\documentclass[12pt,a4paper]{report}
\usepackage{amsfonts}
\usepackage{amsmath}
\usepackage{amssymb}
\usepackage{geometry}
\usepackage{setspace}
\usepackage{subfigure}
\usepackage{ragged2e}
\usepackage{inputenc}
\usepackage{indentfirst}
\usepackage{graphicx}
\usepackage{hyperref}
\usepackage{fancyhdr}
\usepackage{slashed}

\fancypagestyle{plain}{
    \fancyhead{}      }

\begin{document}

\thispagestyle{empty}

\begin{center}
{\LARGE Centro Brasileiro de Pesquisas F\'{\i}sicas - CBPF}

\begin{figure}[h]
\centering
\includegraphics[scale=0.2]{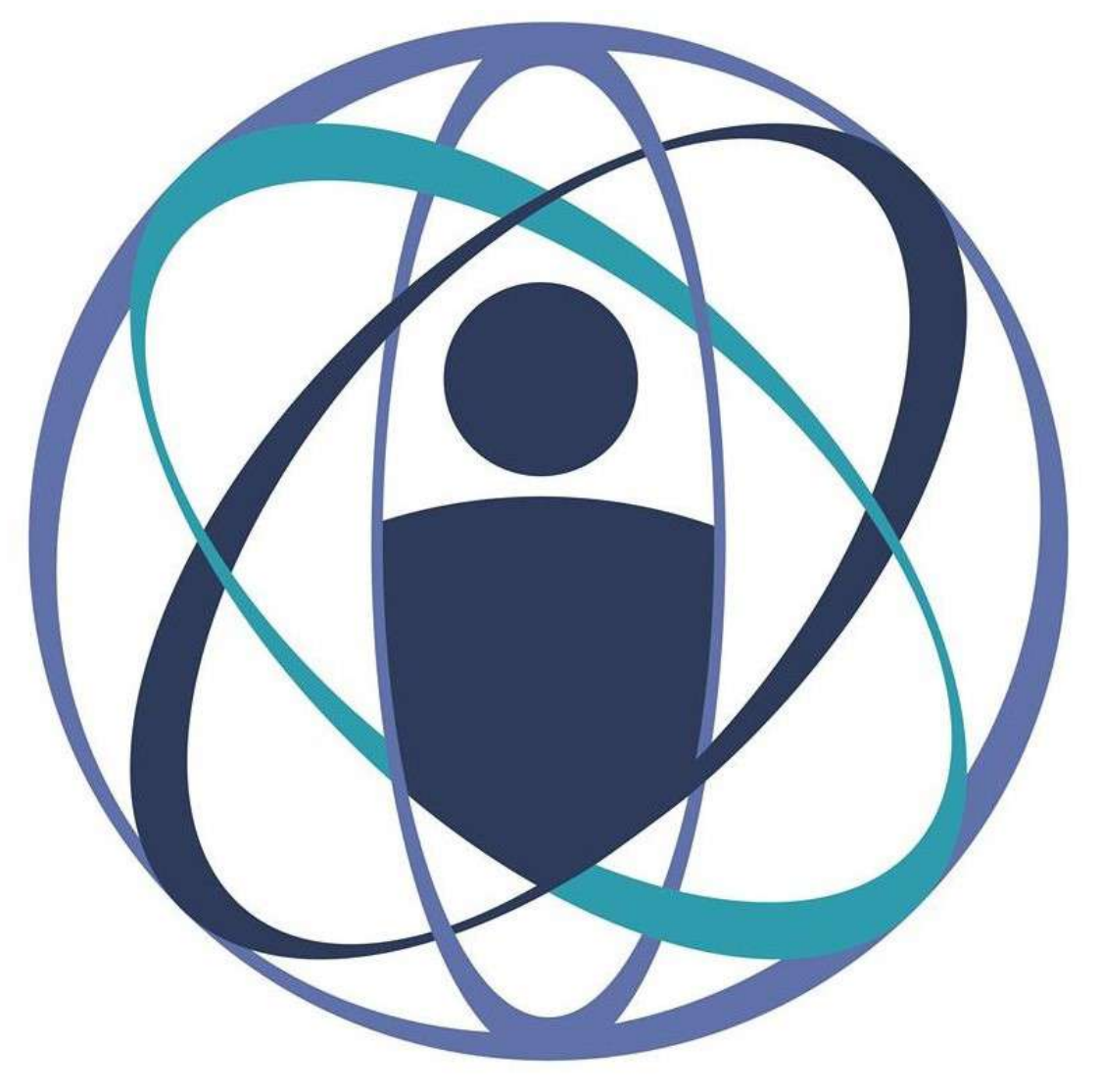}
\end{figure}

\bigskip

\bigskip

\bigskip

\bigskip

\bigskip

\bigskip

\bigskip

\bigskip

\bigskip

\bigskip

\bigskip

\bigskip

{\Huge A study of anomalies using functional integration and perturbative
calculations.}

\bigskip

\bigskip

\bigskip

\bigskip

\bigskip

\bigskip

\bigskip

\bigskip

\bigskip

{\LARGE {Thalis Jos\'{e} Girardi}}

\bigskip

\bigskip

\bigskip

\bigskip

\bigskip

\bigskip

\bigskip

\bigskip

\bigskip \bigskip

\bigskip

\bigskip \bigskip

\bigskip

\bigskip
\end{center}

\begin{flushright}
PhD Thesis

\bigskip

Advisor: Prof. Dr. Sebasti\~{a}o Alves Dias
\end{flushright}

\begin{flushleft}
\bigskip

\bigskip

\bigskip

\bigskip

\bigskip

\bigskip
\end{flushleft}

\begin{center}
Rio de Janeiro, RJ

2023

\newpage

\vfill

\baselineskip=18pt
\end{center}

\section*{Agradecimentos\protect\footnote{%
I chose to write acknowledgements in my first language, Portuguese.}}

\pagenumbering{roman}

A contribui\c{c}\~{a}o de diversas pessoas e organiza\c{c}\~{o}es, de forma
direta ou indireta, permitiu que eu pudesse concluir esta tese. Uso esse
texto para expressar minha gratid\~{a}o a eles.

Come\c{c}o agradecendo \`{a} Capes, cujo financiamento foi essencial para o
desenvolvimento dessa tese. Tamb\'{e}m agrade\c{c}o ao CBPF e todos que
fazem parte dele. \`{A} M\^{o}nica, que felizmente me convenceu a participar
como orientador no Provoc. \`{A} Edna, Cl\'{a}udia, Bete, Denise, Larissa e
outras personalidades que fazem esse centro de pesquisa.

Aos educadores que contribu\'{\i}ram para a minha trajet\'{o}ria, desde a
educa\c{c}\~{a}o b\'{a}sica at\'{e} o doutorado. Agrade\c{c}o por criarem as
possibilidades para que eu pudesse expressar minha individualidade,
tornando-me quem sou. Agrade\c{c}o por estimularem minhas inquietudes,
mostrando que o conhecimento n\~{a}o est\'{a} atrelado ao passado, mas \'{e}
um processo constante de inven\c{c}\~{a}o e reinven\c{c}\~{a}o.

Ao professor Sebasti\~{a}o Alves Dias por sua orienta\c{c}\~{a}o e pela
serenidade que trouxe em momentos dif\'{\i}ceis. Ao professor Orimar
Battistel pelos primeiros passos na F\'{\i}sica, pelas discuss\~{o}es que
seguem e levam ao ato constante de construir conhecimento. Ao professor Jos%
\'{e} Helay\"{e}l-Neto por sua presen\c{c}a encorajadora e reconfortante.
Agrade\c{c}o aos tr\^{e}s pelos excelentes cursos que enriqueceram minha
forma\c{c}\~{a}o acad\^{e}mica e por influenciarem minha paix\~{a}o por ci%
\^{e}ncia e educa\c{c}\~{a}o.

\`{A} Luciana e ao Jos\'{e}, que est\~{a}o sempre pr\'{o}ximos
compartilhando felicidades, ang\'{u}stias e saberes. \`{A} Grecia, que
tornou meus dias mais leves com longas conversas que parecem durar poucos
minutos. Ao Jo\~{a}o, pelas mem\'{o}rias compartilhadas e pelo carinho que
sempre haver\'{a} entre n\'{o}s. Aos colegas Erich, Guilherme e F\'{a}bio
pelo privil\'{e}gio e prazer de convivermos. A todos que fizeram parte da
minha vida no Rio de Janeiro: Neusa, Ivana, Jade, Judismar, Yuri, Gustavo,
Miguel, Pablo, Guga, Manoel, Liliana, Alexandre e tantos outros. Aos amigos
que est\~{a}o por perto desde a gradua\c{c}\~{a}o: Thiago, Giuliano, H\'{e}%
lio e Carine.

\`{A} minha fam\'{\i}lia, cujo apoio foi fundamental para que eu pudesse
concluir essa etapa. Agrade\c{c}o, em especial, aos meus pais por estarem
sempre ao meu lado. Gostaria de finalizar com um agradecimento \`{a}queles
que tiveram profunda influ\^{e}ncia na minha vida e se foram nesses anos. 
\`{A} minha av\'{o}, que amo tanto e cujas lembran\c{c}as sempre me
confortam. Ao meu av\^{o}, cujas provoca\c{c}\~{o}es sempre far\~{a}o falta. 
\`{A} Maria, uma pessoa t\~{a}o querida que ser\'{a} sempre lembrada pelo
carinho e cuidado que nos trouxe.

\newpage

\section*{Abstract}

We present two lines of investigation involving anomalies. First, we review
mechanisms behind the classical and quantum conservation of symmetries using
functional integration. This discussion clarifies conditions for quantum
violations, as acknowledged in chiral theories. Then, we elucidate the
subject of gauge anomaly cancellation when all fields are quantized. Such an
outcome requires gauge invariance of the bosonic measure, so our first
object is proving this invariance within Fujikawa's approach. Second, we
investigate anomalies in fermionic perturbative amplitudes using Implicit
Regularization. The discussion of the single-axial triangle fundaments this
analysis, bringing the elements necessary to approach the single-axial box.
When organizing their mathematical structure, we highlight the role of
traces involving the chiral matrix. Choosing a specific expression for them
reflects on the position of symmetry violations, which has implications
regarding the linearity of integration. Power counting and tensor structure
imply the presence of surface terms related to momenta ambiguities. We
present the results without computing these surface terms. In this neutral
perspective, we explore possibilities achieved under different prescriptions.

Keywords: Gauge and Chiral Anomalies. Divergences. Implicit Regularization.

\section*{Resumo}

N\'{o}s apresentamos duas linhas de investiga\c{c}\~{a}o envolvendo
anomalias. Primeiro, revisamos mecanismos por tr\'{a}s da conserva\c{c}\~{a}%
o cl\'{a}ssica e qu\^{a}ntica de simetrias usando integra\c{c}\~{a}o
funcional. Essa discuss\~{a}o clarifica condi\c{c}\~{o}es para a viola\c{c}%
\~{a}o qu\^{a}ntica, como reconhecido em teorias quirais. Em seguida,
elucidamos o assunto de cancelamento da anomalia de calibre quando todos os
campos s\~{a}o quantizados. Isso requer a invari\^{a}ncia de calibre da
medida bos\^{o}nica, ent\~{a}o nosso primeiro objetivo \'{e} provar essa
invari\^{a}ncia atrav\'{e}s do m\'{e}todo de Fujikawa. Segundo, investigamos
anomalias em amplitudes perturbativas fermi\^{o}nicas usando Regulariza\c{c}%
\~{a}o Impl\'{\i}cita. A discuss\~{a}o do tri\^{a}ngulo com um v\'{e}rtice
axial fundamenta essa an\'{a}lise, trazendo os elementos necess\'{a}rios
para abordar o \textit{box} com um v\'{e}rtice axial. Ao organizarmos suas
estruturas matem\'{a}ticas, destacamos o papel de tra\c{c}os envolvendo a
matriz quiral. Escolher uma express\~{a}o espec\'{\i}fica para eles reflete
na posi\c{c}\~{a}o de viola\c{c}\~{o}es de simetria, trazendo implica\c{c}%
\~{o}es quanto \`{a} linearidade da integra\c{c}\~{a}o. Contagem de pot\^{e}%
ncias e estrutura tensorial implicam na presen\c{c}a de termos de superf%
\'{\i}cie relacionados a combina\c{c}\~{o}es amb\'{\i}guas de momenta.
Apresentamos esses resultados sem calcular termos de superf\'{\i}cie. Nesta
perspectiva neutra, exploramos possibilidades encontradas em prescri\c{c}%
\~{o}es diferentes.

Palavras-chave: Anomalias de calibre e quiral. Diverg\^{e}ncias. Regulariza%
\c{c}\~{a}o Impl\'{\i}cita.

\baselineskip=18pt

\tableofcontents%
\thispagestyle{empty}

\newpage

\chapter{Introduction}

\setcounter{page}{1}\pagenumbering{arabic}When building an interacting model
through a quantum field theory, one starts by constructing a functional of
free fields whose interaction dynamics one aims to describe. In general, the
original functional exhibits invariance under global transformations, with
parameters that do not depend on the space-time position. Then, modifying
this functional promotes the symmetry to a local one. The new functional
emerges after introducing a set of fields (called gauge fields), with
transformations chosen to ensure invariance when parameters depend on the
spacetime. These local transformations of the fields are called gauge
transformations, and the corresponding symmetry is called gauge invariance.
The main consequence is the generation of interaction between the previously
free fields. This proposed interaction generates previsions (i.e., decay
rates or cross sections) capable of being compared with experimentally
measured quantities.

Quantum Electrodynamics is a well-known example of this construction,
corresponding to the quantum field theory for electromagnetic interaction.
The first step is to build the Dirac action, describing free spin-(1/2)
fermions (such as electrons and positrons). Although this functional
exhibits global $U(1)$ invariance, the presence of derivatives prevents
gauge invariance. The solution comes when substituting conventional
derivatives with covariant ones, which induces coupling with a gauge field
(interpreted as the photon field). This field arises from the only generator
of the Abelian symmetry group $U(1)$. There are interacting terms involving
gauge and matter fields in the modified action, so all mentioned
contributions constitute a locally invariant object. One adds a gauge
invariant term involving only the gauge field (the Maxwell action) to
furnish dynamics for the photon.

Something analogous occurs when developing Quantum Chromodynamics to
describe the strong interaction. The outset is on the Dirac action, now
built with free quarks, symmetric under global $SU(3)$ symmetry. Promoting
it to be local generates interaction terms involving gauge and matter
fields. As $SU(3)$ has eight generators, eight gluons emerge as gauge
fields. The difference from the abelian case resides in the non-commutative
character of the algebra, which implies self-interacting gauge fields. The
Yang-Mills functional is introduced to provide dynamics for gauge fields.

Regarding Electroweak Theory, the symmetry group is $SU\left( 2\right)
\times U\left( 1\right) $. As this theory unifies electromagnetic and weak
interactions, it adds new gauge bosons ($W^{\pm }$, $Z$) besides that
corresponding to the photon. The main difference is that these new fields
are massive, while gauge invariance does not admit this type of contribution
to the action. The strategy to deal with this problem is to start from a
massless theory, with the Higgs mechanism generating masses. That means
quarks and leptons are seen as massless Weyl fermions (with defined
chiralities) instead of Dirac fermions. Under these circumstances, the
functional displays gauge invariance before spontaneous symmetry breaking.
That is crucial for the renormalization of the theory. The masses are
generated for all the experimentally known massive fields without spoiling
the renormalizability. This mechanism is extended to the group $SU\left(
3\right) \times SU\left( 2\right) \times U\left( 1\right) $, defining the
Standard Model (SM), which unifies the three mentioned interactions.

There is another issue to be faced by the SM, the so-called \textit{anomalies%
}. They are quantum violations of symmetries originally present at the
action. They have a vast history, initiated by Johnson's discovery of the
two-dimensional chiral anomaly \cite{Johnson:1963vz}. A few years later,
this subject received prominence due to the Adler-Bell-Jackiw anomaly \cite%
{Adler:1969gk, Bell:1969ts}. Both refer to the quantum breaking of\ the
global (constant parameter) chiral symmetry, present in theories with
massless fermions. It also became clear the impossibility of simultaneous
maintenance of chiral and gauge symmetries at the quantum level \cite%
{Jackiw-Johnson}. These symmetries are mixed in the SM before spontaneous
symmetry breaking, which means that gauge invariance is apparently broken at
the quantum level. This phenomenon is known as gauge anomaly. In the SM,
gauge invariance is only achieved through a careful adjustment of the group
representation where one puts the three families of quarks and leptons so
that anomalous contributions from both sectors cancel each other. Meanwhile,
gauge invariance is necessary to ensure renormalizability and unitarity to
the theory. Gauge anomalies modify Slavnov-Taylor identities, preventing one
from relating distinct renormalization constants with each other and
canceling infinities systematically to all orders of perturbation theory 
\cite{Bardeen:1969md, Cheng:1985bj}. We end with an uncomfortable situation
where the SM is a superposition of apparently inconsistent theories, which
result in a consistent one by a very peculiar arrangement.

This situation motivated investigations on gauge-anomalous theories. Jackiw
and Rajaraman \cite{Jackiw, Jackiw2} showed that chiral Quantum
Electrodynamics in two dimensions is consistent and unitary. Furthermore%
\textbf{,} the gauge field, initially massless in the classical action,
became massive after radiative corrections without needing a Higgs
mechanism. Faddeev and Shatashvili clarified the quantization of this type
of theory \cite{Faddeev}. They introduced new quantum degrees of freedom
that provided an equivalent gauge theory (without anomalies). In addition,
Harada and Tsutsui \cite{Harada} and Babelon, Schaposnik and Viallet \cite%
{Babelon} observed that these new degrees could be obtained naturally
through the employment of the Faddeev-Popov procedure. That allowed them to
express the effective action as a gauge scalar for any space-time dimension.
These results suggested that theories with gauge anomalies could be
consistent.

By taking into account gauge invariance of the gauge field measure, in the
context of functional integrals, a recent investigation \cite%
{Gabriel-Rafael-Tiao} showed the vanishing of the insertion of the anomaly
operator in any correlator of gauge invariant operators. This result
suggested that the anomaly vanishes in the part of the Hilbert space
associated with physical states. That motivated us to investigate gauge
invariance of the boson measure in more detail. We do this in Chapter (\ref%
{Sec1}), providing explicit proof of this fact that is, up to our knowledge,
absent from the literature.

We continue to investigate symmetries in the quantum context through an
approach known as \textit{Implicit Regularization} (IReg), a procedure to
identify and separate the divergent part of Feynman diagrams by manipulating
the integrands before integration. The study of an amplitude associated with
the neutral pion decay (the single axial triangle) establishes the
foundations for this analysis. Afterward, we examine the possibility of one
amplitude with an analogous mathematical structure (the single axial box)
exhibiting the same characteristics. Hence, surveying features shared by
these processes highlights new aspects of the anomalies. That corresponds to
the second part of this thesis, whose development occurs in Chapter (\ref%
{Sec2}).

Conclusions will be presented separately for Chapters (\ref{Sec1}) and (\ref%
{Sec2}) since they use different methodologies to approach the subject of
anomalies.

\chapter{Gauge Anomaly and Invariance of the Bosonic Measure}

\label{Sec1}Investigating the consequences of gauge symmetry in classical
and quantum theories is the general objective of this chapter. Starting with
the classical discussion in Section (\ref{ClassSym}), we use arguments
involving action invariance to achieve current conservation. These
preliminary calculations work as a guide to explorations at the quantum
level, made in Section (\ref{QuanSym}). After finding requirements for
quantum invariance, the source of violations in functional integration is
discussed in Section (\ref{GaugeAno}). With the mathematical structure of
the anomaly in our hands, we use a simple procedure to show that its
expected value vanishes when quantizing all theory fields.

The gauge invariance of the gauge field measure is central to this
argumentation. This property has several usages in the literature, as in
investigations involving the Faddeev-Popov method. Since there is (up to our
knowledge) an absence of explicit demonstration of this invariance, our
first contribution is to provide proof of it. To do so, we use general
functional integral arguments to show that the Jacobian associated\ with the
measure has to be 1 (one) when inserted in correlation functions of
gauge-invariant operators. Performing the same analysis for general
operators would complete this demonstration. Since this step brings
complications, we employ a Fujikawa-like approach to calculate this Jacobian
explicitly and show that it is 1 in general.

\section{Classical Symmetry}

\label{ClassSym}This section aims for a preliminary understanding of gauge
theories, emphasizing current conservation at the classical level. It is
also the moment to introduce notations, which follow the material from R.
Jackiw's course in reference \cite{DEWITT} and G. L. S. Lima's works \cite%
{Gabriel, Gabriel2}.

Throughout the Introduction, we mentioned some aspects of theories employed
to describe fundamental interactions. The starting point was the functional
associated with the dynamics of free matter fields. This object is not
invariant under local transformations since it contains derivatives. So, the
idea was to implement this symmetry by making the derivative covariant. The
price paid is inducing terms of interaction with gauge fields. In other
words, gauge symmetry generates dynamics among the fields described by a
theory \cite{Cheng:1985bj}. A contribution associated with free gauge fields
is also necessary. Below, we write the action with these two sectors
separated, so it is clear that each part is invariant by itself:%
\begin{equation}
S\left[ \psi ,\overline{\psi },A_{\mu }\right] =S_{G}\left[ A_{\mu }\right]
+S_{M}\left[ \psi ,\overline{\psi },A_{\mu }\right] .
\end{equation}%
The vector $A_{\mu }=A_{\mu }^{a}T_{a}$ represents the gauge fields with $%
T_{a}$ being generators of the gauge group, while ($\psi $, $\overline{\psi }
$) represent fermionic matter fields.

Saying that the action is invariant means no changes occur when fields
modify through a given set of transformations. Our concern is with gauge
theories, in which case these transformations belong to special unitary
groups $SU\left( N\right) $. Its generators satisfy commutation relations
like%
\begin{equation}
\left[ T^{a},T^{b}\right] =if^{abc}T_{c},
\end{equation}%
along with the normalization%
\begin{equation}
\mathrm{tr}\left( T^{a}T^{b}\right) =-\frac{1}{2}\delta ^{ab}.
\end{equation}%
The symbol $f^{abc}$ represents the structure constants, which have the
property of total antisymmetry through index permutations. Indices denoted
by Latin letters refer to internal degrees of freedom, ranging over the
group dimension (equivalent to the number of generators). As gauge fields
take values on the Lie algebra of the symmetry group, there is one field for
each generator. Greek letters in the indices refer to Minkowski space-time
in the chosen theory.

To analyze current conservation, let us adopt an arbitrary element $%
g=e^{i\theta \left( x\right) }$ to perform a transformation. The parameters
depend on the space-time position $\theta \left( x\right) =\theta ^{a}\left(
x\right) T_{a}$, characterizing a local transformation. As mentioned, the
action is invariant under simultaneous changes of boson and fermion fields%
\begin{eqnarray}
A_{\mu } &\rightarrow &A_{\mu }^{g}=gA_{\mu }g^{-1}+\frac{i}{e}\left(
\partial _{\mu }g\right) g^{-1},  \label{a1} \\
\psi &\rightarrow &\psi ^{g}=g\psi ,  \label{a3} \\
\overline{\psi } &\rightarrow &\overline{\psi }^{g}=\overline{\psi }g^{-1}.
\label{a4}
\end{eqnarray}

\vspace{3cm}By considering small values for the parameter, we take its
first-order contribution to obtain infinitesimal transformations%
\begin{eqnarray}
A_{\mu } &\rightarrow &A_{\mu }^{g}=A_{\mu }-\frac{1}{e}\mathcal{D}_{\mu
}\theta ,  \label{a2} \\
\psi &\rightarrow &\psi ^{g}=\psi +i\theta \psi , \\
\overline{\psi } &\rightarrow &\overline{\psi }^{g}=\overline{\psi }-i%
\overline{\psi }\theta .  \label{a6}
\end{eqnarray}%
We define the covariant derivative of Lie algebra valued quantities through
the mathematical expression%
\begin{equation}
\mathcal{D}_{\mu }\theta =\partial _{\mu }\theta +ie\left[ A_{\mu },\theta %
\right] ,
\end{equation}%
so using the commutation relations allows specifying its components%
\begin{equation}
\mathcal{D}_{\mu }\theta =T^{a}\left( \partial _{\mu }\delta
^{ac}-ef^{abc}A_{\mu }^{b}\right) \theta ^{c}\equiv T^{a}\mathcal{D}_{\mu
}^{ac}\theta ^{c}.
\end{equation}

Since the action is invariant under local transformations, it is also
invariant under global transformations. As the parameter is constant in the
second case, the derivative $\partial _{\mu }\theta $ cancels out within the
vector field transformation. Then, by reversing this line of reasoning,
starting from global transformations and imposing dependence on the position
is feasible. In such a case, the absence of the inhomogeneous term implies
symmetry is lost. That means the following variation must be proportional to
derivatives of the parameter%
\begin{equation}
\delta S_{M}=S_{M}\left[ \psi ^{g},\overline{\psi }^{g},gA_{\mu }g^{-1}%
\right] -S_{M}\left[ \psi ,\overline{\psi },A_{\mu }\right] =\int \mathrm{dx}%
\text{ }\partial _{\mu }\theta ^{a}\left( x\right) J_{a}^{\mu }\left(
x\right) ,  \label{dsm}
\end{equation}%
where we introduce the vector $J_{a}^{\mu }\left( x\right) $, determined by
the fields present in the model. Recalling that both sectors of the action
are invariant when considered by themselves, we focus exclusively on the
matter action. On the other hand, an infinitesimal transformation over the
action leads to another form for the same variation:%
\begin{equation}
\delta S_{M}=\int \mathrm{dx}\text{ }\theta ^{a}\left( x\right) \left[ \frac{%
\delta S_{M}}{\delta \psi \left( x\right) }iT^{a}\psi \left( x\right) -i%
\overline{\psi }\left( x\right) T^{a}\frac{\delta S_{M}}{\delta \overline{%
\psi }\left( x\right) }+f^{abc}A_{\mu }^{b}\left( x\right) \frac{\delta S_{M}%
}{\delta A_{\mu }^{c}\left( x\right) }\right] .  \label{dsm2}
\end{equation}

Equating both expressions to produce one identity is feasible. By performing
an integration by parts on the contribution from (\ref{dsm}), the parameter $%
\theta ^{a}$ factorizes inside the integration sign:%
\begin{equation}
\int \mathrm{dx}\text{ }\theta ^{a}\left( x\right) \left[ \partial _{\mu
}J_{a}^{\mu }\left( x\right) +\frac{\delta S_{M}}{\delta \psi \left(
x\right) }iT^{a}\psi \left( x\right) -i\overline{\psi }\left( x\right) T^{a}%
\frac{\delta S_{M}}{\delta \overline{\psi }\left( x\right) }+f^{abc}A_{\mu
}^{b}\left( x\right) \frac{\delta S_{M}}{\delta A_{\mu }^{c}\left( x\right) }%
\right] =0.
\end{equation}%
Hence, the arbitrariness of this object implies that the structure in
squared brackets vanishes regardless of the integration 
\begin{equation}
\partial _{\mu }J_{a}^{\mu }\left( x\right) +\frac{\delta S_{M}}{\delta \psi
\left( x\right) }iT^{a}\psi \left( x\right) -i\overline{\psi }\left(
x\right) T^{a}\frac{\delta S_{M}}{\delta \overline{\psi }\left( x\right) }%
+f^{abc}A_{\mu }^{b}\left( x\right) \frac{\delta S_{M}}{\delta A_{\mu
}^{c}\left( x\right) }=0.  \label{dj}
\end{equation}%
As we have not considered local symmetry up to this point, such a result is
a consequence of global invariance.

Next, observe that equations of motion associated with fermion fields fall
over the matter action%
\begin{eqnarray}
\frac{\delta S}{\delta \psi \left( x\right) } &=&\frac{\delta S_{M}}{\delta
\psi \left( x\right) }=0,  \label{eq1} \\
\frac{\delta S}{\delta \overline{\psi }\left( x\right) } &=&\frac{\delta
S_{M}}{\delta \overline{\psi }\left( x\right) }=0.  \label{eq2}
\end{eqnarray}%
Hence, replacing them in Equation (\ref{dj}) cancels out some contributions,
which leads to the simplified version\footnote{%
As structure constants cancel out in the Abelian theory, the conservation of
the current comes directly from this equation. That means it is unnecessary
to consider gauge transformations at any point in the calculations.}%
\begin{equation}
\partial _{\mu }J_{a}^{\mu }\left( x\right) +f^{abc}A_{\mu }^{b}\left(
x\right) \frac{\delta S_{M}}{\delta A_{\mu }^{c}\left( x\right) }=0.
\label{c}
\end{equation}

Now, we consider gauge transformations as the final step before achieving
conservation. Invariance of the action establishes the relation%
\begin{equation}
S_{M}\left[ \psi ^{g},\overline{\psi }^{g},A_{\mu }\right] =S_{M}\left[ \psi
,\overline{\psi },A_{\mu }^{g^{-1}}\right] =S_{M}\left[ \psi ,\overline{\psi 
},g^{-1}A_{\mu }g+\frac{i}{e}\left( \partial _{\mu }g^{-1}\right) g\right] .
\end{equation}%
By adopting the configuration for the gauge field $A_{\mu }^{\prime
}=gA_{\mu }g^{-1}$, the variation of $S_{M}$ is achievable again. To that
end, rewrite the relation above by considering the infinitesimal form of the
transformation%
\begin{equation}
S_{M}\left[ \psi ^{g},\overline{\psi }^{g},gA_{\mu }g^{-1}\right] =S_{M}%
\left[ \psi ,\overline{\psi },A_{\mu }+\frac{1}{e}\partial _{\mu }\theta %
\right] .
\end{equation}%
The mentioned variation emerges through an expansion over the parameter%
\begin{equation}
\delta S_{M}=\frac{1}{e}\int \mathrm{dx}\text{ }\partial _{\mu }\theta
^{a}\left( x\right) \frac{\partial S_{M}}{\partial A_{\mu }^{a}\left(
x\right) }.
\end{equation}%
Therefore, a comparison between this form and Equation (\ref{dsm}) generates
the following result%
\begin{equation}
\int \mathrm{dx}\text{ }\partial _{\mu }\theta ^{a}\left( x\right) \left[
J_{a}^{\mu }\left( x\right) -\frac{1}{e}\frac{\partial S_{M}}{\partial
A_{\mu }^{a}\left( x\right) }\right] =0.
\end{equation}%
Again, the quantity in squared brackets has to vanish by itself as
transformation parameters are arbitrary. That produces the relation%
\begin{equation}
J_{a}^{\mu }\left( x\right) =\frac{1}{e}\frac{\partial S_{M}}{\partial
A_{\mu }^{a}\left( x\right) },  \label{corrente}
\end{equation}%
whose replacement within Equation (\ref{c}) allows recognizing the covariant
derivative introduced in the gauge field transformation%
\begin{equation}
\mathcal{D}_{\mu }^{ac}J_{c}^{\mu }\left( x\right) =\left( \partial _{\mu
}\delta ^{ac}-ef^{abc}A_{\mu }^{b}\right) J_{c}^{\mu }\left( x\right) =0.
\end{equation}%
We identify the vector $J_{a}^{\mu }\left( x\right) $ as a current, while
the last equation represents its covariant conservation. Two ingredients
were necessary to achieve this outcome: local gauge invariance of the matter
action and equations of motion for fermions.

\section{Quantum Symmetry}

\label{QuanSym}Since we finalized exploring manifestations of gauge symmetry
in classical theories, let us extend this discussion to the quantum context.
To accomplish this goal, we start by introducing the effective action $W%
\left[ A_{\mu }\right] $ through the functional integral%
\begin{equation}
e^{iW\left[ A_{\mu }\right] }=\int d\psi d\overline{\psi }\exp \left( iS%
\left[ \psi ,\overline{\psi },A_{\mu }\right] \right) .  \label{effective}
\end{equation}%
Since gauge fields are considered external classical fields, the integration
occurs exclusively over (quantized) fermion fields.

Following the same reasoning from the previous section, we consider global
transformation and impose that parameters depend on the position. When
applying infinitesimal transformations (\ref{a2})-(\ref{a6}), the changed
expression for the exponential follows%
\begin{equation}
e^{iW^{\prime }}=\int d\psi d\overline{\psi }\exp \left( iS_{M}\left[ \psi
+i\theta \psi ,\overline{\psi }-i\overline{\psi }\theta ,A_{\mu }-i\left[
A_{\mu },\theta \right] \right] +iS_{G}\left[ A_{\mu }\right] \right) ,
\end{equation}%
with the gauge action invariant. Although gauge fields change through the
covariant derivative, only the contribution on the commutator concerns
global invariance. Recognizing the exponential argument as the action plus a
variation allows detaching both parts%
\begin{equation}
e^{iW^{\prime }}=\int d\psi d\overline{\psi }\exp \left( i\delta
S_{M}\right) \exp \left( iS\left[ \psi ,\overline{\psi },A_{\mu }\right]
\right) .
\end{equation}%
Hence, an expansion on the infinitesimal parameter leads to the exponential
variation of the effective action%
\begin{equation}
e^{iW^{\prime }}-e^{iW}=\int d\psi d\overline{\psi }\left( i\delta
S_{M}\right) \exp \left( iS\left[ \psi ,\overline{\psi },A_{\mu }\right]
\right) .
\end{equation}

As this result depends on the action variation, let us recall the
information obtained. On the one hand, we reasoned that it is proportional
to the derivative of the parameter and the current $J_{a}^{\mu }\left(
x\right) $; see Equation (\ref{dsm}). In the quantum context, that leads to
the expression%
\begin{equation}
e^{iW^{\prime }}-e^{iW}=-\int d\psi d\overline{\psi }\exp \left( iS\left[
\psi ,\overline{\psi },A_{\mu }\right] \right) \left[ i\int \mathrm{dx}\text{
}\theta ^{a}\left( x\right) \partial _{\mu }J_{a}^{\mu }\left( x\right) %
\right] ,
\end{equation}%
where integration by parts changes the derivative position. On the other
hand, the infinitesimal transformation produced result (\ref{dsm2}), which
reflects on the form%
\begin{eqnarray}
e^{iW^{\prime }}-e^{iW} &=&\int d\psi d\overline{\psi }\exp \left( iS\left[
\psi ,\overline{\psi },A_{\mu }\right] \right) \times \\
&&\times i\int \mathrm{dx}\text{ }\theta ^{a}\left( x\right) \left[ \frac{%
\delta S_{M}}{\delta \psi \left( x\right) }iT^{a}\psi \left( x\right) -i%
\overline{\psi }\left( x\right) T^{a}\frac{\delta S_{M}}{\delta \overline{%
\psi }\left( x\right) }+ef^{abc}A_{\mu }^{b}\left( x\right) J_{c}^{\mu
}\left( x\right) \right] .  \notag
\end{eqnarray}%
We already used the association (\ref{corrente}) to recognize the current
within this equation.

Since there are two forms for the same object, let us equate them to produce
an identity. Due to the arbitrariness of the transformation parameter, the
relation applies regardless of space-time integration. We emphasize that
this does not occur if the parameter is constant, as it would factor from
the integration sign without further simplifications. By identifying the
covariant derivative, the variation produces the result%
\begin{eqnarray}
&&\int d\psi d\overline{\psi }\exp \left( iS\left[ \psi ,\overline{\psi }%
,A_{\mu }\right] \right) \left[ D_{\mu }^{ab}J_{b}^{\mu }\left( x\right) %
\right]  \notag \\
&=&\int d\psi d\overline{\psi }\exp \left( iS\left[ \psi ,\overline{\psi }%
,A_{\mu }\right] \right) \left[ i\overline{\psi }\left( x\right) T^{a}\frac{%
\delta S_{M}}{\delta \overline{\psi }\left( x\right) }-\frac{\delta S_{M}}{%
\delta \psi \left( x\right) }iT^{a}\psi \left( x\right) \right] .
\label{eqq}
\end{eqnarray}

In the classical discussion, the conservation law arose posteriorly to
employing equations of motion for fermions in an analogous equation. We
would expect Dyson-Schwinger equations to perform this task here, as they
embody the equations of motion within this context. In that case, current
conservation would result from the translational invariance of the fermion
measure \cite{Rivers}. Nonetheless, gauge invariance emerges as a condition
at the quantum level. Let us integrate an arbitrary functional and explore
its transformation to understand the consequences:%
\begin{eqnarray}
\int d\psi d\overline{\psi }\text{ }F\left[ \psi ,\overline{\psi },A_{\mu }%
\right] &=&\int d\psi ^{g}d\overline{\psi }^{g}\text{ }F\left[ \psi ^{g},%
\overline{\psi }^{g},A_{\mu }\right] \\
&=&\int d\psi d\overline{\psi }\text{ }F\left[ \psi ,\overline{\psi },A_{\mu
}\right] +\int d\psi d\overline{\psi }\int \mathrm{dx}\left[ \frac{\delta F}{%
\delta \psi }\delta \psi +\frac{\delta F}{\delta \overline{\psi }}\delta 
\overline{\psi }\right] .  \notag
\end{eqnarray}%
Under the hypothesis of gauge-invariance of the fermion measure%
\begin{equation}
d\psi ^{g}d\overline{\psi }^{g}=d\psi d\overline{\psi },
\end{equation}%
the condition applies%
\begin{equation}
\int d\psi d\overline{\psi }\int \mathrm{dx}\left[ \frac{\delta F}{\delta
\psi }\delta \psi +\delta \overline{\psi }\frac{\delta F}{\delta \overline{%
\psi }}\right] =0.
\end{equation}%
Disregarding space-time integration, observe that this object cancels out
the right-hand side of Equation (\ref{eqq}) when we set the functional.
Hence, the referred equation turns into the quantum version of the gauge
current covariant conservation:%
\begin{equation}
\int d\psi d\overline{\psi }\left[ D_{\mu }^{ab}J_{b}^{\mu }\left( x\right) %
\right] \exp \left( iS\left[ \psi ,\overline{\psi },A_{\mu }\right] \right)
=0.
\end{equation}

Such an argumentation shows that gauge invariance of the fermion measure is
enough for current conservation. Invariance of the matter action does not
guarantee symmetry maintenance within quantum theory, even if it guarantees
classical conservation.

\section{Gauge Anomaly}

\label{GaugeAno}After shedding light on conditions for quantum conservation,
we aim to inquire about situations characterized by violations. The
literature on functional integrals recognizes non-trivial Jacobians for the
fermion measure as the cause of symmetry breaking \cite{Fujikawa}. This
non-invariance is typical of investigations involving chiral fermions, as in
the Standard Model before spontaneous symmetry breaking.

We approach this subject by introducing the fermion measure Jacobian as
follows%
\begin{equation}
d\psi ^{g}d\overline{\psi }^{g}=J\left[ g,A_{\mu }\right] d\psi d\overline{%
\psi }
\end{equation}%
while considering the possibility of dependence on gauge fields. Although
that is unreasonable for usual integration, this type of contribution might
arise through regularization techniques when dealing with divergent objects
associated with functional integrals \cite{Fujikawa79, Fujikawa80}. That
means integrals and functional derivatives do not necessarily commute,
requiring extra care to avoid inconsistent results.

Given the structure of calculations developed in the previous section,
expressing the Jacobian as the exponential of another functional is
convenient%
\begin{equation}
J\left[ g,A_{\mu }\right] =\exp \left( i\alpha _{1}\left[ g,A_{\mu }\right]
\right) .
\end{equation}%
Thence, writing the Jacobian associated with the inverse transformation is
straightforward%
\begin{equation}
J\left[ g^{-1},A_{\mu }\right] =\exp \left( i\alpha _{1}\left[ g^{-1},A_{\mu
}\right] \right) =\exp \left( -i\alpha _{1}\left[ g,A_{\mu }\right] \right) ,
\label{jac}
\end{equation}%
and so is the property attributed to the exponential argument%
\begin{equation}
\alpha _{1}\left[ g^{-1},A_{\mu }\right] =-\alpha _{1}\left[ g,A_{\mu }%
\right] .
\end{equation}%
Besides, we consider first-order contributions on the infinitesimal
transformation parameter to build the expansion%
\begin{equation}
\alpha _{1}\left[ g,A_{\mu }\right] =\alpha _{1}\left[ 1,A_{\mu }\right]
+\int \mathrm{dx}\text{ }\theta ^{a}\left( x\right) \left. \frac{\delta
\alpha _{1}\left[ g,A_{\mu }\right] }{\delta \theta ^{a}\left( x\right) }%
\right\vert _{\theta =0}.  \label{alpha}
\end{equation}%
As the first term represents the case without transformation, the Jacobian
corresponds to the identity $J\left[ 1,A_{\mu }\right] =1$ and implies the
vanishing argument $\alpha _{1}\left[ 1,A_{\mu }\right] =0$.

Since we discussed how fermionic variables change, let us explore the
implications for the effective action introduced in Equation (\ref{effective}%
). By relabeling fermion fields as $\psi \rightarrow \psi ^{g^{-1}}$ and $%
\overline{\psi }\rightarrow \overline{\psi }^{g^{-1}}$, we get the modified
expression%
\begin{equation}
e^{iW\left[ A_{\mu }\right] }=\int d\psi ^{g^{-1}}d\overline{\psi }%
^{g^{-1}}\exp \left( iS\left[ \psi ^{g^{-1}},\overline{\psi }%
^{g^{-1}},A_{\mu }\right] \right) .
\end{equation}%
After employing action invariance and inserting the Jacobian for the inverse
(\ref{jac}), we achieve another form:%
\begin{equation}
e^{iW\left[ A_{\mu }\right] }=\exp \left( -i\alpha _{1}\left[ g,A_{\mu }%
\right] \right) \int d\psi d\overline{\psi }\exp \left( iS\left[ \psi ,%
\overline{\psi },A_{\mu }^{g}\right] \right) .
\end{equation}%
The Jacobian factors out of the integral sign as it does not depend on
quantized fermion fields. This integral corresponds to the effective action
with modified gauge fields, so expressing the Jacobian through the effective
action is feasible%
\begin{equation}
\exp \left( i\alpha _{1}\left[ g,A_{\mu }\right] \right) =\exp \left( iW%
\left[ A_{\mu }^{g}\right] -iW\left[ A_{\mu }\right] \right) .
\end{equation}%
Taking the logarithm on both sides emphasizes that the effective action is
not invariant under this type of transformation:%
\begin{equation}
\alpha _{1}\left[ g,A_{\mu }\right] =W\left[ A_{\mu }^{g}\right] -W\left[
A_{\mu }\right] .  \label{aWW}
\end{equation}

By recalling the gauge field transformation (\ref{a2}), we expand $W\left[
A_{\mu }^{g}\right] $ to the first order on the infinitesimal parameter.
That allows writing the variation of the effective action through the
integral%
\begin{equation}
W\left[ A_{\mu }^{g}\right] -W\left[ A_{\mu }\right] =\int \mathrm{dx}\text{ 
}\theta ^{c}\mathcal{D}_{\mu }^{ac}\left( \frac{1}{e}\frac{\delta W\left[
A_{\mu }\right] }{\delta A_{\mu }^{a}}\right) .  \label{w'}
\end{equation}%
But Equation (\ref{aWW}) links this structure to the functional $\alpha _{1}$%
, whose expansion is (\ref{alpha}). Given the parameter arbitrariness,
comparing both equations establishes the relation%
\begin{equation}
\left. \frac{\delta \alpha _{1}\left[ g,A_{\mu }\right] }{\delta \theta
\left( x\right) }\right\vert _{\theta =0}=\mathcal{D}_{\mu }\left( \frac{1}{e%
}\frac{\delta W\left[ A_{\mu }\right] }{\delta A_{\mu }}\right) ,
\label{ano0}
\end{equation}%
where the notation involving components is omitted.

For the last step of the current discussion, we recall that both effective
action and action itself are Lorentz scalars. That means the commutation
between these objects and the covariant derivative does not bring
complications. Hence, multiplying the relation above and the exponential of
the effective action leads to the mathematical expression%
\begin{eqnarray}
&&\left. \frac{\delta \alpha _{1}\left[ g,A_{\mu }\right] }{\delta \theta
\left( x\right) }\right\vert _{\theta =0}\int d\psi d\overline{\psi }\exp
\left( iS\left[ \psi ,\overline{\psi },A_{\mu }\right] \right)  \notag \\
&=&\mathcal{D}_{\mu }\left\{ -\frac{i}{e}\frac{\delta }{\delta A_{\mu }}\int
d\psi d\overline{\psi }\exp \left( iS\left[ \psi ,\overline{\psi },A_{\mu }%
\right] \right) \right\} .
\end{eqnarray}%
Since the gauge action is invariant, the functional derivative acts
exclusively on the matter action%
\begin{eqnarray}
&&\left. \frac{\delta \alpha _{1}\left[ g,A_{\mu }\right] }{\delta \theta
\left( x\right) }\right\vert _{\theta =0}\int d\psi d\overline{\psi }\exp
\left( iS\left[ \psi ,\overline{\psi },A_{\mu }\right] \right)  \notag
\label{da} \\
&=&\int d\psi d\overline{\psi }\text{ }\mathcal{D}_{\mu }\left( \frac{1}{e}%
\frac{\delta S_{M}\left[ \psi ,\overline{\psi },A_{\mu }\right] }{\delta
A_{\mu }}\right) \exp \left( iS\left[ \psi ,\overline{\psi },A_{\mu }\right]
\right) .
\end{eqnarray}%
As the term in parenthesis is precisely the current identified in the
classical discussion (\ref{corrente}), the relation applies%
\begin{equation}
\left. \frac{\delta \alpha _{1}\left[ g,A_{\mu }\right] }{\delta \theta
_{a}\left( x\right) }\right\vert _{\theta =0}=\frac{\int d\psi d\overline{%
\psi }\left( \mathcal{D}_{\mu }^{ab}J_{b}^{\mu }\right) e^{iS\left[ \psi ,%
\overline{\psi },A_{\mu }\right] }}{\int d\psi d\overline{\psi }\text{ }e^{iS%
\left[ \psi ,\overline{\psi },A_{\mu }\right] }}.
\end{equation}%
We transposed the effective action to the right-hand side to identify this
structure as the vacuum expectation value of the covariant divergence of the
current. The non-vanishing of this expression characterizes the so-called 
\textit{gauge anomaly}:%
\begin{equation}
\mathcal{A}_{a}\left( A_{\mu }\right) =\left. \frac{\delta \alpha _{1}\left[
g,A_{\mu }\right] }{\delta \theta _{a}\left( x\right) }\right\vert _{\theta
=0}\neq 0.  \label{ano1}
\end{equation}%
This condition is what characterizes the theory as \textit{gauge anomalous}.
We stress that this happens when gauge bosons are external classical fields
interacting with quantum fermion fields.

Further explorations show that the expectation value for the gauge anomaly
cancels out for the fully-quantized theory. To verify that, let us define
the generating functional%
\begin{equation}
Z\left[ \eta ,\overline{\eta },j_{a}^{\mu }\right] =\int d\psi d\overline{%
\psi }dA_{\mu }\text{ }\exp \left( iS\left[ \psi ,\overline{\psi },A_{\mu }%
\right] +i\int \mathrm{dx}\left[ \overline{\eta }\psi +\overline{\psi }\eta
+j_{a}^{\mu }A_{\mu }^{a}\right] \right) .  \label{z}
\end{equation}%
Since our concern relates to vacuum expectation value, contributions
associated with external sources are unnecessary. The notation simplifies
under these circumstances, being viable to express this equation in terms of
the effective action 
\begin{equation}
Z\left[ 0,0,0\right] =\int dA_{\mu }\text{ }e^{iW\left[ A_{\mu }\right] }.
\label{z1}
\end{equation}%
Following a strategy similar to previous cases, we start by relabeling the
structure above through $A_{\mu }\rightarrow A_{\mu }^{g}$. The changed
version for the effective action corresponds to the original plus a
variation. After replacing the result from the previous section (\ref{w'}),
we split the exponential argument. Then, expanding the variation part on the
infinitesimal parameter produces the equation%
\begin{equation}
Z\left[ 0,0,0\right] =\int dA_{\mu }^{g}\text{ }e^{iW\left[ A_{\mu }\right] }%
\left[ 1+i\int \mathrm{dx}\text{ }\theta ^{c}\mathcal{D}_{\mu }^{ac}\left( 
\frac{1}{e}\frac{\delta W\left[ A_{\mu }\right] }{\delta A_{\mu }^{a}}%
\right) \right] .
\end{equation}%
The difference between the generating functional and the first term on the
right-hand side resides in the integration variable; thus, they coincide if
the bosonic measure is gauge-invariant $dA_{\mu }=dA_{\mu }^{g}$. The second
functional integral must be zero under this condition. Since the
arbitrariness of the transformation parameter allows dropping the space-time
integral, the relation emerges%
\begin{equation}
\int dA_{\mu }\text{ }e^{iW\left[ A_{\mu }\right] }\mathcal{D}_{\mu
}^{ac}\left( \frac{1}{e}\frac{\delta W\left[ A_{\mu }\right] }{\delta A_{\mu
}^{a}}\right) =0.
\end{equation}%
At this point, we recall Equations (\ref{ano0}) and (\ref{ano1}) to
recognize the anomaly. Hence, by making the dependence on the fermionic
variables explicit, we showed that its vacuum expectation value is zero for
the fully quantized theory:%
\begin{equation}
\int d\psi d\overline{\psi }dA_{\mu }\text{ }\mathcal{A}_{a}\left( A_{\mu
}\right) \exp \left( iS\left[ \psi ,\overline{\psi },A_{\mu }\right] \right)
=\left\langle 0\right\vert \mathcal{A}_{a}\left( A_{\mu }\right) \left\vert
0\right\rangle =0.
\end{equation}

In addition to its role in the demonstration above, we stress that the
bosonic measure invariance has other applications in investigations in this
area. Even so, we did not find proof of this property in the literature. The
primary objective of this part of the thesis is to provide one, which is our
next subject.

\section{Gauge Invariance of the Bosonic Measure}

\label{GaugeMeasure}This section investigates the behavior of the bosonic
measure under gauge transformations. To this end, we display a preparatory
argument by considering the generating functional for correlators of
gauge-invariant operators $O_{i}\left( A_{\mu }^{g}\right) =O_{i}\left(
A_{\mu }\right) $ in the pure Yang-Mills theory (without chiral fermions):%
\begin{equation}
Z\left[ \lambda ^{i}\right] =\int dA_{\mu }\exp i\int \mathrm{dx}\text{ tr}%
\left( \frac{1}{2}F_{\mu \nu }F^{\mu \nu }+\lambda ^{i}O_{i}\left[ A_{\mu }%
\right] \right) .
\end{equation}%
The quantities $\lambda _{i}$ are currents, and functional derivatives with
respect to them yield the $n$-point correlators%
\begin{equation}
\left. \frac{\delta ^{n}}{\delta \lambda ^{1}\left( x_{1}\right) ...\delta
\lambda ^{n}\left( x_{n}\right) }Z\left[ \lambda ^{i}\right] \right\vert
_{\lambda ^{i}=0}=\left\langle 0\right\vert T\left( O_{1}\left( A_{\mu
}\right) \left( x_{1}\right) ...O_{n}\left( A_{\mu }\right) \left(
x_{n}\right) \right) \left\vert 0\right\rangle .
\end{equation}

Considering the integration over $A_{\mu }$ and also over its gauge
transformed version $A_{\mu }^{g}$, we develop the comparison%
\begin{align}
Z\left[ \lambda ^{i}\right] & =\int dA_{\mu }\exp i\int \mathrm{dx}\text{ tr}%
\left[ \frac{1}{2}F_{\mu \nu }F^{\mu \nu }+\lambda ^{i}O_{i}\left( A_{\mu
}\right) \right]  \notag \\
& =\int dA_{\mu }^{g}\exp i\int \mathrm{dx}\text{ tr}\left[ \frac{1}{2}%
\left( F_{\mu \nu }F^{\mu \nu }\right) ^{g}+\lambda ^{i}O_{i}\left( A_{\mu
}^{g}\right) \right]  \notag \\
& =\int dA_{\mu }\text{ }J\left[ A_{\mu },g\right] \exp i\int \mathrm{dx}%
\text{ tr}\left[ \frac{1}{2}F_{\mu \nu }F^{\mu \nu }+\lambda ^{i}O_{i}\left(
A_{\mu }\right) \right] ,
\end{align}%
where the potential presence of a Jacobian $J\left[ A_{\mu },g\right] $ for
the gauge transformation of the measure is allowed. Thus, we obtain the
correlators associated with both expressions for the generating functional
as follows%
\begin{align}
& \left\langle 0\right\vert T\left( J\left[ A_{\mu },g\right] O_{1}\left(
A_{\mu }\right) \left( x_{1}\right) ...O_{n}\left( A_{\mu }\right) \left(
x_{n}\right) \right) \left\vert 0\right\rangle  \notag \\
& =\left\langle 0\right\vert T\left( O_{1}\left( A_{\mu }\right) \left(
x_{1}\right) ...O_{n}\left( A_{\mu }\right) \left( x_{n}\right) \right)
\left\vert 0\right\rangle .
\end{align}%
Translated into words, that means all correlators involving the Jacobian $J%
\left[ A_{\mu },g\right] $ with gauge invariant operators are the same as
those involving the identity. Thus, in the physical Hilbert space of the
theory, both operators are the same.

This argument does not generalize to arbitrary operators that are not
gauge-invariant, as required to recover the entire Hilbert space. However,
an explicit calculation can solve this problem. Let us use the usual
prescription of defining the bosonic measure through a complete set of
orthonormal eigenfunctions $\left\{ \phi _{n}\right\} $ of a hermitian
operator $\bar{D}$:%
\begin{equation}
\bar{D}\phi _{n}=\lambda _{n}\phi _{n},
\end{equation}%
with the conditions%
\begin{equation}
\int \mathrm{dx}\text{ }\phi _{n}^{\dag }\phi _{m}=\delta _{nm}\text{ and }%
\sum_{n}\phi _{n}\left( x\right) \phi _{n}^{\dag }\left( y\right) =\delta
\left( x-y\right) .
\end{equation}%
Posteriorly to expanding the bosonic field, we build the connection with the
measure as follows%
\begin{equation}
A_{\mu }^{a}\left( x\right) =\sum_{n}a_{\mu ,n}^{a}\phi _{n}\left( x\right)
\rightarrow dA_{\mu }=\prod_{a,\mu ,n}da_{\mu ,n}^{a}.
\end{equation}

Next, we put the changed field into this prescription. By introducing
coefficients $\bar{a}$ to the new expansion, let us rewrite the
infinitesimal gauge transformation (\ref{a2}):%
\begin{align}
A_{\mu }^{g}& =\sum_{n}\bar{a}_{\mu ,n}^{a}T_{a}\phi _{n}\left( x\right)
=\sum_{n}a_{\mu ,n}^{a}T_{a}\phi _{n}\left( x\right) -\frac{i}{e}\mathcal{D}%
_{\mu }\theta  \notag \\
& =\left[ \sum_{n}\left( a_{\mu ,n}^{a}+ia_{\mu ,n}^{b}f_{abc}\theta
^{c}\right) \phi _{n}\left( x\right) -\frac{i}{e}\partial _{\mu }\theta ^{a}%
\right] T_{a}.
\end{align}%
Then, after decomposing parameters $\theta ^{a}$ in terms of the same
eigenfunctions of $\bar{D}$%
\begin{equation}
-\frac{i}{e}\partial _{\mu }\theta ^{a}\left( x\right) =\sum_{n}\tilde{a}%
_{\mu ,n}^{a}\phi _{n}\left( x\right) ,
\end{equation}%
obtaining a transformation rule to coefficients is feasible%
\begin{equation}
\bar{a}_{\mu ,n}^{a}=\sum_{m}\left( \delta _{ab}\delta _{nm}+\int \mathrm{dx}%
\text{ }\phi _{n}^{\dag }\left( x\right) if_{abc}\theta ^{c}\left( x\right)
\phi _{m}\left( x\right) \right) a_{\mu ,m}^{b}+\tilde{a}_{\mu ,n}^{a}.
\end{equation}%
That reflects on the transformation linked to the bosonic measure%
\begin{equation}
\prod_{a,\mu ,n}d\bar{a}_{\mu ,n}^{a}=\det \left[ \delta _{ab}\delta
_{nm}+\int \mathrm{dx}\text{ }\phi _{n}^{\dag }\left( x\right)
if_{abc}\theta ^{c}\left( x\right) \phi _{m}\left( x\right) \right]
\prod_{a,\mu ,n}da_{\mu ,n}^{a},
\end{equation}%
where the term $\tilde{a}_{\mu ,n}^{a}$ does not contribute because of the
translational invariance of each measure $da_{\mu ,n}^{a}$.

Following the steps of Fujikawa \cite{Fujikawa}, we get the expression for
the Jacobian:%
\begin{equation}
J\left[ A_{\mu },\theta \right] =\exp \left[ \sum_{n}\left( \text{tr}\int 
\mathrm{dx}\text{ }\phi _{n}^{\dag }\left( x\right) if_{abc}\theta
^{c}\left( x\right) \phi _{n}\left( x\right) \right) \right] .
\end{equation}%
This trace acts over Lie algebra indices, which cancels out the total
antisymmetric structure constant $f_{abc}$. Meanwhile, we recognize the
product of fields taken at the same point $\phi _{n}\left( x\right) \phi
_{n}^{\dag }\left( x\right) $. When putting both pieces of information
together, it is easy to see that the Jacobian expression is indefinite:%
\begin{align}
\sum_{n}\left( \text{tr}\int \mathrm{dx}\text{ }\phi _{n}^{\dag }\left(
x\right) if_{abc}\theta ^{c}\left( x\right) \phi _{n}\left( x\right) \right)
& =\text{tr}\int \mathrm{dx}\text{ }if_{abc}\theta ^{c}\left( x\right)
\sum_{n}\phi _{n}\left( x\right) \phi _{n}^{\dag }\left( x\right)  \notag \\
& =\int \mathrm{dx}\text{ }if_{aac}\theta ^{c}\left( x\right) \delta \left(
0\right) =0\times \infty .
\end{align}%
Thus, let us regularize this object by introducing eigenvalues of the
operator $\bar{D}$ as%
\begin{align}
J\left[ A_{\mu },\theta \right] & \equiv \exp \left[ \lim_{M^{2}\rightarrow
\infty }\sum_{n}\left( \text{tr}\int \mathrm{dx}\text{ }\phi _{n}^{\dag
}\left( x\right) if_{abc}\theta ^{c}\left( x\right) \exp \left( -\frac{%
\lambda _{n}^{2}}{M^{\alpha }}\right) \phi _{n}\left( x\right) \right) %
\right]  \notag \\
& =\exp \left[ \lim_{M^{2}\rightarrow \infty }\sum_{n}\left( \text{tr}\int 
\mathrm{dx}\text{ }\phi _{n}^{\dag }\left( x\right) if_{abc}\theta
^{c}\left( x\right) \exp \left( -\frac{\bar{D}^{2}}{M^{\alpha }}\right) \phi
_{n}\left( x\right) \right) \right] ,
\end{align}%
where $\alpha $ is chosen so the exponential argument is dimensionless.

The choice of operator $\bar{D}$ usually considers the requisites of
naturally appearing in the theory, being gauge invariant, and having real
eigenvalues. Furthermore, our choice of coefficients $a_{\mu ,n}^{a}$
carrying all the dependence on $\mu $ and $a$ implies that the $\phi _{n}$
must be eigenfunctions of a scalar colorless operator; therefore, a good
choice is%
\begin{equation}
\bar{D}=\text{tr}\left( \mathcal{D}_{\mu }\mathcal{D}^{\mu }\right) ,
\end{equation}%
where the trace is taken only over color indices. We see that the sum is
regularized under these conditions, so proceeding with the evaluation of the
Jacobian is possible. Since no additional dependence on color indices comes
from the exponential argument $\bar{D}^{2}/M^{4}$, the trace can be
immediately taken, yielding the unity%
\begin{align}
J\left[ A_{\mu },\theta \right] & =\exp \left[ \lim_{M^{2}\rightarrow \infty
}\sum_{n}\left( if_{aac}\int \mathrm{dx}\text{ }\phi _{n}^{\dag }\left(
x\right) \theta ^{c}\left( x\right) \exp \left( -\frac{\bar{D}^{2}}{M^{4}}%
\right) \phi _{n}\left( x\right) \right) \right]  \notag \\
& =\exp \left( 0\right) =1.
\end{align}

Such a result accomplishes our objective of furnishing proof for the
invariance of the bosonic measure. Of course, one could choose other
strategies so a result different from 1 could arise. Nevertheless, the
\textquotedblleft gauge anomaly\textquotedblright\ coming from this
\textquotedblleft non-trivial\textquotedblright\ Jacobian could be removed
by an adequate choice of counterterms. To say this more precisely, we can
use what we know from the fact that Yang-Mills theories are renormalizable.
In fact, 't Hooft's proof \cite{tHooft} shows that it is possible to
preserve gauge invariance at every order in perturbation theory, which is
crucial for demonstrating that the theory is renormalizable. Algebraic
renormalization results confirm this by noticing that the cohomology of the
Slavnov-Taylor operator is trivial for a Yang-Mills theory \cite{refalgren}.
Then, even if we would regularize the theory with non-gauge invariant
regulators (obtaining a non-trivial Jacobian), a change in the
renormalization scheme could restore gauge invariance and set the Jacobian
as the unity.

The results in this chapter are the main part of our published work \cite%
{GTS}.

\section{Final Remarks and Conclusions}

\label{Conc1}In the second chapter, we checked aspects related to gauge
symmetry maintenance in gauge theories. At the classical level, current
conservation arose after implementing local invariance in the theory action.
Equations of motion for fermion fields were necessary to achieve this
result. This part of the discussion established a route to follow in the
quantum theory.

With this in mind, it would be reasonable for Dyson-Schwinger equations to
play a role in the current conservation due to their analogy with classical
equations of motion. It would be a consequence of the translational
invariance of the fermion measure, which is a condition to obtain the
mentioned equations. Nevertheless, we saw that gauge invariance of the
fermion measure is the new requirement for conservation.

Once the panorama was clear, we focused on gauge-anomalous theories. For
them, considering external gauge fields, the presence of a Jacobian to the
fermion measure implies a non-zero result to the expectation value of the
covariant derivative of the current (the anomaly). We saw that, when
quantizing the gauge field, the expectation value vanishes in a simple way.
This outcome is a direct consequence of considering the boson measure
invariant, and the properties of the fermion measure were unnecessary. There
is no gauge anomaly preventing current conservation in the fully quantized
theory. That does not affect the topological interpretation of the gauge
anomaly since it is present when we do not consider the integration on the
gauge field.

Although our argumentation depends on gauge measure invariance, we took this
property for granted. That is usual in the literature but not explicitly
proved. This proof was achieved by G. de Lima e Silva, T.J. Girardi, and S.
A. Dias and published in reference \cite{GTS}. Such a result completes the
theoretical setup for our claim that the vacuum expectation value of the
gauge anomaly vanishes. The natural course of this investigation is to
define a chiral theory perturbatively, aiming at a detailed analysis of its
renormalizability and unitarity.

\chapter{Anomalies in Fermionic Amplitudes}

\label{Sec2}This chapter refers to another line of investigation in this
thesis, which concerns the occurrence of anomalies in fermionic amplitudes.
As mentioned, the single axial triangle ($AVV$) establishes the foundations
for this analysis. Although this process is largely explored in the
literature, our perspective shows new aspects of anomalies while emphasizing
patterns related to their tensor structures. The single axial box ($AVVV$)
exhibits similar elements in a more complex scene, substantiating this
investigation.

Both correlators depend on traces involving the chiral matrix, which lead to
products between the Levi-Civita symbol and metric tensors. In addition to
its manifestation in anomalous amplitudes, this type of structure is common
in chiral theories and investigations developed in odd space-time
dimensions. That is part of the motivation for this work and emphasizes the
significance of mathematical resources developed throughout our calculations.

Integrals in perturbative calculus usually exhibit diverging content, which
requires using regularization techniques in intermediate steps of
calculations \cite{Gnendiger:2017pys}. These prescriptions make mathematical
structures finite, so manipulations problematic to the original expressions
become valid. That implies modifying amplitudes through the introduction of
non-physical parameters. Results independent of regularizations emerge after
renormalization \cite{Pittau:2012zd}. Then, establishing predictions to
compare with experimental data becomes feasible.

Choosing a specific regularization scheme brings consequences to the
interpretation of results. To clarify this aspect, we get back to the
impossibility of preserving chiral and gauge symmetry simultaneously \cite%
{Jackiw-Johnson}. This time, however, we emphasize the issue of the
maintenance of Ward identities for the single axial triangle. This amplitude
unavoidably exhibits dependence on a diverging surface term \cite%
{Treiman:1986ep}, so choosing a prescription that eliminates this object
preserves some Ward identities (but not all). Methods that allow shifts in
the integration variable accomplish this task, e.g., Dimensional
Regularization \cite{Bollini:1972bi, tHooft:1972tcz, Ashmore:1972uj}. Other
prescriptions do not lead to this outcome.

Even though there is an inclination towards preserving gauge symmetry, there
are other possibilities. The reason for such is the presence of divergent
Feynman integrals having a divergence degree higher than the logarithmic
one. For them, a shift in the integration variable requires compensation
through (non-zero) surface terms to maintain the connection with the
original expression \cite{Cheng:1985bj, Treiman:1986ep, Bertlmann:1996xk}.
That implies the existence of different versions for perturbative
contributions involving loops, which differ by these surface terms after
integration. This situation is a manifestation of internal momenta
arbitrariness, although they relate to external momenta through
energy-momentum conservation \cite{Sterman:1994ce}. We will illustrate that
choices occur when taking Dirac traces, leading to one version with a\
specific behavior regarding symmetries; i.e., choosing one form sets the
position of violating terms typical of anomalous amplitudes.

The mentioned aspects motivate the perspective adopted in this investigation
and, therefore, the employment of Implicit Regularization (IReg) \cite%
{ORIMAR-TESE}. Its main feature is avoiding the evaluation of divergent
structures. That means we only integrate finite contributions without
modifying ill-defined objects. Our analysis falls on the accessible values
for these divergences within final expressions for amplitudes. We also avoid
choices for the internal momenta, adopting arbitrary routings for internal
lines of the graphs. This arbitrariness is intrinsic to the perturbative
calculus and received attention in recent works \cite{Ferreira:2011cv,
Vieira:2015fra, Viglioni:2016nqc}. The study of schemes to compute traces
involving chiral matrices also received attention from the authors \cite%
{Tsai:2009it, Tsai:2009hp, Bruque:2018bmy}. This concept characterizes
another class of possibilities for anomalous amplitudes.

The discussion is organized as follows. Section (\ref{Model}) introduces the
model and the correlators that concern this investigation. We also comment
on expectations about symmetries (through Ward identities) and their
relation with the linearity of integration. Section (\ref{Integrands}) looks
into integrands of amplitudes, highlighting tensor arrangements associated
with structures that compound the intended organization. This feature is
part of the IReg, approached in Section (\ref{Strategy}). We also introduce
the elements used to describe diverging quantities and finite functions.
With our perspective clear, Section (\ref{Integral}) focus on the explicit
integration of amplitudes while providing a preliminary discussion of these
results. A careful analysis occurs in Section (\ref{An}), where we inquire
about relations involving amplitudes and the consequences of different
prescriptions to evaluate divergent objects. Section (\ref{Final}) discusses
important aspects of the investigation while presenting the conclusions.

\newpage

\section{Model and Definitions}

\label{Model}We consider a $\left( 1+3\right) $-dimensional model where
massive spin $1/2$ fields interact with different types of bosons. The
corresponding couplings\footnote{%
Although some couplings do not concern this investigation at first glance,
perturbative corrections bring all these possibilities.} are listed in the
interacting Lagrangian%
\begin{eqnarray}
\mathcal{L}_{I} &=&e_{S}\left( \bar{\psi}\psi \right) \phi +e_{P}\left( \bar{%
\psi}\gamma _{5}\psi \right) \pi -e_{V}\left( \bar{\psi}\gamma ^{\mu }\psi
\right) V_{\mu }  \notag \\
&&-e_{A}\left( \bar{\psi}\gamma _{5}\gamma ^{\mu }\psi \right) A_{\mu
}+e_{T}\left( \bar{\psi}\gamma _{5}\sigma ^{\mu \nu }\psi \right) H_{\mu \nu
},  \label{Lag}
\end{eqnarray}%
where elements belonging to the set $\left\{ \phi ,\pi ,V_{\mu },A_{\mu
},H_{\mu \nu }\right\} $ are respectively scalar, pseudoscalar, vector,
axial, and pseudotensor boson fields, while $\psi $ corresponds to Dirac
fermions. As coupling constants $\left\{
e_{S},e_{P},e_{V},e_{A},e_{T}\right\} $ do not concern the intended
discussion, we set them as the unity.

The remaining structures emerge in the context of the four-dimensional
Clifford algebra. The objects $\gamma ^{\mu }$ are Dirac matrices, whose
commutator is denoted as $\left[ \gamma ^{\mu },\gamma ^{\nu }\right]
=2\sigma ^{\mu \nu }$. Since establishing a chiral matrix that anticommutes
with all gamma matrices is feasible in even dimensions, we introduce the
definition employed within this context $\gamma _{5}=\frac{i}{4!}\varepsilon
_{\mu \nu \alpha \beta }\gamma ^{\mu }\gamma ^{\nu }\gamma ^{\alpha }\gamma
^{\beta }$. Even though omitted, the identity $\boldsymbol{1}$ appears
within the scalar coupling.

Those structures in parentheses within the Lagrangian correspond to Noether
currents, which couple to boson fields. Current conservation establishes
relations involving these quantities. Although violations are expected for
anomalous amplitudes, we discuss preliminary expectations here. In a case
involving fermions with different masses, the vector current divergence
would be proportional to the scalar one with a coefficient depending on the
difference between masses. Nevertheless, the vector current is conserved as
we delimit this investigation to the equal masses context%
\begin{equation}
\partial _{\mu }\left( \bar{\psi}\gamma ^{\mu }\psi \right) =0.  \label{dV}
\end{equation}%
That suggests implications at the quantum level through Ward identities for
correlators involving vector vertices. The result should vanish whenever we
contract an external momentum with an index corresponding to this vertex
type. On the other hand, the axial current divergence is classically
proportional to the pseudoscalar one%
\begin{equation}
\partial _{\mu }\left( \bar{\psi}\gamma _{5}\gamma ^{\mu }\psi \right)
=2m\left( \bar{\psi}\gamma _{5}\psi \right) .  \label{dA}
\end{equation}%
Such relation leads to Ward identities involving similar amplitudes that
differ by the corresponding vertices. Establishing an analogous association
involving the pseudotensor current is not possible.

Our objective is on the next-to-leading order corrections for processes
involving external bosons, which produces purely fermionic loops. We
introduce them in two steps. First, we employ Feynman rules to construct
graphs for a single value of the unrestricted (loop) momentum. Hence, we
inspect them and survey expectations without worrying about ill-defined
mathematical quantities. This problem arises when implementing the last
Feynman rule, which consists of momenta integration. We only consider this
operation (in the second step) after discussing a strategy to deal with the
mentioned problem. Upper and lower case letters distinguish these two
versions of amplitudes%
\begin{equation}
T^{\Gamma _{i}\Gamma _{j}\cdots \Gamma _{l}}=\int \frac{d^{4}k}{\left( 2\pi
\right) ^{4}}t^{\Gamma _{i}\Gamma _{j}\cdots \Gamma _{l}}.  \label{T}
\end{equation}%
Such notation is extended to other integrals that emerge throughout this
work.

The general form of amplitudes for a single value of the loop momentum is%
\begin{eqnarray}
&&t^{\Gamma _{i}\Gamma _{j}\cdots \Gamma _{l}}\left( k_{1},k_{2},\ldots
,k_{n}\right)  \notag \\
&=&\text{tr}\left\{ \Gamma _{i}\left[ S_{F}\left( k+k_{1};m\right) \right]
\Gamma _{j}\left[ S_{F}\left( k+k_{2};m\right) \right] \cdots \Gamma _{l}%
\left[ S_{F}\left( k+k_{n};m\right) \right] \right\} ,  \label{t}
\end{eqnarray}%
whose argument is omitted unless it associates with configurations different
from $\left( k_{1},k_{2},\ldots ,k_{n}\right) $. This structure depends on
fermion propagators $S_{F}$ and vertex operators $\Gamma _{l}$. We express
the propagator of a Dirac fermion carrying momentum $K_{n}=k+k_{n}$ and mass 
$m$ through the structure%
\begin{equation}
S_{F}\left( k+k_{n},m\right) =\frac{1}{\slashed{F}_{n}}=\frac{1}{\left( %
\slashed{k}+\slashed{k}_{n}\right) -m}=\frac{\left( \slashed{k}+\slashed{k}%
_{n}\right) +m}{D_{n}}.  \label{fermion}
\end{equation}%
Although we use the form $\slashed{F}_{n}^{-1}$ to introduce perturbative
amplitudes and derive relations among them, employing the denominator $%
D_{n}=\left( k+k_{n}\right) ^{2}-m^{2}$ is useful to the integration. Due to
the adopted simplifications, vertices have the following structures%
\begin{equation}
\Gamma _{l}=\left\{ \Gamma _{S},\Gamma _{P},\Gamma _{V},\Gamma _{A},\Gamma _{%
\tilde{T}}\right\} =\left\{ \boldsymbol{1},\gamma _{5},\gamma ^{\mu },\gamma
^{\mu }\gamma _{5},\gamma _{5}\sigma ^{\mu \nu }\right\} .  \label{vertex}
\end{equation}%
Capital Latin subindices denote the nature of each object. They correspond
respectively to scalar, pseudoscalar, vector, axial, and pseudotensor
vertices. We extend this notation to perturbative amplitudes, where these
labels indicate the vertex content and the specific position of each
operator.

Loop corrections to processes involving two, three, and four external bosons
arise within this discussion. The Figure \ref{F} shows representations
through Feynman diagrams associated with these amplitudes. We have yet to
specify the vertex content, but the momenta configuration is set. Although
routings $k_{i}$ do not have physical meaning by themselves, conservation
laws on the vertices connect external (physical) momenta with differences
between routings. The conventions adopted allow summarizing these relations
into the object $p_{i}=k_{1}-k_{i}$, whose accessible values are the
following%
\begin{equation*}
\begin{array}{ccc}
p_{2}=k_{1}-k_{2}=p, & p_{3}=k_{1}-k_{3}=q, & p_{4}=k_{1}-k_{4}=r.%
\end{array}%
\end{equation*}

In order to proceed with definitions, we must cast the processes that
concern this investigation (by setting the vertex configurations). This
subject is covered in the sequence.

\begin{center}
{\small 
\begin{figure}[h]
{\small 
\begin{equation*}
\includegraphics[scale=0.8]{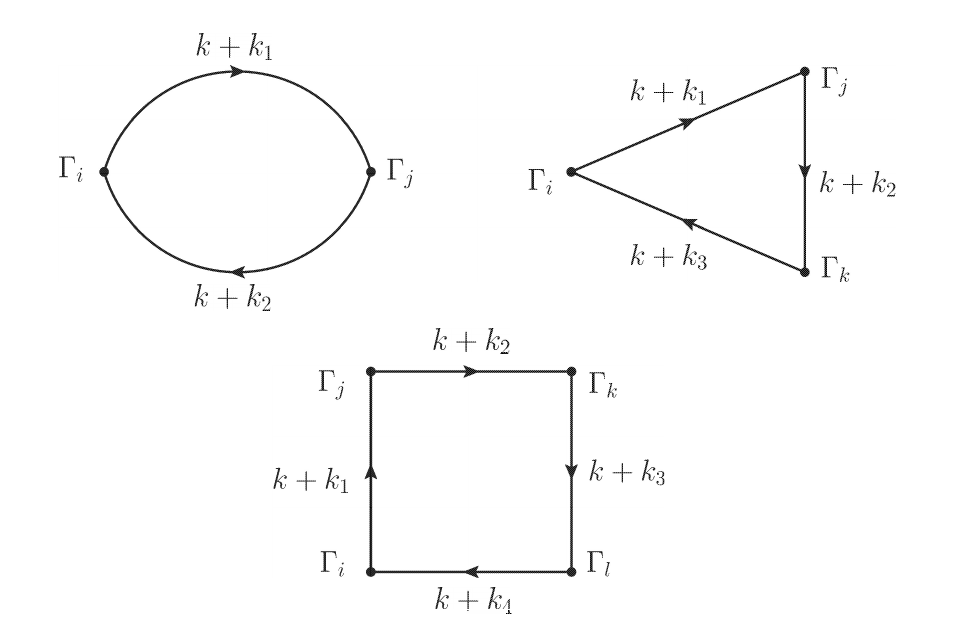}
\end{equation*}
}
\caption{One-loop corrections to processes described by external bosons.}
\label{F}
\end{figure}
}
\end{center}

\subsection{Perturbative Amplitudes}

\label{Amplitudes}Since resources required to construct any fermionic
amplitude are at our disposal, let us delimit those of interest and how they
relate to each other. We consider constraints coming from their mathematical
structure and symmetry implications in this process.

The neutral pion decay in two photons has a remarkable role in studies on
anomalies, so we take the single axial triangle amplitude as the first
laboratory. It is described by one axial and two vector vertices, assuming
the form%
\begin{equation}
t_{\mu \nu \alpha }^{AVV}=\text{tr}\left\{ \gamma _{\mu }\gamma _{5}\frac{1}{%
\slashed{F}_{1}}\gamma _{\nu }\frac{1}{\slashed{F}_{2}}\gamma _{\alpha }%
\frac{1}{\slashed{F}_{3}}\right\} .
\end{equation}%
From now on, we use labels to refer to a particular perturbative amplitude,
i.e., we designate this one as $AVV$. Within the IReg perspective, the
systematization of results highlights certain features concerning the
mathematical structure of the $AVV$ that relate to its anomalous character.
They motivate us to pursue one higher-order amplitude exhibiting similar
elements: the single axial box amplitude%
\begin{equation}
t_{\mu \nu \alpha \beta }^{AVVV}=\text{tr}\left\{ \gamma _{\mu }\gamma _{5}%
\frac{1}{\slashed{F}_{1}}\gamma _{\nu }\frac{1}{\slashed{F}_{2}}\gamma
_{\alpha }\frac{1}{\slashed{F}_{3}}\gamma _{\beta }\frac{1}{\slashed{F}_{4}}%
\right\} ,
\end{equation}%
denominated $AVVV$. Even though its evaluation is complex, all operations
involved are analogous to those performed in the triangle context. Thus, we
consider the first process as a guide for analyzing the second.

These amplitudes are the central elements of this work. Nevertheless, as
acknowledged in the discussion about Noether currents (\ref{dV})-(\ref{dA}),
relations among amplitudes could be derived through contractions with the
external momenta. Thus, we explore this operation for the integrands above
to introduce the remaining correlators while discussing potential
constraints on the results.

For such purpose, let us express contractions involving physical momenta and
Dirac matrices in terms of fermion propagators (\ref{fermion}):%
\begin{equation}
\slashed{k}_{i}-\slashed{k}_{j}=\slashed{F}_{i}-\slashed{F}_{j}.  \label{id0}
\end{equation}%
Now, consider the specific contraction on the index associated with the
first vector vertex of the triangle amplitude. Posteriorly to the
implementation of this identity, trace linearity leads to the result%
\begin{equation*}
p^{\nu }t_{\mu \nu \alpha }^{AVV}=\text{tr}\left\{ \gamma _{\mu }\gamma _{5}%
\frac{1}{\slashed{F}_{2}}\gamma _{\alpha }\frac{1}{\slashed{F}_{3}}\right\} -%
\text{tr}\left\{ \gamma _{\mu }\gamma _{5}\frac{1}{\slashed{F}_{1}}\gamma
_{\alpha }\frac{1}{\slashed{F}_{3}}\right\} ,
\end{equation*}%
where a difference between $AV$ two-point amplitudes is identified%
\begin{equation}
p^{\nu }t_{\mu \nu \alpha }^{AVV}=t_{\mu \alpha }^{AV}\left(
k_{2},k_{3}\right) -t_{\mu \alpha }^{AV}\left( k_{1},k_{3}\right) .
\label{pavv2}
\end{equation}%
An analogous relation arises for the second vector vertex through the same
steps%
\begin{equation}
\left( q-p\right) ^{\alpha }t_{\mu \nu \alpha }^{AVV}=t_{\mu \nu
}^{AV}\left( k_{1},k_{3}\right) -t_{\mu \nu }^{AV}\left( k_{1},k_{2}\right) .
\label{pavv3}
\end{equation}%
As for contractions with axial vertices, we multiply the identity (\ref{id0}%
) by the chiral matrix. Permuting its position is necessary to allow
identifications%
\begin{equation}
\left( \slashed{k}_{i}-\slashed{k}_{j}\right) \gamma _{5}=\slashed{F}%
_{i}\gamma _{5}+\gamma _{5}\slashed{F}_{j}+2m\gamma _{5}.
\end{equation}%
Besides the difference between $AV$ amplitudes, one additional term
corresponds to the $PVV$ amplitude%
\begin{equation}
q^{\mu }t_{\mu \nu \alpha }^{AVV}=t_{\nu \alpha }^{AV}\left(
k_{2},k_{3}\right) -t_{\alpha \nu }^{AV}\left( k_{1},k_{2}\right) -2mt_{\nu
\alpha }^{PVV}.  \label{pavv1}
\end{equation}

Concerning the $AVVV$ box amplitude, contractions follow the same procedure
and yield the results:%
\begin{eqnarray}
r^{\mu }t_{\mu \nu \alpha \beta }^{AVVV} &=&t_{\nu \alpha \beta
}^{AVV}\left( k_{2},k_{3},k_{4}\right) -t_{\beta \nu \alpha }^{AVV}\left(
k_{1},k_{2},k_{3}\right) -2mt_{\nu \alpha \beta }^{PVVV},  \label{pavvv1} \\
p^{\nu }t_{\mu \nu \alpha \beta }^{AVVV} &=&t_{\mu \alpha \beta
}^{AVV}\left( k_{2},k_{3},k_{4}\right) -t_{\mu \alpha \beta }^{AVV}\left(
k_{1},k_{3},k_{4}\right) ,  \label{pavvv2} \\
\left( q-p\right) ^{\alpha }t_{\mu \nu \alpha \beta }^{AVVV} &=&t_{\mu \nu
\beta }^{AVV}\left( k_{1},k_{3},k_{4}\right) -t_{\mu \nu \beta }^{AVV}\left(
k_{1},k_{2},k_{4}\right) ,  \label{pavvv3} \\
\left( r-q\right) ^{\beta }t_{\mu \nu \alpha \beta }^{AVVV} &=&t_{\mu \nu
\alpha }^{AVV}\left( k_{1},k_{2},k_{3}\right) -t_{\mu \nu \alpha
}^{AVV}\left( k_{1},k_{2},k_{4}\right) .  \label{pavvv4}
\end{eqnarray}%
Although all operations lead to the difference between $AVV$ triangles, the $%
PVVV$ four-point amplitude appears as the extra contribution in the axial
contraction.

Obtaining these relations considers only the mathematical structure of
integrands, which consist of identities at this level. Their validity after
integration represents a manifestation of linearity. Nonetheless, we will
see that the anomalous character of involved amplitudes might affect these
prospects. Then, if their verification is successful, proper relations among
Green functions (GF) are established. Expectations for contractions with the 
$AVV$ triangle are the following%
\begin{eqnarray}
q^{\mu }T_{\mu \nu \alpha }^{AVV} &\rightarrow &T_{\nu \alpha }^{AV}\left(
k_{2},k_{3}\right) -T_{\alpha \nu }^{AV}\left( k_{1},k_{2}\right) -2mT_{\nu
\alpha }^{PVV},  \label{ragf1} \\
p^{\nu }T_{\mu \nu \alpha }^{AVV} &\rightarrow &T_{\mu \alpha }^{AV}\left(
k_{2},k_{3}\right) -T_{\mu \alpha }^{AV}\left( k_{1},k_{3}\right) ,
\label{ragf2} \\
\left( q-p\right) ^{\alpha }T_{\mu \nu \alpha }^{AVV} &\rightarrow &T_{\mu
\nu }^{AV}\left( k_{1},k_{3}\right) -T_{\mu \nu }^{AV}\left(
k_{1},k_{2}\right) ,  \label{ragf3}
\end{eqnarray}%
while contractions involving the $AVVV$ box yield%
\begin{eqnarray}
r^{\mu }T_{\mu \nu \alpha \beta }^{AVVV} &\rightarrow &T_{\nu \alpha \beta
}^{AVV}\left( k_{2},k_{3},k_{4}\right) -T_{\beta \nu \alpha }^{AVV}\left(
k_{1},k_{2},k_{3}\right) -2mT_{\nu \alpha \beta }^{PVVV},  \label{ragf4} \\
p^{\nu }T_{\mu \nu \alpha \beta }^{AVVV} &\rightarrow &T_{\mu \alpha \beta
}^{AVV}\left( k_{2},k_{3},k_{4}\right) -T_{\mu \alpha \beta }^{AVV}\left(
k_{1},k_{3},k_{4}\right) ,  \label{ragf5} \\
\left( q-p\right) ^{\alpha }T_{\mu \nu \alpha \beta }^{AVVV} &\rightarrow
&T_{\mu \nu \beta }^{AVV}\left( k_{1},k_{3},k_{4}\right) -T_{\mu \nu \beta
}^{AVV}\left( k_{1},k_{2},k_{4}\right) ,  \label{ragf6} \\
\left( r-q\right) ^{\beta }T_{\mu \nu \alpha \beta }^{AVVV} &\rightarrow
&T_{\mu \nu \alpha }^{AVV}\left( k_{1},k_{2},k_{3}\right) -T_{\mu \nu \alpha
}^{AVV}\left( k_{1},k_{2},k_{4}\right) .  \label{ragf7}
\end{eqnarray}

Previously, we stated that current conservation (\ref{dV})-(\ref{dA})
generates implications over quantum corrections. Ward identities (WIs)
relate to momenta contractions over perturbative amplitudes. In the
hypothesis that one relation among GF applies, the maintenance of the
corresponding WI requires the cancellation of differences between amplitudes
above ($AV$s in the first set and $AVV$s in the second). We cast these
expectations in the sequence, where the required sum of channels is implicit
in the notation $\mathcal{T}$. Nevertheless, we will see that our analysis
applies channel by channel. The identities for the $AVV$ amplitude are%
\begin{eqnarray}
q^{\mu }\mathcal{T}_{\mu \nu \alpha }^{AVV} &\rightarrow &-2m\mathcal{T}%
_{\nu \alpha }^{PVV},  \label{qAVV1} \\
p^{\nu }\mathcal{T}_{\mu \nu \alpha }^{AVV} &\rightarrow &0, \\
\left( q-p\right) ^{\alpha }\mathcal{T}_{\mu \nu \alpha }^{AVV} &\rightarrow
&0,  \label{qAVV3}
\end{eqnarray}%
while those for the $AVVV$ amplitude are%
\begin{eqnarray}
r^{\mu }\mathcal{T}_{\mu \nu \alpha \beta }^{AVVV} &\rightarrow &-2m\mathcal{%
T}_{\nu \alpha \beta }^{PVVV}, \\
p^{\nu }\mathcal{T}_{\mu \nu \alpha \beta }^{AVVV} &\rightarrow &0, \\
\left( q-p\right) ^{\alpha }\mathcal{T}_{\mu \nu \alpha \beta }^{AVVV}
&\rightarrow &0, \\
\left( r-p\right) ^{\beta }\mathcal{T}_{\mu \nu \alpha \beta }^{AVVV}
&\rightarrow &0.
\end{eqnarray}%
Given the impossibility of simultaneous satisfaction of gauge and axial
symmetries, these are also preliminary prospects.

Through this argumentation, we connected concepts of integral linearity and
symmetry implications. If relations among GF are identically satisfied,
canceling those differences on their right-hand side also satisfies WIs.
Nevertheless, the fact that these amplitudes exhibit diverging power
counting is problematic when testing these expectations. That is
particularly important in the anomalies context. We will return to this
discussion after exploring the perturbative amplitudes at the integrand
level.

\newpage

\section{Structure of Perturbative Amplitudes}

\label{Integrands}This work implements Feynman rules in two parts, starting
with obtaining perturbative amplitudes for a single value of the
unrestricted (loop) momentum. Thus, organizing and examining their content
without worries about the divergences that come with integration is
attainable. We begin by introducing an example illustrating the elements
required for this task. Subsequently, we inquire about two, three, and
four-point functions concerning this investigation.

\subsection{Two-Point Amplitudes - Preliminary Notions}

\label{Notion}This analysis uses a simple example to familiarize with
calculations while producing tools for more complex scenes. Soon we will
come across extensive mathematical expressions that might seem vague.
Thereby, designing mechanisms to compact them and systematizing operations
is part of our task.

The next-to-leading order correction to processes involving external bosons
corresponds to pure fermionic loops. We denoted these amplitudes using
uppercase letters (\ref{T}), while their integrands use lowercase letters (%
\ref{t}). These structures contain traces of vertex operators $\Gamma _{i}$
and fermion propagators $\slashed{F}_{n}^{-1}$, as seen in the example of
two-point functions:%
\begin{equation}
t^{\Gamma _{i}\Gamma _{j}}=\text{tr}\left( \Gamma _{i}\frac{1}{\slashed{F}%
_{1}}\Gamma _{j}\frac{1}{\slashed{F}_{2}}\right) .
\end{equation}%
After rewriting the propagator (\ref{fermion}), using the linearity of the
trace makes its matrix content explicit%
\begin{eqnarray}
t^{\Gamma _{i}\Gamma _{j}} &=&\text{tr}\left( \Gamma _{i}\gamma _{A}\Gamma
_{j}\gamma _{B}\right) \frac{K_{1}^{A}K_{2}^{B}}{D_{12}}+m^{2}\text{tr}%
\left( \Gamma _{i}\Gamma _{j}\right) \frac{1}{D_{12}}  \notag \\
&&+m\text{tr}\left( \Gamma _{i}\gamma _{A}\Gamma _{j}\right) \frac{K_{1}^{A}%
}{D_{12}}+m\text{tr}\left( \Gamma _{i}\Gamma _{j}\gamma _{B}\right) \frac{%
K_{2}^{B}}{D_{12}}.  \label{tgg}
\end{eqnarray}

As several notations appear within this context, let us explain them
subsequently. We introduced compact products as that of the denominator $%
D_{ij}=D_{i}D_{j}$ for propagator-like objects $D_{i}=\left( k+k_{i}\right)
^{2}-m^{2}$. Our goal in this section is to express integrands through
combinations depending on these structures%
\begin{equation*}
\frac{1}{D_{i}},\frac{\left[ 1,k_{\mu },k_{\mu \nu }\right] }{D_{ij}},\frac{%
\left[ 1,k_{\mu },k_{\mu \nu },k_{\mu \nu \alpha }\right] }{D_{ijk}},\frac{%
\left[ 1,k_{\mu },k_{\mu \nu },k_{\mu \nu \alpha },k_{\mu \nu \alpha \beta }%
\right] }{D_{ijkl}},
\end{equation*}%
which leads to identifying Feynman integrals in Section (\ref{Strategy}).
That means the usage of the symbol $K_{i}=k+k_{i}$ is limited to the current
analysis, being another artifice to reduce expressions. We also introduced
compact notations for products of momenta or routings:%
\begin{equation*}
\begin{array}{ccccccc}
k_{\mu \nu }=k_{\mu }k_{\nu }, &  &  & p_{\mu \nu }=p_{\mu }p_{\nu }, &  & 
& k_{1\mu \nu }=k_{1\mu }k_{1\nu }.%
\end{array}%
\end{equation*}

The second type of notation consists of (the possibility of) adopting
uppercase Latin letters for summed indices and neglecting their covariant or
contravariant character. This resource facilitates the recognition of
sectors with analogous index configurations inside tensor amplitudes, making
substructures promptly noticeable. Hence, identifying other amplitudes
inside the original only requires sign comparisons among options.
Furthermore, other terms receive a suitable organization through standard
tensors. We also use this notation to emphasize symmetry properties.

Since we know these tools and ideas, we implement them in the mentioned
example. It consists of the double-vector function $VV$, which associates
with the photon self-energy in the Quantum Electrodynamics context. The
replacement of Dirac matrices as vertex operators ($\Gamma _{i}=\gamma _{\mu
}$ and $\Gamma _{j}=\gamma _{\nu }$) on the integrand above generates the
expression%
\begin{eqnarray}
t_{\mu \nu }^{VV} &=&\text{tr}\left( \gamma _{\mu }\gamma _{A}\gamma _{\nu
}\gamma _{B}\right) \frac{K_{1}^{A}K_{2}^{B}}{D_{12}}+m^{2}\text{tr}\left(
\gamma _{\mu }\gamma _{\nu }\right) \frac{1}{D_{12}}  \notag \\
&&+m\text{tr}\left( \gamma _{\mu }\gamma _{A}\gamma _{\nu }\right) \frac{%
K_{1}^{A}}{D_{12}}+m\text{tr}\left( \gamma _{\mu }\gamma _{\nu }\gamma
_{B}\right) \frac{K_{2}^{B}}{D_{12}}.
\end{eqnarray}

Even though Dirac traces are common ingredients, we discuss them to ground
future calculations. The property of anticommutation followed by Dirac
matrices is the outset%
\begin{equation}
\gamma _{\mu }\gamma _{\nu }+\gamma _{\nu }\gamma _{\mu }=2g_{\mu \nu }.
\end{equation}%
By taking the trace on both sides, linearity and invariance under cyclic
permutations lead to the equation%
\begin{equation}
\text{tr}\left( \gamma _{\mu }\gamma _{\nu }\right) =g_{\mu \nu }\text{tr}%
\left( \mathbf{1}\right) =4g_{\mu \nu }.  \label{trace1}
\end{equation}%
Any other trace involving an even number of Dirac matrices could be reduced
to this one. For instance, we use the anticommutation property to express
the four matrices trace as the following combination%
\begin{eqnarray}
\text{tr}\left( \gamma _{\mu }\gamma _{A}\gamma _{\nu }\gamma _{B}\right) &=&%
\text{tr}\left( 2g_{\mu A}\gamma _{\nu }\gamma _{B}-2g_{\mu \nu }\gamma
_{A}\gamma _{B}+2g_{\mu B}\gamma _{A}\gamma _{\nu }-\gamma _{A}\gamma _{\nu
}\gamma _{B}\gamma _{\mu }\right)  \notag \\
&=&4g_{\mu A}g_{\nu B}-4g_{\mu \nu }g_{AB}+4g_{\mu B}g_{A\nu }.
\label{trace2}
\end{eqnarray}

As for products involving an odd number of Dirac matrices, trace operation
vanishes. To prove this statement, introduce the identity $\mathbf{1}=\gamma
_{5}^{2}$ inside the argument. Using (respectively) the fact that the chiral
matrix anticommutes with any Dirac matrix and the cyclicity, we show that
these traces are equal to their negative and, therefore, vanish. To
illustrate, take the trace of one single Dirac matrix%
\begin{equation*}
\text{tr}\left( \gamma _{\mu }\right) =\text{tr}\left( \gamma _{5}\gamma
_{5}\gamma _{\mu }\right) =-\text{tr}\left( \gamma _{5}\gamma _{\mu }\gamma
_{5}\right) =-\text{tr}\left( \gamma _{5}\gamma _{5}\gamma _{\mu }\right) =-%
\text{tr}\left( \gamma _{\mu }\right) .
\end{equation*}

When replacing these results on the $VV$ amplitude and rearranging it, the
sorting of free indices shows two sectors%
\begin{equation}
t_{\mu \nu }^{VV}=4\frac{K_{1\mu }K_{2\nu }+K_{1\nu }K_{2\mu }}{D_{12}}%
+g_{\mu \nu }\left[ -\text{tr}\left( \gamma _{A}\gamma _{B}\right) \frac{%
K_{1}^{A}K_{2}^{B}}{D_{12}}+m^{2}\text{tr}\left( \mathbf{1}\right) \frac{1}{%
D_{12}}\right] .  \label{exa}
\end{equation}%
The first corresponds to the symmetric version of the following standard
tensor%
\begin{equation}
t_{2\mu \nu }^{\left( s\right) }\left( k_{i},k_{j}\right) =\frac{\left(
k+k_{i}\right) _{\mu }\left( k+k_{j}\right) _{\nu }+s\left( k+k_{j}\right)
_{\mu }\left( k+k_{i}\right) _{\nu }}{D_{12}}.  \label{t2s}
\end{equation}%
This general definition admits a numerical subindex, characterizing the
number of propagator-like objects in the denominator (two in this case $%
D_{12}=D_{1}D_{2}$), and it allows different signs $s=\pm 1$. Since this
expression is a combination of structures previously mentioned, it does not
require further analysis.

As for the sector proportional to the metric tensor $g_{\mu \nu }$, we
recognized traces involving fewer matrices. They associate with a scalar
amplitude from two possibilities: $SS$ and $PP$. Thus, replace the
corresponding vertices on Equation (\ref{tgg}) to determine their integrands:%
\begin{eqnarray}
t^{SS} &=&\text{tr}\left( \gamma _{A}\gamma _{B}\right) \frac{%
K_{1}^{A}K_{2}^{B}}{D_{12}}+m^{2}\text{tr}\left( \mathbf{1}\right) \frac{1}{%
D_{12}},  \label{ss1} \\
t^{PP} &=&-\text{tr}\left( \gamma _{A}\gamma _{B}\right) \frac{%
K_{1}^{A}K_{2}^{B}}{D_{12}}+m^{2}\text{tr}\left( \mathbf{1}\right) \frac{1}{%
D_{12}}.  \label{pp1}
\end{eqnarray}%
Since we did not rename any index, the precise identification occurs by
comparing signs, and we achieve the organization%
\begin{equation}
t_{\mu \nu }^{VV}=4t_{2\mu \nu }^{\left( +\right) }\left( k_{1},k_{2}\right)
+g_{\mu \nu }t^{PP}.  \label{vv}
\end{equation}

Exploring the $PP$ structure is still necessary, so we draw attention to its
dependence on the objects%
\begin{equation}
2K_{ij}\rightarrow 2\left( K_{i}\cdot K_{j}-m^{2}\right) =D_{i}+D_{j}-\left(
k_{i}-k_{j}\right) ^{2}.  \label{k.k}
\end{equation}%
This identity brings propagator-like objects to numerators, which reflects
on reductions of denominators within the amplitude integrand%
\begin{equation}
t^{PP}=-2\left[ \frac{1}{D_{1}}+\frac{1}{D_{2}}-p^{2}\frac{1}{D_{12}}\right]
,  \label{pp}
\end{equation}%
where we identified the external momentum $p=k_{1}-k_{2}$. The recurrent
application of this resource throughout this investigation justifies generic
indices. Notice that, with the momenta integration, this identity reduces
part of the Feynman integrals to those involving one less propagator.

We do not integrate these amplitudes in the future since they are not part
of this work. Even so, take them as a guide to calculations performed from
now on.

\subsection{Two-Point Amplitudes - $AV$}

\label{2pt}Given the general expression for two-point amplitudes (\ref{tgg}%
), we replace vertex operators to write the integrand of the axial-vector
amplitude%
\begin{eqnarray}
t_{\mu \nu }^{AV} &=&\text{tr}\left( \gamma _{\mu }\gamma _{5}\gamma
_{A}\gamma _{\nu }\gamma _{B}\right) \frac{K_{1}^{A}K_{2}^{B}}{D_{12}}+m^{2}%
\text{tr}\left( \gamma _{\mu }\gamma _{5}\gamma _{\nu }\right) \frac{1}{%
D_{12}}  \notag \\
&&+m\text{tr}\left( \gamma _{\mu }\gamma _{5}\gamma _{A}\gamma _{\nu
}\right) \frac{K_{1}^{A}}{D_{12}}+m\text{tr}\left( \gamma _{\mu }\gamma
_{5}\gamma _{\nu }\gamma _{B}\right) \frac{K_{2}^{B}}{D_{12}},
\end{eqnarray}%
where numerators depend on $K_{i}=k+k_{i}$ and denominators are $%
D_{12}=D_{1}D_{2}$. We refer to this structure as $AV$, which specifies the
first vertex as an axial $\Gamma _{i}=\gamma _{\mu }\gamma _{5}$ and the
second as a vector $\Gamma _{j}=\gamma _{\nu }$. Although these traces
contain the chiral matrix, replacing its definition $\gamma _{5}=\frac{i}{4!}%
\varepsilon _{\mu \nu \alpha \beta }\gamma ^{\mu }\gamma ^{\nu }\gamma
^{\alpha }\gamma ^{\beta }$ suppresses this dependence. That adds four extra
Dirac matrices to the argument while introducing a global factor through the
Levi-Civita symbol. Within this perspective, we must compute even traces
following steps seen in the previous subsection and then perform
contractions.

Immediately, occurrences involving an odd number of Dirac matrices plus the
chiral one vanish. That also happens in the case involving two Dirac
matrices since it leads to contractions between symmetric and antisymmetric
tensors. Hence, the only non-zero trace involves four Dirac matrices, whose
computation leads to the Levi-Civita symbol%
\begin{equation}
\text{tr}\left( \gamma _{5}\gamma _{\mu }\gamma _{A}\gamma _{\nu }\gamma
_{B}\right) =4i\varepsilon _{\mu A\nu B}.  \label{tr1}
\end{equation}%
When replacing it, symmetry properties allow identifying the antisymmetric
version of the standard tensor (\ref{t2s}):%
\begin{equation}
t_{\mu \nu }^{AV}=2i\varepsilon _{\mu \nu XY}t_{2XY}^{\left( -\right)
}\left( k_{1},k_{2}\right) .  \label{av}
\end{equation}

One would expect two ingredients to compound the integrated substructure:
metric tensor and external momentum $p=k_{1}-k_{2}$. Since they combine
exclusively into symmetric quantities ($g_{XY}$ and $p_{XY}=p_{X}p_{Y}$),
the contraction should cancel out. Nevertheless, two-point functions exhibit
quadratic power counting in the physical dimension. Therefore, these
integrals are not invariant under translations, admitting the emergence of
non-physical momenta associated with surface terms. That provides another
vector to build up the substructure: the sum of arbitrary routings $%
k_{1}+k_{2}$. Hence, we expect the integrated $AV$ amplitude to have the
following form%
\begin{equation}
T_{\mu \nu }^{AV}\rightarrow \varepsilon _{\mu \nu XY}\left(
k_{1}-k_{2}\right) ^{X}\left( k_{1}+k_{2}\right) ^{Y}G_{0},  \label{exp1}
\end{equation}%
where $G_{0}$ represents a surface term that is logarithmically divergent to
adjust with mass dimension.

Such dependence characterizes an ambiguity, a quantity depending on
arbitrary choices. Momenta conservation sets differences between labels as
external momenta; however, it does not attribute a particular meaning to
routings themselves or their sum. As proposed before, this arbitrariness is
preserved throughout this investigation.

\subsection{Three-Point Amplitudes - $PVV$}

Previously, we used lowercase letters to denote the integrand of fermionic
amplitudes (\ref{t}). They correspond to traces containing vertex operators $%
\Gamma _{i}$ and fermion propagators $\slashed{F}_{n}^{-1}$, as seen for the
particular case of three-point functions:%
\begin{equation}
t^{\Gamma _{i}\Gamma _{j}\Gamma _{k}}=\text{tr}\left( \Gamma _{i}\frac{1}{%
\slashed{F}_{1}}\Gamma _{j}\frac{1}{\slashed{F}_{2}}\Gamma _{k}\frac{1}{%
\slashed{F}_{3}}\right) .  \label{t30}
\end{equation}%
Rewriting the propagators (\ref{fermion}) emphasizes the coefficients as
Dirac traces%
\begin{eqnarray}
t^{\Gamma _{i}\Gamma _{j}\Gamma _{k}} &=&\text{tr}\left( \Gamma _{i}\gamma
_{A}\Gamma _{j}\gamma _{B}\Gamma _{k}\gamma _{C}\right) \frac{%
K_{1}^{A}K_{2}^{B}K_{3}^{C}}{D_{123}}+m\text{tr}\left( \Gamma _{i}\Gamma
_{j}\gamma _{B}\Gamma _{k}\gamma _{C}\right) \frac{K_{2}^{B}K_{3}^{C}}{%
D_{123}}  \notag \\
&&+m\text{tr}\left( \Gamma _{i}\gamma _{A}\Gamma _{j}\Gamma _{k}\gamma
_{C}\right) \frac{K_{1}^{A}K_{3}^{C}}{D_{123}}+m\text{tr}\left( \Gamma
_{i}\gamma _{A}\Gamma _{j}\gamma _{B}\Gamma _{k}\right) \frac{%
K_{1}^{A}K_{2}^{B}}{D_{123}}  \notag \\
&&+m^{2}\text{tr}\left( \Gamma _{i}\gamma _{A}\Gamma _{j}\Gamma _{k}\right) 
\frac{K_{1}^{A}}{D_{123}}+m^{2}\text{tr}\left( \Gamma _{i}\Gamma _{j}\gamma
_{B}\Gamma _{k}\right) \frac{K_{2}^{B}}{D_{123}}  \notag \\
&&+m^{2}\text{tr}\left( \Gamma _{i}\Gamma _{j}\Gamma _{k}\gamma _{C}\right) 
\frac{K_{3}^{C}}{D_{123}}+m^{3}\text{tr}\left( \Gamma _{i}\Gamma _{j}\Gamma
_{k}\right) \frac{1}{D_{123}},  \label{t3}
\end{eqnarray}%
where numerators depend on $K_{i}=k+k_{i}$ and denominators are $%
D_{123}=D_{1}D_{2}D_{3}$.

To study the structure of a specific amplitude, we set its vertex content
and evaluate corresponding traces. For the $PVV$ case, the first vertex
indicates a pseudoscalar $\Gamma _{i}=\gamma _{5}$ while the others indicate
vectors $\Gamma _{j}=\gamma _{\nu }$ and $\Gamma _{k}=\gamma _{\alpha }$.
Its non-zero contributions are the following%
\begin{eqnarray}
t_{\nu \alpha }^{PVV} &=&m\text{tr}\left( \gamma _{5}\gamma _{\nu }\gamma
_{B}\gamma _{\alpha }\gamma _{C}\right) \frac{K_{2}^{B}K_{3}^{C}}{D_{123}}+m%
\text{tr}\left( \gamma _{5}\gamma _{A}\gamma _{\nu }\gamma _{\alpha }\gamma
_{C}\right) \frac{K_{1}^{A}K_{3}^{C}}{D_{123}}  \notag \\
&&+m\text{tr}\left( \gamma _{5}\gamma _{A}\gamma _{\nu }\gamma _{B}\gamma
_{\alpha }\right) \frac{K_{1}^{A}K_{2}^{B}}{D_{123}}.
\end{eqnarray}%
They are proportional to the Levi-Civita symbol (\ref{tr1}), leading to the
antisymmetric version of the standard tensor%
\begin{equation}
t_{3\mu \nu }^{\left( s\right) }\left( k_{i},k_{j}\right) =\frac{\left(
k+k_{i}\right) _{\mu }\left( k+k_{j}\right) _{\nu }+s\left( k+k_{j}\right)
_{\mu }\left( k+k_{i}\right) _{\nu }}{D_{123}},  \label{t3uv}
\end{equation}%
Such an object is analogous to the previous one (\ref{t2s}); however, it
depends on three propagators embodied in $D_{123}$ as indicated by the
numerical subindex. With these identifications, the integrand of the
amplitude exhibits the form%
\begin{equation}
t_{\nu \alpha }^{PVV}=-2im\varepsilon _{\nu \alpha XY}\left[ t_{3XY}^{\left(
-\right) }\left( k_{2},k_{3}\right) +t_{3XY}^{\left( -\right) }\left(
k_{3},k_{1}\right) +t_{3XY}^{\left( -\right) }\left( k_{1},k_{2}\right) %
\right] .  \label{pvv}
\end{equation}

Observe the analogy between the $PVV$ structure and that of the $AV$ (\ref%
{av}); both are 2nd-order tensors contracted with the Levi-Civita symbol.
Nonetheless, expectations are different now. Even though three-point
functions exhibit linear power counting, contributions involving diverging
surface terms are prohibited since only finite contributions adjust to the
correct mass dimension. On the other hand, after integration, two external
momenta ($p=k_{1}-k_{2}$ and $q=k_{1}-k_{3}$) are available to build up the
tensor structure%
\begin{equation}
T_{\nu \alpha }^{PVV}\rightarrow \varepsilon _{\nu \alpha XY}p^{X}q^{Y}F_{0}.
\label{exp2}
\end{equation}%
The object $F_{0}=F_{0}\left( p_{i}\cdot p_{j}\right) $ represents a finite
scalar function depending on momenta bilinears $p_{i}\cdot p_{j}=\left\{
p^{2},\text{ }q^{2},\text{ }p\cdot q\right\} $.

\subsection{Three-Point Amplitudes - $AVV$}

\label{AVVint}The $AVV$ integrand emerges by replacing the corresponding
vertex operators within Equation (\ref{t3}); they are axial $\Gamma
_{i}=\gamma _{\mu }\gamma _{5}$, vector $\Gamma _{j}=\gamma _{\nu }$ and
vector $\Gamma _{k}=\gamma _{\alpha }$. Leaving null contributions aside, we
cast its initial structure:%
\begin{eqnarray}
t_{\mu \nu \alpha }^{AVV} &=&\text{tr}\left( \gamma _{\mu }\gamma _{5}\gamma
_{A}\gamma _{\nu }\gamma _{B}\gamma _{\alpha }\gamma _{C}\right) \frac{%
K_{1}^{A}K_{2}^{B}K_{3}^{C}}{D_{123}}+m^{2}\text{tr}\left( \gamma _{\mu
}\gamma _{5}\gamma _{A}\gamma _{\nu }\gamma _{\alpha }\right) \frac{K_{1}^{A}%
}{D_{123}}  \notag \\
&&+m^{2}\text{tr}\left( \gamma _{\mu }\gamma _{5}\gamma _{\nu }\gamma
_{B}\gamma _{\alpha }\right) \frac{K_{2}^{B}}{D_{123}}+m^{2}\text{tr}\left(
\gamma _{\mu }\gamma _{5}\gamma _{\nu }\gamma _{\alpha }\gamma _{C}\right) 
\frac{K_{3}^{C}}{D_{123}},  \label{AVV0}
\end{eqnarray}%
where numerators depend on $K_{i}=k+k_{i}$ and denominators are $%
D_{123}=D_{1}D_{2}D_{3}$. Terms associated with the squared mass are already
known, being proportional to the Levi-Civita symbol (\ref{tr1}).

Our next task is to take the trace involving six Dirac matrices plus the
chiral one. Nevertheless, different ways to perform this operation attribute
different expressions for it. Although all forms attributed to one trace are
linked through identities, the divergent character of perturbative
calculations affects these relations after integration. Clarifying these
aspects is essential to this investigation, so we are very detailed in this
discussion.

To introduce these ideas, we use the chiral matrix anticommutation in
studying two possibilities%
\begin{equation*}
\text{tr}\left( \gamma _{\mu }\gamma _{5}\gamma _{A}\gamma _{\nu }\gamma
_{B}\gamma _{\alpha }\gamma _{C}\right) =-\text{tr}\left( \gamma _{5}\gamma
_{\mu }\gamma _{A}\gamma _{\nu }\gamma _{B}\gamma _{\alpha }\gamma
_{C}\right) .
\end{equation*}%
After replacing the definition $\gamma _{5}=\frac{i}{4!}\varepsilon _{\mu
\nu \alpha \beta }\gamma ^{\mu }\gamma ^{\nu }\gamma ^{\alpha }\gamma
^{\beta }$ in these specific places, we obtain a trace involving only Dirac
matrices. Its computation yields combinations of the metric tensors, which
are contracted with the Levi-Civita symbol. The expression obtained through
the first path is%
\begin{eqnarray}
&&\text{tr}\left( \gamma _{\mu }\gamma _{5}\gamma _{A}\gamma _{\nu }\gamma
_{B}\gamma _{\alpha }\gamma _{C}\right)  \notag \\
&=&-4i\left[ \varepsilon _{\mu A\nu B}g_{\alpha C}-\varepsilon _{\mu A\nu
\alpha }g_{BC}+\varepsilon _{\mu A\nu C}g_{B\alpha }+\varepsilon _{\mu
AB\alpha }g_{\nu C}-\varepsilon _{\mu ABC}g_{\nu \alpha }\right.  \notag \\
&&+\varepsilon _{\mu A\alpha C}g_{\nu B}-\varepsilon _{\mu \nu B\alpha
}g_{AC}+\varepsilon _{\mu \nu BC}g_{A\alpha }-\varepsilon _{\mu \nu \alpha
C}g_{AB}+\varepsilon _{\mu B\alpha C}g_{A\nu }  \notag \\
&&\left. -\varepsilon _{A\nu B\alpha }g_{\mu C}+\varepsilon _{A\nu BC}g_{\mu
\alpha }-\varepsilon _{A\nu \alpha C}g_{\mu B}+\varepsilon _{AB\alpha
C}g_{\mu \nu }-\varepsilon _{\nu B\alpha C}g_{\mu A}\right] ,  \label{form2}
\end{eqnarray}%
while the other is%
\begin{eqnarray}
&&-\text{tr}\left( \gamma _{5}\gamma _{\mu }\gamma _{A}\gamma _{\nu }\gamma
_{B}\gamma _{\alpha }\gamma _{C}\right)  \notag \\
&=&-4i\left[ \varepsilon _{\mu A\nu B}g_{\alpha C}-\varepsilon _{\mu A\nu
\alpha }g_{BC}+\varepsilon _{\mu A\nu C}g_{B\alpha }+\varepsilon _{\mu
AB\alpha }g_{\nu C}-\varepsilon _{\mu ABC}g_{\nu \alpha }\right.  \notag \\
&&+\varepsilon _{\mu A\alpha C}g_{\nu B}-\varepsilon _{\mu \nu B\alpha
}g_{AC}+\varepsilon _{\mu \nu BC}g_{A\alpha }-\varepsilon _{\mu \nu \alpha
C}g_{AB}+\varepsilon _{\mu B\alpha C}g_{A\nu }  \notag \\
&&\left. +\varepsilon _{A\nu B\alpha }g_{\mu C}-\varepsilon _{A\nu BC}g_{\mu
\alpha }+\varepsilon _{A\nu \alpha C}g_{\mu B}-\varepsilon _{AB\alpha
C}g_{\mu \nu }+\varepsilon _{\nu B\alpha C}g_{\mu A}\right] .  \label{form1}
\end{eqnarray}

Although there are other strategies to compute them, one reason to choose
this path is that the results contain all contributions with non-equivalent
tensor configurations. This feature is convenient for the organization
developed throughout this section, which is part of IReg. Furthermore, the
reason for replacing the chiral matrix definition in these specific
positions (adjacent to $\gamma _{\mu }$) is to induce a simplification.

The layout of these (equivalent) expressions highlights that they only
differ by signs on the last row, characterizing one identity:%
\begin{equation}
g_{\mu C}\varepsilon _{A\nu B\alpha }-g_{\mu \alpha }\varepsilon _{A\nu
BC}+g_{\mu B}\varepsilon _{A\nu \alpha C}-g_{\mu \nu }\varepsilon _{AB\alpha
C}+g_{\mu A}\varepsilon _{\nu B\alpha C}=0.  \label{Schouten}
\end{equation}%
From another perspective, note that this tensor is antisymmetric in five
indices ($\mu $ fixed); therefore, identically zero for a four-dimensional
setting. Achieving this identity is not a coincidence but a direct
consequence of comparing positions adjacent to the $\mu $-index. Finding
similar identities where other free indices play this role is within reach.
That is only the first example seen here of the so-called Schouten
identities.

With this argumentation, we developed the know-how to find the same
resources in more complex expressions from four-point amplitudes. Although
that significantly reduces our efforts in these calculations, there is no
damage in ignoring these identities. We verified that these contributions
produce null integrals when evaluating perturbative amplitudes.

As a brief comment on this subject, suppose we achieve three trace
expressions corresponding to each vertex position represented by free
indices ($\mu $, $\nu $, and $\alpha $). They are equivalent since their
obtainment comes from pure algebraic manipulations. Nevertheless, due to
their divergent content, their connection might not apply after integrating
the amplitude. We attribute a central role to the $\mu $-index for now, but
Subsection (\ref{Lin}) extends this notion. The author, L. Ebani, and J. F.
Thuorst develop a broad investigation of the behavior of different versions
of odd-tensor correlators in reference \cite{Arxiv22}.

Returning to the $AVV$ triangle, replacing traces leads to its integrand%
\begin{eqnarray}
t_{\mu \nu \alpha }^{AVV} &=&4i\left( g_{\nu A}\varepsilon _{\mu \alpha
BC}-g_{\nu B}\varepsilon _{\mu \alpha CA}-g_{\nu C}\varepsilon _{\mu \alpha
AB}\right) \frac{K_{1}^{A}K_{2}^{B}K_{3}^{C}}{D_{123}}  \notag \\
&&+4i\left( -g_{\alpha A}\varepsilon _{\mu \nu BC}-g_{B\alpha }\varepsilon
_{\mu \nu CA}+g_{\alpha C}\varepsilon _{\mu \nu AB}\right) \frac{%
K_{1}^{A}K_{2}^{B}K_{3}^{C}}{D_{123}}  \notag \\
&&+4ig_{\nu \alpha }\varepsilon _{\mu ABC}\frac{K_{1}^{A}K_{2}^{B}K_{3}^{C}}{%
D_{123}}  \notag \\
&&+\varepsilon _{\mu \nu \alpha \beta }\left[ \text{tr}\left( \gamma _{\beta
}\gamma _{A}\gamma _{B}\gamma _{C}\right) \frac{K_{1}^{A}K_{2}^{B}K_{3}^{C}}{%
D_{123}}-m^{2}\text{tr}\left( \gamma _{\beta }\gamma _{A}\right) \frac{%
K_{1}^{A}}{D_{123}}\right.  \notag \\
&&\left. +m^{2}\text{tr}\left( \gamma _{\beta }\gamma _{B}\right) \frac{%
K_{2}^{B}}{D_{123}}-m^{2}\text{tr}\left( \gamma _{\beta }\gamma _{C}\right) 
\frac{K_{3}^{C}}{D_{123}}\right] .  \label{1}
\end{eqnarray}%
We already split sectors corresponding to different tensor configurations
and identified less complex traces. As terms with the free index $\mu $
within the metric compound the identity (\ref{Schouten}), we disregarded
them.

Following the reasoning established in example (\ref{exa}), trace content
suggests that the last term above consists of a vector subamplitude\footnote{%
The trace structure indicates this subamplitude has one Lorentz index, which
links to one axial or vector vertex. Other vertices might be scalar or
pseudoscalar combined to produce an even trace. That leads to amplitudes
corresponding to vectors: $VPP$, $VSS$, $APS$, and their permutations.}. If
one maintains the notations for summed indices, comparing signs is enough to
identify the $VPP$ among all possibilities. Meanwhile, the antisymmetric
character of the Levi-Civita symbol allows rewriting the remaining terms
through a new standard tensor characterized by three momenta on the numerator%
\begin{equation}
t_{3\mu ;\nu \alpha }^{\left( s\right) }\left( k_{l};k_{i},k_{j}\right) =%
\frac{\left( k+k_{l}\right) _{\mu }\left[ \left( k+k_{i}\right) _{\nu
}\left( k+k_{j}\right) _{\alpha }+s\left( k+k_{i}\right) _{\alpha }\left(
k+k_{j}\right) _{\nu }\right] }{D_{123}}.  \label{t33s}
\end{equation}%
Following previous notations, the superindex $s$ indicates a sign choice,
and the numerical subindex indicates the association with three propagators
through the denominator $D_{123}$. Hence, we achieve the final organization%
\begin{eqnarray}
t_{\mu \nu \alpha }^{AVV} &=&2i\varepsilon _{\mu \alpha XY}\left[ -t_{3\nu
;XY}^{\left( -\right) }\left( k_{3};k_{1},k_{2}\right) -t_{3\nu ;XY}^{\left(
-\right) }\left( k_{2};k_{3},k_{1}\right) +t_{3\nu ;XY}^{\left( -\right)
}\left( k_{1};k_{2},k_{3}\right) \right]  \notag \\
&&+2i\varepsilon _{\mu \nu XY}\left[ t_{3\alpha ;XY}^{\left( -\right)
}\left( k_{3};k_{1},k_{2}\right) -t_{3\alpha ;XY}^{\left( -\right) }\left(
k_{2};k_{3},k_{1}\right) -t_{3\alpha ;XY}^{\left( -\right) }\left(
k_{1};k_{2},k_{3}\right) \right]  \notag \\
&&+2g_{\nu \alpha }\varepsilon _{\mu XYZ}t_{3}^{\left( -\right) X;YZ}\left(
k_{1};k_{2},k_{3}\right) -i\varepsilon _{\mu \nu \alpha \beta }t_{\beta
}^{VPP}.  \label{avv1}
\end{eqnarray}

After replacing the corresponding vertices\footnote{%
There are three vertices: one vector $\gamma _{\beta }$ followed by two
pseudoscalars $\gamma _{5}$.} in the original integrand (\ref{t3}) and
taking traces, we study the vector subamplitude%
\begin{eqnarray}
t_{\beta }^{VPP} &=&-4\left( g_{\beta A}g_{BC}-g_{\beta B}g_{AC}+g_{\beta
C}g_{AB}\right) \frac{K_{1}^{A}K_{2}^{B}K_{3}^{C}}{D_{123}}  \notag \\
&&+4m^{2}\left[ \frac{K_{1\beta }}{D_{123}}-\frac{K_{2\beta }}{D_{123}}+%
\frac{K_{3\beta }}{D_{123}}\right] .
\end{eqnarray}%
Scalar products on the momenta emerge with the contraction, which leads to
reducing bilinears in analogy with scalar functions used as example (\ref%
{k.k}). Then, some manipulations produce the structure%
\begin{eqnarray}
t_{\beta }^{VPP} &=&-2p_{\beta }\frac{1}{D_{12}}-4\frac{k_{\beta }}{D_{13}}%
-2\left( k_{1}+k_{3}\right) _{\beta }\frac{1}{D_{13}}+2\left( q-p\right)
_{\beta }\frac{1}{D_{23}}  \notag \\
&&+2\left( q-p\right) ^{2}\frac{\left( k+k_{1}\right) _{\beta }}{D_{123}}%
-2q^{2}\frac{\left( k+k_{2}\right) _{\beta }}{D_{123}}+2p^{2}\frac{\left(
k+k_{3}\right) _{\beta }}{D_{123}}.  \label{vpp}
\end{eqnarray}

Lastly, we recall the $AV$ discussion to infer expectations regarding
integration. The objective was to compose a 2nd-order antisymmetric tensor
with available tools, namely, external and ambiguous momenta ($k_{i}-k_{j}$
and $k_{i}+k_{j}$). The only possibility was to employ them both, which
necessarily implies the presence of diverging surface terms. For this to be
consistent with the quadratic power counting, these surface terms must be
logarithmically divergent.

We find similar circumstances for any 3rd-order tensor exhibiting the
property of total antisymmetry. At least three different vectors are
necessary to compound it, which requires the presence of ambiguous momenta.
This structure brings diverging surface terms, which prevents obtaining the
correct mass dimension. As a consequence, 3rd-order antisymmetric tensors
are zero under these circumstances.

The most immediate event of this type is the (three-index) contraction
between the Levi-Civita symbol and the standard tensor. For it to be
non-zero, the tensor must have a total-antisymmetric component. As this
leads to the argumentation above, we expect its cancellation%
\begin{equation}
\varepsilon _{\nu XYZ}T_{3}^{\left( -\right) X;YZ}\left(
k_{1};k_{2},k_{3}\right) \rightarrow 0.  \label{exp3}
\end{equation}%
Furthermore, we combine all non-equivalent momenta configurations to produce
an identity involving this tensor%
\begin{equation}
T_{3\mu ;\nu \alpha }^{\left( -\right) }\left( k_{1};k_{2},k_{3}\right)
+T_{3\mu ;\nu \alpha }^{\left( -\right) }\left( k_{2};k_{3},k_{1}\right)
+T_{3\mu ;\nu \alpha }^{\left( -\right) }\left( k_{3};k_{1},k_{2}\right)
\rightarrow 0.  \label{exp4}
\end{equation}%
If these expectations realize, simplifications apply to the integrated
amplitude, yielding the expression:%
\begin{equation}
T_{\mu \nu \alpha }^{AVV}\rightarrow 4i\varepsilon _{\mu \alpha XY}T_{3\nu
;XY}^{\left( -\right) }\left( k_{1};k_{2},k_{3}\right) +4i\varepsilon _{\mu
\nu XY}T_{3\alpha ;XY}^{\left( -\right) }\left( k_{3};k_{1},k_{2}\right)
-i\varepsilon _{\mu \nu \alpha \beta }T_{\beta }^{VPP}.  \label{exp5}
\end{equation}%
We stress that the $\mu $-index appears exclusively within the Levi-Civita
symbol as a direct consequence of its prioritized role when taking the
traces; simplification only made this clear.

\subsection{Four-Point Amplitudes - $PVVV$}

We still have to look into four-point amplitudes, whose integrands assume
the form%
\begin{equation}
t^{\Gamma _{i}\Gamma _{j}\Gamma _{k}\Gamma _{l}}=\text{tr}\left( \Gamma _{i}%
\frac{1}{\slashed{F}_{1}}\Gamma _{j}\frac{1}{\slashed{F}_{2}}\Gamma _{k}%
\frac{1}{\slashed{F}_{3}}\Gamma _{l}\frac{1}{\slashed{F}_{4}}\right) .
\end{equation}%
After replacing fermion propagators (\ref{fermion}), linearity makes the
matrix content evident within Dirac traces:%
\begin{eqnarray}
&&t^{\Gamma _{i}\Gamma _{j}\Gamma _{k}\Gamma _{l}}  \notag \\
&=&\text{tr}\left( \Gamma _{i}\gamma _{A}\Gamma _{j}\gamma _{B}\Gamma
_{k}\gamma _{C}\Gamma _{l}\gamma _{D}\right) \frac{%
K_{1}^{A}K_{2}^{B}K_{3}^{C}K_{4}^{D}}{D_{1234}}+m^{4}\text{tr}\left( \Gamma
_{i}\Gamma _{j}\Gamma _{k}\Gamma _{l}\right) \frac{1}{D_{1234}}  \notag \\
&&+m^{2}\text{tr}\left( \Gamma _{i}\gamma _{A}\Gamma _{j}\gamma _{B}\Gamma
_{k}\Gamma _{l}\right) \frac{K_{1}^{A}K_{2}^{B}}{D_{1234}}+m^{2}\text{tr}%
\left( \Gamma _{i}\gamma _{A}\Gamma _{j}\Gamma _{k}\gamma _{C}\Gamma
_{l}\right) \frac{K_{1}^{A}K_{3}^{C}}{D_{1234}}  \notag \\
&&+m^{2}\text{tr}\left( \Gamma _{i}\gamma _{A}\Gamma _{j}\Gamma _{k}\Gamma
_{l}\gamma _{D}\right) \frac{K_{1}^{A}K_{4}^{D}}{D_{1234}}+m^{2}\text{tr}%
\left( \Gamma _{i}\Gamma _{j}\gamma _{B}\Gamma _{k}\gamma _{C}\Gamma
_{l}\right) \frac{K_{2}^{B}K_{3}^{C}}{D_{1234}}  \notag \\
&&+m^{2}\text{tr}\left( \Gamma _{i}\Gamma _{j}\gamma _{B}\Gamma _{k}\Gamma
_{l}\gamma _{D}\right) \frac{K_{2}^{B}K_{4}^{D}}{D_{1234}}+m^{2}\text{tr}%
\left( \Gamma _{i}\Gamma _{j}\Gamma _{k}\gamma _{C}\Gamma _{l}\gamma
_{D}\right) \frac{K_{3}^{C}K_{4}^{D}}{D_{1234}}  \notag \\
&&+m\text{tr}\left( \Gamma _{i}\Gamma _{j}\gamma _{B}\Gamma _{k}\gamma
_{C}\Gamma _{l}\gamma _{D}\right) \frac{K_{2}^{B}K_{3}^{C}K_{4}^{D}}{D_{1234}%
}+m^{3}\text{tr}\left( \Gamma _{i}\gamma _{A}\Gamma _{j}\Gamma _{k}\Gamma
_{l}\right) \frac{K_{1}^{A}}{D_{1234}}  \notag \\
&&+m\text{tr}\left( \Gamma _{i}\gamma _{A}\Gamma _{j}\Gamma _{k}\gamma
_{C}\Gamma _{l}\gamma _{D}\right) \frac{K_{1}^{A}K_{3}^{C}K_{4}^{D}}{D_{1234}%
}+m^{3}\text{tr}\left( \Gamma _{i}\Gamma _{j}\gamma _{B}\Gamma _{k}\Gamma
_{l}\right) \frac{K_{2}^{B}}{D_{1234}}  \notag \\
&&+m\text{tr}\left( \Gamma _{i}\gamma _{A}\Gamma _{j}\gamma _{B}\Gamma
_{k}\Gamma _{l}\gamma _{D}\right) \frac{K_{1}^{A}K_{2}^{B}K_{4}^{D}}{D_{1234}%
}+m^{3}\text{tr}\left( \Gamma _{i}\Gamma _{j}\Gamma _{k}\gamma _{C}\Gamma
_{l}\right) \frac{K_{3}^{C}}{D_{1234}}  \notag \\
&&+m\text{tr}\left( \Gamma _{i}\gamma _{A}\Gamma _{j}\gamma _{B}\Gamma
_{k}\gamma _{C}\Gamma _{l}\right) \frac{K_{1}^{A}K_{2}^{B}K_{3}^{C}}{D_{1234}%
}+m^{3}\text{tr}\left( \Gamma _{i}\Gamma _{j}\Gamma _{k}\Gamma _{l}\gamma
_{D}\right) \frac{K_{4}^{D}}{D_{1234}},  \label{t4}
\end{eqnarray}%
where numerators depend on $K_{i}=k+k_{i}$ and denominators are $%
D_{1234}=D_{1}D_{2}D_{3}D_{4}$.

Obtaining a specific function requires replacing the corresponding vertex
operators within this expression. For the case of $PVVV$ amplitude, we use
one pseudoscalar vertex ($\Gamma _{i}=\gamma _{5}$) followed by vector ones (%
$\Gamma _{j}=\gamma _{\nu }$, $\Gamma _{k}=\gamma _{\alpha }$, and $\Gamma
_{l}=\gamma _{\beta }$), achieving the non-zero contributions%
\begin{eqnarray}
t_{\nu \alpha \beta }^{PVVV} &=&m\text{tr}\left( \gamma _{5}\gamma _{\nu
}\gamma _{B}\gamma _{\alpha }\gamma _{C}\gamma _{\beta }\gamma _{D}\right) 
\frac{K_{2}^{B}K_{3}^{C}K_{4}^{D}}{D_{1234}}+m\text{tr}\left( \gamma
_{5}\gamma _{A}\gamma _{\nu }\gamma _{\alpha }\gamma _{C}\gamma _{\beta
}\gamma _{D}\right) \frac{K_{1}^{A}K_{3}^{C}K_{4}^{D}}{D_{1234}}  \notag \\
&&+m\text{tr}\left( \gamma _{5}\gamma _{A}\gamma _{\nu }\gamma _{B}\gamma
_{\alpha }\gamma _{\beta }\gamma _{D}\right) \frac{%
K_{1}^{A}K_{2}^{B}K_{4}^{D}}{D_{1234}}+m\text{tr}\left( \gamma _{5}\gamma
_{A}\gamma _{\nu }\gamma _{B}\gamma _{\alpha }\gamma _{C}\gamma _{\beta
}\right) \frac{K_{1}^{A}K_{2}^{B}K_{3}^{C}}{D_{1234}}  \notag \\
&&+m^{3}\text{tr}\left( \gamma _{5}\gamma _{A}\gamma _{\nu }\gamma _{\alpha
}\gamma _{\beta }\right) \frac{K_{1}^{A}}{D_{1234}}+m^{3}\text{tr}\left(
\gamma _{5}\gamma _{\nu }\gamma _{B}\gamma _{\alpha }\gamma _{\beta }\right) 
\frac{K_{2}^{B}}{D_{1234}}  \notag \\
&&+m^{3}\text{tr}\left( \gamma _{5}\gamma _{\nu }\gamma _{\alpha }\gamma
_{C}\gamma _{\beta }\right) \frac{K_{3}^{C}}{D_{1234}}+m^{3}\text{tr}\left(
\gamma _{5}\gamma _{\nu }\gamma _{\alpha }\gamma _{\beta }\gamma _{D}\right) 
\frac{K_{4}^{D}}{D_{1234}}.
\end{eqnarray}%
All traces are known and can be consulted in Equations (\ref{tr1}) and (\ref%
{form1}). Posteriorly to their employment, our task is to group terms that
share their index configuration to recognize subamplitudes or standard
tensors. We consider each of these sectors separately since their
mathematical expressions are more extensive now.

Finding those terms where the metric tensor has exclusively free indices, we
identify the first sector:%
\begin{eqnarray}
\left[ t_{\nu \alpha \beta }^{PVVV}\right] _{1} &=&-4im\left( g_{\nu \alpha
}\varepsilon _{\beta BCD}-g_{\nu \beta }\varepsilon _{\alpha BCD}+g_{\alpha
\beta }\varepsilon _{\nu BCD}\right) \frac{K_{2}^{B}K_{3}^{C}K_{4}^{D}}{%
D_{1234}}  \notag \\
&&+4im\left( g_{\nu \alpha }\varepsilon _{\beta ACD}-g_{\nu \beta
}\varepsilon _{\alpha ACD}+g_{\alpha \beta }\varepsilon _{\nu ACD}\right) 
\frac{K_{1}^{A}K_{3}^{C}K_{4}^{D}}{D_{1234}}  \notag \\
&&-4im\left( g_{\nu \alpha }\varepsilon _{\beta ABD}-g_{\nu \beta
}\varepsilon _{\alpha ABD}+g_{\alpha \beta }\varepsilon _{\nu ABD}\right) 
\frac{K_{1}^{A}K_{2}^{B}K_{4}^{D}}{D_{1234}}  \notag \\
&&+4im\left( g_{\nu \alpha }\varepsilon _{\beta ABC}-g_{\nu \beta
}\varepsilon _{\alpha ABC}+g_{\alpha \beta }\varepsilon _{\nu ABC}\right) 
\frac{K_{1}^{A}K_{2}^{B}K_{3}^{C}}{D_{1234}}.
\end{eqnarray}%
By following the same procedure from previous cases, axial vector amplitudes
would be achievable. Nevertheless, since quantities in parenthesis are
alike, we rename summed indices to compact them into a single object%
\begin{equation}
\left[ t_{\nu \alpha \beta }^{PVVV}\right] _{1}=-4im\left( g_{\kappa \nu
}g_{\alpha \beta }-g_{\kappa \alpha }g_{\nu \beta }+g_{\kappa \beta }g_{\nu
\alpha }\right) f_{4\kappa }.
\end{equation}%
The introduced object has the following structure%
\begin{eqnarray}
f_{4\kappa } &=&\varepsilon _{\kappa XYZ}t_{4X;YZ}^{\left( -\right) }\left(
k_{2};k_{3},k_{4}\right) -\varepsilon _{\kappa XYZ}t_{4X;YZ}^{\left(
-\right) }\left( k_{1};k_{3},k_{4}\right)  \notag \\
&&+\varepsilon _{\kappa XYZ}t_{4X;YZ}^{\left( -\right) }\left(
k_{1};k_{2},k_{4}\right) -\varepsilon _{\kappa XYZ}t_{4X;YZ}^{\left(
-\right) }\left( k_{1};k_{2},k_{3}\right) ,  \label{f4}
\end{eqnarray}%
which depends on the new standard tensor%
\begin{equation}
t_{4\mu ;\nu \alpha }^{\left( s\right) }\left( k_{l};k_{i},k_{j}\right) =%
\frac{\left( k+k_{l}\right) _{\mu }\left[ \left( k+k_{i}\right) _{\nu
}\left( k+k_{j}\right) _{\alpha }+s\left( k+k_{i}\right) _{\alpha }\left(
k+k_{j}\right) _{\nu }\right] }{D_{1234}}.  \label{t34s}
\end{equation}%
Although this object is analogous to that defined in Equation (\ref{t33s}),
the numerical subindex indicates the association with four propagators
through the denominator $D_{1234}$.

For the second sector, let us group components where all free indices appear
within the Levi-Civita symbol, including traces of four Dirac matrices. We
introduce a summed index $\kappa $ to isolate a global factor and recognize
less complex traces%
\begin{eqnarray}
\left[ t_{\nu \alpha \beta }^{PVVV}\right] _{2} &=&i\varepsilon _{\nu \alpha
\beta \kappa }\left[ -m\text{tr}\left( \gamma _{B}\gamma _{C}\gamma
_{D}\gamma _{\kappa }\right) \frac{K_{2}^{B}K_{3}^{C}K_{4}^{D}}{D_{1234}}+m%
\text{tr}\left( \gamma _{A}\gamma _{C}\gamma _{D}\gamma _{\kappa }\right) 
\frac{K_{1}^{A}K_{3}^{C}K_{4}^{D}}{D_{1234}}\right.  \notag \\
&&-m\text{tr}\left( \gamma _{A}\gamma _{B}\gamma _{D}\gamma _{\kappa
}\right) \frac{K_{1}^{A}K_{2}^{B}K_{4}^{D}}{D_{1234}}+m\text{tr}\left(
\gamma _{A}\gamma _{B}\gamma _{C}\gamma _{\kappa }\right) \frac{%
K_{1}^{A}K_{2}^{B}K_{3}^{C}}{D_{1234}}  \notag \\
&&-m^{3}\text{tr}\left( \gamma _{A}\gamma _{\kappa }\right) \frac{K_{1}^{A}}{%
D_{1234}}+m^{3}\text{tr}\left( \gamma _{B}\gamma _{\kappa }\right) \frac{%
K_{2}^{B}}{D_{1234}}  \notag \\
&&\left. -m^{3}\text{tr}\left( \gamma _{C}\gamma _{\kappa }\right) \frac{%
K_{3}^{C}}{D_{1234}}+m^{3}\text{tr}\left( \gamma _{D}\gamma _{\kappa
}\right) \frac{K_{4}^{D}}{D_{1234}}\right] .
\end{eqnarray}%
This structure associates with a vector subamplitude; thus, comparing signs
among the possibilities leads to the $APPP$ function\footnote{%
There are four vertices: one vector $\gamma _{\kappa }$ followed by three
pseudoscalars $\gamma _{5}$.}:%
\begin{equation}
\left[ t_{\nu \alpha \beta }^{PVVV}\right] _{2}=-i\varepsilon _{\kappa \nu
\alpha \beta }t_{\kappa }^{APPP}.
\end{equation}

As bilinears arise from traces within this subamplitude, we reduce them
through identity (\ref{k.k}). The loop momentum from numerators cancels out
with this operation. Hence, the integrand associated with this function has
the final structure%
\begin{eqnarray}
t_{\kappa }^{APPP} &=&4mp_{\kappa }\frac{1}{D_{124}}+4m\left( r-q\right)
_{\kappa }\frac{1}{D_{134}}  \notag \\
&&-4m\left[ \left( q^{2}-q\cdot r\right) p_{\kappa }-\left( p^{2}-p\cdot
r\right) q_{\kappa }+\left( p^{2}-p\cdot q\right) r_{\kappa }\right] \frac{1%
}{D_{1234}}.  \label{appp}
\end{eqnarray}%
All external momenta arose within this expression: $p=k_{1}-k_{2}$, $%
q=k_{1}-k_{3}$, and $r=k_{1}-k_{4}$.

Lastly, consider those terms that mix free and summed indices%
\begin{eqnarray}
\left[ t_{\nu \alpha \beta }^{PVVV}\right] _{3} &=&4im\left( g_{\nu
B}\varepsilon _{\alpha C\beta D}+g_{\nu C}\varepsilon _{B\alpha \beta
D}+g_{\nu D}\varepsilon _{B\alpha C\beta }+g_{B\alpha }\varepsilon _{\nu
C\beta D}+g_{B\beta }\varepsilon _{\nu \alpha CD}\right.  \notag \\
&&\left. +g_{\alpha C}\varepsilon _{\nu B\beta D}+g_{\alpha D}\varepsilon
_{\nu BC\beta }+g_{C\beta }\varepsilon _{\nu B\alpha D}+g_{\beta
D}\varepsilon _{\nu B\alpha C}\right) \frac{K_{2}^{B}K_{3}^{C}K_{4}^{D}}{%
D_{1234}}  \notag \\
&&+4im\left( g_{A\nu }\varepsilon _{\alpha C\beta D}-g_{A\alpha }\varepsilon
_{\nu C\beta D}-g_{A\beta }\varepsilon _{\nu \alpha CD}-g_{\nu C}\varepsilon
_{A\alpha \beta D}-g_{\nu D}\varepsilon _{A\alpha C\beta }\right.  \notag \\
&&\left. +g_{\alpha C}\varepsilon _{A\nu \beta D}+g_{\alpha D}\varepsilon
_{A\nu C\beta }+g_{C\beta }\varepsilon _{A\nu \alpha D}+g_{\beta
D}\varepsilon _{A\nu \alpha C}\right) \frac{K_{1}^{A}K_{3}^{C}K_{4}^{D}}{%
D_{1234}}  \notag \\
&&+4im\left( g_{A\nu }\varepsilon _{B\alpha \beta D}+g_{A\alpha }\varepsilon
_{\nu B\beta D}-g_{A\beta }\varepsilon _{\nu B\alpha D}+g_{\nu B}\varepsilon
_{A\alpha \beta D}-g_{\nu D}\varepsilon _{AB\alpha \beta }\right.  \notag \\
&&\left. +g_{B\alpha }\varepsilon _{A\nu \beta D}-g_{B\beta }\varepsilon
_{A\nu \alpha D}-g_{\alpha D}\varepsilon _{A\nu B\beta }+g_{\beta
D}\varepsilon _{A\nu B\alpha }\right) \frac{K_{1}^{A}K_{2}^{B}K_{4}^{D}}{%
D_{1234}}  \notag \\
&&+4im\left( g_{A\nu }\varepsilon _{B\alpha C\beta }+g_{A\alpha }\varepsilon
_{\nu BC\beta }+g_{A\beta }\varepsilon _{\nu B\alpha C}+g_{\nu B}\varepsilon
_{A\alpha C\beta }+g_{\nu C}\varepsilon _{AB\alpha \beta }\right.  \notag \\
&&\left. +g_{B\alpha }\varepsilon _{A\nu C\beta }+g_{B\beta }\varepsilon
_{A\nu \alpha C}+g_{\alpha C}\varepsilon _{A\nu B\beta }+g_{C\beta
}\varepsilon _{A\nu B\alpha }\right) \frac{K_{1}^{A}K_{2}^{B}K_{3}^{C}}{%
D_{1234}}.
\end{eqnarray}%
Once again, using the antisymmetric character of the Levi-Civita symbol, we
recognize combinations of the standard tensor (\ref{t34s}).Then, this sector
leads to the following tensor by factorizing $2im$:%
\begin{eqnarray}
f_{4\nu \alpha \beta } &=&-\left( \varepsilon _{\alpha \beta XY}g_{\nu
Z}-\varepsilon _{\nu \beta XY}g_{\alpha Z}+\varepsilon _{\nu \alpha
XY}g_{\beta Z}\right) t_{4Z;XY}^{\left( -\right) }\left(
k_{1};k_{3},k_{4}\right)  \notag \\
&&+\left( \varepsilon _{\alpha \beta XY}g_{\nu Z}-\varepsilon _{\nu \beta
XY}g_{\alpha Z}+\varepsilon _{\nu \alpha XY}g_{\beta Z}\right)
t_{4Z;XY}^{\left( -\right) }\left( k_{1};k_{2},k_{4}\right)  \notag \\
&&-\left( \varepsilon _{\alpha \beta XY}g_{\nu Z}-\varepsilon _{\nu \beta
XY}g_{\alpha Z}+\varepsilon _{\nu \alpha XY}g_{\beta Z}\right)
t_{4Z;XY}^{\left( -\right) }\left( k_{1};k_{2},k_{3}\right)  \notag \\
&&-\left( \varepsilon _{\alpha \beta XY}g_{\nu Z}+\varepsilon _{\nu \beta
XY}g_{\alpha Z}-\varepsilon _{\nu \alpha XY}g_{\beta Z}\right)
t_{4Z;XY}^{\left( -\right) }\left( k_{2};k_{3},k_{4}\right)  \notag \\
&&+\left( \varepsilon _{\alpha \beta XY}g_{\nu Z}+\varepsilon _{\nu \beta
XY}g_{\alpha Z}-\varepsilon _{\nu \alpha XY}g_{\beta Z}\right)
t_{4Z;XY}^{\left( -\right) }\left( k_{2};k_{1},k_{4}\right)  \notag \\
&&-\left( \varepsilon _{\alpha \beta XY}g_{\nu Z}+\varepsilon _{\nu \beta
XY}g_{\alpha Z}-\varepsilon _{\nu \alpha XY}g_{\beta Z}\right)
t_{4Z;XY}^{\left( -\right) }\left( k_{2};k_{1},k_{3}\right)  \notag \\
&&+\left( \varepsilon _{\alpha \beta XY}g_{\nu Z}-\varepsilon _{\nu \beta
XY}g_{\alpha Z}-\varepsilon _{\nu \alpha XY}g_{\beta Z}\right)
t_{4Z;XY}^{\left( -\right) }\left( k_{3};k_{2},k_{4}\right)  \notag \\
&&-\left( \varepsilon _{\alpha \beta XY}g_{\nu Z}-\varepsilon _{\nu \beta
XY}g_{\alpha Z}-\varepsilon _{\nu \alpha XY}g_{\beta Z}\right)
t_{4Z;XY}^{\left( -\right) }\left( k_{3};k_{1},k_{4}\right)  \notag \\
&&+\left( \varepsilon _{\alpha \beta XY}g_{\nu Z}-\varepsilon _{\nu \beta
XY}g_{\alpha Z}-\varepsilon _{\nu \alpha XY}g_{\beta Z}\right)
t_{4Z;XY}^{\left( -\right) }\left( k_{3};k_{1},k_{2}\right)  \notag \\
&&-\left( \varepsilon _{\alpha \beta XY}g_{\nu Z}-\varepsilon _{\nu \beta
XY}g_{\alpha Z}+\varepsilon _{\nu \alpha XY}g_{\beta Z}\right)
t_{4Z;XY}^{\left( -\right) }\left( k_{4};k_{2},k_{3}\right)  \notag \\
&&+\left( \varepsilon _{\alpha \beta XY}g_{\nu Z}-\varepsilon _{\nu \beta
XY}g_{\alpha Z}+\varepsilon _{\nu \alpha XY}g_{\beta Z}\right)
t_{4Z;XY}^{\left( -\right) }\left( k_{4};k_{1},k_{3}\right)  \notag \\
&&-\left( \varepsilon _{\alpha \beta XY}g_{\nu Z}-\varepsilon _{\nu \beta
XY}g_{\alpha Z}+\varepsilon _{\nu \alpha XY}g_{\beta Z}\right)
t_{4Z;XY}^{\left( -\right) }\left( k_{4};k_{1},k_{2}\right) .  \label{f4vab}
\end{eqnarray}%
The most significant difference between both occurrences of this tensor is
in the contraction. Whereas all indices were contracted with the Levi-Civita
symbol in the previous case, only those that show the antisymmetry property
are contracted this time.

With all sectors explored, we write the $PVVV$ final form%
\begin{equation}
t_{\nu \alpha \beta }^{PVVV}=-4im\left( g_{\kappa \nu }g_{\alpha \beta
}-g_{\kappa \alpha }g_{\nu \beta }+g_{\kappa \beta }g_{\nu \alpha }\right)
f_{4\kappa }+2imf_{4\nu \alpha \beta }-i\varepsilon _{\kappa \nu \alpha
\beta }t_{\kappa }^{APPP}.  \label{pvvv}
\end{equation}%
It depends on two main structures: the vector subamplitude $APPP$ and the
standard tensor with three momenta on the numerator. Even though four-point
functions have logarithmic power counting, mass dimension analysis suggests
that integrals within this particular amplitude are finite.

\subsection{Four-Point Amplitudes - $AVVV$}

\label{box1}The last correlator concerning this investigation is the $AVVV$
box, whose structure contains one axial vertex ($\Gamma _{i}=\gamma _{\mu
}\gamma _{5}$) and three vector vertices ($\Gamma _{j}=\gamma _{\nu }$, $%
\Gamma _{k}=\gamma _{\alpha }$, and $\Gamma _{l}=\gamma _{\beta }$). We
obtain its initial structure by replacing the corresponding vertices on the
general integrand of four-point functions (\ref{t4}):%
\begin{eqnarray}
t_{\mu \nu \alpha \beta }^{AVVV} &=&\text{tr}\left( \gamma _{\mu }\gamma
_{5}\gamma _{A}\gamma _{\nu }\gamma _{B}\gamma _{\alpha }\gamma _{C}\gamma
_{\beta }\gamma _{D}\right) \frac{K_{1}^{A}K_{2}^{B}K_{3}^{C}K_{4}^{D}}{%
D_{1234}}  \notag \\
&&+m^{2}\text{tr}\left( \gamma _{\mu }\gamma _{5}\gamma _{A}\gamma _{\nu
}\gamma _{B}\gamma _{\alpha }\gamma _{\beta }\right) \frac{K_{1}^{A}K_{2}^{B}%
}{D_{1234}}+m^{2}\text{tr}\left( \gamma _{\mu }\gamma _{5}\gamma _{A}\gamma
_{\nu }\gamma _{\alpha }\gamma _{C}\gamma _{\beta }\right) \frac{%
K_{1}^{A}K_{3}^{C}}{D_{1234}}  \notag \\
&&+m^{2}\text{tr}\left( \gamma _{\mu }\gamma _{5}\gamma _{A}\gamma _{\nu
}\gamma _{\alpha }\gamma _{\beta }\gamma _{D}\right) \frac{K_{1}^{A}K_{4}^{D}%
}{D_{1234}}+m^{2}\text{tr}\left( \gamma _{\mu }\gamma _{5}\gamma _{\nu
}\gamma _{B}\gamma _{\alpha }\gamma _{C}\gamma _{\beta }\right) \frac{%
K_{2}^{B}K_{3}^{C}}{D_{1234}}  \notag \\
&&+m^{2}\text{tr}\left( \gamma _{\mu }\gamma _{5}\gamma _{\nu }\gamma
_{B}\gamma _{\alpha }\gamma _{\beta }\gamma _{D}\right) \frac{%
K_{2}^{B}K_{4}^{D}}{D_{1234}}+m^{2}\text{tr}\left( \gamma _{\mu }\gamma
_{5}\gamma _{\nu }\gamma _{\alpha }\gamma _{C}\gamma _{\beta }\gamma
_{D}\right) \frac{K_{3}^{C}K_{4}^{D}}{D_{1234}}  \notag \\
&&+m^{4}\text{tr}\left( \gamma _{\mu }\gamma _{5}\gamma _{\nu }\gamma
_{\alpha }\gamma _{\beta }\right) \frac{1}{D_{1234}}.  \label{avvv1}
\end{eqnarray}

This subsection deals with numerous contributions that might compromise the
visualization and understanding of mathematical expressions. For this
reason, we introduce a compact notation for products of gamma matrices,
e.g., $\gamma _{\mu }\gamma _{5}\gamma _{\nu }\gamma _{\alpha }\gamma
_{\beta }=\gamma _{\mu 5\nu \alpha \beta }$. That is a temporary resource
employed exclusively in $AVVV$ calculations.

Most traces above are known and can be consulted in Equations (\ref{tr1})
and (\ref{form1}). We also identify the presence of a trace involving eight
Dirac matrices plus the chiral one, which leads to products involving the
Levi-Civita symbol and metric tensors. This type of structure admits
equivalent expressions distinguished in their tensor structure.
Nevertheless, this connection is not guaranteed for perturbative amplitudes
due to their divergent character. That is analogous to the $AVV$ case and
motivated us to choose the $AVVV$ as an extension of our discussion.

Evaluating this trace follows the same procedure adopted in previous cases:
replace the chiral matrix definition, take the new trace, and perform
contractions with the Levi-Civita symbol. This strategy leads to a result
exhibiting all non-equivalent tensor contributions, which makes the
existence of identities clear. Even so, to allow a careful analysis, we
chose to approach this subject after the complete organization of the
amplitude. Thus, let us directly introduce the trace expression prioritizing
the $\mu $-index:%
\begin{eqnarray}
&&i\text{tr}\left( \gamma _{5\mu A\nu B\alpha C\beta D}\right)  \notag \\
&=&-\varepsilon _{\mu A\nu B}\text{tr}\left( \gamma _{\alpha C\beta
D}\right) +\varepsilon _{\mu A\nu \alpha }\text{tr}\left( \gamma _{BC\beta
D}\right) -\varepsilon _{\mu A\nu C}\text{tr}\left( \gamma _{B\alpha \beta
D}\right) +\varepsilon _{\mu A\nu \beta }\text{tr}\left( \gamma _{B\alpha
CD}\right)  \notag \\
&&-\varepsilon _{\mu A\nu D}\text{tr}\left( \gamma _{B\alpha C\beta }\right)
-\varepsilon _{\mu AB\alpha }\text{tr}\left( \gamma _{\nu C\beta D}\right)
+\varepsilon _{\mu ABC}\text{tr}\left( \gamma _{\nu \alpha \beta D}\right)
-\varepsilon _{\mu AB\beta }\text{tr}\left( \gamma _{\nu \alpha CD}\right) 
\notag \\
&&+\varepsilon _{\mu ABD}\text{tr}\left( \gamma _{\nu \alpha C\beta }\right)
-\varepsilon _{\mu A\alpha C}\text{tr}\left( \gamma _{\nu B\beta D}\right)
+\varepsilon _{\mu A\alpha \beta }\text{tr}\left( \gamma _{\nu BCD}\right)
-\varepsilon _{\mu A\alpha D}\text{tr}\left( \gamma _{\nu BC\beta }\right) 
\notag \\
&&-\varepsilon _{\mu AC\beta }\text{tr}\left( \gamma _{\nu B\alpha D}\right)
+\varepsilon _{\mu ACD}\text{tr}\left( \gamma _{\nu B\alpha \beta }\right)
-\varepsilon _{\mu A\beta D}\text{tr}\left( \gamma _{\nu B\alpha C}\right)
+\varepsilon _{\mu \nu B\alpha }\text{tr}\left( \gamma _{AC\beta D}\right) 
\notag \\
&&-\varepsilon _{\mu \nu BC}\text{tr}\left( \gamma _{A\alpha \beta D}\right)
+\varepsilon _{\mu \nu B\beta }\text{tr}\left( \gamma _{A\alpha CD}\right)
-\varepsilon _{\mu \nu BD}\text{tr}\left( \gamma _{A\alpha C\beta }\right)
+\varepsilon _{\mu \nu \alpha C}\text{tr}\left( \gamma _{AB\beta D}\right) 
\notag \\
&&-\varepsilon _{\mu \nu \alpha \beta }\text{tr}\left( \gamma _{ABCD}\right)
+\varepsilon _{\mu \nu \alpha D}\text{tr}\left( \gamma _{ABC\beta }\right)
+\varepsilon _{\mu \nu C\beta }\text{tr}\left( \gamma _{AB\alpha D}\right)
-\varepsilon _{\mu \nu CD}\text{tr}\left( \gamma _{AB\alpha \beta }\right) 
\notag \\
&&+\varepsilon _{\mu \nu \beta D}\text{tr}\left( \gamma _{AB\alpha C}\right)
-\varepsilon _{\mu B\alpha C}\text{tr}\left( \gamma _{A\nu \beta D}\right)
+\varepsilon _{\mu B\alpha \beta }\text{tr}\left( \gamma _{A\nu CD}\right)
-\varepsilon _{\mu B\alpha D}\text{tr}\left( \gamma _{A\nu C\beta }\right) 
\notag \\
&&-\varepsilon _{\mu BC\beta }\text{tr}\left( \gamma _{A\nu \alpha D}\right)
+\varepsilon _{\mu BCD}\text{tr}\left( \gamma _{A\nu \alpha \beta }\right)
-\varepsilon _{\mu B\beta D}\text{tr}\left( \gamma _{A\nu \alpha C}\right)
+\varepsilon _{\mu \alpha C\beta }\text{tr}\left( \gamma _{A\nu BD}\right) 
\notag \\
&&-\varepsilon _{\mu \alpha CD}\text{tr}\left( \gamma _{A\nu B\beta }\right)
+\varepsilon _{\mu \alpha \beta D}\text{tr}\left( \gamma _{A\nu BC}\right)
-\varepsilon _{\mu C\beta D}\text{tr}\left( \gamma _{A\nu B\alpha }\right)
-\varepsilon _{A\nu B\alpha }\text{tr}\left( \gamma _{\mu C\beta D}\right) 
\notag \\
&&+\varepsilon _{A\nu BC}\text{tr}\left( \gamma _{\mu \alpha \beta D}\right)
-\varepsilon _{A\nu B\beta }\text{tr}\left( \gamma _{\mu \alpha CD}\right)
+\varepsilon _{A\nu BD}\text{tr}\left( \gamma _{\mu \alpha C\beta }\right)
-\varepsilon _{A\nu \alpha C}\text{tr}\left( \gamma _{\mu B\beta D}\right) 
\notag \\
&&+\varepsilon _{A\nu \alpha \beta }\text{tr}\left( \gamma _{\mu BCD}\right)
-\varepsilon _{A\nu \alpha D}\text{tr}\left( \gamma _{\mu BC\beta }\right)
-\varepsilon _{A\nu C\beta }\text{tr}\left( \gamma _{\mu B\alpha D}\right)
+\varepsilon _{A\nu CD}\text{tr}\left( \gamma _{\mu B\alpha \beta }\right) 
\notag \\
&&-\varepsilon _{A\nu \beta D}\text{tr}\left( \gamma _{\mu B\alpha C}\right)
+\varepsilon _{AB\alpha C}\text{tr}\left( \gamma _{\mu \nu \beta D}\right)
-\varepsilon _{AB\alpha \beta }\text{tr}\left( \gamma _{\mu \nu CD}\right)
+\varepsilon _{AB\alpha D}\text{tr}\left( \gamma _{\mu \nu C\beta }\right) 
\notag \\
&&+\varepsilon _{ABC\beta }\text{tr}\left( \gamma _{\mu \nu \alpha D}\right)
-\varepsilon _{ABCD}\text{tr}\left( \gamma _{\mu \nu \alpha \beta }\right)
+\varepsilon _{AB\beta D}\text{tr}\left( \gamma _{\mu \nu \alpha C}\right)
-\varepsilon _{A\alpha C\beta }\text{tr}\left( \gamma _{\mu \nu BD}\right) 
\notag \\
&&+\varepsilon _{A\alpha CD}\text{tr}\left( \gamma _{\mu \nu B\beta }\right)
-\varepsilon _{A\alpha \beta D}\text{tr}\left( \gamma _{\mu \nu BC}\right)
+\varepsilon _{AC\beta D}\text{tr}\left( \gamma _{\mu \nu B\alpha }\right)
-\varepsilon _{\nu B\alpha C}\text{tr}\left( \gamma _{\mu A\beta D}\right) 
\notag \\
&&+\varepsilon _{\nu B\alpha \beta }\text{tr}\left( \gamma _{\mu ACD}\right)
-\varepsilon _{\nu B\alpha D}\text{tr}\left( \gamma _{\mu AC\beta }\right)
-\varepsilon _{\nu BC\beta }\text{tr}\left( \gamma _{\mu A\alpha D}\right)
+\varepsilon _{\nu BCD}\text{tr}\left( \gamma _{\mu A\alpha \beta }\right) 
\notag \\
&&-\varepsilon _{\nu B\beta D}\text{tr}\left( \gamma _{\mu A\alpha C}\right)
+\varepsilon _{\nu \alpha C\beta }\text{tr}\left( \gamma _{\mu ABD}\right)
-\varepsilon _{\nu \alpha CD}\text{tr}\left( \gamma _{\mu AB\beta }\right)
+\varepsilon _{\nu \alpha \beta D}\text{tr}\left( \gamma _{\mu ABC}\right) 
\notag \\
&&-\varepsilon _{\nu D\beta D}\text{tr}\left( \gamma _{\mu AB\alpha }\right)
-\varepsilon _{B\alpha C\beta }\text{tr}\left( \gamma _{\mu A\nu D}\right)
+\varepsilon _{B\alpha CD}\text{tr}\left( \gamma _{\mu A\nu \beta }\right)
-\varepsilon _{B\alpha \beta D}\text{tr}\left( \gamma _{\mu A\nu C}\right) 
\notag \\
&&+\varepsilon _{BC\beta D}\text{tr}\left( \gamma _{\mu A\nu \alpha }\right)
-\varepsilon _{\alpha C\beta D}\text{tr}\left( \gamma _{\mu A\nu B}\right) .
\label{tr3}
\end{eqnarray}

Since numerous components exist, we split this analysis\footnote{%
Although all vertex operators appear within this context, we only comment on
cases that remain in the final form.} into sectors grouping terms where free
indices play similar roles. This line of reasoning extends to all parts of
the initial integrand (\ref{avvv1}). Thus, we will call upon the term
proportional to $K_{1}^{A}K_{2}^{B}$ to illustrate a trace involving six
Dirac matrices%
\begin{eqnarray}
\text{tr}\left( \gamma _{\mu 5A\nu B\alpha \beta }\right) &=&-4i\left[
g_{\mu A}\varepsilon _{\nu B\alpha \beta }-g_{\mu \nu }\varepsilon
_{AB\alpha \beta }+g_{\mu B}\varepsilon _{A\nu \alpha \beta }-g_{\mu \alpha
}\varepsilon _{A\nu B\beta }+g_{\mu \beta }\varepsilon _{A\nu B\alpha
}\right.  \notag \\
&&+g_{A\nu }\varepsilon _{\mu B\alpha \beta }-g_{AB}\varepsilon _{\mu \nu
\alpha \beta }+g_{A\alpha }\varepsilon _{\mu \nu B\beta }-g_{A\beta
}\varepsilon _{\mu \nu B\alpha }+g_{\nu B}\varepsilon _{\mu A\alpha \beta } 
\notag \\
&&\left. -g_{\nu \alpha }\varepsilon _{\mu AB\beta }+g_{\nu \beta
}\varepsilon _{\mu AB\alpha }+g_{B\alpha }\varepsilon _{\mu A\nu \beta
}-g_{B\beta }\varepsilon _{\mu A\nu \alpha }+g_{\alpha \beta }\varepsilon
_{\mu A\nu B}\right] .  \label{tr3m}
\end{eqnarray}

Our first step is to find those terms depending on the metric tensor with
free indices. The artifice of using uppercase Latin letters on summed
indices makes this process a lot easier. For the equation above, the
following components interest us%
\begin{equation*}
\text{tr}\left( \gamma _{\mu 5A\nu B\alpha \beta }\right) \rightarrow g_{\mu
\nu }\varepsilon _{AB\alpha \beta }+g_{\mu \alpha }\varepsilon _{A\nu B\beta
}-g_{\mu \beta }\varepsilon _{A\nu B\alpha }+g_{\nu \alpha }\varepsilon
_{\mu AB\beta }-g_{\nu \beta }\varepsilon _{\mu AB\alpha }-g_{\alpha \beta
}\varepsilon _{\mu A\nu B},
\end{equation*}%
where the Levi-Civita symbols are recognized as less complex traces.
Extending this idea to the complete amplitude, we have all contributions
belonging to this sector:%
\begin{eqnarray}
\left[ t_{\mu \nu \alpha \beta }^{AVVV}\right] _{1} &=&-\left[ g_{\alpha
\beta }\text{tr}\left( \gamma _{\mu 5A\nu BCD}\right) +g_{\nu \beta }\text{tr%
}\left( \gamma _{\mu 5AB\alpha CD}\right) +g_{\nu \alpha }\text{tr}\left(
\gamma _{\mu 5ABC\beta D}\right) \right.  \notag \\
&&+g_{\mu \beta }\text{tr}\left( \gamma _{A\nu B\alpha C5D}\right) +g_{\mu
\alpha }\text{tr}\left( \gamma _{A\nu B5C\beta D}\right) +g_{\mu \nu }\text{%
tr}\left( \gamma _{A5B\alpha C\beta D}\right)  \notag \\
&&\left. +\left( g_{\mu \nu }g_{\alpha \beta }-g_{\mu \alpha }g_{\nu \beta
}+g_{\mu \beta }g_{\nu \alpha }\right) \text{tr}\left( \gamma
_{A5BCD}\right) \right] \frac{K_{1}^{A}K_{2}^{B}K_{3}^{C}K_{4}^{D}}{D_{1234}}
\notag \\
&&+m^{2}\left[ -g_{\alpha \beta }\text{tr}\left( \gamma _{5\mu A\nu
B}\right) -g_{\nu \beta }\text{tr}\left( \gamma _{5\mu AB\alpha }\right)
+g_{\nu \alpha }\text{tr}\left( \gamma _{5\mu AB\beta }\right) \right. 
\notag \\
&&\left. -g_{\mu \beta }\text{tr}\left( \gamma _{5A\nu B\alpha }\right)
+g_{\mu \alpha }\text{tr}\left( \gamma _{5A\nu B\beta }\right) +g_{\mu \nu }%
\text{tr}\left( \gamma _{5AB\alpha \beta }\right) \right] \frac{%
K_{1}^{A}K_{2}^{B}}{D_{1234}}  \notag \\
&&+m^{2}\left[ g_{\alpha \beta }\text{tr}\left( \gamma _{5\mu A\nu C}\right)
-g_{\nu \beta }\text{tr}\left( \gamma _{5\mu A\alpha C}\right) -g_{\nu
\alpha }\text{tr}\left( \gamma _{5\mu AC\beta }\right) \right.  \notag \\
&&\left. -g_{\mu \beta }\text{tr}\left( \gamma _{5A\nu \alpha C}\right)
-g_{\mu \alpha }\text{tr}\left( \gamma _{5A\nu C\beta }\right) +g_{\mu \nu }%
\text{tr}\left( \gamma _{5A\alpha C\beta }\right) \right] \frac{%
K_{1}^{A}K_{3}^{C}}{D_{1234}}  \notag \\
&&+m^{2}\left[ -g_{\alpha \beta }\text{tr}\left( \gamma _{5\mu A\nu
D}\right) +g_{\nu \beta }\text{tr}\left( \gamma _{5\mu A\alpha D}\right)
-g_{\nu \alpha }\text{tr}\left( \gamma _{5\mu A\beta D}\right) \right. 
\notag \\
&&\left. g_{\mu \beta }\text{tr}\left( \gamma _{5A\nu \alpha D}\right)
-g_{\mu \alpha }\text{tr}\left( \gamma _{5A\nu \beta D}\right) +g_{\mu \nu }%
\text{tr}\left( \gamma _{5A\alpha \beta D}\right) \right] \frac{%
K_{1}^{A}K_{4}^{D}}{D_{1234}}  \notag \\
&&+m^{2}\left[ g_{\alpha \beta }\text{tr}\left( \gamma _{5\mu \nu BC}\right)
+g_{\nu \beta }\text{tr}\left( \gamma _{5\mu B\alpha C}\right) +g_{\nu
\alpha }\text{tr}\left( \gamma _{5\mu BC\beta }\right) \right.  \notag \\
&&\left. -g_{\mu \beta }\text{tr}\left( \gamma _{5\nu B\alpha C}\right)
-g_{\mu \alpha }\text{tr}\left( \gamma _{5\nu BC\beta }\right) -g_{\mu \nu }%
\text{tr}\left( \gamma _{5B\alpha C\beta }\right) \right] \frac{%
K_{2}^{B}K_{3}^{C}}{D_{1234}}  \notag \\
&&+m^{2}\left[ -g_{\alpha \beta }\text{tr}\left( \gamma _{5\mu \nu
BD}\right) -g_{\nu \beta }\text{tr}\left( \gamma _{5\mu B\alpha D}\right)
+g_{\nu \alpha }\text{tr}\left( \gamma _{5\mu B\beta D}\right) \right. 
\notag \\
&&\left. +g_{\mu \beta }\text{tr}\left( \gamma _{5\nu B\alpha D}\right)
-g_{\mu \alpha }\text{tr}\left( \gamma _{5\nu B\beta D}\right) -g_{\mu \nu }%
\text{tr}\left( \gamma _{5B\alpha \beta D}\right) \right] \frac{%
K_{2}^{B}K_{4}^{D}}{D_{1234}}  \notag \\
&&+m^{2}\left[ g_{\alpha \beta }\text{tr}\left( \gamma _{5\mu \nu CD}\right)
-g_{\nu \beta }\text{tr}\left( \gamma _{5\mu \alpha CD}\right) -g_{\nu
\alpha }\text{tr}\left( \gamma _{5\mu C\beta D}\right) \right.  \notag \\
&&\left. +g_{\mu \beta }\text{tr}\left( \gamma _{5\nu \alpha CD}\right)
+g_{\mu \alpha }\text{tr}\left( \gamma _{5\nu C\beta D}\right) -g_{\mu \nu }%
\text{tr}\left( \gamma _{5\alpha C\beta D}\right) \right] \frac{%
K_{3}^{C}K_{4}^{D}}{D_{1234}}.
\end{eqnarray}

The final part of this task is identifying substructures by noticing that
these traces correspond to odd amplitudes that are 2nd-order tensors. Since
indices are unchanged, our work reduces to replacing vertices within
Equation (\ref{t4}) and comparing sign differences among all possibilities.
Ultimately, this part contains exclusively odd amplitudes%
\begin{eqnarray}
\left[ t_{\mu \nu \alpha \beta }^{AVVV}\right] _{1} &=&\left[ g_{\alpha
\beta }t_{\mu \nu }^{AVPP}+g_{\nu \beta }t_{\mu \alpha }^{APVP}+g_{\nu
\alpha }t_{\mu \beta }^{APPV}\right]  \notag \\
&&-\left[ g_{\mu \beta }t_{\nu \alpha }^{SVVP}+g_{\mu \alpha }t_{\nu \beta
}^{SVPV}+g_{\mu \nu }t_{\alpha \beta }^{SPVV}\right]  \notag \\
&&+\left( g_{\mu \nu }g_{\alpha \beta }-g_{\mu \alpha }g_{\nu \beta }+g_{\mu
\beta }g_{\nu \alpha }\right) t^{SPPP}.
\end{eqnarray}%
The $SPPP$ numeric factor changes because this amplitude also appears inside
the others.

In the second sector, we group those terms where the Levi-Civita symbol has
three or four free indices. Let us return to expression (\ref{tr3m}) to
illustrate the analysis of these components%
\begin{eqnarray*}
\text{tr}\left( \gamma _{\mu 5A\nu B\alpha \beta }\right) &\rightarrow
&\varepsilon _{\mu \nu B\alpha }g_{A\beta }+\varepsilon _{\mu A\nu \alpha
}g_{B\beta }-\varepsilon _{\mu \nu B\beta }g_{A\alpha }-\varepsilon _{\mu
A\nu \beta }g_{B\alpha }-\varepsilon _{\mu B\alpha \beta }g_{A\nu } \\
&&-\varepsilon _{\mu A\alpha \beta }g_{\nu B}-\varepsilon _{\nu B\alpha
\beta }g_{\mu A}-\varepsilon _{A\nu \alpha \beta }g_{\mu B}+\varepsilon
_{\mu \nu \alpha \beta }g_{AB}.
\end{eqnarray*}%
Our objective is finding substructures, which requires combining monomials
with the same index arrangement. To do so, we introduce a new index $\kappa $
to generate metric products corresponding to less complex traces%
\begin{eqnarray*}
\text{tr}\left( \gamma _{\mu 5A\nu B\alpha \beta }\right) &\rightarrow
&\varepsilon _{\mu \nu \alpha \kappa }\text{tr}\left( \gamma _{\kappa
AB\beta }\right) -\varepsilon _{\mu \nu \beta \kappa }\text{tr}\left( \gamma
_{\kappa AB\alpha }\right) -\varepsilon _{\mu \alpha \beta \kappa }\text{tr}%
\left( \gamma _{\kappa A\nu B}\right) \\
&&-\varepsilon _{\nu \alpha \beta \kappa }\text{tr}\left( \sigma _{\kappa
\mu }\gamma _{AB}\right) -2\varepsilon _{\mu \nu \alpha \beta }\text{tr}%
\left( \gamma _{AB}\right) .
\end{eqnarray*}%
Note that the performed manipulations changed the last numerical
coefficient. The traces below are recognized when extending this discussion
to the remaining cases:%
\begin{eqnarray}
\left[ t_{\mu \nu \alpha \beta }^{AVVV}\right] _{2} &=&i\left[ \varepsilon
_{\nu \alpha \beta \kappa }\text{tr}\left( \sigma _{\kappa \mu }\gamma
_{ABCD}\right) -\varepsilon _{\mu \alpha \beta \kappa }\text{tr}\left(
\gamma _{\kappa A\nu BCD}\right) -\varepsilon _{\mu \nu \beta \kappa }\text{%
tr}\left( \gamma _{\kappa AB\alpha CD}\right) \right.  \notag \\
&&\left. -\varepsilon _{\mu \nu \alpha \kappa }\text{tr}\left( \gamma
_{\kappa ABC\beta D}\right) +2\varepsilon _{\mu \nu \alpha \beta }\text{tr}%
\left( \gamma _{ABCD}\right) \right] \frac{%
K_{1}^{A}K_{2}^{B}K_{3}^{C}K_{4}^{D}}{D_{1234}}  \notag \\
&&+im^{2}\left[ \varepsilon _{\mu \nu \alpha \kappa }\text{tr}\left( \gamma
_{\kappa AB\beta }\right) -\varepsilon _{\mu \nu \beta \kappa }\text{tr}%
\left( \gamma _{\kappa AB\alpha }\right) -\varepsilon _{\mu \alpha \beta
\kappa }\text{tr}\left( \gamma _{\kappa A\nu B}\right) \right.  \notag \\
&&\left. -\varepsilon _{\nu \alpha \beta \kappa }\text{tr}\left( \sigma
_{\kappa \mu }\gamma _{AB}\right) -2\varepsilon _{\mu \nu \alpha \beta }%
\text{tr}\left( \gamma _{AB}\right) \right] \frac{K_{1}^{A}K_{2}^{B}}{%
D_{1234}}  \notag \\
&&+im^{2}\left[ -\varepsilon _{\mu \nu \alpha \kappa }\text{tr}\left( \gamma
_{\kappa AC\beta }\right) -\varepsilon _{\mu \nu \beta \kappa }\text{tr}%
\left( \gamma _{\kappa A\alpha C}\right) +\varepsilon _{\mu \alpha \beta
\kappa }\text{tr}\left( \gamma _{\kappa A\nu C}\right) \right.  \notag \\
&&\left. +\varepsilon _{\nu \alpha \beta \kappa }\text{tr}\left( \sigma
_{\kappa \mu }\gamma _{AC}\right) +2\varepsilon _{\mu \nu \alpha \beta }%
\text{tr}\left( \gamma _{AC}\right) \right] \frac{K_{1}^{A}K_{3}^{C}}{%
D_{1234}}  \notag \\
&&+im^{2}\left[ -\varepsilon _{\mu \nu \alpha \kappa }\text{tr}\left( \gamma
_{\kappa A\beta D}\right) +\varepsilon _{\mu \nu \beta \kappa }\text{tr}%
\left( \gamma _{\kappa A\alpha D}\right) -\varepsilon _{\mu \alpha \beta
\kappa }\text{tr}\left( \gamma _{\kappa A\nu D}\right) \right.  \notag \\
&&\left. -\varepsilon _{\nu \alpha \beta \kappa }\text{tr}\left( \sigma
_{\kappa \mu }\gamma _{AD}\right) -2\varepsilon _{\mu \nu \alpha \beta }%
\text{tr}\left( \gamma _{AD}\right) \right] \frac{K_{1}^{A}K_{4}^{D}}{%
D_{1234}}  \notag \\
&&+im^{2}\left[ \varepsilon _{\mu \nu \alpha \kappa }\text{tr}\left( \gamma
_{\kappa BC\beta }\right) +\varepsilon _{\mu \nu \beta \kappa }\text{tr}%
\left( \gamma _{\kappa B\alpha C}\right) +\varepsilon _{\mu \alpha \beta
\kappa }\text{tr}\left( \gamma _{\kappa \nu BC}\right) \right.  \notag \\
&&\left. -\varepsilon _{\nu \alpha \beta \kappa }\text{tr}\left( \sigma
_{\kappa \mu }\gamma _{BC}\right) -2\varepsilon _{\mu \nu \alpha \beta }%
\text{tr}\left( \gamma _{BC}\right) \right] \frac{K_{2}^{B}K_{3}^{C}}{%
D_{1234}}  \notag \\
&&+im^{2}\left[ \varepsilon _{\mu \nu \alpha \kappa }\text{tr}\left( \gamma
_{\kappa B\beta D}\right) -\varepsilon _{\mu \nu \beta \kappa }\text{tr}%
\left( \gamma _{\kappa B\alpha D}\right) -\varepsilon _{\mu \alpha \beta
\kappa }\text{tr}\left( \gamma _{\kappa \nu BD}\right) \right.  \notag \\
&&\left. +\varepsilon _{\nu \alpha \beta \kappa }\text{tr}\left( \sigma
_{\kappa \mu }\gamma _{BD}\right) +2\varepsilon _{\mu \nu \alpha \beta }%
\text{tr}\left( \gamma _{BD}\right) \right] \frac{K_{2}^{B}K_{4}^{D}}{%
D_{1234}}  \notag \\
&&+im^{2}\left[ -\varepsilon _{\mu \nu \alpha \kappa }\text{tr}\left( \gamma
_{\kappa C\beta D}\right) -\varepsilon _{\mu \nu \beta \kappa }\text{tr}%
\left( \gamma _{\kappa \alpha CD}\right) +\varepsilon _{\mu \alpha \beta
\kappa }\text{tr}\left( \gamma _{\kappa \nu CD}\right) \right.  \notag \\
&&\left. -\varepsilon _{\nu \alpha \beta \kappa }\text{tr}\left( \sigma
_{\kappa \mu }\gamma _{CD}\right) -2\varepsilon _{\mu \nu \alpha \beta }%
\text{tr}\left( \gamma _{CD}\right) \right] \frac{K_{3}^{C}K_{4}^{D}}{%
D_{1234}}  \notag \\
&&-im^{4}\text{tr}\left( \gamma _{5\mu \nu \alpha \beta }\right) \frac{1}{%
D_{1234}}.
\end{eqnarray}

This time, traces correspond to even amplitudes that are 2nd-order tensors.
We write the ensuing organization when examining differences among all
possibilities:%
\begin{equation}
\left[ t_{\mu \nu \alpha \beta }^{AVVV}\right] _{2}=i\varepsilon _{\nu
\alpha \beta \kappa }t_{\kappa \mu }^{\widetilde{T}PPP}-i\left[ \varepsilon
_{\mu \alpha \beta \kappa }t_{\kappa \nu }^{VVPP}+\varepsilon _{\mu \nu
\beta \kappa }t_{\kappa \alpha }^{VPVP}+\varepsilon _{\mu \nu \alpha \kappa
}t_{\kappa \beta }^{VPPV}\right] +2i\varepsilon _{\mu \nu \alpha \beta
}t^{PPPP}.
\end{equation}%
Observe that the commutator $\sigma _{\kappa \mu }$ appeared throughout
calculations and now reflects on the emergence of the pseudo-tensor vertex $%
\widetilde{T}$. Since the scalar function $PPPP$ appears inside other terms,
one must adjust its coefficient adequately.

The last sector comprehends all remaining contributions, which are
combinations of standard tensors with four momenta in the numerator. Without
performing manipulations, we group terms according to their index arrangement%
\begin{eqnarray}
\left[ t_{\mu \nu \alpha \beta }^{AVVV}\right] _{3} &=&4i\left[ \varepsilon
_{\mu \nu XY}t_{XY\alpha \beta }^{\left( 12\right) }+\varepsilon _{\mu
\alpha XY}t_{XY\nu \beta }^{\left( 13\right) }+\varepsilon _{\mu \beta
XY}t_{XY\nu \alpha }^{\left( 14\right) }\right.  \notag \\
&&\left. +\varepsilon _{\nu \alpha XY}t_{XY\mu \beta }^{\left( 23\right)
}+\varepsilon _{\nu \beta XY}t_{XY\mu \alpha }^{\left( 24\right)
}+\varepsilon _{\alpha \beta XY}t_{XY\mu \nu }^{\left( 34\right) }\right] .
\end{eqnarray}%
We will provide an adequate definition of these tensors eventually, so
consider the direct associations introduced in the sequence for now.%
\begin{eqnarray}
&&\varepsilon _{\mu \nu XY}t_{XY\alpha \beta }^{\left( 12\right) }  \notag \\
&\rightarrow &-\left[ \varepsilon _{\mu A\nu B}\left( g_{\alpha C}g_{\beta
D}+g_{\alpha D}g_{C\beta }\right) +\varepsilon _{\mu A\nu C}\left(
g_{B\alpha }g_{\beta D}-g_{B\beta }g_{\alpha D}\right) \right.  \notag \\
&&+\varepsilon _{\mu A\nu D}\left( g_{B\alpha }g_{C\beta }+g_{B\beta
}g_{\alpha C}\right) +\varepsilon _{\mu \nu BC}\left( g_{A\alpha }g_{\beta
D}-g_{A\beta }g_{\alpha D}\right)  \notag \\
&&\left. +\varepsilon _{\mu \nu BD}\left( g_{A\alpha }g_{C\beta }+g_{A\beta
}g_{\alpha C}\right) +\varepsilon _{\mu \nu CD}\left( g_{A\beta }g_{B\alpha
}-g_{A\alpha }g_{B\beta }\right) \right] \frac{%
K_{1}^{A}K_{2}^{B}K_{3}^{C}K_{4}^{D}}{D_{1234}}  \label{12a}
\end{eqnarray}%
\begin{eqnarray}
&&\varepsilon _{\mu \alpha XY}t_{XY\nu \beta }^{\left( 13\right) }  \notag \\
&\rightarrow &-\left[ \varepsilon _{\mu AB\alpha }\left( g_{\nu C}g_{\beta
D}+g_{\nu D}g_{C\beta }\right) +\varepsilon _{\mu A\alpha C}\left( g_{\nu
B}g_{\beta D}+g_{\nu D}g_{B\beta }\right) \right.  \notag \\
&&+\varepsilon _{\mu A\alpha D}\left( g_{\nu B}g_{C\beta }-g_{\nu
C}g_{B\beta }\right) +\varepsilon _{\mu B\alpha C}\left( g_{A\nu }g_{\beta
D}-g_{A\beta }g_{\nu D}\right)  \notag \\
&&\left. +\varepsilon _{\mu B\alpha D}\left( g_{A\nu }g_{C\beta }+g_{A\beta
}g_{\nu C}\right) +\varepsilon _{\mu \alpha CD}\left( g_{A\nu }g_{B\beta
}+g_{A\beta }g_{\nu B}\right) \right] \frac{%
K_{1}^{A}K_{2}^{B}K_{3}^{C}K_{4}^{D}}{D_{1234}}
\end{eqnarray}%
\begin{eqnarray}
&&\varepsilon _{\mu \beta XY}t_{XY\nu \alpha }^{\left( 14\right) }  \notag \\
&\rightarrow &-\left[ \varepsilon _{\mu AB\beta }\left( -g_{\nu C}g_{\alpha
D}+g_{\nu D}g_{\alpha C}\right) +\varepsilon _{\mu AC\beta }\left( g_{\nu
B}g_{\alpha D}+g_{\nu D}g_{B\alpha }\right) \right.  \notag \\
&&+\varepsilon _{\mu A\beta D}\left( g_{\nu B}g_{\alpha C}+g_{\nu
C}g_{B\alpha }\right) +\varepsilon _{\mu BC\beta }\left( g_{A\nu }g_{\alpha
D}-g_{A\alpha }g_{\nu D}\right)  \notag \\
&&\left. +\varepsilon _{\mu B\beta D}\left( g_{A\nu }g_{\alpha C}-g_{A\alpha
}g_{\nu C}\right) +\varepsilon _{\mu C\beta D}\left( g_{A\nu }g_{B\alpha
}+g_{A\alpha }g_{\nu B}\right) \right] \frac{%
K_{1}^{A}K_{2}^{B}K_{3}^{C}K_{4}^{D}}{D_{1234}}  \label{14a}
\end{eqnarray}%
\begin{eqnarray}
&&\varepsilon _{\nu \alpha XY}t_{XY\mu \beta }^{\left( 23\right) }  \notag \\
&\rightarrow &-\left[ \varepsilon _{A\nu B\alpha }\left( g_{\mu C}g_{\beta
D}+g_{\mu D}g_{C\beta }\right) +\varepsilon _{A\nu \alpha C}\left( g_{\mu
B}g_{\beta D}+g_{\mu D}g_{B\beta }\right) \right.  \notag \\
&&+\varepsilon _{A\nu \alpha D}\left( g_{\mu B}g_{C\beta }-g_{\mu
C}g_{B\beta }\right) +\varepsilon _{\nu B\alpha C}\left( g_{\mu A}g_{\beta
D}+g_{\mu D}g_{A\beta }\right)  \notag \\
&&\left. +\varepsilon _{\nu B\alpha D}\left( g_{\mu A}g_{C\beta }-g_{\mu
C}g_{A\beta }\right) +\varepsilon _{\nu \alpha CD}\left( g_{\mu A}g_{B\beta
}-g_{\mu B}g_{A\beta }\right) \right] \frac{%
K_{1}^{A}K_{2}^{B}K_{3}^{C}K_{4}^{D}}{D_{1234}}
\end{eqnarray}%
\begin{eqnarray}
&&\varepsilon _{\nu \beta XY}t_{XY\mu \alpha }^{\left( 24\right) }  \notag \\
&\rightarrow &-\left[ \varepsilon _{A\nu B\beta }\left( -g_{\mu C}g_{\alpha
D}+g_{\mu D}g_{\alpha C}\right) +\varepsilon _{A\nu C\beta }\left( g_{\mu
B}g_{\alpha D}+g_{\mu D}g_{B\alpha }\right) \right.  \notag \\
&&+\varepsilon _{A\nu \beta D}\left( g_{\mu B}g_{\alpha C}+g_{\mu
C}g_{B\alpha }\right) +\varepsilon _{\nu BC\beta }\left( g_{\mu A}g_{\alpha
D}+g_{\mu D}g_{A\alpha }\right)  \notag \\
&&\left. +\varepsilon _{\nu B\beta D}\left( g_{\mu A}g_{\alpha C}+g_{\mu
C}g_{A\alpha }\right) +\varepsilon _{\nu C\beta D}\left( g_{\mu A}g_{B\alpha
}-g_{\mu B}g_{A\alpha }\right) \right] \frac{%
K_{1}^{A}K_{2}^{B}K_{3}^{C}K_{4}^{D}}{D_{1234}}
\end{eqnarray}%
\begin{eqnarray}
&&\varepsilon _{\alpha \beta XY}t_{XY\mu \nu }^{\left( 34\right) }  \notag \\
&\rightarrow &-\left[ \varepsilon _{AB\alpha \beta }\left( -g_{\mu C}g_{\nu
D}+g_{\mu D}g_{\nu C}\right) +\varepsilon _{A\alpha C\beta }\left( -g_{\mu
B}g_{\nu D}+g_{\mu D}g_{\nu B}\right) \right.  \notag \\
&&+\varepsilon _{A\alpha \beta D}\left( -g_{\mu B}g_{\nu C}+g_{\mu C}g_{\nu
B}\right) +\varepsilon _{B\alpha C\beta }\left( g_{\mu A}g_{\nu D}+g_{\mu
D}g_{A\nu }\right)  \notag \\
&&\left. +\varepsilon _{B\alpha \beta D}\left( g_{\mu A}g_{\nu C}+g_{\mu
C}g_{A\nu }\right) +\varepsilon _{\alpha C\beta D}\left( g_{\mu A}g_{\nu
B}+g_{\mu B}g_{A\nu }\right) \right] \frac{%
K_{1}^{A}K_{2}^{B}K_{3}^{C}K_{4}^{D}}{D_{1234}}
\end{eqnarray}

Once all pieces are known, the $AVVV$ integrand assumes the following form%
\begin{eqnarray}
t_{\mu \nu \alpha \beta }^{AVVV} &=&4i\left[ \varepsilon _{\mu \nu
XY}t_{XY\alpha \beta }^{\left( 12\right) }+\varepsilon _{\mu \alpha
XY}t_{XY\nu \beta }^{\left( 13\right) }+\varepsilon _{\mu \beta XY}t_{XY\nu
\alpha }^{\left( 14\right) }\right]  \notag \\
&&+\left[ g_{\alpha \beta }t_{\mu \nu }^{AVPP}+g_{\nu \beta }t_{\mu \alpha
}^{APVP}+g_{\nu \alpha }t_{\mu \beta }^{APPV}\right] +2i\varepsilon _{\mu
\nu \alpha \beta }t^{PPPP}  \notag \\
&&-i\left[ \varepsilon _{\mu \alpha \beta \kappa }t_{\kappa \nu
}^{VVPP}+\varepsilon _{\mu \nu \beta \kappa }t_{\kappa \alpha
}^{VPVP}+\varepsilon _{\mu \nu \alpha \kappa }t_{\kappa \beta }^{VPPV}\right]
\notag \\
&&+4i\left[ \varepsilon _{\nu \alpha XY}t_{XY\mu \beta }^{\left( 23\right)
}+\varepsilon _{\nu \beta XY}t_{XY\mu \alpha }^{\left( 24\right)
}+\varepsilon _{\alpha \beta XY}t_{XY\mu \nu }^{\left( 34\right) }\right] 
\notag \\
&&+i\varepsilon _{\nu \alpha \beta \kappa }t_{\kappa \mu }^{\widetilde{T}%
PPP}-\left[ g_{\mu \beta }t_{\nu \alpha }^{SVVP}+g_{\mu \alpha }t_{\nu \beta
}^{SVPV}+g_{\mu \nu }t_{\alpha \beta }^{SPVV}\right]  \notag \\
&&+\left( g_{\mu \nu }g_{\alpha \beta }-g_{\mu \alpha }g_{\nu \beta }+g_{\mu
\beta }g_{\nu \alpha }\right) t^{SPPP}.  \notag
\end{eqnarray}%
We reiterate that expressions adopted for traces contain all non-equivalent
tensor configurations, which was convenient for identifying substructures.
As this task is over, let us pursue simplifications in the same fashion as
the triangle discussion. There, we acknowledged the presence of a Schouten
identity\ with the trace-defining index fixed. In other words, when
replacing the chiral matrix definition adjacent to the matrix $\gamma _{\mu
} $, an identity with $\mu $ fixed arose (\ref{Schouten}). This feature also
applies to the box amplitude; thus, let us look closer at terms having this
index outside the Levi-Civita symbol to verify that each coefficient
vanishes identically (last three rows of the equation above). We do not
compact products involving Dirac matrices from this point on.

Following this reasoning, we check over terms proportional to the squared
mass. Notwithstanding the 2nd-order tensor amplitudes count with these
contributions, the following combination does not exhibit such dependence:%
\begin{eqnarray}
&&i\varepsilon _{\nu \alpha \beta \kappa }t_{\kappa \mu }^{\widetilde{T}PPP}-%
\left[ g_{\mu \beta }t_{\nu \alpha }^{SVVP}+g_{\mu \alpha }t_{\nu \beta
}^{SVPV}+g_{\mu \nu }t_{\alpha \beta }^{SPVV}\right]  \notag \\
&=&\left[ i\varepsilon _{\nu \alpha \beta \kappa }\text{tr}\left( \sigma
_{\kappa \mu }\gamma _{A}\gamma _{B}\gamma _{C}\gamma _{D}\right) -g_{\mu
\beta }\text{tr}\left( \gamma _{A}\gamma _{\nu }\gamma _{B}\gamma _{\alpha
}\gamma _{C}\gamma _{5}\gamma _{D}\right) \right.  \notag \\
&&\left. -g_{\mu \alpha }\text{tr}\left( \gamma _{A}\gamma _{\nu }\gamma
_{B}\gamma _{5}\gamma _{C}\gamma _{\beta }\gamma _{D}\right) -g_{\mu \nu }%
\text{tr}\left( \gamma _{A}\gamma _{5}\gamma _{B}\gamma _{\alpha }\gamma
_{C}\gamma _{\beta }\gamma _{D}\right) \right] \frac{%
K_{1}^{A}K_{2}^{B}K_{3}^{C}K_{4}^{D}}{D_{1234}}.  \label{2}
\end{eqnarray}%
To prove this result, we recall the coefficient associated with $%
K_{1}^{A}K_{2}^{B}$ in (\ref{tr3m}). The first row of the referred equation
are monomials having $\mu $ within the metric tensor. Since it is a tensor
antisymmetric in five indices, it cancels out identically in a
four-dimensional setting%
\begin{equation}
g_{\mu A}\varepsilon _{\nu B\alpha \beta }-g_{\mu \nu }\varepsilon
_{AB\alpha \beta }+g_{\mu B}\varepsilon _{A\nu \alpha \beta }-g_{\mu \alpha
}\varepsilon _{A\nu B\beta }+g_{\mu \beta }\varepsilon _{A\nu B\alpha }=0%
\text{.}
\end{equation}%
Alternatively, one generates this result by performing successive
permutations of the matrix $\gamma _{\mu }$ within a more complex trace tr$%
\left( \gamma _{5}\gamma _{\mu }\gamma _{\nu }\gamma _{\alpha }\gamma
_{\beta }\gamma _{A}\gamma _{B}\right) $; observe the form: 
\begin{eqnarray}
g_{\mu A}\text{tr}\left( \gamma _{5}\gamma _{\nu }\gamma _{B}\gamma _{\alpha
}\gamma _{\beta }\right) -g_{\mu \nu }\text{tr}\left( \gamma _{5}\gamma
_{A}\gamma _{B}\gamma _{\alpha }\gamma _{\beta }\right) +g_{\mu B}\text{tr}%
\left( \gamma _{5}\gamma _{A}\gamma _{\nu }\gamma _{\alpha }\gamma _{\beta
}\right) &&  \notag \\
-g_{\mu \alpha }\text{tr}\left( \gamma _{5}\gamma _{A}\gamma _{\nu }\gamma
_{B}\gamma _{\beta }\right) +g_{\mu \beta }\text{tr}\left( \gamma _{5}\gamma
_{A}\gamma _{\nu }\gamma _{B}\gamma _{\alpha }\right) &=&0\text{.}
\end{eqnarray}%
We find the same outcome when examining other coefficients; therefore,
completing this part of the demonstration.

As a primary ingredient to examine tensor contributions, we follow the ideas
seen in the previous case to derive the identities:%
\begin{eqnarray}
&&g_{\mu A}\text{tr}\left( \gamma _{5}\gamma _{\nu }\gamma _{B}\gamma
_{\alpha }\gamma _{C}\gamma _{\beta }\gamma _{D}\right) +g_{\mu B}\text{tr}%
\left( \gamma _{5}\gamma _{A}\gamma _{\nu }\gamma _{\alpha }\gamma
_{C}\gamma _{\beta }\gamma _{D}\right)  \notag \\
&&+g_{\mu C}\text{tr}\left( \gamma _{5}\gamma _{A}\gamma _{\nu }\gamma
_{B}\gamma _{\alpha }\gamma _{\beta }\gamma _{D}\right) +g_{\mu D}\text{tr}%
\left( \gamma _{5}\gamma _{A}\gamma _{\nu }\gamma _{B}\gamma _{\alpha
}\gamma _{C}\gamma _{\beta }\right)  \notag \\
&=&-g_{\mu \nu }\text{tr}\left( \gamma _{A}\gamma _{5}\gamma _{B}\gamma
_{\alpha }\gamma _{C}\gamma _{\beta }\gamma _{D}\right) -g_{\mu \alpha }%
\text{tr}\left( \gamma _{A}\gamma _{\nu }\gamma _{B}\gamma _{5}\gamma
_{C}\gamma _{\beta }\gamma _{D}\right)  \notag \\
&&-g_{\mu \beta }\text{tr}\left( \gamma _{A}\gamma _{\nu }\gamma _{B}\gamma
_{\alpha }\gamma _{C}\gamma _{5}\gamma _{D}\right)
\end{eqnarray}%
and%
\begin{eqnarray}
g_{\mu \nu }\text{tr}\left( \gamma _{5}\gamma _{A}\gamma _{B}\gamma
_{C}\gamma _{D}\right) &=&g_{\mu A}\text{tr}\left( \gamma _{5}\gamma _{\nu
}\gamma _{B}\gamma _{C}\gamma _{D}\right) -g_{\mu B}\text{tr}\left( \gamma
_{5}\gamma _{\nu }\gamma _{A}\gamma _{C}\gamma _{D}\right)  \notag \\
&&+g_{\mu C}\text{tr}\left( \gamma _{5}\gamma _{\nu }\gamma _{A}\gamma
_{B}\gamma _{D}\right) -g_{\mu D}\text{tr}\left( \gamma _{5}\gamma _{\nu
}\gamma _{A}\gamma _{B}\gamma _{C}\right) .  \label{id1}
\end{eqnarray}%
They perform the task of permuting index positions in Equation (\ref{2}),
leading directly to the expected identifications%
\begin{eqnarray}
&&\left[ g_{\mu \beta }t_{\nu \alpha }^{SVVP}+g_{\mu \alpha }t_{\nu \beta
}^{SVPV}+g_{\mu \nu }t_{\alpha \beta }^{SPVV}\right] -\varepsilon _{\nu
\alpha \beta \kappa }t_{\kappa \mu }^{\widetilde{T}PPP}  \notag \\
&=&4\left[ \varepsilon _{\nu \alpha XY}t_{XY\mu \beta }^{\left( 23\right)
}+\varepsilon _{\nu \beta XY}t_{XY\mu \alpha }^{\left( 24\right)
}+\varepsilon _{\alpha \beta XY}t_{XY\mu \nu }^{\left( 34\right) }\right] 
\notag \\
&&+\left( g_{\mu \nu }g_{\alpha \beta }-g_{\mu \alpha }g_{\nu \beta }+g_{\mu
\beta }g_{\nu \alpha }\right) t^{SPPP}.  \label{idbox}
\end{eqnarray}%
With this identity, we achieve a much simpler view of the $AVVV$ integrand%
\begin{eqnarray}
t_{\mu \nu \alpha \beta }^{AVVV} &=&f_{4\mu \nu \alpha \beta }-\left[
\varepsilon _{\mu \alpha \beta \kappa }t_{\kappa \nu }^{VVPP}+\varepsilon
_{\mu \nu \beta \kappa }t_{\kappa \alpha }^{VPVP}+\varepsilon _{\mu \nu
\alpha \kappa }t_{\kappa \beta }^{VPPV}\right]  \notag \\
&&+\left[ g_{\alpha \beta }t_{\mu \nu }^{AVPP}+g_{\nu \beta }t_{\mu \alpha
}^{APVP}+g_{\nu \alpha }t_{\mu \beta }^{APPV}\right] +2\varepsilon _{\mu \nu
\alpha \beta }t^{PPPP},  \label{avvv}
\end{eqnarray}%
where $f_{4\mu \nu \alpha \beta }$ represents the tensor sector (we clarify
this object below).

Inquiring about each object structure is the final part of this exploration,
which occurs in the subsequent topics.

\subsubsection{Fourth-Order Tensors}

First, we inspect pure tensor contributions grouped into the structure%
\begin{equation}
f_{4\mu \nu \alpha \beta }=4\varepsilon _{\mu \nu XY}t_{XY\alpha \beta
}^{\left( 12\right) }+4\varepsilon _{\mu \alpha XY}t_{XY\nu \beta }^{\left(
13\right) }+4\varepsilon _{\mu \beta XY}t_{XY\nu \alpha }^{\left( 14\right)
}.  \label{f41}
\end{equation}%
After performing index contractions in the original expressions (\ref{12a})-(%
\ref{14a}), our goal is to relabel summed indices and factorize the
Levi-Civita symbol. Using the antisymmetric character of tensors is
recurrent throughout this process. Thus, we introduce the following
organization of the parts%
\begin{eqnarray}
2t_{XY\alpha \beta }^{\left( 12\right) } &=&t_{4XY;\alpha \beta }^{\left(
-;+\right) }+t_{4X\alpha ;Y\beta }^{\left( -;+\right) }-t_{4X\beta ;Y\alpha
}^{\left( -;-\right) }+t_{4\alpha Y;\beta X}^{\left( -;+\right) }+t_{4\beta
Y;\alpha X}^{\left( -;-\right) }+t_{4\alpha \beta ;XY}^{\left( -;-\right) },
\\
2t_{XY\nu \beta }^{\left( 13\right) } &=&t_{4Y\beta ;\nu X}^{\left(
-;-\right) }-t_{4XY;\nu \beta }^{\left( -;+\right) }-t_{4\nu Y;\beta
X}^{\left( +;+\right) }-t_{4\beta X;Y\nu }^{\left( -;-\right) }+t_{4\nu
X;Y\beta }^{\left( +;+\right) }-t_{4\nu \beta ;XY}^{\left( +;-\right) }, \\
2t_{XY\nu \alpha }^{\left( 14\right) } &=&t_{4XY;\nu \alpha }^{\left(
-;-\right) }+t_{4\alpha Y;\nu X}^{\left( -;-\right) }-t_{4\nu Y;\alpha
X}^{\left( +;-\right) }+t_{4\alpha X;Y\nu }^{\left( -;-\right) }-t_{4\nu
X;Y\alpha }^{\left( +;-\right) }+t_{4\nu \alpha ;XY}^{\left( +;-\right) },
\label{f42}
\end{eqnarray}%
where a new standard tensor arises%
\begin{eqnarray}
&&t_{4\mu \nu ;\alpha \beta }^{\left( s_{1};s_{2}\right) }\left(
k_{i},k_{j};k_{l},k_{m}\right)  \notag \\
&=&\left[ \left( k+k_{i}\right) _{\mu }\left( k+k_{j}\right) _{\nu
}+s_{1}\left( k+k_{j}\right) _{\mu }\left( k+k_{i}\right) _{\nu }\right]
\times  \notag \\
&&\times \left[ \left( k+k_{l}\right) _{\alpha }\left( k+k_{m}\right)
_{\beta }+s_{2}\left( k+k_{m}\right) _{\alpha }\left( k+k_{l}\right) _{\beta
}\right] \frac{1}{D_{1234}}.  \label{t44s}
\end{eqnarray}%
This notation employs a numerical subindex to indicate dependence on four
internal momenta and admits two sign choices: $s_{1}$ and $s_{2}$. We omit
arguments in occurrences exhibiting the momenta hierarchy $t_{4\mu \nu
;\alpha \beta }^{\left( s_{1};s_{2}\right) }=t_{4\mu \nu ;\alpha \beta
}^{\left( s_{1};s_{2}\right) }\left( k_{1},k_{2};k_{3},k_{4}\right) $.

Such an organization allows reducing our efforts to computing a single
object, which consists of the simplified version%
\begin{equation}
t_{4\mu \nu \alpha \beta }\left( k_{i},k_{j},k_{l},k_{m}\right) =\frac{%
\left( k+k_{i}\right) _{\mu }\left( k+k_{j}\right) _{\nu }\left(
k+k_{l}\right) _{\alpha }\left( k+k_{m}\right) _{\beta }}{D_{1234}}.
\label{t4+}
\end{equation}%
Besides appearing by itself inside some amplitudes, redefining indices of
this tensor to build up the standard version is attainable%
\begin{eqnarray}
t_{4\mu \nu ;\alpha \beta }^{\left( s_{1},s_{2}\right) }\left(
k_{i},k_{j};k_{l},k_{m}\right) &=&t_{4\mu \nu \alpha \beta }\left(
k_{i},k_{j},k_{l},k_{m}\right) +s_{1}t_{4\mu \nu \alpha \beta }\left(
k_{j},k_{i},k_{l},k_{m}\right)  \notag \\
&&+s_{2}t_{4\mu \nu \alpha \beta }\left( k_{i},k_{j},k_{m},k_{l}\right)
+s_{1}s_{2}t_{4\mu \nu ;\alpha \beta }\left( k_{j},k_{i},k_{m},k_{l}\right) .
\label{t4comp}
\end{eqnarray}

\subsubsection{Even Amplitudes - $VVPP$, $VPVP$, and $VPPV$}

Second, we inspect even amplitudes that are 2nd-order tensors:\ $VVPP$, $%
VPVP $, and $VPPV$. For convenience, we check over these possibilities
together. Therefore, replacing their vertex operators\footnote{%
There are two vector vertices denoted by $\gamma _{\mu }$ and $\gamma _{\nu
} $; and two pseudoscalar vertices $\gamma _{5}$. Different configurations
produce the acknowledged sign differences.} on the general integrand (\ref%
{t4}) leads to the form%
\begin{eqnarray}
t_{\mu \nu }^{\Gamma _{i}\Gamma _{j}\Gamma _{k}\Gamma _{l}}
&=&+4s_{2}K_{34}\left( K_{1\mu }K_{2\nu }+s_{3}K_{1\nu }K_{2\mu }\right) 
\frac{1}{D_{1234}}  \notag \\
&&+4s_{3}K_{24}\left( K_{1\mu }K_{3\nu }-s_{2}K_{1\nu }K_{3\mu }\right) 
\frac{1}{D_{1234}}  \notag \\
&&+4s_{1}K_{23}\left( K_{1\mu }K_{4\nu }+K_{1\nu }K_{4\mu }\right) \frac{1}{%
D_{1234}}  \notag \\
&&-4s_{3}K_{14}\left( K_{2\mu }K_{3\nu }+s_{1}K_{2\nu }K_{3\mu }\right) 
\frac{1}{D_{1234}}  \notag \\
&&-4s_{1}K_{13}\left( K_{2\mu }K_{4\nu }-s_{3}K_{2\nu }K_{4\mu }\right) 
\frac{1}{D_{1234}}  \notag \\
&&+4s_{1}K_{12}\left( K_{3\mu }K_{4\nu }+s_{2}K_{3\nu }K_{4\mu }\right) 
\frac{1}{D_{1234}}  \notag \\
&&-s_{1}g_{\mu \nu }t^{PPPP},
\end{eqnarray}%
where bilinears $K_{ij}=K_{i}\cdot K_{j}-m^{2}$ appear. Each vertex
configuration $\Gamma _{i}\Gamma _{j}\Gamma _{k}\Gamma _{l}$ considers three
signs $s_{i}$, so we obtain the $VVPP$ function by fixing $s_{i}=\left(
-1,-1,+1\right) $, the $VPVP$ by fixing $s_{i}=\left( +1,-1,-1\right) $, and
the $VPPV$ by fixing $s_{i}=\left( -1,+1,-1\right) $. The $PPPP$ scalar
function appears as a subamplitude here.

When reducing bilinears with the aid of identity (\ref{k.k}), identifying
2nd-order standard tensors is straightforward. Their systematization
remembers the version depending on three internal momenta (\ref{t3uv}) and
introduces the analogous involving four momenta%
\begin{equation}
t_{4\mu \nu }^{\left( s\right) }\left( k_{i},k_{j}\right) =\frac{\left(
k+k_{i}\right) _{\mu }\left( k+k_{j}\right) _{\nu }+s\left( k+k_{j}\right)
_{\mu }\left( k+k_{i}\right) _{\nu }}{D_{1234}}.  \label{t24s}
\end{equation}%
By performing these identifications and grouping terms with the same
denominator, we achieve the structure:%
\begin{eqnarray}
t_{\mu \nu }^{\Gamma _{i}\Gamma _{j}\Gamma _{k}\Gamma _{l}} &=&2s_{1}\left[
s_{3}t_{3\mu \nu }^{\left( s_{3}\right) }\left( k_{1},k_{2}\right)
+s_{2}t_{3\mu \nu }^{\left( -s_{2}\right) }\left( k_{1},k_{3}\right)
-s_{2}t_{3\mu \nu }^{\left( s_{1}\right) }\left( k_{2},k_{3}\right) \right] 
\notag \\
&&+2s_{1}\left[ s_{3}t_{3\mu \nu }^{\left( s_{3}\right) }\left(
k_{1},k_{2}\right) +t_{3\mu \nu }^{\left( +\right) }\left(
k_{1},k_{4}\right) -t_{3\mu \nu }^{\left( -s_{3}\right) }\left(
k_{2},k_{4}\right) \right] ^{\prime }  \notag \\
&&+2s_{1}\left[ s_{2}t_{3\mu \nu }^{\left( -s_{2}\right) }\left(
k_{1},k_{3}\right) +t_{3\mu \nu }^{\left( +\right) }\left(
k_{1},k_{4}\right) +t_{3\mu \nu }^{\left( s_{2}\right) }\left(
k_{3},k_{4}\right) \right] ^{\prime \prime }  \notag \\
&&+2s_{1}\left[ -s_{2}t_{3\mu \nu }^{\left( s_{1}\right) }\left(
k_{2},k_{3}\right) -t_{3\mu \nu }^{\left( -s_{3}\right) }\left(
k_{2},k_{4}\right) +t_{3\mu \nu }^{\left( s_{2}\right) }\left(
k_{3},k_{4}\right) \right] ^{\prime \prime \prime }  \notag \\
&&-2s_{1}\left[ s_{3}\left( q-r\right) ^{2}t_{4\mu \nu }^{\left(
s_{3}\right) }\left( k_{1},k_{2}\right) +s_{2}\left( p-r\right) ^{2}t_{4\mu
\nu }^{\left( -s_{2}\right) }\left( k_{1},k_{3}\right) \right.  \notag \\
&&+\left( p-q\right) ^{2}t_{4\mu \nu }^{\left( +\right) }\left(
k_{1},k_{4}\right) -s_{2}r^{2}t_{4\mu \nu }^{\left( s_{1}\right) }\left(
k_{2},k_{3}\right)  \notag \\
&&\left. -q^{2}t_{4\mu \nu }^{\left( -s_{3}\right) }\left(
k_{2},k_{4}\right) +p^{2}t_{4\mu \nu }^{\left( s_{2}\right) }\left(
k_{3},k_{4}\right) \right] +s_{1}g_{\mu \nu }t^{PPPP}.  \label{vvpp}
\end{eqnarray}

Objects typical of three-point amplitudes arose, bringing different momenta
configurations with them. We introduced the associations below to
distinguish these possibilities.%
\begin{equation}
\begin{array}{ccc}
\frac{1}{D_{123}}\rightarrow \left[ \text{structure}\right] &  & \frac{1}{%
D_{134}}\rightarrow \left[ \text{structure}\right] ^{\prime \prime } \\ 
\frac{1}{D_{124}}\rightarrow \left[ \text{structure}\right] ^{\prime } &  & 
\frac{1}{D_{234}}\rightarrow \left[ \text{structure}\right] ^{\prime \prime
\prime }%
\end{array}
\label{line}
\end{equation}

\subsubsection{Odd Amplitudes - $AVPP$, $APVP$, and $APPV$}

Third, we inspect odd amplitudes that are 2nd-order tensors:\ $AVPP$, $APVP$%
, and $APPV$. By replacing the corresponding vertex operators\footnote{%
There are four vertices: one axial $\gamma _{\mu }\gamma _{5}$, one vector $%
\gamma _{\nu }$, and two pseudoscalars $\gamma _{5}$. Different
configurations produce sign differences.} on the general integrand (\ref{t4}%
), we approach all possibilities together%
\begin{eqnarray}
s_{1}t_{\mu \nu }^{\Gamma _{i}\Gamma _{j}\Gamma _{k}\Gamma _{l}} &=&4\left(
-\varepsilon _{ABCD}g_{\mu \nu }-\varepsilon _{\nu BCD}g_{\mu A}-\varepsilon
_{\mu BCD}g_{\nu A}+\varepsilon _{\nu ACD}g_{\mu B}+s_{2}\varepsilon _{\mu
ACD}g_{\nu B}\right.  \notag \\
&&\left. -\varepsilon _{\nu ABD}g_{\mu C}+s_{3}\varepsilon _{\mu ABD}g_{\nu
C}+\varepsilon _{\nu ABC}g_{\mu D}-\varepsilon _{\mu ABC}g_{\nu D}\right) 
\frac{K_{1}^{A}K_{2}^{B}K_{3}^{C}K_{4}^{D}}{D_{1234}}  \notag \\
&&+4\varepsilon _{\mu \nu CD}K_{12}\frac{K_{3}^{C}K_{4}^{D}}{D_{1234}}%
-4\varepsilon _{\mu \nu BD}K_{13}\frac{K_{2}^{B}K_{4}^{D}}{D_{1234}}%
+4\varepsilon _{\mu \nu BC}K_{14}\frac{K_{2}^{B}K_{3}^{C}}{D_{1234}}  \notag
\\
&&+4\varepsilon _{\mu \nu AD}K_{23}\frac{K_{1}^{A}K_{4}^{D}}{D_{1234}}%
-4\varepsilon _{\mu \nu AC}K_{24}\frac{K_{1}^{A}K_{3}^{C}}{D_{1234}}%
+4\varepsilon _{\mu \nu AB}K_{34}\frac{K_{1}^{A}K_{2}^{B}}{D_{1234}},
\end{eqnarray}%
where bilinears $K_{ij}=K_{i}\cdot K_{j}-m^{2}$ appear. Each vertex
configuration $\Gamma _{i}\Gamma _{j}\Gamma _{k}\Gamma _{l}$ considers three
signs $s_{i}$, so we obtain the $AVPP$ function by fixing $s_{i}=\left(
-1,-1,+1\right) $, the $APVP$ by fixing $s_{i}=\left( +1,+1,+1\right) $, and
the $APPV$ by fixing $s_{i}=\left( -1,+1,-1\right) $.

Since there is an evident distinction between both parts of these
amplitudes, we rename summed indices to emphasize them%
\begin{equation}
t_{\mu \nu }^{\Gamma _{i}\Gamma _{j}\Gamma _{k}\Gamma _{l}}=s_{1}\varepsilon
_{\mu XYZ}f_{4\nu XYZ}^{\left( s_{2},s_{3}\right) }+s_{1}\varepsilon _{\mu
\nu XY}f_{4XY}.  \label{avpp}
\end{equation}%
The first depends on the simplified version of the 4th-order tensor (\ref%
{t4+}). Taking a closer look at its coefficients, observe that contributions
having the index $\mu $ on the metric compound the Schouten identity (\ref%
{id1}). Hence, these terms cancel out, and this sector assumes the form%
\begin{equation}
\varepsilon _{\mu XYZ}f_{4\nu XYZ}^{\left( s_{2},s_{3}\right) }=4\left(
-\varepsilon _{\mu BCD}g_{\nu A}+s_{2}\varepsilon _{\mu ACD}g_{\nu
B}+s_{3}\varepsilon _{\mu ABD}g_{\nu C}-\varepsilon _{\mu ABC}g_{\nu
D}\right) t_{4ABCD}.  \label{sector}
\end{equation}%
Analogously to what happened with even amplitudes, bilinear reductions in
the second part lead to 2nd-order standard tensors: (\ref{t3uv}) and (\ref%
{t24s}). This time, however, all objects are antisymmetric tensors:%
\begin{eqnarray}
f_{4XY} &=&\left[ t_{3XY}^{\left( -\right) }\left( k_{2},k_{3}\right)
-t_{3XY}^{\left( -\right) }\left( k_{1},k_{3}\right) +t_{3XY}^{\left(
-\right) }\left( k_{1},k_{2}\right) \right]  \notag \\
&&+\left[ -t_{3XY}^{\left( -\right) }\left( k_{2},k_{4}\right)
+t_{3XY}^{\left( -\right) }\left( k_{1},k_{4}\right) +t_{3XY}^{\left(
-\right) }\left( k_{1},k_{2}\right) \right] ^{\prime }  \notag \\
&&+\left[ t_{3XY}^{\left( -\right) }\left( k_{3},k_{4}\right)
+t_{3XY}^{\left( -\right) }\left( k_{1},k_{4}\right) -t_{3XY}^{\left(
-\right) }\left( k_{1},k_{3}\right) \right] ^{\prime \prime }  \notag \\
&&+\left[ t_{3XY}^{\left( -\right) }\left( k_{3},k_{4}\right)
-t_{3XY}^{\left( -\right) }\left( k_{2},k_{4}\right) +t_{3XY}^{\left(
-\right) }\left( k_{2},k_{3}\right) \right] ^{\prime \prime \prime }  \notag
\\
&&-p^{2}t_{4XY}^{\left( -\right) }\left( k_{3},k_{4}\right)
+q^{2}t_{4XY}^{\left( -\right) }\left( k_{2},k_{4}\right)
-r^{2}t_{4XY}^{\left( -\right) }\left( k_{2},k_{3}\right)  \notag \\
&&-\left( p-q\right) ^{2}t_{4XY}^{\left( -\right) }\left( k_{1},k_{4}\right)
+\left( p-r\right) ^{2}t_{4XY}^{\left( -\right) }\left( k_{1},k_{3}\right) 
\notag \\
&&-\left( q-r\right) ^{2}t_{4XY}^{\left( -\right) }\left( k_{1},k_{2}\right)
.
\end{eqnarray}

\subsubsection{Scalar Amplitude $PPPP$}

Fourth, we inspect the scalar amplitude $PPPP$. Our task is to explore the
structure achieved by replacing the chiral matrix as vertices in the
original integrand (\ref{t4}):%
\begin{eqnarray}
t^{PPPP} &=&\left[ K_{1}^{A}K_{2}^{B}K_{3}^{C}K_{4}^{D}\text{tr}\left(
\gamma _{A}\gamma _{B}\gamma _{C}\gamma _{D}\right) -m^{2}K_{1}^{A}K_{2}^{B}%
\text{tr}\left( \gamma _{A}\gamma _{B}\right) \right.  \notag \\
&&+m^{2}K_{1}^{A}K_{3}^{C}\text{tr}\left( \gamma _{A}\gamma _{C}\right)
-m^{2}K_{1}^{A}K_{4}^{D}\text{tr}\left( \gamma _{A}\gamma _{D}\right)
-m^{2}K_{2}^{B}K_{3}^{C}\text{tr}\left( \gamma _{B}\gamma _{C}\right)  \notag
\\
&&\left. +m^{2}K_{2}^{B}K_{4}^{D}\text{tr}\left( \gamma _{B}\gamma
_{D}\right) -m^{2}K_{3}^{C}K_{4}^{D}\text{tr}\left( \gamma _{C}\gamma
_{D}\right) +m^{4}\text{tr}\left( \mathbf{1}\right) \right] \frac{1}{D_{1234}%
},
\end{eqnarray}%
where sign changes come from permutations. All traces contain exclusively
Dirac matrices, so results depend on the metric tensor in such a way that
bilinears $K_{ij}=K_{i}\cdot K_{j}-m^{2}$ appear:%
\begin{equation}
t^{PPPP}=4\left( K_{12}K_{34}-K_{13}K_{24}+K_{14}K_{23}\right) \frac{1}{%
D_{1234}}.
\end{equation}%
Once again, rewriting them through identity (\ref{k.k}) reduces the
dependence on propagator-like objects $D_{i}$. Since there are two
reductions this time, quantities typical of two and three-point functions
emerge. In the end, we obtain the $PPPP$ final organization%
\begin{eqnarray}
t^{PPPP} &=&2\left[ \frac{1}{D_{24}}+\frac{1}{D_{13}}\right] -2\left(
p^{2}-p\cdot q\right) \frac{1}{D_{123}}-2\left( p\cdot r\right) \frac{1}{%
D_{124}}  \notag \\
&&-2\left( r^{2}-q\cdot r\right) \frac{1}{D_{134}}+2\left( p-q\right) \cdot
\left( q-r\right) \frac{1}{D_{234}}  \notag \\
&&+\left[ p^{2}\left( r-q\right) ^{2}-q^{2}\left( p-r\right)
^{2}+r^{2}\left( p-q\right) ^{2}\right] \frac{1}{D_{1234}}.  \label{pppp}
\end{eqnarray}

\subsection{Comments}

After implementing the first part of Feynman rules, we analyzed integrands
of amplitudes relevant to this investigation. The grouping of components
that share similar sorting of indices allowed the identification of less
complex correlators and standard tensors inside them. Ultimately, each piece
corresponds to a combination of rational functions having propagator-like
quantities in denominators with loop momentum products on numerators. This
subsection briefly comments on them while introducing one-loop Feynman
integrals.

The general integrand of two-point amplitudes (\ref{tgg}) indicates they are
combinations of structures having denominators $D_{ij}=D_{i}D_{j}$ and
numerators $\left[ 1,k_{\mu },k_{\mu \nu }\right] $. Nevertheless, the $AV$ (%
\ref{av}) is an antisymmetric tensor and does not admit dependence on the
symmetric numerator. These objects also appear inside higher-order
amplitudes due to reducing bilinears, in which cases discriminating the
arguments is fundamental. That is the case of the vector $VPP$ (\ref{vpp})
and the scalar $PPPP$ (\ref{pppp}). They lead to two-propagator Feynman
integrals%
\begin{equation}
\left[ I_{2},I_{2\alpha }\right] =\int \frac{d^{4}k}{\left( 2\pi \right) ^{4}%
}\frac{\left[ 1,k_{\alpha }\right] }{D_{ij}}.  \label{I2}
\end{equation}%
Although power counting indicates quadratic divergences for two-point
amplitudes, the integrals above exhibit logarithmic and linear power
counting, respectively.

Extending this reasoning indicates that three-point amplitudes (\ref{t3})
are combinations of structures with denominators $D_{ijk}=D_{i}D_{j}D_{k}$
and numerators $\left[ 1,k_{\alpha },k_{\alpha \beta },k_{\alpha \beta \rho }%
\right] $. Again, the property of antisymmetry prohibits the emergence of
the last numerator. Hence, the following three-propagator Feynman integrals
manifest within this investigation:%
\begin{equation}
\left[ I_{3},I_{3\alpha },I_{3\alpha \beta }\right] =\int \frac{d^{4}k}{%
\left( 2\pi \right) ^{4}}\frac{\left[ 1,k_{\alpha },k_{\alpha \beta }\right] 
}{D_{ijk}}.  \label{I3}
\end{equation}%
Besides appearing within $PVV$ (\ref{pvv}) and $AVV$ (\ref{avv1}), bilinear
reductions bring these structures to all subamplitudes belonging to box
amplitudes. In this cases, we enforce the need for a notation to avoid
confusion (\ref{line}). Even though three-point amplitudes exhibit linear
power counting, only the 2nd-order integral is (logarithmically) divergent.

As for four-point amplitudes, the general integrand indicates the need to
compute the following Feynman integrals%
\begin{equation}
\left[ I_{4},I_{4\alpha },I_{4\alpha \beta },I_{4\alpha \beta \rho
},I_{4\alpha \beta \rho \sigma }\right] =\int \frac{d^{4}k}{\left( 2\pi
\right) ^{4}}\frac{\left[ 1,k_{\alpha },k_{\alpha \beta },k_{\alpha \beta
\rho },k_{\alpha \beta \rho \sigma }\right] }{D_{1234}}.  \label{I4}
\end{equation}%
Only the last one indicates a logarithmic divergence in this case.
Meanwhile, note that this contribution appears exclusively within 4th-order
standard tensors%
\begin{eqnarray}
&&t_{4\mu \nu ;\alpha \beta }^{\left( s_{1};s_{2}\right) }\left(
k_{i},k_{j};k_{l},k_{m}\right)  \notag \\
&=&\left[ \left( k+k_{i}\right) _{\mu }\left( k+k_{j}\right) _{\nu
}+s_{1}\left( k+k_{j}\right) _{\mu }\left( k+k_{i}\right) _{\nu }\right]
\times  \notag \\
&&\times \left[ \left( k+k_{l}\right) _{\alpha }\left( k+k_{m}\right)
_{\beta }+s_{2}\left( k+k_{m}\right) _{\alpha }\left( k+k_{l}\right) _{\beta
}\right] \frac{1}{D_{1234}},
\end{eqnarray}%
which is contracted with the Levi-Civita symbol within the $AVVV$ box, see
Equations (\ref{f41}) and (\ref{sector}). That simplifies some
contributions, so tensors symmetric in four indices might not appear in this
work.

We still want to comment on other standard tensors appearing throughout this
section. They emphasize patterns followed by tensor amplitudes at the
integrand level, which continues to occur after integration. This reasoning
is essential to this work, particularly for 3rd-order tensors involving
three and four propagators ($n=3,4$)%
\begin{equation}
t_{n\mu ;\nu \alpha }^{\left( s\right) }\left( k_{l};k_{i},k_{j}\right) =%
\frac{\left( k+k_{l}\right) _{\mu }\left[ \left( k+k_{i}\right) _{\nu
}\left( k+k_{j}\right) _{\alpha }+s\left( k+k_{i}\right) _{\alpha }\left(
k+k_{j}\right) _{\nu }\right] }{D_{a_{1}a_{2}\ldots a_{n}}}
\end{equation}%
and for 2nd-order tensors involving two, three, and four propagators ($%
n=2,3,4$)%
\begin{equation}
t_{n\mu \nu }^{\left( s\right) }\left( k_{i},k_{j}\right) =\frac{\left(
k+k_{i}\right) _{\mu }\left( k+k_{j}\right) _{\nu }+s\left( k+k_{j}\right)
_{\mu }\left( k+k_{i}\right) _{\nu }}{D_{a_{1}a_{2}\ldots a_{n}}}.
\end{equation}

\newpage

\section{Strategy to Handle Divergences}

\label{Strategy}As Feynman integrals are necessary to compound perturbative
amplitudes, our objective becomes their explicit computation. Thus, it is
crucial to adopt a strategy to deal with the divergences acknowledged above.
We employ the Implicit Regularization (IReg), proposed and developed by O.
A. Battistel in his Ph.D. thesis \cite{ORIMAR-TESE}. This strategy has
several applications in the anomalies subject \cite{Ebani2018:gre,
Battistel:2018rqe, Battistel:2012zz}, including cases involving the
single-axial triangle \cite{Battistel:2002dm, Battistel:2002ve}. We also
draw attention to works developed in (odd and even) extra dimensions \cite%
{FONSECA:2013eoa, Battistel:2014vea, Fonseca:2014cba} since they relate to
more complex tensor structures, as it occurs for the box amplitude.

The central ingredient of IReg is a representation of the propagator (\ref%
{fermion}) capable of splitting ill-defined mathematical structures from
finite contributions of integrals. The finite part is univocal, and its
evaluation employs usual methods of perturbative calculations. Without
choosing a prescription to compute diverging objects, we organize them and
study properties relevant to the intended discussion. This view allows a
transparent connection among mathematical expressions attributed to a
perturbative amplitude in different stages of calculations.

Following this strategy, one writes the mentioned representation through an
identity with the property that the power counting decreases from term to
term. The performed operations are purely algebraic; therefore, this
strategy has no restrictions on applicability. Besides, as such identity
consists of a summation, the only requirement for its implementation is that
linearity applies to Feynman integrals.

Let us use the object $D_{n}^{-1}$ as a study case to illustrate the
procedure. By introducing an arbitrary parameter $\lambda $, we construct
the identity%
\begin{equation}
\int \frac{d^{4}k}{\left( 2\pi \right) ^{4}}\frac{1}{D_{n}}=\int \frac{d^{4}k%
}{\left( 2\pi \right) ^{4}}\left[ \frac{1}{\left( k^{2}-\lambda ^{2}\right) }%
-\frac{2k_{n}\cdot k+k_{n}^{2}+\lambda ^{2}-m^{2}}{\left( k^{2}-\lambda
^{2}\right) D_{n}}\right] .
\end{equation}%
Although the power counting exhibited by the first term on the right-hand
side remains the same as the original integral, this term does not depend on
physical parameters. Meanwhile, the power counting of the second term
decreases by one. That compels successive implementations, so finite
integrals emerge eventually. Here, three iterations are enough to achieve
this perspective. In the end, the separation comes as follows%
\begin{equation}
\int \frac{d^{4}k}{\left( 2\pi \right) ^{4}}\frac{1}{D_{n}}=\int \frac{d^{4}k%
}{\left( 2\pi \right) ^{4}}\left[ \frac{1}{D_{\lambda }}-\frac{A_{n}}{%
D_{\lambda }^{2}}+\frac{A_{n}^{2}}{D_{\lambda }^{3}}-\frac{A_{n}^{3}}{%
D_{\lambda }^{3}D_{n}}\right] ,
\end{equation}%
where notations were introduced:%
\begin{equation}
D_{\lambda }=k^{2}-\lambda ^{2}\text{ \ and \ }A_{n}=2k_{n}\cdot
k+k_{n}^{2}+\lambda ^{2}-m^{2}.  \label{Dl}
\end{equation}%
Even though we could keep repeating this process, nothing new would occur.
Only redundant finite integrals would emerge, generating extra effort with
their inspection. Observe that $\lambda $ works as a scale connecting finite
and ill-defined structures. Furthermore, the final results must not depend
on it since it is an outsider to the theory.

At this point, we induce a general representation for the propagator (\ref%
{fermion}), capable of splitting successfully any structure of interest. It
assumes the form of the identity%
\begin{equation}
\frac{1}{D_{n}}=\sum_{j=0}^{N}\frac{\left( -1\right) ^{j}A_{n}^{j}}{%
D_{\lambda }^{j+1}}+\frac{\left( -1\right) ^{N+1}A_{n}^{N+1}}{D_{\lambda
}^{N+1}D_{n}},  \label{id}
\end{equation}%
with $N$ being equal to or higher than the superficial degree of divergence
of the aimed integral. This condition guarantees that at least the last term
leads to a finite structure.

By itself, the systematization proposed by the IReg is very useful as a tool
in this type of calculation. The subsequent discussion brings ingredients
from references \cite{Battistel:2006zq, Battistel:2012qpm}, introducing
mathematical structures necessary to express the amplitudes investigated
here. They are standard divergent objects and finite structure functions.
Further information on the implementation of this strategy is elucidated in
Section (\ref{Integral}).

\subsection{Standard Divergent Objects}

\label{Div}For the separation to be effective, the last term of identity (%
\ref{id}) must be finite. That implies any diverging object is shaped
accordingly to the elements from the summation sign. Expanding the powers $%
A_{n}^{j}$ shows that this sector combines structures depending on the loop
momentum and the scale:%
\begin{equation*}
\frac{A_{n}^{j}}{D_{\lambda }^{j+1}}\rightarrow \frac{1}{D_{\lambda
}^{\alpha }},\frac{k_{\mu _{1}}k_{\mu _{2}}}{D_{\lambda }^{\alpha +1}},\frac{%
k_{\mu _{1}}k_{\mu _{2}}k_{\mu _{3}}k_{\mu _{4}}}{D_{\lambda }^{\alpha +2}}%
,\ldots ,\frac{k_{\mu _{1}}k_{\mu _{2}}k_{\mu _{3}}k_{\mu _{4}}\ldots k_{\mu
_{2n-1}}k_{\mu _{2n}}}{D_{\lambda }^{\alpha +n}}.
\end{equation*}%
Exclusively even terms are cast since odd ones do not generate non-zero
contributions after integration.

Some configurations of parameters lead to quantities whose integration is
finite. Therefore, establishing a restriction is needed since our targets
are divergent quantities. Given that the investigation occurs in the
physical space-time dimension, we come across the constraint $N=2-\alpha
\geq 0$. Nevertheless, we saw that integrals with quadratic power counting
cancel out, and linearly diverging objects are not allowed as they associate
with odd integrands. That delimits our discussion to logarithmically
divergent quantities, and we set $\alpha =2$.

Our specific goal is to organize them into standard objects. When studying
structures of amplitudes, we established expectations towards the emergence
of surface terms. As we adjusted all structures above so that their
integrals share the power counting, putting them together to obtain these
terms is straightforward. Starting with a 2nd-order tensor, we combine the
first two integrals into the object\footnote{%
Even though we are introducing these objects beforehand, their arising is
automatic when employing resources from IReg. For instance, if one uses the
identity to separate the $AV$ integrand, it will find precisely the
2nd-order surface term when performing the loop integration. Such an outcome
requires only algebraic operations at the integrand level.}%
\begin{equation}
-\int \frac{d^{4}k}{\left( 2\pi \right) ^{4}}\frac{\partial }{\partial
k^{\mu }}\frac{k_{\nu }}{D_{\lambda }^{2}}=\int \frac{d^{4}k}{\left( 2\pi
\right) ^{4}}\left[ \frac{4k_{\mu \nu }}{D_{\lambda }^{3}}-g_{\mu \nu }\frac{%
1}{D_{\lambda }^{2}}\right] .
\end{equation}%
Following the same reasoning, we have the 4th-order tensor%
\begin{equation}
-\int \frac{d^{4}k}{\left( 2\pi \right) ^{4}}\frac{\partial }{\partial
k^{\mu }}\frac{4k_{\nu \alpha \beta }}{D_{\lambda }^{3}}=\int \frac{d^{4}k}{%
\left( 2\pi \right) ^{4}}\left[ \frac{24k_{\mu \nu \alpha \beta }}{%
D_{\lambda }^{4}}-g_{\mu \nu }\frac{4k_{\alpha \beta }}{D_{\lambda }^{3}}%
-g_{\mu \alpha }\frac{4k_{\nu \beta }}{D_{\lambda }^{3}}-g_{\mu \beta }\frac{%
4k_{\nu \alpha }}{D_{\lambda }^{3}}\right] ,
\end{equation}%
where the global numerical factor is an adjustment related to the first
object. We recall the notation introduced to products involving momenta $%
k_{\alpha \beta }=k_{\alpha }k_{\beta }$.

These two cases comprise all elements that arise throughout this
investigation, so we introduce the proper definitions. Concerning the
4th-order tensor, observe that the $\mu $-index has a privileged role with
respect to other indices. We prefer a symmetrized version, taking all
different index configurations into account to introduce the surface term:%
\begin{eqnarray}
\square _{\mu \nu \alpha \beta }\left( \lambda ^{2}\right) &=&\int \frac{%
d^{4}k}{\left( 2\pi \right) ^{4}}\left[ \frac{24k_{\mu \nu \alpha \beta }}{%
D_{\lambda }^{4}}-\frac{1}{2}g_{\mu \nu }\frac{4k_{\alpha \beta }}{%
D_{\lambda }^{3}}-\frac{1}{2}g_{\mu \alpha }\frac{4k_{\nu \beta }}{%
D_{\lambda }^{3}}\right.  \notag \\
&&\left. -\frac{1}{2}g_{\mu \beta }\frac{4k_{\nu \alpha }}{D_{\lambda }^{3}}-%
\frac{1}{2}g_{\nu \alpha }\frac{4k_{\mu \beta }}{D_{\lambda }^{3}}-\frac{1}{2%
}g_{\nu \beta }\frac{4k_{\mu \alpha }}{D_{\lambda }^{3}}-\frac{1}{2}%
g_{\alpha \beta }\frac{4k_{\mu \nu }}{D_{\lambda }^{3}}\right] .  \label{box}
\end{eqnarray}%
Tensors that share the power counting connect, generating one irreducible
object at the end of the process. That means we find 2nd-order tensors
inside the expression above%
\begin{equation}
\Delta _{\mu \nu }\left( \lambda ^{2}\right) =\int \frac{d^{4}k}{\left( 2\pi
\right) ^{4}}\left[ \frac{4k_{\mu \nu }}{D_{\lambda }^{3}}-g_{\mu \nu }\frac{%
1}{D_{\lambda }^{2}}\right]  \label{delta3}
\end{equation}%
and ultimately the mentioned irreducible object arises%
\begin{equation}
I_{\log }\left( \lambda ^{2}\right) =\int \frac{d^{4}k}{\left( 2\pi \right)
^{4}}\frac{1}{D_{\lambda }^{2}}.  \label{Ilog}
\end{equation}%
We omit the argument of divergent objects in the calculations for simplicity
since variations do not appear.

The objects above represent the mathematically ill-defined part of the
results. Differently from finite integrals, we do not evaluate them. In
possession of amplitudes expressions, the analysis of results reflects on
possibilities for these structures. From this perspective, it is feasible to
investigate different prescriptions for their evaluation and the
consequences they bring.

As an example of this reasoning, suppose our aim is computing a specific
amplitude contraction. In general, most relations among Green functions (GF)
arise from pure algebraic operations without further conditions.
Nevertheless, for anomalous amplitudes, one or more relations might depend
on the prescription for evaluating divergences. Beyond that, choosing a
prescription affects the maintenance (or not) of Ward identities (WIs)
corresponding to the same contractions. We aim to clarify how these
constraints relate, inquiring about the role played by divergent objects.

\subsection{Finite Structure Functions - Part I}

\label{Finite1}The systematization involving finite functions is a
fundamental ingredient of the IReg that makes it easier to visualize and
interpret results even in the face of extensive mathematical expressions. We
discuss this subject in three parts directed to structures typical of two,
three, and four-point amplitudes.

Firstly, we focus on objects related to Feynman integrals depending on two
internal lines (\ref{I2}). In any space-time dimension, one-loop
calculations for theories involving equal masses lead to the following
polynomial on the parameter $z$\footnote{%
We deal with a particular form of the polynomial with different masses 
\begin{equation}
Q\left( z\right) =p^{2}z\left( 1-z\right) +\left( m_{1}^{2}-m_{2}^{2}\right)
-m_{1}^{2}.
\end{equation}%
}:%
\begin{equation}
Q\left( z\right) =p^{2}z\left( 1-z\right) -m^{2}.  \label{Q1}
\end{equation}%
Subsection (\ref{Two}) is very detailed in evaluating finite contributions,
clarifying how this polynomial emerges after adopting a Feynman
parametrization. As two-propagator integrals have divergent power counting
in the physical dimension, one initially acknowledges dependence on
non-physical quantities, i.e., arbitrary labels $k_{i}$ and the scale $%
\lambda $. They cancel out in the integration, so only dependence on
physical parameters ultimately remains. For this case, the polynomial
carries the external momentum $p=k_{1}-k_{2}$ and the mass.

The specific family of functions that concern four-dimensional calculations
is%
\begin{equation}
\xi _{a}^{\left( 0\right) }\left( p^{2},m^{2};\lambda ^{2}\right) =\xi
_{a}^{\left( 0\right) }\left( p\right) =\int_{0}^{1}dz\text{ }z^{a}\ln \frac{%
Q\left( z\right) }{-\lambda ^{2}}.  \label{xi0}
\end{equation}%
Since momentum is the only parameter that changes throughout this
investigation, we omit the others from the argument. We even suppress this
information when the dependence is undoubtedly clear.

For our purposes, the integral representation of finite functions is enough.
Nevertheless, if needed, computing them is doable. The first stage of this
task is integrating the function with the lowest parameter power ($a=0$),
which yields%
\begin{equation}
\xi _{0}^{\left( 0\right) }\left( p\right) =\ln \frac{m^{2}}{\lambda ^{2}}-2-%
\frac{1}{2p^{2}}h\left( p^{2},m^{2}\right) .
\end{equation}%
The object $h\left( p^{2},m^{2}\right) $ admits three different
representations depending on the squared momentum value:

\begin{enumerate}
\item In the region where $p^{2}<0$%
\begin{equation}
h\left( p^{2},m^{2}\right) =2\sqrt{4m^{2}-p^{2}}\sqrt{-p^{2}}\ln \left[ 
\frac{\sqrt{4m^{2}-p^{2}}+\sqrt{-p^{2}}}{\sqrt{4m^{2}-p^{2}}-\sqrt{-p^{2}}}%
\right]
\end{equation}

\item In the region where $0<p^{2}<4m^{2}$%
\begin{equation}
h\left( p^{2},m^{2}\right) =-4\sqrt{4m^{2}-p^{2}}\sqrt{p^{2}}\tan ^{-1}\left[
\frac{\sqrt{p^{2}}}{\sqrt{4m^{2}-p^{2}}}\right]
\end{equation}

\item In the region where $p^{2}>4m^{2}$%
\begin{eqnarray}
h\left( p^{2},m^{2}\right) &=&2\sqrt{p^{2}-4m^{2}}\sqrt{p^{2}}\ln \left\{ 
\frac{\sqrt{p^{2}-4m^{2}}+\sqrt{p^{2}}}{\sqrt{p^{2}}-\sqrt{p^{2}-4m^{2}}}%
\right\}  \notag \\
&&+2i\pi \sqrt{p^{2}-4m^{2}}\sqrt{p^{2}}.
\end{eqnarray}
\end{enumerate}

Instead of integrating more complex elements, the idea is to reduce them to
those already known. The main ingredient for such is one identity that
expresses the parameter in terms of the $Q$-polynomial derivative%
\begin{equation}
z=\frac{1}{2}\left[ 1-\frac{1}{p^{2}}\frac{\partial Q\left( z\right) }{%
\partial z}\right] .
\end{equation}%
When replacing this structure within a finite function, the first term
represents another function with decreased parameter power, while the second
corresponds to compensating terms evaluated posteriorly to integration by
parts.

The closest example of this reasoning resides in the element defined by $a=1$%
. Whereas the first term reduces the parameter power to $a=0$, the rest is a
total derivative that vanishes by considering both integration limits%
\begin{equation}
\xi _{1}^{\left( 0\right) }\left( p\right) =\frac{1}{2}\xi _{0}^{\left(
0\right) }\left( p\right) .  \label{red}
\end{equation}%
These instructions also lead to a general expression reducing any
higher-order function ($a\geq 2$) to most elementary ones%
\begin{equation}
\xi _{a}^{\left( 0\right) }\left( p\right) =\frac{a}{a+1}\xi _{a-1}^{\left(
0\right) }\left( p\right) -\frac{a-1}{a+1}\frac{m^{2}}{p^{2}}\xi
_{a-2}^{\left( 0\right) }\left( p\right) +\frac{1}{a+1}\frac{m^{2}}{p^{2}}%
\ln \frac{m^{2}}{\lambda ^{2}}-\frac{1}{a}\frac{a-1}{\left( a+1\right) ^{2}}.
\end{equation}

\subsection{Finite Structure Functions - Part II}

\label{Finite2}The second part of this discussion studies structures related
to Feynman integrals depending on three internal lines (\ref{I3}). Although
two different families arise in this investigation%
\begin{eqnarray}
\xi _{ab}^{\left( -1\right) }\left( p,q,m^{2}\right) &=&\xi _{ab}^{\left(
-1\right) }\left( p,q\right) =\int_{0}^{1}dz\int_{0}^{1-z}dy\text{ }%
y^{b}z^{a}\frac{1}{Q\left( y,z\right) },  \label{xii1} \\
\xi _{ab}^{\left( 0\right) }\left( p,q,m^{2};\lambda ^{2}\right) &=&\xi
_{ab}^{\left( 0\right) }\left( p,q\right) =\int_{0}^{1}dz\int_{0}^{1-z}dy%
\text{ }y^{b}z^{a}\ln \frac{Q\left( y,z\right) }{-\lambda ^{2}},
\label{xii0}
\end{eqnarray}%
the second type appears exclusively for Feynman integrals that are 2nd-order
tensors. Again, we suppress their argument if this is transparent throughout
the discussion. Meanwhile, since different momenta configurations appear
within the box exploration, we resort to the line notation wherever
necessary (\ref{line}). These functions manifest dependence on a polynomial
on Feynman parameters $\left\{ z\text{, }y\right\} $:%
\begin{equation}
Q\left( y,z\right) =p^{2}y\left( 1-y\right) +q^{2}z\left( 1-z\right)
-2\left( p\cdot q\right) yz-m^{2},  \label{Q2}
\end{equation}%
where $p=k_{1}-k_{2}$ and $q=k_{1}-k_{3}$.

Our focus is understanding how to reduce parameter powers in analogy with
the $\xi _{a}$ cases in Section (\ref{Finite1}). By examining both
derivatives, we establish the following relations%
\begin{eqnarray}
2\left[ p^{2}y+\left( p\cdot q\right) z\right] &=&p^{2}-\frac{\partial
Q\left( y,z\right) }{\partial y},  \label{dyq} \\
2\left[ \left( p\cdot q\right) y+q^{2}z\right] &=&q^{2}-\frac{\partial
Q\left( y,z\right) }{\partial z}.  \label{dzq}
\end{eqnarray}%
Notice that both parameters appear together, which indicates reductions
concern the sum of powers $a+b$. When computing Feynman integrals, we will
see this is part of a pattern: finite structure functions do not emerge
randomly but in packages having $a+b$ fixed.

Then, starting with the constraint $a+b=1$, let us examine how functions $%
\xi _{10}^{\left( -1\right) }$ and $\xi _{01}^{\left( -1\right) }$ combine.
When multiplying both sides of the first relation by $Q^{-1}$ and applying
the integration, identifications are straightforward%
\begin{equation}
2\left[ p^{2}\xi _{10}^{\left( -1\right) }+\left( p\cdot q\right) \xi
_{01}^{\left( -1\right) }\right] =p^{2}\xi _{00}^{\left( -1\right)
}-\int_{0}^{1}dz\int_{0}^{1-z}dy\text{ }\frac{\partial }{\partial y}\ln 
\frac{Q\left( y,z\right) }{-\lambda ^{2}}.
\end{equation}%
As it is the objective, parameter powers decreased and now $a+b=0$. The
compensating term is a total derivative; thus, considering the integration
limits allows recognizing two-point finite functions%
\begin{equation}
2\left[ p^{2}\xi _{10}^{\left( -1\right) }+\left( p\cdot q\right) \xi
_{01}^{\left( -1\right) }\right] =p^{2}\xi _{00}^{\left( -1\right) }-\xi
_{0}^{\left( 0\right) }\left( p-q\right) +\xi _{0}^{\left( 0\right) }\left(
q\right) .  \label{p10}
\end{equation}%
Since different momenta configurations are concomitant, their distinction is
crucial.

From the second relation (\ref{dzq}), using the derivative with respect to
the $z$ variable generates an analogous structure%
\begin{equation}
2\left[ \left( p\cdot q\right) \xi _{10}^{\left( -1\right) }+q^{2}\xi
_{01}^{\left( -1\right) }\right] =q^{2}\xi _{00}^{\left( -1\right)
}-\int_{0}^{1}dz\int_{0}^{1-z}dy\text{ }\frac{\partial }{\partial z}\ln 
\frac{Q\left( y,z\right) }{-\lambda ^{2}}.  \label{8}
\end{equation}%
The novelty is that exchanging positions of integral (in $y$) and derivative
(in $z$) is needed before computing the last term. Nevertheless,
difficulties emerge due to the $z$ parameter presence in the integration
limit. Under adequate continuity conditions, Leibniz rule for
differentiation under the integral sign applies%
\begin{equation}
\frac{d}{dz}\int_{a\left( z\right) }^{b\left( z\right) }dy\text{ }f\left(
y,z\right) =\int_{a\left( z\right) }^{b\left( z\right) }dy\text{ }\frac{%
\partial }{\partial z}f\left( y,z\right) -f\left( a\left( z\right) ,z\right) 
\frac{d}{dz}a\left( z\right) +f\left( b\left( z\right) ,z\right) \frac{d}{dz}%
b\left( z\right) .  \label{Leibniz}
\end{equation}%
For the specific case from Equation (\ref{8}), we set limits of integration
and perform their derivatives. When integrating with respect to the $z$
variable, the rule is established%
\begin{equation}
\int_{0}^{1}dz\int_{0}^{1-z}dy\text{ }\frac{\partial }{\partial z}f\left(
y,z\right) =\int_{0}^{1}dz\text{ }f\left( 1-z,z\right) -\int_{0}^{1}dy\text{ 
}f\left( y,0\right) .
\end{equation}%
Lastly, we establish the second reduction after replacing the corresponding
integrand%
\begin{equation}
2\left[ \left( p\cdot q\right) \xi _{10}^{\left( -1\right) }+q^{2}\xi
_{01}^{\left( -1\right) }\right] =q^{2}\xi _{00}^{\left( -1\right) }-\xi
_{0}^{\left( 0\right) }\left( p-q\right) +\xi _{0}^{\left( 0\right) }\left(
p\right) .  \label{q01}
\end{equation}

One might think these reductions have some redundancies since they involve
the same functions, so introducing all of them would be unnecessary. In
truth, they correspond to different properties attributed to the same
object. That is an important feature we will address when computing the
Feynman integrals. To illustrate, observe that both cases derived above
consist of momenta contractions over one vector:%
\begin{eqnarray*}
p^{\mu }\left( p_{\mu }\xi _{10}^{\left( -1\right) }+q_{\mu }\xi
_{01}^{\left( -1\right) }\right) &\rightarrow &\left[ p^{2}\xi _{10}^{\left(
-1\right) }+\left( p\cdot q\right) \xi _{01}^{\left( -1\right) }\right] , \\
q^{\mu }\left( p_{\mu }\xi _{10}^{\left( -1\right) }+q_{\mu }\xi
_{01}^{\left( -1\right) }\right) &\rightarrow &\left[ \left( p\cdot q\right)
\xi _{10}^{\left( -1\right) }+q^{2}\xi _{01}^{\left( -1\right) }\right] .
\end{eqnarray*}

Aiming for reductions involving $a+b=2$, multiply each relation from
Equations (\ref{dyq})-(\ref{dzq}) by each Feynman parameter. Hence,
following the line of reasoning employed in previous cases yields%
\begin{eqnarray}
2\left[ p^{2}\xi _{20}^{\left( -1\right) }+\left( p\cdot q\right) \xi
_{11}^{\left( -1\right) }\right] &=&p^{2}\xi _{10}^{\left( -1\right) }+\xi
_{00}^{\left( 0\right) }-\frac{1}{2}\xi _{0}^{\left( 0\right) }\left(
p-q\right) ,  \label{p20} \\
2\left[ p^{2}\xi _{11}^{\left( -1\right) }+\left( p\cdot q\right) \xi
_{02}^{\left( -1\right) }\right] &=&p^{2}\xi _{01}^{\left( -1\right) }-\frac{%
1}{2}\xi _{0}^{\left( 0\right) }\left( p-q\right) +\frac{1}{2}\xi
_{0}^{\left( 0\right) }\left( q\right) ,  \label{p11} \\
2\left[ \left( p\cdot q\right) \xi _{20}^{\left( -1\right) }+q^{2}\xi
_{11}^{\left( -1\right) }\right] &=&q^{2}\xi _{10}^{\left( -1\right) }-\frac{%
1}{2}\xi _{0}^{\left( 0\right) }\left( p-q\right) +\frac{1}{2}\xi
_{0}^{\left( 0\right) }\left( p\right) , \\
2\left[ \left( p\cdot q\right) \xi _{11}^{\left( -1\right) }+q^{2}\xi
_{02}^{\left( -1\right) }\right] &=&q^{2}\xi _{01}^{\left( -1\right) }+\xi
_{00}^{\left( 0\right) }-\frac{1}{2}\xi _{0}^{\left( 0\right) }\left(
p-q\right) .  \label{q02}
\end{eqnarray}%
Although the function $\xi _{1}^{\left( 0\right) }$ appears with this
procedure, we have already performed its reduction (\ref{red}).

Observe that the $\xi _{00}^{\left( 0\right) }$ emerged as compensation for
integration by parts. Starting from the expression%
\begin{equation*}
\frac{1}{2}=\int_{0}^{1}dz\int_{0}^{1-z}dy\frac{Q\left( y,z\right) }{Q\left(
y,z\right) },
\end{equation*}%
let us expand the numerator to identify some reductions above. Their
replacement produces a similar relation%
\begin{equation}
2\xi _{00}^{\left( 0\right) }=p^{2}\xi _{10}^{\left( -1\right) }+q^{2}\xi
_{01}^{\left( -1\right) }+\xi _{0}^{\left( 0\right) }\left( p-q\right)
-2m^{2}\xi _{00}^{\left( -1\right) }-1.  \label{x00}
\end{equation}%
We stress two unusual contributions here, i.e., the term proportional to the
squared mass and the constant. They will play relevant roles in this
investigation, so we return to them eventually.

\subsection{Finite Structure Functions - Part III}

\label{Finite3}The last part of this discussion surveys structures related
to Feynman integrals depending on four internal lines (\ref{I4}), which
consist of three families of finite functions:%
\begin{eqnarray}
\xi _{abc}^{\left( -2\right) }\left( p,q,r\right)
&=&\int_{0}^{1}dz\int_{0}^{1-z}dy\int_{0}^{1-y-z}dx\text{ }x^{c}y^{b}z^{a}%
\frac{1}{\left[ Q\left( x,y,z\right) \right] ^{2}},  \label{xiii2} \\
\xi _{abc}^{\left( -1\right) }\left( p,q,r\right)
&=&\int_{0}^{1}dz\int_{0}^{1-z}dy\int_{0}^{1-y-z}dx\text{ }x^{c}y^{b}z^{a}%
\frac{1}{Q\left( x,y,z\right) },  \label{xiii1} \\
\xi _{abc}^{\left( 0\right) }\left( p,q,r\right)
&=&\int_{0}^{1}dz\int_{0}^{1-z}dy\int_{0}^{1-y-z}dx\text{ }%
x^{c}y^{b}z^{a}\ln \frac{Q\left( x,y,z\right) }{-\lambda ^{2}}.
\label{xiii0}
\end{eqnarray}%
Although they depend on the mass and possibly the scale, our notation omits
this information. These functions contain a new polynomial depending on
three Feynman parameters $\left\{ z\text{, }y\text{, }x\right\} $, whose
expression is%
\begin{eqnarray}
Q\left( x,y,z\right) &=&p^{2}x\left( 1-x\right) +q^{2}y\left( 1-y\right)
+r^{2}z\left( 1-z\right)  \notag \\
&&-2\left( p\cdot q\right) xy-2\left( q\cdot r\right) yz-2\left( p\cdot
r\right) xz-m^{2},  \label{Q3}
\end{eqnarray}%
where $p=k_{1}-k_{2}$, $q=k_{1}-k_{3}$, and $r=k_{1}-k_{4}$.

Since reducing combinations of finite functions is our primary objective, we
perform derivatives of this polynomial to build up relations\ among
parameters%
\begin{eqnarray}
\left[ p^{2}x+\left( p\cdot q\right) y+\left( p\cdot r\right) z\right] &=&%
\frac{1}{2}p^{2}-\frac{1}{2}\frac{\partial Q\left( x,y,z\right) }{\partial x}%
,  \label{rela1} \\
\left[ \left( p\cdot q\right) x+q^{2}y+\left( q\cdot r\right) z\right] &=&%
\frac{1}{2}q^{2}-\frac{1}{2}\frac{\partial Q\left( x,y,z\right) }{\partial y}%
, \\
\left[ \left( p\cdot r\right) x+\left( q\cdot r\right) y+r^{2}z\right] &=&%
\frac{1}{2}r^{2}-\frac{1}{2}\frac{\partial Q\left( x,y,z\right) }{\partial z}%
.  \label{rela3}
\end{eqnarray}%
These fundamental elements shape results when inserting the adequate
multiplicative factors and performing the integration. The first term on the
right-hand side represents a decrease in the parameter power.
Identifications of four-point functions are straightforward in this
procedure, even when they come from integration by parts.

On the other hand, evaluating the other terms might require permutations
among derivatives and integrals. The Leibniz rule for differentiation under
the integral sign (\ref{Leibniz}) applies in these cases. Beforehand, we
summarize these possibilities through the following set of rules:%
\begin{eqnarray}
&&\int_{0}^{1}dz\int_{0}^{1-z}dy\int_{0}^{1-y-z}dx\text{ }\frac{\partial }{%
\partial x}f\left( x,y,z\right)  \notag \\
&=&\int_{0}^{1}dz\int_{0}^{1-z}dy\text{ }f\left( 1-y-z,y,z\right)
-\int_{0}^{1}dz\int_{0}^{1-z}dy\text{ }f\left( 0,y,z\right) ,
\end{eqnarray}%
\begin{eqnarray}
&&\int_{0}^{1}dz\int_{0}^{1-z}dy\int_{0}^{1-y-z}dx\text{ }\frac{\partial }{%
\partial y}f\left( x,y,z\right)  \notag \\
&=&\int_{0}^{1}dz\int_{0}^{1-z}dy\text{ }f\left( 1-y-z,y,z\right)
-\int_{0}^{1}dz\int_{0}^{1-z}dx\text{ }f\left( x,0,z\right) ,
\end{eqnarray}%
\begin{eqnarray}
&&\int_{0}^{1}dz\int_{0}^{1-z}dy\int_{0}^{1-y-z}dx\text{ }\frac{\partial }{%
\partial z}f\left( x,y,z\right)  \notag \\
&=&\int_{0}^{1}dz\int_{0}^{1-z}dy\text{ }f\left( 1-y-z,y,z\right)
-\int_{0}^{1}dy\int_{0}^{1-y}dx\text{ }f\left( x,y,0\right) .
\end{eqnarray}

We use them to suppress derivatives and then find quantities considered
typical of calculations related to three-point amplitudes. Even though they
only admit dependence on two external momenta, three are available in the
box context. That means different momenta configurations appear mixed inside
each reduction. Below, we reintroduce the line notation (\ref{line}) by
considering these new ingredients. All equations above exhibit the same type
of object as the first term on the right-hand side, whose identifications
lead to $\xi _{ab}^{\prime \prime \prime }$-like functions. In contrast, the
second term varies; the associations occur respectively with $\xi
_{ab}^{\prime \prime }$, $\xi _{ab}^{\prime }$, and $\xi _{ab}$.%
\begin{equation}
\left\{ 
\begin{array}{l}
\xi _{ab}\rightarrow \xi _{ab}\left( p,q\right) \\ 
\xi _{ab}^{\prime }\rightarrow \xi _{ab}\left( p,r\right) \\ 
\xi _{ab}^{\prime \prime }\rightarrow \xi _{ab}\left( q,r\right) \\ 
\xi _{ab}^{\prime \prime \prime }\rightarrow \xi _{ab}\left( q-p,r-p\right)%
\end{array}%
\right.  \label{line2}
\end{equation}

Without further delay, we cast the required reductions in the sequence.
Their presentation is divided accordingly with the sum of parameter powers,
while subdivisions indicate the relation used in each calculation (\ref%
{rela1})-(\ref{rela3}).

\begin{itemize}
\item Constraint $a+b+c=1$ - Functions $\xi _{abc}^{\left( -2\right) }$ are
typical of four-dimensional calculations, appearing within all Feynman
integrals involving four propagators. For this constraint, one considers the
structure $Q^{-2}$ in the relations.
\end{itemize}

\textbf{First relation}%
\begin{equation}
2\left[ p^{2}\xi _{100}^{\left( -2\right) }+\left( p\cdot q\right) \xi
_{010}^{\left( -2\right) }+\left( p\cdot r\right) \xi _{001}^{\left(
-2\right) }\right] =p^{2}\xi _{000}^{\left( -2\right) }+\left[ \xi
_{00}^{\left( -1\right) }\right] ^{\prime \prime \prime }-\left[ \xi
_{00}^{\left( -1\right) }\right] ^{\prime \prime }  \label{rel1}
\end{equation}

\textbf{Second relation}%
\begin{equation}
2\left[ \left( p\cdot q\right) \xi _{100}^{\left( -2\right) }+q^{2}\xi
_{010}^{\left( -2\right) }+\left( q\cdot r\right) \xi _{001}^{\left(
-2\right) }\right] =q^{2}\xi _{000}^{\left( -2\right) }+\left[ \xi
_{00}^{\left( -1\right) }\right] ^{\prime \prime \prime }-\left[ \xi
_{00}^{\left( -1\right) }\right] ^{\prime }
\end{equation}

\textbf{Third relation}%
\begin{equation}
2\left[ \left( p\cdot r\right) \xi _{100}^{\left( -2\right) }+\left( q\cdot
r\right) \xi _{010}^{\left( -2\right) }+r^{2}\xi _{001}^{\left( -2\right) }%
\right] =r^{2}\xi _{000}^{\left( -2\right) }+\left[ \xi _{00}^{\left(
-1\right) }\right] ^{\prime \prime \prime }-\left[ \xi _{00}^{\left(
-1\right) }\right]  \label{rel3}
\end{equation}

\begin{itemize}
\item Constraint $a+b+c=2$ - Besides the structure $Q^{-2}$, multiplicative
factors also consider each of the Feynman parameters $\left\{ x\text{, }y%
\text{, }z\right\} $. We adopt the symbol $\xi _{\text{\textbf{one}}%
}^{\left( -1\right) }=\xi _{00}^{\left( -1\right) }-\xi _{10}^{\left(
-1\right) }-\xi _{01}^{\left( -1\right) }$ to improve the visualization.
\end{itemize}

\textbf{First relation}%
\begin{eqnarray}
2\left[ p^{2}\xi _{200}^{\left( -2\right) }+\left( p\cdot q\right) \xi
_{110}^{\left( -2\right) }+\left( p\cdot r\right) \xi _{101}^{\left(
-2\right) }\right] &=&p^{2}\xi _{100}^{\left( -2\right) }-\xi _{000}^{\left(
-1\right) }+\left[ \xi _{\text{\textbf{one}}}^{\left( -1\right) }\right]
^{\prime \prime \prime } \\
2\left[ p^{2}\xi _{110}^{\left( -2\right) }+\left( p\cdot q\right) \xi
_{020}^{\left( -2\right) }+\left( p\cdot r\right) \xi _{011}^{\left(
-2\right) }\right] &=&p^{2}\xi _{010}^{\left( -2\right) }+\left[ \xi
_{10}^{\left( -1\right) }\right] ^{\prime \prime \prime }-\left[ \xi
_{10}^{\left( -1\right) }\right] ^{\prime \prime } \\
2\left[ p^{2}\xi _{101}^{\left( -2\right) }+\left( p\cdot q\right) \xi
_{011}^{\left( -2\right) }+\left( p\cdot r\right) \xi _{002}^{\left(
-2\right) }\right] &=&p^{2}\xi _{001}^{\left( -2\right) }+\left[ \xi
_{01}^{\left( -1\right) }\right] ^{\prime \prime \prime }-\left[ \xi
_{01}^{\left( -1\right) }\right] ^{\prime \prime }
\end{eqnarray}

\textbf{Second relation}%
\begin{eqnarray}
2\left[ \left( p\cdot q\right) \xi _{200}^{\left( -2\right) }+q^{2}\xi
_{110}^{\left( -2\right) }+\left( q\cdot r\right) \xi _{101}^{\left(
-2\right) }\right] &=&q^{2}\xi _{100}^{\left( -2\right) }-\left[ \xi
_{10}^{\left( -1\right) }\right] ^{\prime }+\left[ \xi _{\text{\textbf{one}}%
}^{\left( -1\right) }\right] ^{\prime \prime \prime } \\
2\left[ \left( p\cdot q\right) \xi _{110}^{\left( -2\right) }+q^{2}\xi
_{020}^{\left( -2\right) }+\left( q\cdot r\right) \xi _{011}^{\left(
-2\right) }\right] &=&q^{2}\xi _{010}^{\left( -2\right) }-\xi _{000}^{\left(
-1\right) }+\left[ \xi _{10}^{\left( -1\right) }\right] ^{\prime \prime
\prime } \\
2\left[ \left( p\cdot q\right) \xi _{101}^{\left( -2\right) }+q^{2}\xi
_{011}^{\left( -2\right) }+\left( q\cdot r\right) \xi _{002}^{\left(
-2\right) }\right] &=&q^{2}\xi _{001}^{\left( -2\right) }+\left[ \xi
_{01}^{\left( -1\right) }\right] ^{\prime \prime \prime }-\left[ \xi
_{01}^{\left( -1\right) }\right] ^{\prime }
\end{eqnarray}

\textbf{Third relation}%
\begin{eqnarray}
2\left[ \left( p\cdot r\right) \xi _{200}^{\left( -2\right) }+\left( q\cdot
r\right) \xi _{110}^{\left( -2\right) }+r^{2}\xi _{101}^{\left( -2\right) }%
\right] &=&r^{2}\xi _{100}^{\left( -2\right) }+\left[ \xi _{\text{\textbf{one%
}}}^{\left( -1\right) }\right] ^{\prime \prime \prime }-\left[ \xi
_{10}^{\left( -1\right) }\right] \\
2\left[ \left( p\cdot r\right) \xi _{110}^{\left( -2\right) }+\left( q\cdot
r\right) \xi _{020}^{\left( -2\right) }+r^{2}\xi _{011}^{\left( -2\right) }%
\right] &=&r^{2}\xi _{010}^{\left( -2\right) }+\left[ \xi _{10}^{\left(
-1\right) }\right] ^{\prime \prime \prime }-\left[ \xi _{01}^{\left(
-1\right) }\right] \\
2\left[ \left( p\cdot r\right) \xi _{101}^{\left( -2\right) }+\left( q\cdot
r\right) \xi _{011}^{\left( -2\right) }+r^{2}\xi _{002}^{\left( -2\right) }%
\right] &=&r^{2}\xi _{001}^{\left( -2\right) }-\xi _{000}^{\left( -1\right)
}+\left[ \xi _{01}^{\left( -1\right) }\right] ^{\prime \prime \prime }
\end{eqnarray}

\begin{itemize}
\item Constraint $a+b+c=3$ - Besides the structure $Q^{-2}$, multiplicative
factors also consider each of the combinations $\left\{ x^{2}\text{, }xy%
\text{, }xz\text{, }y^{2}\text{, }yz\text{, }z^{2}\right\} $. Since they
appear when computing tensor integrals, $\xi _{abc}^{\left( -1\right) }$%
-type functions are considered here and correspond to the structure $Q^{-1}$%
. This time, we adopt the symbols $\xi _{\text{\textbf{one}}}^{\left(
-1\right) }=\xi _{00}^{\left( -1\right) }-\xi _{10}^{\left( -1\right) }-\xi
_{01}^{\left( -1\right) }$ and $\xi _{\text{\textbf{two}}}^{\left( -1\right)
}=\xi _{00}^{\left( -1\right) }-2\xi _{10}^{\left( -1\right) }-2\xi
_{01}^{\left( -1\right) }+2\xi _{11}^{\left( -1\right) }+\xi _{20}^{\left(
-1\right) }+\xi _{02}^{\left( -1\right) }$ to improve the visualization.
\end{itemize}

\textbf{First relation}%
\begin{eqnarray}
2\left[ p^{2}\xi _{300}^{\left( -2\right) }+\left( p\cdot q\right) \xi
_{210}^{\left( -2\right) }+\left( p\cdot r\right) \xi _{201}^{\left(
-2\right) }\right] &=&p^{2}\xi _{200}^{\left( -2\right) }-2\xi
_{100}^{\left( -1\right) }+\left[ \xi _{\text{\textbf{two}}}^{\left(
-1\right) }\right] ^{\prime \prime \prime } \\
2\left[ p^{2}\xi _{210}^{\left( -2\right) }+\left( p\cdot q\right) \xi
_{120}^{\left( -2\right) }+\left( p\cdot r\right) \xi _{111}^{\left(
-2\right) }\right] &=&p^{2}\xi _{110}^{\left( -2\right) }-\xi _{010}^{\left(
-1\right) }+\left[ \xi _{\text{\textbf{one}}}^{\left( -1\right) }\right]
^{\prime \prime \prime } \\
2\left[ p^{2}\xi _{201}^{\left( -2\right) }+\left( p\cdot q\right) \xi
_{111}^{\left( -2\right) }+\left( p\cdot r\right) \xi _{102}^{\left(
-2\right) }\right] &=&p^{2}\xi _{101}^{\left( -2\right) }-\xi _{001}^{\left(
-1\right) }+\left[ \xi _{\text{\textbf{one}}}^{\left( -1\right) }\right]
^{\prime \prime \prime } \\
2\left[ p^{2}\xi _{120}^{\left( -2\right) }+\left( p\cdot q\right) \xi
_{030}^{\left( -2\right) }+\left( p\cdot r\right) \xi _{021}^{\left(
-2\right) }\right] &=&p^{2}\xi _{020}^{\left( -2\right) }+\left[ \xi
_{20}^{\left( -1\right) }\right] ^{\prime \prime \prime }-\left[ \xi
_{20}^{\left( -1\right) }\right] ^{\prime \prime } \\
2\left[ p^{2}\xi _{111}^{\left( -2\right) }+\left( p\cdot q\right) \xi
_{021}^{\left( -2\right) }+\left( p\cdot r\right) \xi _{012}^{\left(
-2\right) }\right] &=&p^{2}\xi _{011}^{\left( -2\right) }+\left[ \xi
_{11}^{\left( -1\right) }\right] ^{\prime \prime \prime }-\left[ \xi
_{11}^{\left( -1\right) }\right] ^{\prime \prime } \\
2\left[ p^{2}\xi _{102}^{\left( -2\right) }+\left( p\cdot q\right) \xi
_{012}^{\left( -2\right) }+\left( p\cdot r\right) \xi _{003}^{\left(
-2\right) }\right] &=&p^{2}\xi _{002}^{\left( -2\right) }+\left[ \xi
_{02}^{\left( -1\right) }\right] ^{\prime \prime \prime }-\left[ \xi
_{02}^{\left( -1\right) }\right] ^{\prime \prime } \\
2\left[ p^{2}\xi _{100}^{\left( -1\right) }+\left( p\cdot q\right) \xi
_{010}^{\left( -1\right) }+\left( p\cdot r\right) \xi _{001}^{\left(
-1\right) }\right] &=&p^{2}\xi _{000}^{\left( -1\right) }-\left[ \xi
_{00}^{\left( 0\right) }\right] ^{\prime \prime \prime }+\left[ \xi
_{00}^{\left( 0\right) }\right] ^{\prime \prime }
\end{eqnarray}

\textbf{Second relation}%
\begin{eqnarray}
2\left[ \left( p\cdot q\right) \xi _{300}^{\left( -2\right) }+q^{2}\xi
_{210}^{\left( -2\right) }+\left( r\cdot q\right) \xi _{201}^{\left(
-2\right) }\right] &=&q^{2}\xi _{200}^{\left( -2\right) }+\left[ \xi _{\text{%
\textbf{two}}}^{\left( -1\right) }\right] ^{\prime \prime \prime }-\left[
\xi _{20}^{\left( -1\right) }\right] ^{\prime } \\
2\left[ \left( p\cdot q\right) \xi _{210}^{\left( -2\right) }+q^{2}\xi
_{120}^{\left( -2\right) }+\left( r\cdot q\right) \xi _{111}^{\left(
-2\right) }\right] &=&q^{2}\xi _{110}^{\left( -2\right) }-\xi _{100}^{\left(
-1\right) }+\left[ \xi _{\text{\textbf{one}}}^{\left( -1\right) }\right]
^{\prime \prime \prime } \\
2\left[ \left( p\cdot q\right) \xi _{201}^{\left( -2\right) }+q^{2}\xi
_{111}^{\left( -2\right) }+\left( r\cdot q\right) \xi _{102}^{\left(
-2\right) }\right] &=&q^{2}\xi _{101}^{\left( -2\right) }+\left[ \xi _{\text{%
\textbf{one}}}^{\left( -1\right) }\right] ^{\prime \prime \prime }-\left[
\xi _{11}^{\left( -1\right) }\right] ^{\prime } \\
2\left[ \left( p\cdot q\right) \xi _{120}^{\left( -2\right) }+q^{2}\xi
_{030}^{\left( -2\right) }+\left( r\cdot q\right) \xi _{021}^{\left(
-2\right) }\right] &=&q^{2}\xi _{020}^{\left( -2\right) }-2\xi
_{010}^{\left( -1\right) }+\left[ \xi _{20}^{\left( -1\right) }\right]
^{\prime \prime \prime } \\
2\left[ \left( p\cdot q\right) \xi _{111}^{\left( -2\right) }+q^{2}\xi
_{021}^{\left( -2\right) }+\left( r\cdot q\right) \xi _{012}^{\left(
-2\right) }\right] &=&q^{2}\xi _{011}^{\left( -2\right) }-\xi _{001}^{\left(
-1\right) }+\left[ \xi _{11}^{\left( -1\right) }\right] ^{\prime \prime
\prime } \\
2\left[ \left( p\cdot q\right) \xi _{102}^{\left( -2\right) }+q^{2}\xi
_{012}^{\left( -2\right) }+\left( r\cdot q\right) \xi _{003}^{\left(
-2\right) }\right] &=&q^{2}\xi _{002}^{\left( -2\right) }+\left[ \xi
_{02}^{\left( -1\right) }\right] ^{\prime \prime \prime }-\left[ \xi
_{02}^{\left( -1\right) }\right] ^{\prime } \\
2\left[ \left( p\cdot q\right) \xi _{100}^{\left( -1\right) }+q^{2}\xi
_{010}^{\left( -1\right) }+\left( r\cdot q\right) \xi _{001}^{\left(
-1\right) }\right] &=&q^{2}\xi _{000}^{\left( -1\right) }-\left[ \xi
_{00}^{\left( 0\right) }\right] ^{\prime \prime \prime }+\left[ \xi
_{00}^{\left( 0\right) }\right] ^{\prime }
\end{eqnarray}

\textbf{Third relation}%
\begin{eqnarray}
2\left[ \left( p\cdot r\right) \xi _{300}^{\left( -2\right) }+\left( q\cdot
r\right) \xi _{210}^{\left( -2\right) }+r^{2}\xi _{201}^{\left( -2\right) }%
\right] &=&r^{2}\xi _{200}^{\left( -2\right) }+\left[ \xi _{\text{\textbf{two%
}}}^{\left( -1\right) }\right] ^{\prime \prime \prime }-\left[ \xi
_{20}^{\left( -1\right) }\right] \\
2\left[ \left( p\cdot r\right) \xi _{210}^{\left( -2\right) }+\left( q\cdot
r\right) \xi _{120}^{\left( -2\right) }+r^{2}\xi _{111}^{\left( -2\right) }%
\right] &=&r^{2}\xi _{110}^{\left( -2\right) }+\left[ \xi _{\text{\textbf{one%
}}}^{\left( -1\right) }\right] ^{\prime \prime \prime }-\left[ \xi
_{11}^{\left( -1\right) }\right] \\
2\left[ \left( p\cdot r\right) \xi _{201}^{\left( -2\right) }+\left( q\cdot
r\right) \xi _{111}^{\left( -2\right) }+r^{2}\xi _{102}^{\left( -2\right) }%
\right] &=&r^{2}\xi _{101}^{\left( -2\right) }-\xi _{100}^{\left( -1\right)
}+\left[ \xi _{\text{\textbf{one}}}^{\left( -1\right) }\right] ^{\prime
\prime \prime } \\
2\left[ \left( p\cdot r\right) \xi _{120}^{\left( -2\right) }+\left( q\cdot
r\right) \xi _{030}^{\left( -2\right) }+r^{2}\xi _{021}^{\left( -2\right) }%
\right] &=&r^{2}\xi _{020}^{\left( -2\right) }+\left[ \xi _{20}^{\left(
-1\right) }\right] ^{\prime \prime \prime }-\left[ \xi _{02}^{\left(
-1\right) }\right] \\
2\left[ \left( p\cdot r\right) \xi _{111}^{\left( -2\right) }+\left( q\cdot
r\right) \xi _{021}^{\left( -2\right) }+r^{2}\xi _{012}^{\left( -2\right) }%
\right] &=&r^{2}\xi _{011}^{\left( -2\right) }-\xi _{010}^{\left( -1\right)
}+\left[ \xi _{11}^{\left( -1\right) }\right] ^{\prime \prime \prime } \\
2\left[ \left( p\cdot r\right) \xi _{102}^{\left( -2\right) }+\left( q\cdot
r\right) \xi _{012}^{\left( -2\right) }+r^{2}\xi _{003}^{\left( -2\right) }%
\right] &=&r^{2}\xi _{002}^{\left( -2\right) }-2\xi _{001}^{\left( -1\right)
}+\left[ \xi _{02}^{\left( -1\right) }\right] ^{\prime \prime \prime } \\
2\left[ \left( p\cdot r\right) \xi _{100}^{\left( -1\right) }+\left( q\cdot
r\right) \xi _{010}^{\left( -1\right) }+r^{2}\xi _{001}^{\left( -1\right) }%
\right] &=&r^{2}\xi _{000}^{\left( -1\right) }-\left[ \xi _{00}^{\left(
0\right) }\right] ^{\prime \prime \prime }+\left[ \xi _{00}^{\left( 0\right)
}\right]
\end{eqnarray}

Analogously to the $\xi _{00}^{\left( 0\right) }$, whose analysis was
developed in Equation (\ref{x00}), expressing $\xi _{abc}^{-1}$-like
functions in terms of $\xi _{abc}^{-2}$-like functions is convenient. To
accomplish this task, employ $Q^{-1}=Q^{-2}Q$ as a link between both
families. Following the procedure from the referred case and using $\xi _{%
\text{\textbf{one}}}^{\left( -1\right) }=\xi _{00}^{\left( -1\right) }-\xi
_{10}^{\left( -1\right) }-\xi _{01}^{\left( -1\right) }$, we obtain the
relations that concern this investigation:%
\begin{eqnarray}
\xi _{000}^{\left( -1\right) } &=&2m^{2}\xi _{000}^{\left( -2\right) }-\left[
p^{2}\xi _{100}^{\left( -2\right) }+q^{2}\xi _{010}^{\left( -2\right)
}+r^{2}\xi _{001}^{\left( -2\right) }\right] +\left[ \xi _{00}^{\left(
-1\right) }\right] ^{\prime \prime \prime },  \label{000} \\
2\xi _{100}^{\left( -1\right) } &=&2m^{2}\xi _{100}^{\left( -2\right) }- 
\left[ p^{2}\xi _{200}^{\left( -2\right) }+q^{2}\xi _{110}^{\left( -2\right)
}+r^{2}\xi _{101}^{\left( -2\right) }\right] +\left[ \xi _{\text{\textbf{one}%
}}^{\left( -1\right) }\right] ^{\prime \prime \prime }, \\
2\xi _{010}^{\left( -1\right) } &=&2m^{2}\xi _{010}^{\left( -2\right) }- 
\left[ p^{2}\xi _{110}^{\left( -2\right) }+q^{2}\xi _{020}^{\left( -2\right)
}+r^{2}\xi _{011}^{\left( -2\right) }\right] +\left[ \xi _{10}^{\left(
-1\right) }\right] ^{\prime \prime \prime }, \\
2\xi _{001}^{\left( -1\right) } &=&2m^{2}\xi _{001}^{\left( -2\right) }- 
\left[ p^{2}\xi _{101}^{\left( -2\right) }+q^{2}\xi _{011}^{\left( -2\right)
}+r^{2}\xi _{002}^{\left( -2\right) }\right] +\left[ \xi _{01}^{\left(
-1\right) }\right] ^{\prime \prime \prime }.
\end{eqnarray}

\newpage

\section{Explicit Perturbative Amplitudes}

\label{Integral}After understanding the structure of correlators at the
integrand level, we developed a strategy to deal with divergences associated
with their integration. The objective of this section is to perform this
operation explicitly. For each case, the first step is evaluating Feynman
integrals since these are the fundamental pieces that build up the
investigated objects. Subsequently, we obtain standard tensors and
perturbative amplitudes hitherto identified.

\subsection{Two-Point Amplitudes - Feynman Integrals and $AV$}

\label{Two}Our task is to compute quantities introduced in Subsection (\ref%
{2pt}), with a particular interest in the $AV$ correlator. That is also the
opportunity to elucidate elements related to the strategy. After detailing
the procedure for the separation, we organize ill-defined mathematical
quantities through standard divergent objects. Posteriorly, we evaluate
finite contributions using common tools of perturbative calculations, such
as Feynman parametrizations and finite loop integration. One might consult
further information about these resources in introductory books on quantum
field theories \cite{Weinberg}.

We achieved the $AV$ structure in Equation (\ref{av}) through a contraction
with the standard tensor (\ref{t2s}). Considering the antisymmetric
character of the Levi-Civita symbol, the simplified integrand arises%
\begin{equation}
t_{\mu \nu }^{AV}=4i\varepsilon _{\mu \nu \alpha \beta }\left[ k_{1}^{\alpha
}k_{2}^{\beta }\frac{1}{D_{12}}+\left( k_{1}-k_{2}\right) ^{\alpha }\frac{%
k^{\beta }}{D_{12}}\right] .
\end{equation}%
Denoted by an uppercase letter, the amplitude combines the following
two-propagator Feynman integrals (\ref{I2}):%
\begin{equation}
T_{\mu \nu }^{AV}=4i\varepsilon _{\mu \nu \alpha \beta }\left[ k_{1}^{\alpha
}k_{2}^{\beta }I_{2}+\left( k_{1}-k_{2}\right) ^{\alpha }I_{2}^{\beta }%
\right] .  \label{AVI}
\end{equation}

Since this expression exhibits a divergent power counting, we adopt a
prescription to propagator-like objects $D_{n}$ through identity (\ref{id}).
The separation is successful if the identity considers $N$ as equal to or
higher than the power counting of the integral. Thus, $N=2$ would be a
logical option as two-point amplitudes have quadratic power counting in the
physical dimension. Nevertheless, we acknowledged simplifications due to the
antisymmetric character of the $AV$, which allows using the $N=1$ version.
Although both routes lead to the same outcome, the first generates more
finite contributions and involves more laborious calculations.

Alternatively, one might also evaluate Feynman integrals separately,
adopting versions for the identity as it finds suitable. We opt for this
route because these integrals also emerge within higher-order amplitudes.
For instance, as the $I_{2}$ integral exhibits logarithmic power counting
when integrated, employing the $N=0$ identity version rewrites the
propagator-like structure $D_{1}$ and splits its integrand as follows%
\begin{equation}
\frac{1}{D_{12}}=\left[ \frac{1}{D_{\lambda }}-\frac{A_{1}}{D_{\lambda }D_{1}%
}\right] \frac{1}{D_{2}},
\end{equation}%
where denominators involve $D_{n}=\left( k+k_{n}\right) ^{2}-m^{2}$ and $%
D_{\lambda }=k^{2}-\lambda ^{2}$ and numerators exhibit the object $%
A_{n}=2k_{n}\cdot k+k_{n}^{2}+\lambda ^{2}-m^{2}$.

Power counting decreased as required, so the last contribution will generate
a finite integral. Nevertheless, the first term still shows diverging power
counting (when integrated). Exploring both propagator-like objects is
necessary for this term, so divergent objects depend only on non-physical
quantities, i.e., the loop momentum $k$ and the scale $\lambda ^{2}$. Such a
property is intrinsic to the IReg. Therefore, by employing the identity for $%
D_{2}$ within this specific term, the separation assumes the form%
\begin{equation}
\frac{1}{D_{12}}-\frac{1}{D_{\lambda }^{2}}=-\frac{A_{2}}{D_{\lambda
}^{2}D_{2}}-\frac{A_{1}}{D_{\lambda }D_{12}}.
\end{equation}

This organization puts ill-defined mathematical structures on the left-hand
side of equations after integration, so it is transparent that the
right-hand side leads to a finite quantity. Therefore, by identifying the
irreducible divergent object (\ref{Ilog}), we have the $I_{2}$ integral:%
\begin{equation}
I_{2}\left( k_{1},k_{2}\right) -I_{\log }=-\int \frac{d^{4}k}{\left( 2\pi
\right) ^{4}}\left[ \frac{A_{2}}{D_{\lambda }^{2}D_{2}}+\frac{A_{1}}{%
D_{\lambda }D_{12}}\right] .  \label{i20}
\end{equation}

Our next task is to compute the finite part; however, dealing with products
in the denominators is inconvenient. One generally rewrites these structures
through Feynman parametrizations to avoid such circumstances. This resource
expresses rational functions in terms of an integral representation; observe
the examples:%
\begin{equation}
\frac{1}{ab}=\int_{0}^{1}dz\frac{1}{\left[ \left( b-a\right) z+a\right] ^{2}}%
,  \label{fey1}
\end{equation}%
\begin{equation}
\frac{1}{abc}=2\int_{0}^{1}dz\int_{0}^{1-z}dy\frac{1}{\left[ \left(
b-a\right) y+\left( c-a\right) z+a\right] ^{3}},  \label{fey2}
\end{equation}%
\begin{equation}
\frac{1}{abcd}=6\int_{0}^{1}dz\int_{0}^{1-z}dy\int_{0}^{1-y-z}dx\frac{1}{%
\left[ \left( b-a\right) x+\left( c-a\right) y+\left( d-a\right) z+a\right]
^{4}},  \label{fey3}
\end{equation}%
where $x$, $y$, and $z$ are parameters. Variations employed here emerge
through derivatives with respect to $a$, which increases its power on the
left-hand side.

Let us clarify this subject by adopting a variation of (\ref{fey1}) to
express the first finite contribution from the scalar integral (\ref{i20}).
After replacing $a=D_{\lambda }$ and $b=D_{2}$, we group terms on the loop
momentum by completing the square%
\begin{equation}
-\int \frac{d^{4}k}{\left( 2\pi \right) ^{4}}\frac{A_{2}}{D_{\lambda
}^{2}D_{2}}=-2\int_{0}^{1}dz\left( 1-z\right) \int \frac{d^{4}k}{\left( 2\pi
\right) ^{4}}\frac{A_{2}}{\left[ \left( k+k_{2}z\right) ^{2}+P_{1}\right]
^{3}};
\end{equation}%
therefore, one polynomial dependent on the arbitrary routing arises%
\begin{equation}
P_{1}\left( z\right) =k_{2}^{2}z\left( 1-z\right) +\left( \lambda
^{2}-m^{2}\right) z-\lambda ^{2}\text{.}
\end{equation}%
Performing a shift on the variable $k+k_{2}z\rightarrow k$ makes the
denominator momentum-even while generating an additional term in the
numerator, which allows identifying a derivative of the polynomial:%
\begin{equation}
-\int \frac{d^{4}k}{\left( 2\pi \right) ^{4}}\frac{A_{2}}{D_{\lambda
}^{2}D_{2}}=-2\int_{0}^{1}dz\left( 1-z\right) \int \frac{d^{4}k}{\left( 2\pi
\right) ^{4}}\left[ 2k_{2}^{\rho }k_{\rho }+\frac{\partial P_{1}}{\partial z}%
\right] \frac{1}{\left( k^{2}+P_{1}\right) ^{3}}.
\end{equation}

Any finite integral found in one-loop calculations leads to this type of
structure after parametrization. Nevertheless, derivatives (and their
powers) only appear if the original integral has divergent power counting.
The next step consists of the loop integration, which only produces non-zero
contributions for even integrands; the case above yields:%
\begin{equation}
\int \frac{d^{4}k}{\left( 2\pi \right) ^{4}}\frac{1}{\left(
k^{2}+P_{z}\right) ^{3}}=\frac{i}{\left( 4\pi \right) ^{2}}\frac{1}{2P_{1}}.
\label{CC1}
\end{equation}%
Posteriorly to replacing this result, one must integrate by parts until all
derivatives are eliminated. This case requires a sole operation and leads to
the outcome:%
\begin{eqnarray}
-\int \frac{d^{4}k}{\left( 2\pi \right) ^{4}}\frac{A_{2}}{D_{\lambda
}^{2}D_{2}} &=&-\frac{i}{\left( 4\pi \right) ^{2}}\int_{0}^{1}dz\left[ \frac{%
\partial }{\partial z}\left( 1-z\right) \ln P_{1}+\ln P_{1}\right]  \notag \\
&=&-\frac{i}{\left( 4\pi \right) ^{2}}\int_{0}^{1}dz\ln \frac{P_{1}}{%
-\lambda ^{2}}.
\end{eqnarray}

Finite contributions follow a strong pattern since we departed from a
logarithmically divergent integral. Each step described above has an
analogous form in the second contribution from Equation (\ref{i20}). The
fundamental difference is in the parametrization (\ref{fey3}), which
involves two propagators and leads to another polynomial%
\begin{eqnarray}
P_{2}\left( z,y\right) &=&k_{1}^{2}y\left( 1-y\right) +k_{2}^{2}z\left(
1-z\right) -2k_{1}\cdot k_{2}yz  \notag \\
&&+\left( \lambda ^{2}-m^{2}\right) y+\left( \lambda ^{2}-m^{2}\right)
z-\lambda ^{2}.
\end{eqnarray}%
Observe how this dependence reflects on the integration by parts:%
\begin{eqnarray}
-\int \frac{d^{4}k}{\left( 2\pi \right) ^{4}}\frac{A_{1}}{D_{\lambda }D_{12}}
&=&-\frac{i}{\left( 4\pi \right) ^{2}}\int_{0}^{1}dz\int_{0}^{1-z}dy\frac{%
\partial }{\partial y}\ln P_{2}  \notag \\
&=&-\frac{i}{\left( 4\pi \right) ^{2}}\int_{0}^{1}dz\ln \frac{Q}{P_{1}}.
\end{eqnarray}%
The lower limit of integration (in $y=0$) returns the first polynomial;
hence, this type of term disappears when summing up the entire sector. Even
in more complex cases, finite contributions involving arbitrary routings $%
k_{i}$ cancel out identically in a chain effect. Only the term achieved by
applying the upper limit of integration (in $y=1-z$) contributes in the end.
That leads to the dependence on external momentum $p=k_{1}-k_{2}$
acknowledged in Subsection (\ref{Finite1}), embodied in the polynomial:%
\begin{equation}
Q\left( z\right) =p^{2}z\left( 1-z\right) -m^{2}.
\end{equation}

With both contributions at our disposal, building up the scalar Feynman
integral (\ref{i20}) is possible%
\begin{equation}
I_{2}\left( k_{1},k_{2}\right) -I_{\log }=-\frac{i}{\left( 4\pi \right) ^{2}}%
\xi _{0}^{\left( 0\right) }\left( p\right) ,  \label{I20FIN}
\end{equation}%
where the finite function was identified (\ref{xi0}). Such an expression
clarifies that the parameter $\lambda ^{2}$ plays the role of a scale
connecting finite and ill-defined quantities. That becomes transparent by
setting routings as zero $k_{i}=0$ on the equation above:%
\begin{equation}
I_{\log }\left( m^{2}\right) -I_{\log }\left( \lambda ^{2}\right) =-\frac{i}{%
\left( 4\pi \right) ^{2}}\ln \frac{m^{2}}{\lambda ^{2}}.
\end{equation}%
This type of scale relation is implicit whenever logarithmic functions are
present in this investigation.

After detailing the first case, we directly cast one possible separation
linked to the vector integral $I_{2}^{\beta }$ (\ref{I2}). Since its power
counting indicates linear divergence, let us set $N\leq 1$ in identity (\ref%
{id}) and employ both versions to achieve the structure%
\begin{equation}
\left[ \frac{k^{\beta }}{D_{12}}\right] _{not\text{ }odd}+2\left(
k_{1}+k_{2}\right) _{\rho }\frac{k^{\beta }k^{\rho }}{D_{\lambda }^{3}}=%
\frac{A_{2}\left( A_{1}+A_{2}\right) k^{\beta }}{D_{\lambda }^{3}D_{2}}+%
\frac{A_{1}^{2}k^{\beta }}{D_{\lambda }^{2}D_{12}}.
\end{equation}%
We disregard momentum-odd terms since they vanish with the loop integration.
Again, the adopted arrangement puts ill-defined structures on the left-hand
side:%
\begin{equation}
I_{2}^{\beta }\left( k_{1},k_{2}\right) +\frac{1}{2}\left(
k_{1}+k_{2}\right) _{\rho }\left( \Delta ^{\beta \rho }+g^{\beta \rho
}I_{\log }\right) =\frac{i}{\left( 4\pi \right) ^{2}}\frac{1}{2}\left(
k_{1}+k_{2}\right) ^{\beta }\xi _{0}^{\left( 0\right) }\left( p\right) .
\label{I21FIN}
\end{equation}%
We employed the irreducible object (\ref{Ilog}) and the 2nd-order surface
term (\ref{delta3}) to organize the divergent sector. Finite contributions
lead to the family (\ref{xi0}); we also employed the reduction of finite
functions $\xi _{1}=\frac{1}{2}\xi _{0}$, achieved initially in (\ref{red}).

Lastly, let us employ the achieved integrals to build the $AV$ amplitude (%
\ref{AVI}). Finite contributions and irreducible divergent objects cancel
out identically after using the identity $\varepsilon _{\mu \nu \alpha \beta
}\left( k_{1}-k_{2}\right) ^{\alpha }\left( k_{1}+k_{2}\right) ^{\beta
}=2\varepsilon _{\mu \nu \alpha \beta }k_{1}^{\alpha }k_{2}^{\beta }$.
Hence, the only non-trivial contribution is the following%
\begin{equation}
T_{\mu \nu }^{AV}=-2i\varepsilon _{\mu \nu \alpha \beta }p^{\alpha }\left(
k_{1}+k_{2}\right) _{\rho }\Delta ^{\beta \rho }.  \label{AV}
\end{equation}%
That agrees with the expectation from Equation (\ref{exp1}), i.e., it is a
surface term proportional to an arbitrary momenta combination.

Observing this expression isolated, one might expect that restricting
arbitrary labels (as in $k_{2}=-k_{1}$) would eliminate surface terms and
solve issues approached while exploring symmetry aspects. Nevertheless, that
is not enough when considering the complete discussion. For this reason, we
maintain the arbitrariness associated with labels, so the analysis falls
over values accessible to surface terms.

\subsection{Three-Point Amplitudes - Feynman Integrals}

\label{Fey3}Our next objective is to compute quantities typical of
calculations involving three-point correlators, starting with the
corresponding Feynman integrals (\ref{I3}). Afterward, we evaluate standard
tensors and subamplitudes necessary to build the main targets: $PVV$ and $%
AVV $. Since some ingredients also appear when exploring four-point
structures, we broaden their discussion.

As the first couple of integrals is finite, dependence on external momenta
appears from the beginning. In other words, when employing the Feynman
parametrization (\ref{fey2}) and grouping terms on the loop momentum,
denominators exhibit polynomial (\ref{Q2}):%
\begin{equation}
\frac{\left[ 1,k_{\mu }\right] }{D_{123}}=2\int_{0}^{1}dz\int_{0}^{1-z}dy%
\frac{\left[ 1,k_{\mu }\right] }{\left[ \left( k+k_{1}-py-qz\right)
^{2}+Q\left( y,z\right) \right] ^{3}}.
\end{equation}%
Then, integrating both sides of the scalar version of this equation yields
the first integral. No compensation term appears by shifting the momentum $%
k+k_{1}-py-qz\rightarrow k$; hence, obtaining this result is straightforward%
\begin{equation}
I_{3}=\frac{i}{\left( 4\pi \right) ^{2}}\int_{0}^{1}dz\int_{0}^{1-z}dy\text{ 
}\frac{1}{Q\left( y,z\right) }.
\end{equation}%
Although this reasoning extends to the vector version, the momentum shift
brings parameter powers to its numerator%
\begin{equation}
I_{3\mu }=-\frac{i}{\left( 4\pi \right) ^{2}}\int_{0}^{1}dz\int_{0}^{1-z}dy%
\text{ }\frac{\left( k_{1}-py-qz\right) _{\mu }}{Q\left( y,z\right) }.
\end{equation}

Lastly, the 2nd-order tensor is the only integral exhibiting logarithmically
diverging power counting here. We split its integrand by employing the $N=0$
identity version (\ref{id}) whenever necessary%
\begin{equation}
\frac{k_{\mu \nu }}{D_{123}}-\frac{k_{\mu \nu }}{D_{\lambda }^{3}}=-\frac{%
A_{3}k_{\mu \nu }}{D_{\lambda }^{3}D_{3}}-\frac{A_{2}k_{\mu \nu }}{%
D_{\lambda }^{2}D_{23}}-\frac{A_{1}k_{\mu \nu }}{D_{\lambda }D_{123}}.
\end{equation}%
Terms associated with ill-defined contributions are on the left-hand side,
so we use standard objects introduced in Equations (\ref{delta3})-(\ref{Ilog}%
) to express them without additional manipulations.

Regarding finite contributions, each rational function requires a different
Feynman parametrization. Although they lead to structures similar to those
above, polynomials depend on non-physical parameters this time. Furthermore,
momentum shifts induce derivatives of these polynomials in the numerators,
requiring integrations by parts. When completing this procedure, most
contributions fit perfectly, and only those depending on external momenta
remain:%
\begin{eqnarray}
&&I_{3\mu \nu }-\tfrac{1}{4}\left( \Delta _{\mu \nu }+g_{\mu \nu }I_{\log
}\right)  \notag \\
&=&\frac{i}{\left( 4\pi \right) ^{2}}\int_{0}^{1}dz\int_{0}^{1-z}dy\text{ }%
\left( k_{1}-py-qz\right) _{\mu }\left( k_{1}-py-qz\right) _{\nu }\frac{1}{%
Q\left( y,z\right) }  \notag \\
&&-\frac{i}{\left( 4\pi \right) ^{2}}\frac{1}{2}g_{\mu \nu
}\int_{0}^{1}dz\int_{0}^{1-z}dy\text{ }\ln \frac{Q\left( y,z\right) }{%
-\lambda ^{2}}.
\end{eqnarray}

The final step for evaluating these Feynman integrals is to project finite
contributions in terms of structure functions from the families (\ref{xii1}%
)-(\ref{xii0}). Having two parameters highlights some patterns, which
clarifies that these functions do not appear randomly but in tensors having
well-defined properties. We mentioned them in Subsection (\ref{Finite2}).
Then, following the identifications, we group terms depending exclusively on
external momenta into what we call $J$-tensors. Other contributions
correspond to lower-order Feynman integrals having combinations of the
routing $k_{1}$ as coefficients. Such reasoning materializes in the
following organization%
\begin{eqnarray}
I_{3} &=&J_{3},  \label{I30} \\
I_{3\mu } &=&J_{3\mu }-\left[ k_{1\mu }I_{3}\right] ,  \label{I3u} \\
I_{3\mu \nu }-\tfrac{1}{4}\left( \Delta _{\mu \nu }+g_{\mu \nu }I_{\log
}\right) &=&J_{3\mu \nu }-\left[ k_{1\mu }I_{3\nu }+k_{1\nu }I_{3\mu }\right]
-\left[ k_{1\mu }k_{1\nu }I_{3}\right] ,  \label{I3uv}
\end{eqnarray}%
where $J$-tensors are introduced%
\begin{equation}
J_{3}=\frac{i}{\left( 4\pi \right) ^{2}}\xi _{00}^{\left( -1\right) },
\label{J3}
\end{equation}%
\begin{equation}
J_{3\mu }=\frac{i}{\left( 4\pi \right) ^{2}}\left[ p_{\mu }\xi _{10}^{\left(
-1\right) }+q_{\mu }\xi _{01}^{\left( -1\right) }\right] ,  \label{J3u}
\end{equation}%
\begin{equation}
J_{3\mu \nu }=\frac{i}{\left( 4\pi \right) ^{2}}\left[ p_{\mu }p_{\nu }\xi
_{20}^{\left( -1\right) }+q_{\mu }q_{\nu }\xi _{02}^{\left( -1\right)
}+\left( p_{\mu }q_{\nu }+q_{\mu }p_{\nu }\right) \xi _{11}^{\left(
-1\right) }-\tfrac{1}{2}g_{\mu \nu }\xi _{00}^{\left( 0\right) }\right] .
\label{J3uv}
\end{equation}

Using these tensors to express mathematical structures appearing in
perturbative calculations is already very useful. They are introduced in
reference \cite{Battistel:2012qpm} as part of the systematization from IReg,
where they allow a compact presentation of the quadruple-vector box
amplitude.

Although that is part of their purpose here, we stress their remarkable
value regarding algebraic manipulations and interpretation of results. Since 
$J$-tensors concentrate all contributions on external momenta, they are
enough to describe the finite part of physical amplitudes. We consider this
systematization to propose a new perspective, where $J$-tensors are the
fundamental pieces in this analysis. When computing momenta contractions,
for instance, the discussion resorts to their properties as a generalization
of reductions from Section (\ref{Strategy}). Without further delay, let us
employ these ideas in the study of three-point functions subsequently.

\subsection{Three-Point Amplitudes - $PVV$}

Before computing the $PVV$ amplitude, our first task is integrating the
standard tensor having two momenta in the numerator. By integrating Equation
(\ref{t3uv}), we expand products and identify the following combination of
Feynman integrals: 
\begin{eqnarray}
T_{3\mu \nu }^{\left( s\right) }\left( k_{i},k_{j}\right) &=&\left(
1+s\right) I_{3\mu \nu }+\left( k_{j}+sk_{i}\right) _{\nu }I_{3\mu }  \notag
\\
&&+\left( k_{i}+sk_{j}\right) _{\mu }I_{3\nu }+\left( k_{i\mu }k_{j\nu
}+sk_{i\nu }k_{j\mu }\right) I_{3}.
\end{eqnarray}%
We adopt general structures, admitting choices for signs and routings. This
expression applies to any denominator $D_{klm}$ typical of three-point
calculations and extends other cases (e.g., box) by changing the numerical
subindex. The same pattern manifests in finite tensors, expressing all
finite quantities below. We delimit our focus to the $D_{123}$ case for now.

Since there is an intrinsic idea of hierarchy, we start by replacing the
highest-order integral $I_{3\mu \nu }$ (\ref{I3uv}). Divergent objects and
the 2nd-order $J$-tensor are ready; however, this operation brings new
contributions through lower-order structures. With this, external momenta $%
p_{i}=k_{1}-k_{i}$ appear as multiplicative coefficients of the next
integral $I_{3\mu }$ (\ref{I3u}). Its substitution gives continuity to a
chain effect, and now the last integral $I_{2}$ (\ref{I30}) has this type of
coefficient. Once this procedure is over, we achieve the general form%
\begin{eqnarray}
T_{3\mu \nu }^{\left( s\right) }\left( k_{i},k_{j}\right) &=&\tfrac{1}{4}%
\left( 1+s\right) \left( \Delta _{\mu \nu }+g_{\mu \nu }I_{\log }+4J_{3\mu
\nu }\right)  \notag \\
&&-\left( p_{j}+sp_{i}\right) _{\nu }J_{3\mu }-\left( p_{i}+sp_{j}\right)
_{\mu }J_{3\nu }+\left( p_{i\mu }p_{j\nu }+sp_{i\nu }p_{j\mu }\right) J_{3}.
\label{T3s}
\end{eqnarray}%
This procedure is generic, so we resort to it when examining all standard
tensors. We recall that the object $p_{i}$ produces three possibilities
here: $p_{1}=0$, $p_{2}=k_{1}-k_{2}=p$, and $p_{3}=k_{1}-k_{3}=q$.

We aim to build the $PVV$ (\ref{pvv}) using this tool, so let us reintroduce
its expression by using uppercase letters to characterize the integrated
amplitude%
\begin{equation}
T_{\mu \nu }^{PVV}=-2im\varepsilon _{\mu \nu XY}\left[ T_{3XY}^{\left(
-\right) }\left( k_{2},k_{3}\right) +T_{3XY}^{\left( -\right) }\left(
k_{3},k_{1}\right) +T_{3XY}^{\left( -\right) }\left( k_{1},k_{2}\right) %
\right] .
\end{equation}%
The minus sign reflects in their antisymmetry property, hence, canceling the
first row of the general form (\ref{T3s}). Then, setting the different
momenta arrangements, we cast the required versions:%
\begin{equation}
T_{3\mu \nu }^{\left( -\right) }\left( k_{1},k_{2}\right) =-p_{\nu }J_{3\mu
}+p_{\mu }J_{3\nu },
\end{equation}%
\begin{equation}
T_{3\mu \nu }^{\left( -\right) }\left( k_{3},k_{1}\right) =q_{\nu }J_{3\mu
}-q_{\mu }J_{3\nu },
\end{equation}%
\begin{equation}
T_{3\mu \nu }^{\left( -\right) }\left( k_{2},k_{3}\right) =\left( p-q\right)
_{\nu }J_{3\mu }-\left( p-q\right) _{\mu }J_{3\nu }+\left( p_{\mu }q_{\nu
}-p_{\nu }q_{\mu }\right) J_{3}.
\end{equation}%
It is straightforward to sum them to find that these objects collapse into
the finite function%
\begin{equation}
T_{\mu \nu }^{PVV}=-4im\varepsilon _{\mu \nu XY}p^{X}q^{Y}J_{3},  \label{PVV}
\end{equation}%
which agrees with the expectation from Equation (\ref{exp2}).

\subsection{Three-Point Amplitudes - $AVV$}

Our next target is the $AVV$ triangle (\ref{avv1}), which contains a tensor
sector besides the vector subamplitude $VPP$. Given the procedure introduced
in the previous case, let us begin this discussion by writing the integrated
form of the 3rd-order standard tensor (\ref{t33s}) through Feynman integrals%
\begin{eqnarray}
T_{3\mu ;\nu \alpha }^{\left( -\right) }\left( k_{l};k_{i},k_{j}\right)
&=&\left( k_{j}-k_{i}\right) _{\alpha }I_{3\mu \nu }+\left(
k_{i}-k_{j}\right) _{\nu }I_{3\mu \alpha }  \notag \\
&&+\left( k_{j\alpha }k_{i\nu }-k_{i\alpha }k_{j\nu }\right) I_{3\mu
}+\left( k_{j}-k_{i}\right) _{\alpha }k_{l\mu }I_{3\nu }  \notag \\
&&+\left( k_{i}-k_{j}\right) _{\nu }k_{l\mu }I_{3\alpha }+\left( k_{j\alpha
}k_{i\nu }-k_{i\alpha }k_{j\nu }\right) k_{l\mu }I_{3}.  \label{tensor3}
\end{eqnarray}%
We restricted this equation to the minus sign because only antisymmetric
tensors appear throughout this investigation. That also comprehends the
four-propagator version, achieved by changing numerical subindices.

Replacements start with the highest-order integral and follow a hierarchy
until getting to the lowest. Ultimately, finite contributions depend
exclusively on external momenta $p_{i}=k_{1}-k_{i}$ since terms associated
with $k_{1}$ combinations vanish identically:%
\begin{eqnarray}
T_{3\mu ;\nu \alpha }^{\left( -\right) }\left( k_{l};k_{i},k_{j}\right) &=&-%
\tfrac{1}{4}\left[ \left( p_{j}-p_{i}\right) _{\alpha }\Delta _{\mu \nu
}+\left( p_{i}-p_{j}\right) _{\nu }\Delta _{\mu \alpha }\right]  \notag \\
&&-\tfrac{1}{4}\left[ \left( p_{j}-p_{i}\right) _{\alpha }g_{\mu \nu
}+\left( p_{i}-p_{j}\right) _{\nu }g_{\mu \alpha }\right] I_{\log }  \notag
\\
&&-\left( p_{j}-p_{i}\right) _{\alpha }J_{3\mu \nu }-\left(
p_{i}-p_{j}\right) _{\nu }J_{3\mu \alpha }+\left( p_{i}-p_{j}\right) _{\nu
}p_{l\mu }J_{3\alpha }  \notag \\
&&+\left( p_{j}-p_{i}\right) _{\alpha }p_{l\mu }J_{3\nu }+\left( p_{j\alpha
}p_{i\nu }-p_{i\alpha }p_{j\nu }\right) J_{3\mu }  \notag \\
&&-\left( p_{j\alpha }p_{i\nu }-p_{i\alpha }p_{j\nu }\right) p_{l\mu }J_{3}.
\label{7}
\end{eqnarray}%
That becomes transparent as a consequence of $J$-tensors structures. Even
though we introduced the scalar $J_{3}$ for generality, its coefficient
vanishes here due to the unavoidable presence of $p_{1}=0$. As three
routings are available, three non-equivalent configurations of this tensor
are obtainable:%
\begin{eqnarray}
T_{3\mu ;\nu \alpha }^{\left( -\right) }\left( k_{1};k_{2},k_{3}\right) &=&-%
\tfrac{1}{4}\left[ \left( q-p\right) _{\alpha }\Delta _{\mu \nu }+\left(
p-q\right) _{\nu }\Delta _{\mu \alpha }\right]  \notag \\
&&-\tfrac{1}{4}\left[ \left( q-p\right) _{\alpha }g_{\mu \nu }+\left(
p-q\right) _{\nu }g_{\mu \alpha }\right] I_{\log }  \notag \\
&&-\left( q-p\right) _{\alpha }J_{3\mu \nu }-\left( p-q\right) _{\nu
}J_{3\mu \alpha }+\left( q_{\alpha }p_{\nu }-p_{\alpha }q_{\nu }\right)
J_{3\mu }, \\
T_{3\mu ;\nu \alpha }^{\left( -\right) }\left( k_{2};k_{3},k_{1}\right) &=&%
\tfrac{1}{4}\left( q_{\alpha }\Delta _{\mu \nu }-q_{\nu }\Delta _{\mu \alpha
}\right) +\frac{1}{4}\left( q_{\alpha }g_{\mu \nu }-q_{\nu }g_{\mu \alpha
}\right) I_{\log }  \notag \\
&&+q_{\alpha }J_{3\mu \nu }-q_{\nu }J_{3\mu \alpha }+q_{\nu }p_{\mu
}J_{3\alpha }-q_{\alpha }p_{\mu }J_{3\nu }, \\
T_{3\mu ;\nu \alpha }^{\left( -\right) }\left( k_{3};k_{1},k_{2}\right) &=&%
\tfrac{1}{4}\left( p_{\nu }\Delta _{\mu \alpha }-p_{\alpha }\Delta _{\mu \nu
}\right) +\frac{1}{4}\left( p_{\nu }g_{\mu \alpha }-p_{\alpha }g_{\mu \nu
}\right) I_{\log }  \notag \\
&&-p_{\alpha }J_{3\mu \nu }+p_{\nu }J_{3\mu \alpha }-p_{\nu }q_{\mu
}J_{3\alpha }+p_{\alpha }q_{\mu }J_{3\nu }.
\end{eqnarray}

When looking into integrands, we made expectations regarding these
structures (\ref{exp3})-(\ref{exp5}). The main point is the impossibility of
building a 3rd-order tensor with the property of total antisymmetry in this
particular context. Having all ingredients required for the verifications,
we comment on them in the sequence.

First, all terms vanish by contracting the Levi-Civita symbol with the first
configuration above since they correspond to products between symmetric and
antisymmetric objects. Whereas most cases are straightforward, inspecting
the $J$-vector content (\ref{J3u}) is necessary for completing this
verification%
\begin{eqnarray}
\varepsilon ^{\nu XYZ}T_{3X;YZ}^{\left( -\right) }\left(
k_{1};k_{2},k_{3}\right) &=&2\varepsilon ^{\nu XYZ}p_{Y}q_{Z}J_{3X}  \notag
\\
&\rightarrow &\varepsilon ^{\nu XYZ}p_{Y}q_{Z}\left[ p_{X}\xi _{10}^{\left(
-1\right) }+q_{X}\xi _{01}^{\left( -1\right) }\right] =0.
\end{eqnarray}%
Second, all terms cancel out identically when summing these three
configurations. Again, that requires a closer look inside the $J$-vector%
\begin{eqnarray}
&&T_{3\mu ;\nu \alpha }^{\left( -\right) }\left( k_{1};k_{2},k_{3}\right)
+T_{3\mu ;\nu \alpha }^{\left( -\right) }\left( k_{2};k_{3},k_{1}\right)
+T_{3\mu ;\nu \alpha }^{\left( -\right) }\left( k_{3};k_{1},k_{2}\right) 
\notag \\
&=&\left( q_{\alpha }p_{\nu }-p_{\alpha }q_{\nu }\right) J_{3\mu }+\left(
q_{\nu }p_{\mu }-p_{\nu }q_{\mu }\right) J_{3\alpha }+\left( p_{\alpha
}q_{\mu }-q_{\alpha }p_{\mu }\right) J_{3\nu }=0.
\end{eqnarray}%
As these identities are indeed confirmed, the expectation over the amplitude
also applies 
\begin{equation}
T_{\mu \nu \alpha }^{AVV}=4i\varepsilon _{\mu \alpha XY}T_{3\nu ;XY}^{\left(
-\right) }\left( k_{1};k_{2},k_{3}\right) +4i\varepsilon _{\mu \nu
XY}T_{3\alpha ;XY}^{\left( -\right) }\left( k_{3};k_{1},k_{2}\right)
-i\varepsilon _{\mu \nu \alpha \beta }T_{\beta }^{VPP}.  \label{AVV1}
\end{equation}%
If compared with other free indices, $\mu $ has a distinct function in this
equation. That is a direct consequence of the trace version adopted in the
integrand exploration (\ref{form1})-(\ref{form2}).

Proceeding to the last substructure, we consult Equation (\ref{vpp}) to
express the $VPP$ amplitude as a combination of Feynman integrals%
\begin{eqnarray}
T_{\beta }^{VPP} &=&-2p_{\beta }I_{2\beta }\left( k_{1},k_{2}\right)
-4I_{2\beta }\left( k_{1},k_{3}\right) -2\left( k_{1}+k_{3}\right) _{\beta
}I_{2}\left( k_{1},k_{3}\right)  \notag \\
&&+2\left( q-p\right) _{\beta }I_{2}\left( k_{2},k_{3}\right) +2\left(
q-p\right) ^{2}\left( I_{3\beta }+k_{1\beta }I_{3}\right)  \notag \\
&&-2q^{2}\left( I_{3\beta }+k_{2\beta }I_{3}\right) +2p^{2}\left( I_{3\beta
}+k_{3\beta }I_{3}\right) .
\end{eqnarray}%
Besides results obtained at the outset of the triangle discussion,
two-propagator integrals (\ref{I20FIN})-(\ref{I21FIN}) are also ingredients
needed to build this object. Their replacement leads to the following
mathematical expression:%
\begin{eqnarray}
T_{\beta }^{VPP} &=&2\left( k_{1}+k_{3}\right) ^{\rho }\Delta _{\beta \rho
}-2\left( 2p-q\right) _{\beta }I_{\log }  \notag \\
&&+4\left( p^{2}-p\cdot q\right) J_{3\beta }+2\left( q^{2}p_{\beta
}-p^{2}q_{\beta }\right) J_{3}  \notag \\
&&+i\left( 4\pi \right) ^{-2}\left[ p_{\beta }\xi _{0}^{\left( 0\right)
}\left( p\right) -\left( q-p\right) _{\beta }\xi _{0}^{\left( 0\right)
}\left( p-q\right) \right] .  \label{VPP}
\end{eqnarray}%
Since the dependence on external momenta is not univocal for $\xi _{k}$%
-functions, we must specify their argument.

As we determined all substructures, renaming indices and organizing
contributions is the final task before expressing the $AVV$ amplitude:%
\begin{eqnarray}
T_{\mu \nu \alpha }^{AVV} &=&-2i\varepsilon _{\mu \alpha \rho \sigma }\left(
p-q\right) ^{\rho }\Delta _{\nu }^{\sigma }+2i\varepsilon _{\mu \nu \rho
\sigma }p^{\rho }\Delta _{\alpha }^{\sigma }  \notag \\
&&-2i\varepsilon _{\mu \nu \alpha \rho }\left( k_{1}+k_{3}\right) _{\sigma
}\Delta ^{\rho \sigma }-8i\varepsilon _{\mu \alpha \rho \sigma }\left(
p-q\right) ^{\rho }J_{3\nu }^{\sigma }  \notag \\
&&+8i\varepsilon _{\mu \nu \rho \sigma }p^{\rho }J_{3\alpha }^{\sigma
}-8i\varepsilon _{\mu \nu \rho \sigma }p^{\rho }q_{\alpha }J_{3\sigma
}+8i\varepsilon _{\mu \alpha \rho \sigma }p^{\rho }q^{\sigma }J_{3\nu } 
\notag \\
&&-4i\varepsilon _{\mu \nu \alpha \beta }\left( p^{2}-p\cdot q\right)
J_{3\beta }-2i\varepsilon _{\mu \nu \alpha \beta }\left( q^{2}p^{\beta
}-p^{2}q^{\beta }\right) J_{3}  \notag \\
&&+2\left( 4\pi \right) ^{-2}\varepsilon _{\mu \nu \alpha \beta }\left[
p^{\beta }\xi _{0}^{\left( 0\right) }\left( p\right) -\left( q-p\right)
^{\beta }\xi _{0}^{\left( 0\right) }\left( p-q\right) \right] .  \label{AVV}
\end{eqnarray}%
Some comments on ill-defined quantities are pertinent to conclude this
analysis. Even though it appears when we survey substructures individually,
the irreducible standard object $I_{\log }$ does not appear within the final
expression since the corresponding coefficient vanishes. That implies all
divergences concentrate on surface terms $\Delta _{\rho \sigma }$, whose
coefficient unavoidably depends on a non-physical momenta combination.
Interestingly, this ambiguous contribution comes from the vector function $%
VPP$; standard tensors do not manifest this type of ambiguity.

\subsection{Four-Point Amplitudes - Feynman Integrals}

\label{Fey4}The final task of this section is to compute quantities typical
of calculations involving four-point amplitudes, starting with the
corresponding Feynman integrals (\ref{I4}). Most are finite, therefore,
polynomial (\ref{Q3}) manifests after adopting the Feynman parametrization (%
\ref{fey3}) and grouping terms on the loop momentum%
\begin{equation}
\frac{1}{D_{1234}}=\frac{i}{\left( 4\pi \right) ^{2}}\int_{0}^{1}dz%
\int_{0}^{1-z}dy\int_{0}^{1-y-z}dx\text{ }\frac{1}{\left[ \left( k+L\right)
^{2}+Q\left( x,y,z\right) \right] ^{4}},
\end{equation}%
where $x$, $y$, and $z$ are the parameters. The object $L=k_{1}-px-qy-rz$
corresponds to the quantity shifted posteriorly to applying the integration.
Notations involving it are nothing more than tools to facilitate the
visualization of mathematical expressions, hence suppressed later when
identifying finite functions. Considering these introductions, explicit
integration leads to the following results:%
\begin{eqnarray}
I_{4} &=&\frac{i}{\left( 4\pi \right) ^{2}}\int_{0}^{1}dz\int_{0}^{1-z}dy%
\int_{0}^{1-y-z}dx\text{ }\frac{1}{Q^{2}}, \\
I_{4\mu } &=&-\frac{i}{\left( 4\pi \right) ^{2}}\int_{0}^{1}dz%
\int_{0}^{1-z}dy\int_{0}^{1-y-z}dx\text{ }L_{\mu }\frac{1}{Q^{2}}, \\
I_{4\mu \nu } &=&\frac{i}{\left( 4\pi \right) ^{2}}\int_{0}^{1}dz%
\int_{0}^{1-z}dy\int_{0}^{1-y-z}dx\left[ L_{\mu \nu }\frac{1}{Q^{2}}+\frac{1%
}{2}g_{\mu \nu }\frac{1}{Q}\right] , \\
I_{4\mu \nu \alpha } &=&-\frac{i}{\left( 4\pi \right) ^{2}}%
\int_{0}^{1}dz\int_{0}^{1-z}dy\int_{0}^{1-y-z}dx\left[ L_{\mu \nu \alpha }%
\frac{1}{Q^{2}}+\frac{1}{2}L_{\mu \nu \alpha }^{\prime }\frac{1}{Q}\right] ,
\end{eqnarray}%
where we compact products involving the momentum $L_{\mu \nu }=L_{\mu
}L_{\nu }$ and introduce the combination%
\begin{equation}
L_{\mu \nu \alpha }^{\prime }=L_{\mu }g_{\nu \alpha }+L_{\nu }g_{\mu \alpha
}+L_{\alpha }g_{\mu \nu }.
\end{equation}

We still have to evaluate the 4th-order Feynman integral. Since it exhibits
logarithmic power counting, one form for its separation employs the $N=0$
version of identity (\ref{id}) to write%
\begin{equation}
\frac{k_{\mu \nu \alpha \beta }}{D_{1234}}-\frac{k_{\mu \nu \alpha \beta }}{%
D_{\lambda }^{4}}=-\frac{A_{4}k_{\mu \nu \alpha \beta }}{D_{\lambda
}^{4}D_{4}}-\frac{A_{3}k_{\mu \nu \alpha \beta }}{D_{\lambda }^{3}D_{34}}-%
\frac{A_{2}k_{\mu \nu \alpha \beta }}{D_{\lambda }^{2}D_{234}}-\frac{%
A_{1}k_{\mu \nu \alpha \beta }}{D_{\lambda }D_{1234}}.
\end{equation}%
As this equation follows the developed strategy, integrating the left-hand
side leads to ill-defined quantities. They receive an organization through
symmetric tensors:%
\begin{eqnarray}
&&I_{4\mu \nu \alpha \beta }-\tfrac{1}{24}A_{\mu \nu \alpha \beta }-\tfrac{1%
}{24}g_{\mu \nu \alpha \beta }I_{\log }  \notag \\
&=&-\int \frac{d^{4}k}{\left( 2\pi \right) ^{4}}\left[ \frac{A_{4}k_{\mu \nu
\alpha \beta }}{D_{\lambda }^{4}D_{4}}+\frac{A_{3}k_{\mu \nu \alpha \beta }}{%
D_{\lambda }^{3}D_{34}}+\frac{A_{2}k_{\mu \nu \alpha \beta }}{D_{\lambda
}^{2}D_{234}}+\frac{A_{1}k_{\mu \nu \alpha \beta }}{D_{\lambda }D_{1234}}%
\right] .  \label{I4div}
\end{eqnarray}%
Here, aiming for a cleaner form, we concentrate all surface terms in the
object%
\begin{equation}
A_{\mu \nu \alpha \beta }=\square _{\mu \nu \alpha \beta }+\tfrac{1}{2}%
\left( g_{\mu \nu }\Delta _{\alpha \beta }+g_{\mu \alpha }\Delta _{\nu \beta
}+g_{\mu \beta }\Delta _{\nu \alpha }g_{\nu \alpha }\Delta _{\mu \beta
}+g_{\nu \beta }\Delta _{\mu \alpha }+g_{\alpha \beta }\Delta _{\mu \nu
}\right)
\end{equation}%
while products involving the metric tensor receive a compact notation%
\begin{equation}
g_{\mu \nu \alpha \beta }=g_{\mu \nu }g_{\alpha \beta }+g_{\mu \alpha
}g_{\nu \beta }+g_{\mu \beta }g_{\nu \alpha }.
\end{equation}

Next, proceeding to the finite sector on the right-hand side of this
integral, each rational function requires a different Feynman
parametrization. They differ from the cases above because polynomials depend
on non-physical parameters. This type of contribution cancels out
identically after integrations by parts, which ultimately brings polynomials
dependent on external momenta:%
\begin{eqnarray}
&&-\int \frac{d^{4}k}{\left( 2\pi \right) ^{4}}\left[ \frac{A_{4}k_{\mu \nu
\alpha \beta }}{D_{\lambda }^{4}D_{4}}+\frac{A_{3}k_{\mu \nu \alpha \beta }}{%
D_{\lambda }^{3}D_{34}}+\frac{A_{2}k_{\mu \nu \alpha \beta }}{D_{\lambda
}^{2}D_{234}}+\frac{A_{1}k_{\mu \nu \alpha \beta }}{D_{\lambda }D_{1234}}%
\right] \\
&=&\frac{i}{\left( 4\pi \right) ^{2}}\int_{0}^{1}dz\int_{0}^{1-z}dy%
\int_{0}^{1-y-z}dx\left[ L_{\mu \nu \alpha \beta }\frac{1}{Q^{2}}+\frac{1}{2}%
L_{\mu \nu \alpha \beta }^{\prime \prime }\frac{1}{Q}-\frac{1}{4}g_{\mu \nu
\alpha \beta }\ln \frac{Q}{-\lambda ^{2}}\right] ,  \notag
\end{eqnarray}%
where we introduce the object%
\begin{equation}
L_{\mu \nu \alpha \beta }^{\prime \prime }=L_{\mu \nu }g_{\alpha \beta
}+L_{\mu \alpha }g_{\nu \beta }+L_{\mu \beta }g_{\nu \alpha }+L_{\nu \alpha
}g_{\mu \beta }+L_{\nu \beta }g_{\mu \alpha }+L_{\alpha \beta }g_{\mu \nu }.
\end{equation}%
That completes the expression for the last Feynman integral%
\begin{eqnarray}
&&I_{4\mu \nu \alpha \beta }-\tfrac{1}{24}A_{\mu \nu \alpha \beta }-\tfrac{1%
}{24}g_{\mu \nu \alpha \beta }I_{\log } \\
&=&\frac{i}{\left( 4\pi \right) ^{2}}\int_{0}^{1}dz\int_{0}^{1-z}dy%
\int_{0}^{1-y-z}dx\left[ L_{\mu \nu \alpha \beta }\frac{1}{Q^{2}}+\frac{1}{2}%
L_{\mu \nu \alpha \beta }^{\prime \prime }\frac{1}{Q}-\frac{1}{4}g_{\mu \nu
\alpha \beta }\ln \frac{Q}{-\lambda ^{2}}\right] .  \notag
\end{eqnarray}

To complete the systematization of Feynman integrals, let us identify terms
depending exclusively on external momenta and group them into $J$-tensors.
The remaining terms are proportional to combinations of the arbitrary
routing $k_{1}$, connecting to lower-order Feynman integrals. This process
expands the momentum $L$ and its combinations, so the notations introduced
above are no longer necessary. Nevertheless, we recur to compact notations
to products involving momenta, e.g., $k_{1\mu \nu }=k_{1\mu }k_{1\nu }$ and $%
p_{\mu \nu }=p_{\mu }p_{\nu }$. We cast the final forms for the integrals in
the sequence:%
\begin{eqnarray}
I_{4} &=&J_{4},  \label{I40} \\
I_{4\mu } &=&J_{4\mu }-k_{1\mu }I_{4}, \\
I_{4\mu \nu } &=&J_{4\mu \nu }-\left[ k_{1\mu }I_{4\nu }+k_{1\nu }I_{4\mu }%
\right] -\left[ k_{1\mu \nu }I_{4}\right] , \\
I_{4\mu \nu \alpha } &=&J_{4\mu \nu \alpha }-\left[ k_{1\mu }I_{4\nu \alpha
}+k_{1\nu }I_{4\alpha \mu }+k_{1\alpha }I_{4\mu \nu }\right]  \notag \\
&&-\left[ k_{1\nu \alpha }I_{4\mu }+k_{1\mu \alpha }I_{4\nu }+k_{1\mu \nu
}I_{4\alpha }\right] -\left[ k_{1\mu \nu \alpha }I_{4}\right] ,
\end{eqnarray}%
\begin{eqnarray}
&&I_{4\mu \nu \alpha \beta }-\tfrac{1}{24}A_{\mu \nu \alpha \beta }-\tfrac{1%
}{24}g_{\mu \nu \alpha \beta }I_{\log }  \notag \\
&=&J_{4\mu \nu \alpha \beta }-\left[ k_{1\mu }I_{4\nu \alpha \beta }+k_{1\nu
}I_{4\mu \alpha \beta }+k_{1\alpha }I_{4\mu \nu \beta }+k_{1\beta }I_{4\mu
\nu \alpha }\right]  \notag \\
&&-\left[ k_{1\alpha \beta }I_{4\mu \nu }+k_{1\nu \beta }I_{4\mu \alpha
}+k_{1\nu \alpha }I_{4\mu \beta }+k_{1\mu \beta }I_{4\nu \alpha }+k_{1\mu
\alpha }I_{4\nu \beta }+k_{1\mu \nu }I_{4\alpha \beta }\right]  \notag \\
&&-\left[ k_{1\nu \alpha \beta }I_{4\mu }+k_{1\mu \alpha \beta }I_{4\nu
}+k_{1\mu \nu \beta }I_{4\alpha }+k_{1\mu \nu \alpha }I_{4\beta }\right] -%
\left[ k_{1\mu \nu \alpha \beta }I_{4}\right] .  \label{I4uvab}
\end{eqnarray}%
The $J$-tensors arise as symmetric combinations of finite functions
belonging to the families (\ref{xiii2})-(\ref{xiii0}). All non-equivalent
index permutations compound these objects:%
\begin{eqnarray}
J_{4} &=&i\left( 4\pi \right) ^{-2}\xi _{000}^{\left( -2\right) },
\label{J4} \\
&&  \notag \\
J_{4\mu } &=&i\left( 4\pi \right) ^{-2}\left[ p_{\mu }\xi _{100}^{\left(
-2\right) }+q_{\mu }\xi _{010}^{\left( -2\right) }+r_{\mu }\xi
_{001}^{\left( -2\right) }\right] ,  \label{J4v} \\
&&  \notag \\
J_{4\mu \nu } &=&i\left( 4\pi \right) ^{-2}\left[ p_{\mu \nu }\xi
_{200}^{\left( -2\right) }+q_{\mu \nu }\xi _{020}^{\left( -2\right) }+r_{\mu
\nu }\xi _{002}^{\left( -2\right) }+\left( p_{\mu }q_{\nu }+q_{\mu }p_{\nu
}\right) \xi _{110}^{\left( -2\right) }\right.  \notag \\
&&\left. +\left( p_{\mu }r_{\nu }+r_{\mu }p_{\nu }\right) \xi _{101}^{\left(
-2\right) }+\left( q_{\mu }r_{\nu }+r_{\mu }q_{\nu }\right) \xi
_{011}^{\left( -2\right) }+\tfrac{1}{2}g_{\mu \nu }\xi _{000}^{\left(
-1\right) }\right] ,  \label{J4uv} \\
&&  \notag \\
J_{4\mu \nu \alpha } &=&i\left( 4\pi \right) ^{-2}\left[ p_{\mu \nu \alpha
}\xi _{300}^{\left( -2\right) }+q_{\mu \nu \alpha }\xi _{030}^{\left(
-2\right) }+r_{\mu \nu \alpha }\xi _{003}^{\left( -2\right) }\right.  \notag
\\
&&+\left( p_{\mu \nu }q_{\alpha }+p_{\mu \alpha }q_{\nu }+p_{\nu \alpha
}q_{\mu }\right) \xi _{210}^{\left( -2\right) }+\left( q_{\mu \nu }p_{\alpha
}+q_{\mu \alpha }p_{\nu }+q_{\nu \alpha }p_{\mu }\right) \xi _{120}^{\left(
-2\right) }  \notag \\
&&+\left( p_{\mu \nu }r_{\alpha }+p_{\mu \alpha }r_{\nu }+p_{\nu \alpha
}r_{\mu }\right) \xi _{201}^{\left( -2\right) }+\left( r_{\mu \nu }p_{\alpha
}+r_{\mu \alpha }p_{\nu }+r_{\nu \alpha }p_{\mu }\right) \xi _{102}^{\left(
-2\right) }  \notag \\
&&+\left( q_{\mu \nu }r_{\alpha }+q_{\mu \alpha }r_{\nu }+q_{\nu \alpha
}r_{\mu }\right) \xi _{021}^{\left( -2\right) }+\left( r_{\mu \nu }q_{\alpha
}+r_{\mu \alpha }q_{\nu }+r_{\nu \alpha }q_{\mu }\right) \xi _{012}^{\left(
-2\right) }  \notag \\
&&+\left[ \left( p_{\mu }q_{\nu }+q_{\mu }p_{\nu }\right) r_{\alpha }+\left(
p_{\mu }r_{\nu }+r_{\mu }p_{\nu }\right) q_{\alpha }+\left( q_{\mu }r_{\nu
}+r_{\mu }q_{\nu }\right) p_{\alpha }\right] \xi _{111}^{\left( -2\right) } 
\notag \\
&&+\tfrac{1}{2}\left( g_{\mu \nu }p_{\alpha }+g_{\mu \alpha }p_{\nu }+g_{\nu
\alpha }p_{\mu }\right) \xi _{100}^{\left( -1\right) }+\tfrac{1}{2}\left(
g_{\mu \nu }q_{\alpha }+g_{\mu \alpha }q_{\nu }+g_{\nu \alpha }q_{\mu
}\right) \xi _{010}^{\left( -1\right) }  \notag \\
&&\left. +\tfrac{1}{2}\left( g_{\mu \nu }r_{\alpha }+g_{\mu \alpha }r_{\nu
}+g_{\nu \alpha }r_{\mu }\right) \xi _{001}^{\left( -1\right) }\right] ,
\label{J4uva}
\end{eqnarray}%
\begin{eqnarray}
J_{4\mu \nu \alpha \beta } &=&i\left( 4\pi \right) ^{-2}\left[ p_{\mu \nu
\alpha \beta }\xi _{400}^{\left( -2\right) }+q_{\mu \nu \alpha \beta }\xi
_{040}^{\left( -2\right) }+r_{\mu \nu \alpha \beta }\xi _{004}^{\left(
-2\right) }\right.  \notag \\
&&+p_{\mu \nu \alpha }q_{\beta }\xi _{310}^{\left( -2\right) }+q_{\mu \nu
\alpha }p_{\beta }\xi _{130}^{\left( -2\right) }+p_{\mu \nu \alpha }r_{\beta
}\xi _{301}^{\left( -2\right) }  \notag \\
&&+r_{\mu \nu \alpha }p_{\beta }\xi _{103}^{\left( -2\right) }+q_{\mu \nu
\alpha }r_{\beta }\xi _{031}^{\left( -2\right) }+r_{\mu \nu \alpha }q_{\beta
}\xi _{013}^{\left( -2\right) }  \notag \\
&&+p_{\mu \nu }q_{\alpha \beta }\xi _{220}^{\left( -2\right) }+p_{\mu \nu
}r_{\alpha \beta }\xi _{202}^{\left( -2\right) }+q_{\mu \nu }r_{\alpha \beta
}\xi _{022}^{\left( -2\right) }  \notag \\
&&+p_{\mu }q_{\nu }r_{\alpha \beta }\xi _{112}^{\left( -2\right) }+r_{\mu
}q_{\nu }p_{\alpha \beta }\xi _{211}^{\left( -2\right) }+p_{\mu }r_{\nu
}q_{\alpha \beta }\xi _{121}^{\left( -2\right) }  \notag \\
&&+\tfrac{1}{2}p_{\mu \nu }g_{\alpha \beta }\xi _{200}^{\left( -1\right) }+%
\tfrac{1}{2}q_{\mu \nu }g_{\alpha \beta }\xi _{020}^{\left( -1\right) }+%
\tfrac{1}{2}r_{\mu \nu }g_{\alpha \beta }\xi _{002}^{\left( -1\right) } 
\notag \\
&&+\tfrac{1}{2}p_{\mu }q_{\nu }g_{\alpha \beta }\xi _{110}^{\left( -1\right)
}+\tfrac{1}{2}q_{\mu }r_{\nu }g_{\alpha \beta }\xi _{011}^{\left( -1\right)
}+\tfrac{1}{2}p_{\mu }r_{\nu }g_{\alpha \beta }\xi _{101}^{\left( -1\right) }
\notag \\
&&\left. -\tfrac{1}{4}g_{\mu \nu }g_{\alpha \beta }\xi _{000}^{\left(
0\right) }\right] +\text{\textrm{permutations}}.
\end{eqnarray}

Once the required pieces are at our disposal, the computation of
perturbative amplitudes occurs in the sequence.

\subsection{Four-Point Amplitudes - $PVVV$}

The amplitude $PVVV$ emerges by integrating Equation (\ref{pvvv}), as
symbolized through the adoption of uppercase letters:%
\begin{eqnarray}
T_{\nu \alpha \beta }^{PVVV} &=&-4im\left( g_{\kappa \nu }g_{\alpha \beta
}-g_{\kappa \alpha }g_{\nu \beta }+g_{\kappa \beta }g_{\nu \alpha }\right)
F_{4\kappa }  \notag \\
&&+2imF_{4\nu \alpha \beta }-i\varepsilon _{\kappa \nu \alpha \beta
}T_{\kappa }^{APPP}.  \label{IntPVVV}
\end{eqnarray}%
Its content mirrors the $AVV$ triangle since both are 3rd-order
pseudotensors having a tensor sector and a vector subamplitude. Hence,
operations performed there find their analogs here.

That is particularly evident for standard tensors with three momenta in the
numerator. The four-propagator version follows the structure (\ref{t34s}),
whose integration resembles that with three propagators (\ref{tensor3}). We
must only change the numerical subindex to four to establish the connection.
This same reasoning applies to the result of integration (\ref{7}); however,
there are no divergent objects this time%
\begin{eqnarray}
T_{4\mu ;\nu \alpha }^{\left( -\right) }\left( k_{l};k_{i},k_{j}\right)
&=&-\left( p_{j}-p_{i}\right) _{\alpha }J_{4\mu \nu }-\left(
p_{i}-p_{j}\right) _{\nu }J_{4\mu \alpha }  \notag \\
&&+\left( p_{j\alpha }p_{i\nu }-p_{i\alpha }p_{j\nu }\right) J_{4\mu
}+\left( p_{j}-p_{i}\right) _{\alpha }p_{l\mu }J_{4\nu }  \notag \\
&&+\left( p_{i}-p_{j}\right) _{\nu }p_{l\mu }J_{4\alpha }-\left( p_{j\alpha
}p_{i\nu }-p_{i\alpha }p_{j\nu }\right) p_{l\mu }J_{4}.
\end{eqnarray}%
Now, four routings $k_{i}$ are available and allow twelve non-equivalent
momenta configurations. That generates differences related to external
momenta: $p_{1}=0$, $p_{2}=k_{1}-k_{2}=p$, $p_{3}=k_{1}-k_{3}=q$, and $%
p_{4}=k_{1}-k_{4}=r$. After performing these identifications, we cast the
standard tensors below.%
\begin{eqnarray}
T_{4\mu ;\nu \alpha }^{\left( -\right) }\left( k_{1};k_{2},k_{3}\right)
&=&-\left( q-p\right) _{\alpha }J_{4\mu \nu }-\left( p-q\right) _{\nu
}J_{4\mu \alpha }+\left( q_{\alpha }p_{\nu }-p_{\alpha }q_{\nu }\right)
J_{4\mu } \\
T_{4\mu ;\nu \alpha }^{\left( -\right) }\left( k_{1};k_{2},k_{4}\right)
&=&-\left( r-p\right) _{\alpha }J_{4\mu \nu }-\left( p-r\right) _{\nu
}J_{4\mu \alpha }+\left( r_{\alpha }p_{\nu }-p_{\alpha }r_{\nu }\right)
J_{4\mu } \\
T_{4\mu ;\nu \alpha }^{\left( -\right) }\left( k_{1};k_{3},k_{4}\right)
&=&-\left( r-q\right) _{\alpha }J_{4\mu \nu }-\left( q-r\right) _{\nu
}J_{4\mu \alpha }+\left( r_{\alpha }q_{\nu }-q_{\alpha }r_{\nu }\right)
J_{4\mu } \\
T_{4\mu ;\nu \alpha }^{\left( -\right) }\left( k_{2};k_{1},k_{3}\right)
&=&-q_{\alpha }J_{4\mu \nu }+q_{\nu }J_{4\mu \alpha }+q_{\alpha }p_{\mu
}J_{4\nu }-q_{\nu }p_{\mu }J_{4\alpha } \\
T_{4\mu ;\nu \alpha }^{\left( -\right) }\left( k_{2};k_{1},k_{4}\right)
&=&-r_{\alpha }J_{4\mu \nu }+r_{\nu }J_{4\mu \alpha }+r_{\alpha }p_{\mu
}J_{4\nu }-r_{\nu }p_{\mu }J_{4\alpha } \\
T_{4\mu ;\nu \alpha }^{\left( -\right) }\left( k_{2};k_{3},k_{4}\right)
&=&-\left( r-q\right) _{\alpha }J_{4\mu \nu }-\left( q-r\right) _{\nu
}J_{4\mu \alpha }+\left( r_{\alpha }q_{\nu }-q_{\alpha }r_{\nu }\right)
J_{4\mu }  \notag \\
&&+\left( r-q\right) _{\alpha }p_{\mu }J_{4\nu }+\left( q-r\right) _{\nu
}p_{\mu }J_{4\alpha }-\left( r_{\alpha }q_{\nu }-q_{\alpha }r_{\nu }\right)
p_{\mu }J_{4} \\
T_{4\mu ;\nu \alpha }^{\left( -\right) }\left( k_{3};k_{1},k_{2}\right)
&=&-p_{\alpha }J_{4\mu \nu }+p_{\nu }J_{4\mu \alpha }+p_{\alpha }q_{\mu
}J_{4\nu }-p_{\nu }q_{\mu }J_{4\alpha } \\
T_{4\mu ;\nu \alpha }^{\left( -\right) }\left( k_{3};k_{1},k_{4}\right)
&=&-r_{\alpha }J_{4\mu \nu }+r_{\nu }J_{4\mu \alpha }+r_{\alpha }q_{\mu
}J_{4\nu }-r_{\nu }q_{\mu }J_{4\alpha } \\
T_{4\mu ;\nu \alpha }^{\left( -\right) }\left( k_{3};k_{2},k_{4}\right)
&=&-\left( r-p\right) _{\alpha }J_{4\mu \nu }-\left( p-r\right) _{\nu
}J_{4\mu \alpha }+\left( r_{\alpha }p_{\nu }-p_{\alpha }r_{\nu }\right)
J_{4\mu }  \notag \\
&&+\left( r-p\right) _{\alpha }q_{\mu }J_{4\nu }+\left( p-r\right) _{\nu
}q_{\mu }J_{4\alpha }-\left( r_{\alpha }p_{\nu }-p_{\alpha }r_{\nu }\right)
q_{\mu }J_{4} \\
T_{4\mu ;\nu \alpha }^{\left( -\right) }\left( k_{4};k_{1},k_{2}\right)
&=&-p_{\alpha }J_{4\mu \nu }+p_{\nu }J_{4\mu \alpha }+p_{\alpha }r_{\mu
}J_{4\nu }-p_{\nu }r_{\mu }J_{4\alpha } \\
T_{4\mu ;\nu \alpha }^{\left( -\right) }\left( k_{4};k_{1},k_{3}\right)
&=&-q_{\alpha }J_{4\mu \nu }+q_{\nu }J_{4\mu \alpha }+q_{\alpha }r_{\mu
}J_{4\nu }-q_{\nu }r_{\mu }J_{4\alpha } \\
T_{4\mu ;\nu \alpha }^{\left( -\right) }\left( k_{4};k_{2},k_{4}\right)
&=&-\left( q-p\right) _{\alpha }J_{4\mu \nu }-\left( p-q\right) _{\nu
}J_{4\mu \alpha }+\left( q_{\alpha }p_{\nu }-p_{\alpha }q_{\nu }\right)
J_{4\mu }  \notag \\
&&+\left( q-p\right) _{\alpha }r_{\mu }J_{4\nu }+\left( p-q\right) _{\nu
}r_{\mu }J_{4\alpha }-\left( q_{\alpha }p_{\nu }-p_{\alpha }q_{\nu }\right)
r_{\mu }J_{4}
\end{eqnarray}

Our next step consists of building objects containing these tensors in their
structure. Thus, we start by suiting the notation within the vector $F_{4\mu
}$ (\ref{f4}) to write its integrated version%
\begin{eqnarray}
F_{4\mu } &=&\varepsilon _{\mu \rho \sigma \kappa }\left[ T_{4}^{\left(
-\right) \rho ;\sigma \kappa }\left( k_{2};k_{3},k_{4}\right) -T_{4}^{\left(
-\right) \rho ;\sigma \kappa }\left( k_{1};k_{3},k_{4}\right) \right.  \notag
\\
&&\left. +T_{4}^{\left( -\right) \rho ;\sigma \kappa }\left(
k_{1};k_{2},k_{4}\right) -T_{4}^{\left( -\right) \rho ;\sigma \kappa }\left(
k_{1};k_{2},k_{3}\right) \right] .
\end{eqnarray}%
Whereas contributions on the 2nd-order $J$-tensor cancel out directly due to
symmetry properties in the contraction, the same does not occur for other
sectors%
\begin{eqnarray}
F_{4\mu } &=&\varepsilon _{\mu \rho \sigma \kappa }\left\{ -\left( q^{\kappa
}p^{\sigma }-p^{\kappa }q^{\sigma }\right) J_{4}^{\rho }+\left( r^{\kappa
}p^{\sigma }-p^{\kappa }r^{\sigma }\right) J_{4}^{\rho }\right.  \notag \\
&&\left. +\left( r-q\right) ^{\kappa }p^{\rho }J_{4}^{\sigma }+\left(
q-r\right) ^{\sigma }p^{\rho }J_{4}^{\kappa }-\left( r^{\kappa }q^{\sigma
}-q^{\kappa }r^{\sigma }\right) p^{\rho }J_{4}\right\} .
\end{eqnarray}%
A closer look at the $J$-vector structure (\ref{J4v}) is required to verify
that it also vanishes. Ultimately, all terms disappear for symmetry reasons.
At the end of calculations, only the scalar sector remains%
\begin{equation}
F_{4\mu }=-2\varepsilon _{\mu \rho \sigma \kappa }p^{\rho }q^{\sigma
}r^{\kappa }J_{4}.  \label{F4}
\end{equation}

We need all momenta configurations to assemble the other tensor group (\ref%
{f4vab}) as seen in its integrated version%
\begin{eqnarray}
F_{4\nu \alpha \beta } &=&-\left( \varepsilon _{\alpha \beta XY}g_{\nu
Z}-\varepsilon _{\nu \beta XY}g_{\alpha Z}+\varepsilon _{\nu \alpha
XY}g_{\beta Z}\right) T_{4Z;XY}^{\left( -\right) }\left(
k_{1};k_{3},k_{4}\right)  \notag \\
&&+\left( \varepsilon _{\alpha \beta XY}g_{\nu Z}-\varepsilon _{\nu \beta
XY}g_{\alpha Z}+\varepsilon _{\nu \alpha XY}g_{\beta Z}\right)
T_{4Z;XY}^{\left( -\right) }\left( k_{1};k_{2},k_{4}\right)  \notag \\
&&-\left( \varepsilon _{\alpha \beta XY}g_{\nu Z}-\varepsilon _{\nu \beta
XY}g_{\alpha Z}+\varepsilon _{\nu \alpha XY}g_{\beta Z}\right)
T_{4Z;XY}^{\left( -\right) }\left( k_{1};k_{2},k_{3}\right)  \notag \\
&&-\left( \varepsilon _{\alpha \beta XY}g_{\nu Z}+\varepsilon _{\nu \beta
XY}g_{\alpha Z}-\varepsilon _{\nu \alpha XY}g_{\beta Z}\right)
T_{4Z;XY}^{\left( -\right) }\left( k_{2};k_{3},k_{4}\right)  \notag \\
&&+\left( \varepsilon _{\alpha \beta XY}g_{\nu Z}+\varepsilon _{\nu \beta
XY}g_{\alpha Z}-\varepsilon _{\nu \alpha XY}g_{\beta Z}\right)
T_{4Z;XY}^{\left( -\right) }\left( k_{2};k_{1},k_{4}\right)  \notag \\
&&-\left( \varepsilon _{\alpha \beta XY}g_{\nu Z}+\varepsilon _{\nu \beta
XY}g_{\alpha Z}-\varepsilon _{\nu \alpha XY}g_{\beta Z}\right)
T_{4Z;XY}^{\left( -\right) }\left( k_{2};k_{1},k_{3}\right)  \notag \\
&&+\left( \varepsilon _{\alpha \beta XY}g_{\nu Z}-\varepsilon _{\nu \beta
XY}g_{\alpha Z}-\varepsilon _{\nu \alpha XY}g_{\beta Z}\right)
T_{4Z;XY}^{\left( -\right) }\left( k_{3};k_{2},k_{4}\right)  \notag \\
&&-\left( \varepsilon _{\alpha \beta XY}g_{\nu Z}-\varepsilon _{\nu \beta
XY}g_{\alpha Z}-\varepsilon _{\nu \alpha XY}g_{\beta Z}\right)
T_{4Z;XY}^{\left( -\right) }\left( k_{3};k_{1},k_{4}\right)  \notag \\
&&+\left( \varepsilon _{\alpha \beta XY}g_{\nu Z}-\varepsilon _{\nu \beta
XY}g_{\alpha Z}-\varepsilon _{\nu \alpha XY}g_{\beta Z}\right)
T_{4Z;XY}^{\left( -\right) }\left( k_{3};k_{1},k_{2}\right)  \notag \\
&&-\left( \varepsilon _{\alpha \beta XY}g_{\nu Z}-\varepsilon _{\nu \beta
XY}g_{\alpha Z}+\varepsilon _{\nu \alpha XY}g_{\beta Z}\right)
T_{4Z;XY}^{\left( -\right) }\left( k_{4};k_{2},k_{3}\right)  \notag \\
&&+\left( \varepsilon _{\alpha \beta XY}g_{\nu Z}-\varepsilon _{\nu \beta
XY}g_{\alpha Z}+\varepsilon _{\nu \alpha XY}g_{\beta Z}\right)
T_{4Z;XY}^{\left( -\right) }\left( k_{4};k_{1},k_{3}\right)  \notag \\
&&-\left( \varepsilon _{\alpha \beta XY}g_{\nu Z}-\varepsilon _{\nu \beta
XY}g_{\alpha Z}+\varepsilon _{\nu \alpha XY}g_{\beta Z}\right)
T_{4Z;XY}^{\left( -\right) }\left( k_{4};k_{1},k_{2}\right) .
\end{eqnarray}

Some simplifications are immediate after replacing standard tensors,
yielding the expression%
\begin{eqnarray}
F_{4\nu \alpha \beta } &=&4m\left( -\varepsilon _{\alpha \beta XY}g_{\nu
Z}+\varepsilon _{\nu \beta XY}g_{\alpha Z}-\varepsilon _{\nu \alpha
XY}g_{\beta Z}\right) \left( p_{X}q_{Y}-p_{X}r_{Y}+q_{X}r_{Y}\right) J_{4Z} 
\notag \\
&&+4m\left( -\varepsilon _{\alpha \beta XY}g_{\nu Z}-\varepsilon _{\nu \beta
XY}g_{\alpha Z}+\varepsilon _{\nu \alpha XY}g_{\beta Z}\right)
q_{X}r_{Y}\left( J_{4Z}-p_{Z}J_{4}\right)  \notag \\
&&+4m\left( \varepsilon _{\alpha \beta XY}g_{\nu Z}-\varepsilon _{\nu \beta
XY}g_{\alpha Z}-\varepsilon _{\nu \alpha XY}g_{\beta Z}\right)
p_{X}r_{Y}\left( J_{4Z}-q_{Z}J_{4}\right)  \notag \\
&&+4m\left( -\varepsilon _{\alpha \beta XY}g_{\nu Z}+\varepsilon _{\nu \beta
XY}g_{\alpha Z}-\varepsilon _{\nu \alpha XY}g_{\beta Z}\right)
p_{X}q_{Y}\left( J_{4Z}-r_{Z}J_{4}\right) .
\end{eqnarray}%
As these coefficients are products between the Levi-Civita symbol and the
metric tensor, rearranging indices through Schouten identities (\ref%
{Schouten}) is feasible. Nevertheless, contractions involving external
momenta and $J_{4}$-vectors emerge in this process. From the explicit form (%
\ref{J4v}), we recognize these contractions as reductions obtained in the
strategy context (\ref{rel1})-(\ref{rel3}). Their employment allows
expressing the result as follows%
\begin{eqnarray}
F_{4\nu \alpha \beta } &=&4m\varepsilon _{\nu \alpha \beta \rho }\left[
\left( r-q\right) ^{\rho }\left( J_{3}^{\prime \prime \prime }-J_{3}^{\prime
\prime }\right) +p^{\rho }\left( J_{3}-J_{3}^{\prime }\right) \right]  \notag
\\
&&-8m\varepsilon _{\alpha \beta \rho \sigma }q^{\rho }r^{\sigma }J_{4\nu
}-8m\varepsilon _{\nu \alpha \rho \sigma }p^{\rho }r^{\sigma }J_{4\beta } 
\notag \\
&&+4m\varepsilon _{\nu \alpha \beta \rho }\left[ \left( q^{2}-r^{2}-q\cdot
r\right) p^{\rho }+\left( p\cdot r-p^{2}\right) q^{\rho }+p^{2}r^{\rho }%
\right] J_{4}  \notag \\
&&-4m\varepsilon _{\beta \alpha \rho \sigma }\left( q^{\rho }p_{\nu
}-p^{\rho }q_{\nu }\right) r^{\sigma }J_{4}+4m\varepsilon _{\nu \beta \rho
\sigma }\left( q^{\rho }p_{\alpha }+p^{\rho }q_{\alpha }\right) r^{\sigma
}J_{4}  \notag \\
&&\left. -4m\varepsilon _{\nu \alpha \rho \sigma }\left( q^{\rho }p_{\beta
}-p^{\rho }q_{\beta }\right) \right] r^{\sigma }J_{4},
\end{eqnarray}%
where we identified $J_{3}$-scalars and extended the line notation to them (%
\ref{line2}).

The last ingredient is the vector subamplitude $APPP$ (\ref{appp}), whose
integration leads to the combination%
\begin{eqnarray}
T_{\sigma }^{APPP} &=&4mp_{\sigma }I_{3}\left( k_{1},k_{2},k_{4}\right)
+4m\left( r-q\right) _{\sigma }I_{3}\left( k_{1},k_{3},k_{4}\right)  \notag
\\
&&-4m\left[ \left( q^{2}-q\cdot r\right) p_{\sigma }-\left( p^{2}-p\cdot
r\right) q_{\sigma }+\left( p^{2}-p\cdot q\right) r_{\sigma }\right] I_{4}.
\end{eqnarray}%
Since these Feynman integrals are finite, see Equations (\ref{I30}) and (\ref%
{I40}), the link with the corresponding $J$-scalars is straightforward%
\begin{eqnarray}
T_{\sigma }^{APPP} &=&4mp_{\sigma }J_{3}^{\prime }+4m\left( r-q\right)
_{\sigma }J_{3}^{\prime \prime }-4m\left[ \left( q^{2}-q\cdot r\right)
p_{\sigma }\right.  \notag \\
&&\left. -\left( p^{2}-p\cdot r\right) q_{\sigma }+\left( p^{2}-p\cdot
q\right) r_{\sigma }\right] J_{4}.
\end{eqnarray}

Once all pieces are known, we replace them in the original form (\ref%
{IntPVVV}) to compound the $PVVV$ amplitude:%
\begin{eqnarray}
T_{\nu \alpha \beta }^{PVVV} &=&4im\left( g_{\alpha \beta }\varepsilon _{\nu
\rho \sigma \kappa }-g_{\nu \beta }\varepsilon _{\alpha \rho \sigma \kappa
}+g_{\nu \alpha }\varepsilon _{\beta \rho \sigma \kappa }\right) p^{\rho
}q^{\sigma }r^{\kappa }J_{4}  \notag \\
&&+4im\varepsilon _{\nu \alpha \beta \rho }\left[ \left( r-q\right) ^{\rho
}J_{3}^{\prime \prime \prime }+p^{\rho }J_{3}\right] -8im\varepsilon
_{\alpha \beta \rho \sigma }q^{\rho }r^{\sigma }J_{4\nu }  \notag \\
&&-8im\varepsilon _{\nu \alpha \rho \sigma }p^{\rho }r^{\sigma }J_{4\beta
}+4im\varepsilon _{\nu \alpha \beta \rho }\left[ \left( p\cdot q\right)
r^{\rho }-r^{2}p^{\rho }\right] J_{4}  \notag \\
&&-4im\varepsilon _{\beta \alpha \rho \sigma }\left( q^{\rho }p_{\nu
}-p^{\rho }q_{\nu }\right) r^{\sigma }J_{4}  \notag \\
&&+4im\varepsilon _{\nu \beta \rho \sigma }\left( q^{\rho }p_{\alpha
}+p^{\rho }q_{\alpha }\right) r^{\sigma }J_{4}  \notag \\
&&-4im\varepsilon _{\nu \alpha \rho \sigma }\left( q^{\rho }p_{\beta
}-p^{\rho }q_{\beta }\right) r^{\sigma }J_{4}.  \label{PVVV}
\end{eqnarray}%
As anticipated by the analysis of mass dimension, we found a finite
structure.

\subsection{Four-Point Amplitudes - $AVVV$}

We reach the last correlator that concerns this investigation. From Equation
(\ref{avvv}), we write the integrated version of the $AVVV$ amplitude as%
\begin{eqnarray}
T_{\mu \nu \alpha \beta }^{AVVV} &=&iF_{4\mu \nu \alpha \beta }-i\left[
\varepsilon _{\mu \alpha \beta X}T_{X\nu }^{VVPP}+\varepsilon _{\mu \nu
\beta X}T_{X\alpha }^{VPVP}+\varepsilon _{\mu \nu \alpha X}T_{X\beta }^{VPPV}%
\right]  \notag \\
&&+\left[ g_{\alpha \beta }T_{\mu \nu }^{AVPP}+g_{\nu \beta }T_{\mu \alpha
}^{APVP}+g_{\nu \alpha }T_{\mu \beta }^{APPV}\right] +2i\varepsilon _{\mu
\nu \alpha \beta }T^{PPPP}.  \label{AVVV}
\end{eqnarray}%
Since the involved mathematical expressions are extensive, we focus only on
analyzing substructures without providing the complete object. Although this
presentation follows the same steps from Section (\ref{box1}), we add one
step to discuss 2nd-order tensors before building the corresponding
subamplitudes.

\subsubsection{Fourth-Order Standard Tensors}

First, we compute all required 4th-order tensors starting with the
simplified version. Besides appearing by itself within $AVPP$-like
functions, this object compounds the standard version required to express
the sector $F_{4\mu \nu \alpha \beta }$. These are the only places where the
Feynman integral $I_{4\mu \nu \alpha \beta }$ appears; therefore, containing
all contributions symmetric in four indices\footnote{%
The mentioned structures are a combination of surface terms $A_{\mu \nu
\alpha \beta }$, the irreducible divergent object $I_{\log }$, and the
finite tensor $J_{\mu \nu \alpha \beta }$. Consult Equation (\ref{I4div})
for further information.}. Since most of the involved tensors exhibit
antisymmetry in some indices, we acknowledged the possibility of
cancellation for these contributions. Verifying this prospect is part of our
goal. If this situation indeed occurs, the surface term $\square _{\mu \nu
\alpha \beta }$ and the finite tensor $J_{4\mu \nu \alpha \beta }$ do not
appear in this work.

By expanding products from the numerator of its structure (\ref{t4+}) and
integrating, we recognize the simplified version as a combination of
four-propagator Feynman integrals:%
\begin{eqnarray}
T_{4\mu \nu \alpha \beta }\left( k_{i},k_{j},k_{m},k_{n}\right) &=&I_{4\mu
\nu \alpha \beta }+\left[ k_{i\mu }I_{4\nu \alpha \beta }+k_{j\nu }I_{4\mu
\alpha \beta }+k_{m\alpha }I_{4\mu \nu \beta }+k_{n\beta }I_{4\mu \nu \alpha
}\right]  \notag \\
&&+\left[ k_{i\mu }k_{j\nu }I_{4\alpha \beta }+k_{i\mu }k_{m\alpha }I_{4\nu
\beta }+k_{i\mu }k_{n\beta }I_{4\nu \alpha }+k_{j\nu }k_{m\alpha }I_{4\mu
\beta }\right.  \notag \\
&&\left. +k_{j\nu }k_{n\beta }I_{4\mu \alpha }+k_{m\alpha }k_{n\beta
}I_{4\mu \nu }\right] +\left[ k_{j\nu }k_{m\alpha }k_{n\beta }I_{4\mu
}\right.  \notag \\
&&\left. +k_{i\mu }k_{m\alpha }k_{n\beta }I_{4\nu }+k_{i\mu }k_{j\nu
}k_{n\beta }I_{4\alpha }+k_{i\mu }k_{j\nu }k_{m\alpha }I_{4\beta }\right] 
\notag \\
&&+k_{i\mu }k_{j\nu }k_{m\alpha }k_{n\beta }I_{4}.
\end{eqnarray}%
Next, our task consists of substituting their explicit expressions (\ref{I40}%
)-(\ref{I4uvab}) while obeying the hierarchy observed in previous cases;
consult Equation (\ref{T3s}). This strategy allows writing all finite
structures through $J$-tensors with external momenta $p_{i}=k_{1}-k_{i}$ as
coefficients. Observe that the $J$-scalar does not contribute due to the
unavoidable dependence on $p_{1}=0$. Once these ideas are clear, we
introduce the simplified version%
\begin{eqnarray}
T_{4\mu \nu \alpha \beta }\left( k_{i},k_{j},k_{m},k_{n}\right) &=&\tfrac{1}{%
24}A_{\mu \nu \alpha \beta }+\tfrac{1}{24}g_{\mu \nu \alpha \beta }I_{\log
}+J_{4\mu \nu \alpha \beta }  \notag \\
&&-\left[ p_{i\mu }J_{4\nu \alpha \beta }+p_{j\nu }J_{4\mu \alpha \beta
}+p_{m\alpha }J_{4\mu \nu \beta }+p_{n\beta }J_{4\mu \nu \alpha }\right] 
\notag \\
&&+\left[ p_{i\mu }p_{j\nu }J_{4\alpha \beta }+p_{i\mu }p_{m\alpha }J_{4\nu
\beta }+p_{i\mu }p_{n\beta }J_{4\nu \alpha }+p_{j\nu }p_{m\alpha }J_{4\mu
\beta }\right.  \notag \\
&&\left. +p_{j\nu }p_{n\beta }J_{4\mu \alpha }+p_{m\alpha }p_{n\beta
}J_{4\mu \nu }\right] -\left[ p_{j\nu }p_{m\alpha }p_{n\beta }J_{4\mu
}\right.  \notag \\
&&\left. +p_{i\mu }p_{m\alpha }p_{n\beta }J_{4\nu }+p_{i\mu }p_{j\nu
}p_{n\beta }J_{4\alpha }+p_{i\mu }p_{j\nu }p_{m\alpha }J_{4\beta }\right] ,
\label{ref1}
\end{eqnarray}%
and all necessary momenta configurations%
\begin{eqnarray}
T_{4\mu \nu \alpha \beta }\left( k_{1},k_{2},k_{3},k_{4}\right) &=&\tfrac{1}{%
24}A_{\mu \nu \alpha \beta }+\tfrac{1}{24}g_{\mu \nu \alpha \beta }I_{\log
}+J_{4\mu \nu \alpha \beta }  \notag \\
&&-\left[ p_{\nu }J_{4\mu \alpha \beta }+q_{\alpha }J_{4\mu \nu \beta
}+r_{\beta }J_{4\mu \nu \alpha }\right]  \notag \\
&&+\left[ p_{\nu }q_{\alpha }J_{4\mu \beta }+p_{\nu }r_{\beta }J_{4\mu
\alpha }+q_{\alpha }r_{\beta }J_{4\mu \nu }\right] -p_{\nu }q_{\alpha
}r_{\beta }J_{4\mu },  \label{6}
\end{eqnarray}%
\begin{eqnarray}
T_{4\mu \nu \alpha \beta }\left( k_{1},k_{2},k_{4},k_{3}\right) &=&\tfrac{1}{%
24}A_{\mu \nu \alpha \beta }+\tfrac{1}{24}g_{\mu \nu \alpha \beta }I_{\log
}+J_{4\mu \nu \alpha \beta }  \notag \\
&&-\left[ p_{\nu }J_{4\mu \alpha \beta }+r_{\alpha }J_{4\mu \nu \beta
}+q_{\beta }J_{4\mu \nu \alpha }\right]  \notag \\
&&+\left[ p_{\nu }r_{\alpha }J_{4\mu \beta }+p_{\nu }q_{\beta }J_{4\mu
\alpha }+r_{\alpha }q_{\beta }J_{4\mu \nu }\right] -p_{\nu }r_{\alpha
}q_{\beta }J_{4\mu },
\end{eqnarray}%
\begin{eqnarray}
T_{4\mu \nu \alpha \beta }\left( k_{2},k_{1},k_{3},k_{4}\right) &=&\tfrac{1}{%
24}A_{\mu \nu \alpha \beta }+\tfrac{1}{24}g_{\mu \nu \alpha \beta }I_{\log
}+J_{4\mu \nu \alpha \beta }  \notag \\
&&-\left[ p_{\mu }J_{4\nu \alpha \beta }+q_{\alpha }J_{4\mu \nu \beta
}+r_{\beta }J_{4\mu \nu \alpha }\right]  \notag \\
&&+\left[ p_{\mu }q_{\alpha }J_{4\nu \beta }+p_{\mu }r_{\beta }J_{4\nu
\alpha }+q_{\alpha }r_{\beta }J_{4\mu \nu }\right] -p_{\mu }q_{\alpha
}r_{\beta }J_{4\nu },
\end{eqnarray}%
\begin{eqnarray}
T_{4\mu \nu \alpha \beta }\left( k_{2},k_{1},k_{4},k_{3}\right) &=&\tfrac{1}{%
24}A_{\mu \nu \alpha \beta }+\tfrac{1}{24}g_{\mu \nu \alpha \beta }I_{\log
}+J_{4\mu \nu \alpha \beta }  \notag \\
&&-\left[ p_{\mu }J_{4\nu \alpha \beta }+r_{\alpha }J_{4\mu \nu \beta
}+q_{\beta }J_{4\mu \nu \alpha }\right]  \notag \\
&&+\left[ p_{\mu }r_{\alpha }J_{4\nu \beta }+p_{\mu }q_{\beta }J_{4\nu
\alpha }+r_{\alpha }q_{\beta }J_{4\mu \nu }\right] -p_{\mu }r_{\alpha
}q_{\beta }J_{4\nu }.
\end{eqnarray}

Contributions symmetric in four indices come from the highest-order
integral, appearing in the first row from the equations above. We stress
that version (\ref{6}) appears contracted to the Levi-Civita symbol with $%
AVPP$-type amplitudes; see Equation (\ref{sector}). That implies symmetric
contributions vanish, but we return to this discussion in due time.

With these tools determined, let us obtain the standard version that admits
sign choices (\ref{t44s}). By integrating Equation (\ref{t4comp}), we write
this object through the following combination:%
\begin{eqnarray}
T_{4\mu \nu ;\alpha \beta }^{\left( s_{1},s_{2}\right) } &=&T_{4\mu \nu
\alpha \beta }\left( k_{1},k_{2},k_{3},k_{4}\right) +s_{1}T_{4\mu \nu \alpha
\beta }\left( k_{2},k_{1},k_{3},k_{4}\right)  \notag \\
&&+s_{2}T_{4\mu \nu \alpha \beta }\left( k_{1},k_{2},k_{4},k_{3}\right)
+s_{1}s_{2}T_{4\mu \nu ;\alpha \beta }\left( k_{2},k_{1},k_{4},k_{3}\right) .
\end{eqnarray}%
We omit arguments exhibiting the momenta hierarchy $T_{4\mu \nu ;\alpha
\beta }^{\left( s_{1};s_{2}\right) }=T_{4\mu \nu ;\alpha \beta }^{\left(
s_{1};s_{2}\right) }\left( k_{1},k_{2};k_{3},k_{4}\right) $. Then, our job
consists of replacing expressions attributed to different momenta
configurations. This operation produces the generic form%
\begin{eqnarray}
T_{4\mu \nu ;\alpha \beta }^{\left( s_{1},s_{2}\right) } &=&\left(
1+s_{1}\right) \left( 1+s_{2}\right) \left[ \tfrac{1}{24}A_{\mu \nu \alpha
\beta }+\tfrac{1}{24}g_{\mu \nu \alpha \beta }I_{\log }+J_{4\mu \nu \alpha
\beta }\right]  \notag \\
&&-\left( 1+s_{2}\right) \left[ s_{1}p_{\mu }J_{4\nu \alpha \beta }+p_{\nu
}J_{4\mu \alpha \beta }\right]  \notag \\
&&-\left( 1+s_{1}\right) \left[ \left( q_{\alpha }+s_{2}r_{\alpha }\right)
J_{4\mu \nu \beta }+\left( r_{\beta }+s_{2}q_{\beta }\right) J_{4\mu \nu
\alpha }\right]  \notag \\
&&+\left( 1+s_{1}\right) \left( q_{\alpha }r_{\beta }+s_{2}q_{\beta
}r_{\alpha }\right) J_{4\mu \nu }+\left( r_{\beta }+s_{2}q_{\beta }\right)
\left( p_{\nu }J_{4\mu \alpha }+s_{1}p_{\mu }J_{4\nu \alpha }\right)  \notag
\\
&&+\left( q_{\alpha }+s_{2}r_{\alpha }\right) \left( p_{\nu }J_{4\mu \beta
}+s_{1}p_{\mu }J_{4\nu \beta }\right)  \notag \\
&&-\left( q_{\alpha }r_{\beta }+s_{2}r_{\alpha }q_{\beta }\right) \left(
p_{\nu }J_{4\mu }+s_{1}p_{\mu }J_{4\nu }\right) ;
\end{eqnarray}%
hence, setting the signs leads to four particular forms%
\begin{eqnarray}
T_{4\mu \nu ;\alpha \beta }^{\left( +,+\right) } &=&\tfrac{1}{6}A_{\mu \nu
\alpha \beta }+\tfrac{1}{6}g_{\mu \nu \alpha \beta }I_{\log }+4J_{4\mu \nu
\alpha \beta }  \notag \\
&&-2\left[ p_{\mu }J_{4\nu \alpha \beta }+p_{\nu }J_{4\mu \alpha \beta }%
\right] -2\left[ \left( q_{\alpha }+r_{\alpha }\right) J_{4\mu \nu \beta
}+\left( r_{\beta }+q_{\beta }\right) J_{4\mu \nu \alpha }\right]  \notag \\
&&+2\left( q_{\alpha }r_{\beta }+q_{\beta }r_{\alpha }\right) J_{4\mu \nu
}+\left( r_{\beta }+q_{\beta }\right) \left( p_{\nu }J_{4\mu \alpha }+p_{\mu
}J_{4\nu \alpha }\right)  \notag \\
&&+\left( q_{\alpha }+r_{\alpha }\right) \left( p_{\nu }J_{4\mu \beta
}+p_{\mu }J_{4\nu \beta }\right) -\left( q_{\alpha }r_{\beta }+r_{\alpha
}q_{\beta }\right) \left( p_{\nu }J_{4\mu }+p_{\mu }J_{4\nu }\right) ,
\end{eqnarray}%
\begin{eqnarray}
T_{4\mu \nu ;\alpha \beta }^{\left( +,-\right) } &=&-2\left[ \left(
q_{\alpha }-r_{\alpha }\right) J_{4\mu \nu \beta }+\left( r_{\beta
}-q_{\beta }\right) J_{4\mu \nu \alpha }\right] +2\left( q_{\alpha }r_{\beta
}-q_{\beta }r_{\alpha }\right) J_{4\mu \nu }  \notag \\
&&+\left( r_{\beta }-q_{\beta }\right) \left( p_{\nu }J_{4\mu \alpha
}+p_{\mu }J_{4\nu \alpha }\right) +\left( q_{\alpha }-r_{\alpha }\right)
\left( p_{\nu }J_{4\mu \beta }+p_{\mu }J_{4\nu \beta }\right)  \notag \\
&&-\left( q_{\alpha }r_{\beta }-r_{\alpha }q_{\beta }\right) \left( p_{\nu
}J_{4\mu }+p_{\mu }J_{4\nu }\right) ,
\end{eqnarray}%
\begin{eqnarray}
T_{4\mu \nu ;\alpha \beta }^{\left( -,+\right) } &=&2\left( p_{\mu }J_{4\nu
\alpha \beta }-p_{\nu }J_{4\mu \alpha \beta }\right) +\left( r_{\beta
}+q_{\beta }\right) \left( p_{\nu }J_{4\mu \alpha }-p_{\mu }J_{4\nu \alpha
}\right)  \notag \\
&&+\left( q_{\alpha }+r_{\alpha }\right) \left( p_{\nu }J_{4\mu \beta
}-p_{\mu }J_{4\nu \beta }\right) -\left( q_{\alpha }r_{\beta }+r_{\alpha
}q_{\beta }\right) \left( p_{\nu }J_{4\mu }-p_{\mu }J_{4\nu }\right) ,
\end{eqnarray}%
\begin{eqnarray}
T_{4\mu \nu ;\alpha \beta }^{\left( -,-\right) } &=&\left( r_{\beta
}-q_{\beta }\right) \left( p_{\nu }J_{4\mu \alpha }-p_{\mu }J_{4\nu \alpha
}\right) +\left( q_{\alpha }-r_{\alpha }\right) \left( p_{\nu }J_{4\mu \beta
}-p_{\mu }J_{4\nu \beta }\right)  \notag \\
&&-\left( q_{\alpha }r_{\beta }-r_{\alpha }q_{\beta }\right) \left( p_{\nu
}J_{4\mu }-p_{\mu }J_{4\nu }\right) .
\end{eqnarray}

Lastly, from Equations (\ref{f41})-(\ref{f42}), we aim to determine the
entire sector%
\begin{equation}
F_{4\mu \nu \alpha \beta }=4\varepsilon _{\mu \nu XY}T_{XY\alpha \beta
}^{\left( 12\right) }+4\varepsilon _{\mu \alpha XY}T_{XY\nu \beta }^{\left(
13\right) }+4\varepsilon _{\mu \beta XY}T_{XY\nu \alpha }^{\left( 14\right)
}.  \label{F4uvab}
\end{equation}%
Each of its pieces relates to a combination of standard tensors $T_{4\mu \nu
;\alpha \beta }^{\left( s_{1},s_{2}\right) }=T_{4\mu \nu ;\alpha \beta
}^{\left( s_{1},s_{2}\right) }\left( k_{1},k_{2};k_{3},k_{4}\right) $:%
\begin{eqnarray}
2T_{XY\alpha \beta }^{\left( 12\right) } &=&T_{4XY;\alpha \beta }^{\left(
-;+\right) }+T_{4X\alpha ;Y\beta }^{\left( -;+\right) }-T_{4X\beta ;Y\alpha
}^{\left( -;-\right) }+T_{4\alpha Y;\beta X}^{\left( -;+\right) }+T_{4\beta
Y;\alpha X}^{\left( -;-\right) }+T_{4\alpha \beta ;XY}^{\left( -;-\right) },
\\
2T_{XY\nu \beta }^{\left( 13\right) } &=&-T_{4XY;\nu \beta }^{\left(
-;+\right) }+T_{4Y\beta ;\nu X}^{\left( -;-\right) }-T_{4\nu Y;\beta
X}^{\left( +;+\right) }-T_{4\beta X;Y\nu }^{\left( -;-\right) }+T_{4\nu
X;Y\beta }^{\left( +;+\right) }-T_{4\nu \beta ;XY}^{\left( +;-\right) }, \\
2T_{XY\nu \alpha }^{\left( 14\right) } &=&T_{4XY;\nu \alpha }^{\left(
-;-\right) }+T_{4\alpha Y;\nu X}^{\left( -;-\right) }-T_{4\nu Y;\alpha
X}^{\left( +;-\right) }+T_{4\alpha X;Y\nu }^{\left( -;-\right) }-T_{4\nu
X;Y\alpha }^{\left( +;-\right) }+T_{4\nu \alpha ;XY}^{\left( +;-\right) }.
\end{eqnarray}%
We highlight that the tensor with $s_{1}=s_{2}=+1$ is the only one
containing structures symmetric in four indices; thus, it is straightforward
to verify their cancellation within object $T_{XY\nu \beta }^{\left(
13\right) }$. The immediate consequence is that the entire sector consists
of a finite object. Considering our comment on $AVPP$-like amplitudes, this
result completes the proof that the surface term $\square _{\mu \nu \alpha
\beta }$ and the finite tensor $J_{4\mu \nu \alpha \beta }$ do not appear in
this work.

Since all tensors exhibit the same momenta configuration, no additional
ingredients are necessary for their evaluation. We only have to rename
indices of the particular versions of the standard tensor (with signs set)
and perform the replacements. As the adopted notations emphasize contracted
indices through uppercase Latin letters, simplifications associated with
symmetry properties are evident. After performing them, we present the final
expressions attributed to the tensors below. Arrows indicate that only
non-trivial contributions regarding contractions appear, which is compatible
with Equation (\ref{F4uvab}). 
\begin{eqnarray}
T_{XY\alpha \beta }^{\left( 12\right) } &\rightarrow &4p_{X}J_{4Y\alpha
\beta }-2\left( p_{\alpha }q_{X}+q_{\alpha }p_{X}\right) J_{4Y\beta }  \notag
\\
&&-2\left[ \left( q+r\right) _{\beta }p_{X}-p_{\beta }\left( q-r\right) _{X}%
\right] J_{4Y\alpha }+2r_{X}p_{Y}J_{4\alpha \beta }  \notag \\
&&+\left[ \left( q_{\alpha }r_{\beta }+r_{\alpha }q_{\beta }\right)
p_{X}+\left( r_{\beta }p_{\alpha }-r_{\alpha }p_{\beta }\right) q_{X}+\left(
q_{\beta }p_{\alpha }+q_{\alpha }p_{\beta }\right) r_{X}\right] J_{4Y} 
\notag \\
&&+\left( q_{\beta }p_{X}r_{Y}+r_{\beta }p_{X}q_{Y}+p_{\beta
}r_{X}q_{Y}\right) J_{4\alpha }  \notag \\
&&+\left( q_{\alpha }p_{X}r_{Y}+r_{\alpha }q_{X}p_{Y}+p_{\alpha
}q_{X}r_{Y}\right) J_{4\beta }  \label{12}
\end{eqnarray}%
\begin{eqnarray}
T_{XY\nu \beta }^{\left( 13\right) } &\rightarrow &4\left( q-p\right)
_{X}J_{4Y\nu \beta }+2\left( q_{\nu }p_{X}-p_{\nu }q_{X}\right) J_{4Y\beta }
\notag \\
&&+2\left[ \left( q+r\right) _{\beta }p_{X}-\left( p+r\right) _{\beta
}q_{X}+\left( p-q\right) _{\beta }r_{X}\right] J_{4Y\nu }+2\left( p-q\right)
_{X}r_{Y}J_{4\nu \beta }  \notag \\
&&-\left[ \left( q_{\nu }r_{\beta }+r_{\nu }q_{\beta }\right) p_{X}-\left(
r_{\beta }p_{\nu }+r_{\nu }p_{\beta }\right) q_{X}+\left( q_{\nu }p_{\beta
}-q_{\beta }p_{\nu }\right) r_{X}\right] J_{4Y}  \notag \\
&&+\left( q_{\beta }r_{X}p_{Y}+r_{\beta }q_{X}p_{Y}+q_{X}r_{Y}p_{\beta
}\right) J_{4\nu }  \notag \\
&&+\left( p_{\nu }q_{X}r_{Y}+q_{\nu }r_{X}p_{Y}+r_{\nu }p_{X}q_{Y}\right)
J_{4\beta }  \label{13}
\end{eqnarray}%
\begin{eqnarray}
T_{XY\nu \alpha }^{\left( 14\right) } &\rightarrow &4\left( r-q\right)
_{X}J_{4Y\nu \alpha }-2\left[ \left( q-r\right) _{\nu }p_{X}-p_{\nu }\left(
q-r\right) _{X}\right] J_{4Y\alpha }  \notag \\
&&+2\left[ \left( q-r\right) _{\alpha }p_{X}+\left( p+r\right) _{\alpha
}q_{X}-\left( p+q\right) _{\alpha }r_{X}\right] J_{4Y\nu
}+2q_{X}r_{Y}J_{4\nu \alpha }  \notag \\
&&+\left[ \left( q_{\nu }r_{\alpha }-r_{\nu }q_{\alpha }\right) p_{X}-\left(
r_{\nu }p_{\alpha }+r_{\alpha }p_{\nu }\right) q_{X}+\left( q_{\nu
}p_{\alpha }+q_{\alpha }p_{\nu }\right) r_{X}\right] J_{4Y}  \notag \\
&&+\left( q_{\alpha }r_{X}p_{Y}+r_{\alpha }p_{X}q_{Y}+p_{\alpha
}r_{X}q_{Y}\right) J_{4\nu }  \notag \\
&&+\left( q_{\nu }p_{X}r_{Y}+r_{\nu }q_{X}p_{Y}+p_{\nu }r_{X}q_{Y}\right)
J_{4\alpha }  \label{14}
\end{eqnarray}

\subsubsection{Second-Order Standard Tensors}

\label{Second}Second, we compute the 2nd-order standard tensors required to
build up subamplitudes. Even though we already examined those involving
three propagators, we get back to this subject as the perspective is broader
this time. For this purpose, recall the general form obtained succeeding the
integration (\ref{T3s})%
\begin{eqnarray}
T_{3\mu \nu }^{\left( s\right) }\left( k_{i},k_{j}\right) &=&\tfrac{1}{4}%
\left( 1+s\right) \left( \Delta _{\mu \nu }+g_{\mu \nu }I_{\log }+4J_{3\mu
\nu }\right)  \notag \\
&&-\left( p_{j}+sp_{i}\right) _{\nu }J_{3\mu }-\left( p_{i}+sp_{j}\right)
_{\mu }J_{3\nu }+\left( p_{i\mu }p_{j\nu }+sp_{i\nu }p_{j\mu }\right) J_{3},
\label{citet}
\end{eqnarray}%
where associations with external momenta occur through the relation $%
p_{i}=k_{1}-k_{i}$.

We assigned a special role for the routing $k_{1}$ simply because it is the
first to appear in the adopted ordering. This reasoning was implicit when
evaluating three-point Feynman integrals in Subsection (\ref{Fey3}) and led
to the external momenta $p$ and $q$. The notation for the corresponding
functions is $\xi _{ab}=\xi _{ab}\left( p,q\right) $ and reflects in the
corresponding $J_{3}$-tensors, including the coefficients inside them.

From the first case, let us obtain the second $D_{124}$ through the
transformation $k_{3}\rightarrow k_{4}$. That changes the second external
momentum $q\rightarrow r$, which reflects on the notations for functions $%
\xi _{ab}^{\prime }=\xi _{ab}\left( p,r\right) $ and $J_{3}^{\prime }$%
-tensors. Analogously, the third case $D_{134}$ links to the momenta $q$ and 
$r$ seen in functions $\xi _{ab}^{\prime \prime }=\xi _{ab}\left( q,r\right) 
$ and $J_{3}^{\prime \prime }$-tensors.

Nevertheless, things are different for objects involving the fourth
denominator $D_{234}$. When emphasizing the routing $k_{2}$, these
particular associations come with $p_{i}^{\prime }=k_{2}-k_{i}=p_{i}-p$ and
lead to momenta $q-p$ and $r-p$. The notation for functions $\xi
_{ab}^{\prime \prime \prime }=\xi _{ab}\left( q-p,r-p\right) $ and $%
J_{3}^{\prime \prime \prime }$-tensors follows previous cases; however, the
differences $p_{i}^{\prime }$ generate more structures inside the tensors.
We must consider such information when exploring reductions and other
algebraic manipulations.

The generality brought by $J$-tensors makes extensions of the expression
above direct. Besides changing the versions of these tensors, we recall that
there are no ill-defined contributions for the standard tensor depending on
four propagators. Therefore, the new version is the following%
\begin{equation}
T_{4\mu \nu }^{\left( s\right) }\left( k_{i},k_{j}\right) =\left( 1+s\right)
J_{4\mu \nu }-\left( p_{j}+sp_{i}\right) _{\nu }J_{4\mu }-\left(
p_{i}+sp_{j}\right) _{\mu }J_{4\nu }+\left( p_{i\mu }p_{j\nu }+sp_{i\nu
}p_{j\mu }\right) J_{4},  \label{citett}
\end{equation}%
where the original association $p_{i}=k_{1}-k_{i}$ applies.

Without setting signs, we cast all available momenta configurations for
these objects in the sequence. The line notation (\ref{line2}) is
particularly advantageous in this scene.

$\bullet $ Three propagators $D_{123}$ - $\xi _{ab}=\xi _{ab}\left(
p,q\right) $%
\begin{eqnarray}
\left[ T_{3\mu \nu }^{\left( s\right) }\left( k_{1},k_{2}\right) \right] &=&%
\tfrac{1}{4}\left( 1+s\right) \left( \Delta _{\mu \nu }+g_{\mu \nu }I_{\log
}+4J_{3\mu \nu }\right) -p_{\nu }J_{3\mu }-sp_{\mu }J_{3\nu }  \label{4} \\
\left[ T_{3\mu \nu }^{\left( s\right) }\left( k_{1},k_{3}\right) \right] &=&%
\tfrac{1}{4}\left( 1+s\right) \left( \Delta _{\mu \nu }+g_{\mu \nu }I_{\log
}+4J_{3\mu \nu }\right) -q_{\nu }J_{3\mu }-sq_{\mu }J_{3\nu } \\
\left[ T_{3\mu \nu }^{\left( s\right) }\left( k_{2},k_{3}\right) \right] &=&%
\tfrac{1}{4}\left( 1+s\right) \left( \Delta _{\mu \nu }+g_{\mu \nu }I_{\log
}+4J_{3\mu \nu }\right)  \notag \\
&&-\left( q+sp\right) _{\nu }J_{3\mu }-\left( p+sq\right) _{\mu }J_{3\nu
}+\left( p_{\mu }q_{\nu }+sp_{\nu }q_{\mu }\right) J_{3}
\end{eqnarray}

$\bullet $ Three propagators $D_{124}$ - $\xi _{ab}^{\prime }=\xi
_{ab}\left( p,r\right) $%
\begin{eqnarray}
\left[ T_{3\mu \nu }^{\left( s\right) }\left( k_{1},k_{2}\right) \right]
^{\prime } &=&\tfrac{1}{4}\left( 1+s\right) \left( \Delta _{\mu \nu }+g_{\mu
\nu }I_{\log }+4J_{3\mu \nu }^{\prime }\right) -p_{\nu }J_{3\mu }^{\prime
}-sp_{\mu }J_{3\nu }^{\prime } \\
\left[ T_{3\mu \nu }^{\left( s\right) }\left( k_{1},k_{4}\right) \right]
^{\prime } &=&\tfrac{1}{4}\left( 1+s\right) \left( \Delta _{\mu \nu }+g_{\mu
\nu }I_{\log }+4J_{3\mu \nu }^{\prime }\right) -r_{\nu }J_{3\mu }^{\prime
}-sr_{\mu }J_{3\nu }^{\prime } \\
\left[ T_{3\mu \nu }^{\left( s\right) }\left( k_{2},k_{4}\right) \right]
^{\prime } &=&\tfrac{1}{4}\left( 1+s\right) \left( \Delta _{\mu \nu }+g_{\mu
\nu }I_{\log }+4J_{3\mu \nu }^{\prime }\right)  \notag \\
&&-\left( r+sp\right) _{\nu }J_{3\mu }^{\prime }-\left( p+sr\right) _{\mu
}J_{3\nu }^{\prime }+\left( p_{\mu }r_{\nu }+sp_{\nu }r_{\mu }\right)
J_{3}^{\prime }
\end{eqnarray}

$\bullet $ Three propagators $D_{134}$ - $\xi _{ab}^{\prime \prime }=\xi
_{ab}\left( q,r\right) $%
\begin{eqnarray}
\left[ T_{3\mu \nu }^{\left( s\right) }\left( k_{1},k_{3}\right) \right]
^{\prime \prime } &=&\tfrac{1}{4}\left( 1+s\right) \left( \Delta _{\mu \nu
}+g_{\mu \nu }I_{\log }+4J_{3\mu \nu }^{\prime \prime }\right) -q_{\nu
}J_{3\mu }^{\prime \prime }-sq_{\mu }J_{3\nu }^{\prime \prime } \\
\left[ T_{3\mu \nu }^{\left( s\right) }\left( k_{1},k_{4}\right) \right]
^{\prime \prime } &=&\tfrac{1}{4}\left( 1+s\right) \left( \Delta _{\mu \nu
}+g_{\mu \nu }I_{\log }+4J_{3\mu \nu }^{\prime \prime }\right) -r_{\nu
}J_{3\mu }^{\prime \prime }-sr_{\mu }J_{3\nu }^{\prime \prime } \\
\left[ T_{3\mu \nu }^{\left( s\right) }\left( k_{3},k_{4}\right) \right]
^{\prime \prime } &=&\tfrac{1}{4}\left( 1+s\right) \left( \Delta _{\mu \nu
}+g_{\mu \nu }I_{\log }+4J_{3\mu \nu }^{\prime \prime }\right)  \notag \\
&&-\left( r+sq\right) _{\nu }J_{3\mu }^{\prime \prime }-\left( q+sr\right)
_{\mu }J_{3\nu }^{\prime \prime }+\left( q_{\mu }r_{\nu }+sq_{\nu }r_{\mu
}\right) J_{3}^{\prime \prime }
\end{eqnarray}

$\bullet $ Three propagators $D_{234}$ - $\xi _{ab}^{\prime \prime \prime
}=\xi _{ab}\left( q-p,r-p\right) $ 
\begin{eqnarray}
\left[ T_{3\mu \nu }^{\left( s\right) }\left( k_{2},k_{3}\right) \right]
^{\prime \prime \prime } &=&\tfrac{1}{4}\left( 1+s\right) \left( \Delta
_{\mu \nu }+g_{\mu \nu }I_{\log }+4J_{3\mu \nu }^{\prime \prime \prime
}\right)  \notag \\
&&-\left( q-p\right) _{\nu }J_{3\mu }^{\prime \prime \prime }-s\left(
q-p\right) _{\mu }J_{3\nu }^{\prime \prime \prime } \\
\left[ T_{3\mu \nu }^{\left( s\right) }\left( k_{2},k_{4}\right) \right]
^{\prime \prime \prime } &=&\tfrac{1}{4}\left( 1+s\right) \left( \Delta
_{\mu \nu }+g_{\mu \nu }I_{\log }+4J_{3\mu \nu }^{\prime \prime \prime
}\right)  \notag \\
&&-\left( r-p\right) _{\nu }J_{3\mu }^{\prime \prime \prime }-s\left(
r-p\right) _{\mu }J_{3\nu }^{\prime \prime \prime } \\
\left[ T_{3\mu \nu }^{\left( s\right) }\left( k_{3},k_{4}\right) \right]
^{\prime \prime \prime } &=&\tfrac{1}{4}\left( 1+s\right) \left( \Delta
_{\mu \nu }+g_{\mu \nu }I_{\log }+4J_{3\mu \nu }^{\prime \prime \prime
}\right)  \notag \\
&&-\left[ \left( r-p\right) +s\left( q-p\right) \right] _{\nu }J_{3\mu
}^{\prime \prime \prime }-\left[ \left( q-p\right) +s\left( r-p\right) %
\right] _{\mu }J_{3\nu }^{\prime \prime \prime }  \notag \\
&&+\left[ \left( 1+s\right) p_{\mu }p_{\nu }-\left( p_{\nu }q_{\mu }+sp_{\mu
}q_{\nu }\right) \right.  \notag \\
&&\left. -\left( p_{\mu }r_{\nu }+sp_{\nu }r_{\mu }\right) +\left( q_{\mu
}r_{\nu }+sq_{\nu }r_{\mu }\right) \right] J_{3}^{\prime \prime \prime }
\end{eqnarray}

$\bullet $ Four propagators $D_{1234}$ - $\xi _{abc}=\xi _{abc}\left(
p,q,r\right) $%
\begin{eqnarray}
T_{4\mu \nu }^{\left( s\right) }\left( k_{1},k_{2}\right) &=&\left(
1+s\right) J_{4\mu \nu }-p_{\nu }J_{4\mu }-sp_{\mu }J_{4\nu } \\
T_{4\mu \nu }^{\left( s\right) }\left( k_{1},k_{3}\right) &=&\left(
1+s\right) J_{4\mu \nu }-q_{\nu }J_{4\mu }-sq_{\mu }J_{4\nu } \\
T_{4\mu \nu }^{\left( s\right) }\left( k_{1},k_{4}\right) &=&\left(
1+s\right) J_{4\mu \nu }-r_{\nu }J_{4\mu }-sr_{\mu }J_{4\nu } \\
T_{4\mu \nu }^{\left( s\right) }\left( k_{2},k_{3}\right) &=&\left(
1+s\right) J_{4\mu \nu }-\left( q+sp\right) _{\nu }J_{4\mu } \\
&&-\left( p+sq\right) _{\mu }J_{4\nu }+\left( p_{\mu }q_{\nu }+sp_{\nu
}q_{\mu }\right) J_{4} \\
T_{4\mu \nu }^{\left( s\right) }\left( k_{2},k_{4}\right) &=&\left(
1+s\right) J_{4\mu \nu }-\left( r+sp\right) _{\nu }J_{4\mu }  \notag \\
&&-\left( p+sr\right) _{\mu }J_{4\nu }+\left( p_{\mu }r_{\nu }+sp_{\nu
}r_{\mu }\right) J_{4} \\
T_{4\mu \nu }^{\left( s\right) }\left( k_{3},k_{4}\right) &=&\left(
1+s\right) J_{4\mu \nu }-\left( r+sq\right) _{\nu }J_{4\mu }  \notag \\
&&-\left( q+sr\right) _{\mu }J_{4\nu }+\left( q_{\mu }r_{\nu }+sq_{\nu
}r_{\mu }\right) J_{4}  \label{5}
\end{eqnarray}

\subsubsection{Even Amplitudes - $VVPP$, $VPVP$, and $VPPV$}

Third, we compute even amplitudes that are 2nd-order tensors:\ $VVPP$, $VPVP$%
, and $VPPV$. Taking their general form from Equation (\ref{vvpp}),
integration allows writing%
\begin{eqnarray}
T_{\mu \nu }^{\Gamma _{i}\Gamma _{j}\Gamma _{k}\Gamma _{l}} &=&2s_{1}\left[
s_{3}T_{3\mu \nu }^{\left( s_{3}\right) }\left( k_{1},k_{2}\right)
+s_{2}T_{3\mu \nu }^{\left( -s_{2}\right) }\left( k_{1},k_{3}\right)
-s_{2}T_{3\mu \nu }^{\left( s_{1}\right) }\left( k_{2},k_{3}\right) \right] 
\notag \\
&&+2s_{1}\left[ s_{3}T_{3\mu \nu }^{\left( s_{3}\right) }\left(
k_{1},k_{2}\right) +T_{3\mu \nu }^{\left( +\right) }\left(
k_{1},k_{4}\right) -T_{3\mu \nu }^{\left( -s_{3}\right) }\left(
k_{2},k_{4}\right) \right] ^{\prime }  \notag \\
&&+2s_{1}\left[ s_{2}T_{3\mu \nu }^{\left( -s_{2}\right) }\left(
k_{1},k_{3}\right) +T_{3\mu \nu }^{\left( +\right) }\left(
k_{1},k_{4}\right) +T_{3\mu \nu }^{\left( s_{2}\right) }\left(
k_{3},k_{4}\right) \right] ^{\prime \prime }  \notag \\
&&+2s_{1}\left[ -s_{2}T_{3\mu \nu }^{\left( s_{1}\right) }\left(
k_{2},k_{3}\right) -T_{3\mu \nu }^{\left( -s_{3}\right) }\left(
k_{2},k_{4}\right) +T_{3\mu \nu }^{\left( s_{2}\right) }\left(
k_{3},k_{4}\right) \right] ^{\prime \prime \prime }  \notag \\
&&-2s_{1}\left[ s_{3}\left( q-r\right) ^{2}T_{4\mu \nu }^{\left(
s_{3}\right) }\left( k_{1},k_{2}\right) +s_{2}\left( p-r\right) ^{2}T_{4\mu
\nu }^{\left( -s_{2}\right) }\left( k_{1},k_{3}\right) \right.  \notag \\
&&+\left( p-q\right) ^{2}T_{4\mu \nu }^{\left( +\right) }\left(
k_{1},k_{4}\right) -s_{2}r^{2}T_{4\mu \nu }^{\left( s_{1}\right) }\left(
k_{2},k_{3}\right)  \notag \\
&&\left. -q^{2}T_{4\mu \nu }^{\left( -s_{3}\right) }\left(
k_{2},k_{4}\right) +p^{2}T_{4\mu \nu }^{\left( s_{2}\right) }\left(
k_{3},k_{4}\right) \right] -s_{1}g_{\mu \nu }T^{PPPP},  \label{VVPPs}
\end{eqnarray}%
where we obtain one particular version by setting signs through the
associations: the $VVPP$ function by fixing $s_{i}=\left( -1,-1,+1\right) $,
the $VPVP$ by fixing $s_{i}=\left( +1,-1,-1\right) $, and the $VPPV$ by
fixing $s_{i}=\left( -1,+1,-1\right) $. Replacing standard tensors obtained
in Subsubsection (\ref{Second}) determines the explicit results cast in the
sequence. We anticipate that these are the only substructures effectively
contributing with divergent objects to the $AVVV$.

$\bullet $ The $VVPP$ Amplitude%
\begin{eqnarray}
T_{\mu \nu }^{VVPP} &=&-2\Delta _{\mu \nu }-2g_{\mu \nu }I_{\log }+g_{\mu
\nu }T^{PPPP}  \notag \\
&&-8J_{3\mu \nu }^{\prime }+4\left( p-q\right) _{\mu }J_{3\nu }+4p_{\nu
}J_{3\mu }^{\prime }+4r_{\mu }J_{3\nu }^{\prime }+4\left( r-q\right) _{\nu
}J_{3\mu }^{\prime \prime }  \notag \\
&&-2\left( p_{\mu }q_{\nu }-p_{\nu }q_{\mu }\right) J_{3}+2\left( p_{\mu
}r_{\nu }-p_{\nu }r_{\mu }\right) J_{3}^{\prime }-2\left( q_{\mu }r_{\nu
}-q_{\nu }r_{\mu }\right) J_{3}^{\prime \prime }  \notag \\
&&-2\left[ \left( p_{\mu }q_{\nu }-p_{\nu }q_{\mu }\right) -\left( p_{\mu
}r_{\nu }-p_{\nu }r_{\mu }\right) +\left( q_{\mu }r_{\nu }-q_{\nu }r_{\mu
}\right) \right] J_{3}^{\prime \prime \prime }  \notag \\
&&+8\left( q^{2}-p\cdot q+p\cdot r-q\cdot r\right) J_{4\mu \nu }  \notag \\
&&-4\left[ \left( q^{2}-q\cdot r\right) p_{\nu }+\left( p\cdot
r-p^{2}\right) q_{\nu }+\left( p^{2}-q\cdot p\right) r_{\nu }\right] J_{4\mu
}  \notag \\
&&-4\left[ \left( r^{2}-r\cdot q\right) p_{\mu }+\left( p\cdot
r-r^{2}\right) q_{\mu }+\left( q^{2}-p\cdot q\right) r_{\mu }\right] J_{4\nu
}  \notag \\
&&+2\left[ p^{2}\left( q_{\mu }r_{\nu }-q_{\nu }r_{\mu }\right) -q^{2}\left(
p_{\mu }r_{\nu }-p_{\nu }r_{\mu }\right) +r^{2}\left( p_{\mu }q_{\nu
}-p_{\nu }q_{\mu }\right) \right] J_{4}
\end{eqnarray}

$\bullet $ The $VPVP$ Amplitude%
\begin{eqnarray}
T_{\mu \alpha }^{VPVP} &=&-g_{\mu \alpha }T^{PPPP}-4p_{\mu }J_{3\alpha
}+4p_{\alpha }J_{3\mu }^{\prime }-4\left( r-q\right) _{\alpha }J_{3\mu
}^{\prime \prime }-4\left( q-r\right) _{\mu }J_{3\alpha }^{\prime \prime
\prime }  \notag \\
&&+2\left( p_{\mu }q_{\alpha }+p_{\alpha }q_{\mu }\right) J_{3}-2\left(
p_{\mu }r_{\alpha }+p_{\alpha }r_{\mu }\right) J_{3}^{\prime }+2\left(
q_{\mu }r_{\alpha }-q_{\alpha }r_{\mu }\right) J_{3}^{\prime \prime }  \notag
\\
&&+2\left[ \left( p_{\mu }q_{\alpha }-p_{\alpha }q_{\mu }\right) -\left(
p_{\mu }r_{\alpha }-p_{\alpha }r_{\mu }\right) +\left( q_{\mu }r_{\alpha
}-q_{\alpha }r_{\mu }\right) \right] J_{3}^{\prime \prime \prime }  \notag \\
&&-8\left( p\cdot r-p\cdot q\right) J_{4\mu \alpha }+4\left[ \left(
r^{2}-q\cdot r\right) p_{\mu }+\left( p\cdot r\right) q_{\mu }-\left( p\cdot
q\right) r_{\mu }\right] J_{4\alpha }  \notag \\
&&+4\left[ \left( q\cdot r-q^{2}\right) p_{\alpha }+\left( p\cdot
r-p^{2}\right) q_{\alpha }+\left( p^{2}-p\cdot q\right) r_{\alpha }\right]
J_{4\mu }  \notag \\
&&-2\left[ p^{2}\left( q_{\mu }r_{\alpha }-q_{\alpha }r_{\mu }\right)
-q^{2}\left( p_{\mu }r_{\alpha }+p_{\alpha }r_{\mu }\right) +r^{2}\left(
p_{\mu }q_{\alpha }+p_{\alpha }q_{\mu }\right) \right] J_{4}
\end{eqnarray}

$\bullet $ The $VPPV$ Amplitude%
\begin{eqnarray}
T_{\mu \beta }^{VPPV} &=&-2\Delta _{\mu \beta }-2g_{\mu \beta }I_{\log
}+g_{\mu \beta }T^{PPPP}  \notag \\
&&-8J_{3\mu \beta }^{\prime \prime }-4p_{\beta }J_{3\mu }^{\prime }+4\left(
r+q\right) _{\beta }J_{3\mu }^{\prime \prime }+4r_{\mu }J_{3\beta }^{\prime
\prime }+4\left( q-p\right) _{\mu }J_{3\beta }^{\prime \prime \prime } 
\notag \\
&&+2\left( p_{\mu }q_{\beta }-p_{\beta }q_{\mu }\right) J_{3}+2\left( p_{\mu
}r_{\beta }+p_{\beta }r_{\mu }\right) J_{3}^{\prime }-2\left( q_{\mu
}r_{\beta }+q_{\beta }r_{\mu }\right) J_{3}^{\prime \prime }  \notag \\
&&-2\left[ 2p_{\mu }p_{\beta }-\left( p_{\mu }q_{\beta }+p_{\beta }q_{\mu
}\right) -\left( p_{\mu }r_{\beta }+p_{\beta }r_{\mu }\right) +\left( q_{\mu
}r_{\beta }+q_{\beta }r_{\mu }\right) \right] J_{3}^{\prime \prime \prime } 
\notag \\
&&+8\left( p^{2}-p\cdot q\right) J_{4\mu \beta }+4\left[ \left( q\cdot
r\right) p_{\mu }-\left( p\cdot r\right) q_{\mu }+\left( p\cdot
q-p^{2}\right) r_{\mu }\right] J_{4\beta }  \notag \\
&&+4\left[ \left( q^{2}-q\cdot r\right) p_{\beta }+\left( p\cdot
r-p^{2}\right) q_{\beta }+\left( p\cdot q-p^{2}\right) r_{\beta }\right]
J_{4\mu }  \notag \\
&&+2\left[ p^{2}\left( q_{\mu }r_{\beta }+q_{\beta }r_{\mu }\right)
-q^{2}\left( p_{\mu }r_{\beta }+p_{\beta }r_{\mu }\right) -r^{2}\left(
p_{\mu }q_{\beta }-p_{\beta }q_{\mu }\right) \right] J_{4}
\end{eqnarray}

\subsubsection{Odd Amplitudes - $AVPP$, $APVP$, and $APPV$}

\label{AVPPexp}Forth, we compute odd amplitudes that are 2nd-order tensors:\ 
$AVPP$, $APVP$, and $APPV$. Given the general form (\ref{avpp}), the
integral operation characterizes two sectors corresponding to different
tensor structures:%
\begin{equation}
T_{\mu \nu }^{\Gamma _{i}\Gamma _{j}\Gamma _{k}\Gamma
_{l}}=is_{1}\varepsilon _{\mu XYZ}F_{4\nu XYZ}^{\left( s_{2},s_{3}\right)
}+is_{1}\varepsilon _{\mu \nu XY}F_{4XY}.  \label{AVPPs}
\end{equation}%
We distinguish particular functions when choosing signs through the
association: the $AVPP$ function by fixing $s_{i}=\left( -1,-1,+1\right) $,
the $APVP$ by fixing $s_{i}=\left( +1,+1,+1\right) $, and the $APPV$ by
fixing $s_{i}=\left( -1,+1,-1\right) $.

The first sector is proportional to the simplified version of the 4th-order
standard tensor (\ref{6}):%
\begin{equation}
\varepsilon _{\mu XYZ}F_{4\nu XYZ}^{\left( s_{2},s_{3}\right) }=4\left(
-\varepsilon _{\mu BCD}g_{\nu A}+s_{2}\varepsilon _{\mu ACD}g_{\nu
B}+s_{3}\varepsilon _{\mu ABD}g_{\nu C}-\varepsilon _{\mu ABC}g_{\nu
D}\right) T_{4ABCD}.
\end{equation}%
Following its replacement, symmetry properties bring simplifications so this
product assumes the general form%
\begin{eqnarray}
\varepsilon _{\mu XYZ}F_{4\nu XYZ}^{\left( s_{2},s_{3}\right) }
&=&-4\varepsilon _{\mu XYZ}\left[ \left( 1-s_{2}\right) q_{Y}r_{Z}J_{4\nu
X}+\left( 1+s_{3}\right) p_{X}r_{Z}J_{4\nu Y}\right.  \notag \\
&&+2p_{X}q_{Y}J_{4\nu Z}-p_{X}q_{Y}r_{Z}J_{4\nu }+s_{2}p_{\nu
}q_{Y}r_{Z}J_{4X}  \notag \\
&&\left. -s_{3}q_{\nu }p_{X}r_{Z}J_{4Y}-r_{\nu }p_{X}q_{Y}J_{4Z}\right] .
\label{AVPP}
\end{eqnarray}%
As mentioned before, symmetric objects $J_{4\mu \nu \alpha \beta }$ and $%
\square _{\mu \nu \alpha \beta }$ disappear and do not concern this
investigation. Moving on to the second sector, we have another combination
of 2nd-order standard tensors%
\begin{eqnarray}
F_{4XY} &=&\left[ T_{3XY}^{\left( -\right) }\left( k_{2},k_{3}\right)
-T_{3XY}^{\left( -\right) }\left( k_{1},k_{3}\right) +T_{3XY}^{\left(
-\right) }\left( k_{1},k_{2}\right) \right]  \notag \\
&&+\left[ -T_{3XY}^{\left( -\right) }\left( k_{2},k_{4}\right)
+T_{3XY}^{\left( -\right) }\left( k_{1},k_{4}\right) +T_{3XY}^{\left(
-\right) }\left( k_{1},k_{2}\right) \right] ^{\prime }  \notag \\
&&+\left[ T_{3XY}^{\left( -\right) }\left( k_{3},k_{4}\right)
+T_{3XY}^{\left( -\right) }\left( k_{1},k_{4}\right) -T_{3XY}^{\left(
-\right) }\left( k_{1},k_{3}\right) \right] ^{\prime \prime }  \notag \\
&&+\left[ T_{3XY}^{\left( -\right) }\left( k_{3},k_{4}\right)
-T_{3XY}^{\left( -\right) }\left( k_{2},k_{4}\right) +T_{3XY}^{\left(
-\right) }\left( k_{2},k_{3}\right) \right] ^{\prime \prime \prime }  \notag
\\
&&+\left[ -p^{2}T_{4XY}^{\left( -\right) }\left( k_{3},k_{4}\right)
+q^{2}T_{4XY}^{\left( -\right) }\left( k_{2},k_{4}\right)
-r^{2}T_{4XY}^{\left( -\right) }\left( k_{2},k_{3}\right) \right.  \notag \\
&&-\left( p-q\right) ^{2}T_{4XY}^{\left( -\right) }\left( k_{1},k_{4}\right)
+\left( p-r\right) ^{2}T_{4XY}^{\left( -\right) }\left( k_{1},k_{3}\right) 
\notag \\
&&\left. -\left( q-r\right) ^{2}T_{4XY}^{\left( -\right) }\left(
k_{1},k_{2}\right) \right] .
\end{eqnarray}%
Its structure arises after replacing results from Subsubsection (\ref{Second}%
) and performing simplifications:%
\begin{eqnarray}
F_{4XY} &=&4p_{X}J_{3Y}^{\prime }+4\left( r-q\right) _{X}J_{3Y}^{\prime
\prime }+2p_{X}q_{Y}J_{3}+2r_{X}p_{Y}J_{3}^{\prime }  \notag \\
&&+2q_{X}r_{Y}J_{3}^{\prime \prime }+2\left(
q_{X}r_{Y}+r_{X}p_{Y}+p_{X}q_{Y}\right) J_{3}^{\prime \prime \prime }  \notag
\\
&&-4\left[ \left( q^{2}-q\cdot r\right) p_{X}-\left( p^{2}-p\cdot r\right)
q_{X}+\left( p^{2}-p\cdot q\right) r_{X}\right] J_{4Y}  \notag \\
&&-2\left( p^{2}q_{X}r_{Y}+q^{2}r_{X}p_{Y}+r^{2}p_{X}q_{Y}\right) J_{4}.
\label{T4uv}
\end{eqnarray}

Adjusting signs, we cast the final expressions attributed to odd
perturbative amplitudes below.

$\bullet $ The $AVPP$ Amplitude%
\begin{eqnarray}
T_{\mu \nu }^{AVPP} &=&4i\varepsilon _{\mu XYZ}\left( 2q_{Y}r_{Z}J_{4\nu
X}+2p_{X}r_{Z}J_{4\nu Y}+2p_{X}q_{Y}J_{4\nu Z}-p_{X}q_{Y}r_{Z}J_{4\nu
}\right)  \notag \\
&&+4i\varepsilon _{\mu XYZ}\left( -p_{\nu }q_{Y}r_{Z}J_{4X}-q_{\nu
}p_{X}r_{Z}J_{4Y}-r_{\nu }p_{X}q_{Y}J_{4Z}\right)  \notag \\
&&-i\varepsilon _{\mu \nu XY}F_{4XY}  \label{AVPP1}
\end{eqnarray}

$\bullet $ The $APVP$ Amplitude%
\begin{eqnarray}
T_{\mu \alpha }^{APVP} &=&-4i\varepsilon _{\mu XYZ}\left(
2p_{X}r_{Z}J_{4\alpha Y}+2p_{X}q_{Y}J_{4\alpha Z}-p_{X}q_{Y}r_{Z}J_{4\alpha
}\right)  \notag \\
&&-4i\varepsilon _{\mu XYZ}\left( p_{\alpha }q_{Y}r_{Z}J_{4X}-q_{\alpha
}p_{X}r_{Z}J_{4Y}-r_{\alpha }p_{X}q_{Y}J_{4Z}\right)  \notag \\
&&+i\varepsilon _{\mu \alpha XY}F_{4XY}
\end{eqnarray}

$\bullet $ The $APPV$ Amplitude%
\begin{eqnarray}
T_{\mu \beta }^{APPV} &=&4i\varepsilon _{\mu XYZ}\left( 2p_{X}q_{Y}J_{4\beta
Z}-p_{X}q_{Y}r_{Z}J_{4\beta }\right)  \notag \\
&&+4i\varepsilon _{\mu XYZ}\left( p_{\beta }q_{Y}r_{Z}J_{4X}+q_{\beta
}p_{X}r_{Z}J_{4Y}-r_{\beta }p_{X}q_{Y}J_{4Z}\right)  \notag \\
&&-i\varepsilon _{\mu \beta XY}F_{4XY}  \label{AVPP3}
\end{eqnarray}

\subsubsection{Scalar Amplitude - $PPPP$}

Fifth, we compute the scalar amplitude $PPPP$. The integration of its
structure (\ref{pppp}) allows writing this correlator in terms of scalar
Feynman integrals%
\begin{eqnarray}
T^{PPPP} &=&2\left[ I_{2}\left( k_{2},k_{4}\right) +I_{2}\left(
k_{1},k_{3}\right) \right]  \notag \\
&&-2\left( p^{2}-p\cdot q\right) I_{3}\left( k_{1},k_{2},k_{3}\right)
-2\left( p\cdot r\right) I_{3}\left( k_{1},k_{2},k_{4}\right)  \notag \\
&&-2\left( r^{2}-q\cdot r\right) I_{3}\left( k_{1},k_{3},k_{4}\right)
+2\left( p-q\right) \cdot \left( q-r\right) I_{3}\left(
k_{2},k_{3},k_{4}\right)  \notag \\
&&+\left[ p^{2}\left( r-q\right) ^{2}-q^{2}\left( p-r\right)
^{2}+r^{2}\left( p-q\right) ^{2}\right] I_{4}.
\end{eqnarray}%
The required tools are displayed in Equations (\ref{I20FIN}), (\ref{I30}),
and (\ref{I40}).\ Since structures typical of two and three-point
calculations appear, specifying their momenta content is essential. After
replacing them, we obtain the explicit version of the amplitude%
\begin{eqnarray}
T^{PPPP} &=&4I_{\log }-2i\left( 4\pi \right) ^{-2}\left[ \xi _{0}^{\left(
0\right) }\left( r-p\right) +\xi _{0}^{\left( 0\right) }\left( q\right) %
\right]  \notag \\
&&-2\left( p^{2}-p\cdot q\right) J_{3}-2\left( p\cdot r\right) J_{3}^{\prime
}  \notag \\
&&-2\left( r^{2}-q\cdot r\right) J_{3}^{\prime \prime }+2\left( p-q\right)
\cdot \left( q-r\right) J_{3}^{\prime \prime \prime }  \notag \\
&&+\left[ p^{2}\left( r-q\right) ^{2}-q^{2}\left( p-r\right)
^{2}+r^{2}\left( p-q\right) ^{2}\right] J_{4}.
\end{eqnarray}

\subsection{Comments}

Before proceeding with the analysis of results, let us present a brief
panorama of our calculations. In this section, we have evaluated all
perturbative amplitudes needed for this investigation. Aiming to accomplish
this task, we adopted a strategy to separate ill-defined mathematical
structures from finite contributions of integrals.

After computing finite quantities, we projected them in terms of structure
functions. They do not appear randomly but in particular arrangements named $%
J$-tensors. They stress the exclusive dependence on differences between
routings, i.e., external momenta. Under this new perspective, $J$-tensors'
properties are fundamental ingredients to the intended analysis.

On the other hand, we only organized divergent structures that arose within $%
AV^{n}$-type amplitudes. It is well-known that integrals exhibiting power
counting equal to or higher than linear are not invariant under
translations. Here, this causes the presence of divergent surface terms
inside amplitudes $AV$ and $AVV$. Furthermore, coefficients of these terms
unavoidably carry ambiguous structures materialized into sums of arbitrary
routings $k_{i}$. Interestingly, we acknowledged the same surface term in
logarithmically diverging integrals corresponding to at least 2nd-order
tensors, although coefficients are not ambiguous in these cases. The $AVVV$
box is an example of this type of situation.

To be more precise, only the 2nd-order surface term $\Delta _{\mu \nu }$
effectively concerns this investigation. The 4th-order surface term appears
exclusively inside $AVVV$'s tensor sector but cancels out subsequently. Even
if the irreducible object appears within substructures, it vanishes
identically in the complete amplitudes. Taking a closer look at
contributions from even subamplitudes belonging to the box, we cast its
divergent sector:%
\begin{equation}
\left[ T_{\mu \nu \alpha \beta }^{AVVV}\right] _{div}=2i\left(
\varepsilon _{\mu \alpha \beta \rho }\Delta _{\nu }^{\rho }+\varepsilon
_{\mu \nu \alpha \rho }\Delta _{\beta }^{\rho }\right) .
\end{equation}

As no prescription was adopted to evaluate divergences, expressing them in
the context of a regularization scheme is feasible. Nonetheless, by avoiding
this step, our analysis inquires about the implications of different values
for the surface term $\Delta _{\mu \nu }$. That occurs in the following
section when investigating the connection involving linearity of integration
and symmetries.

\newpage

\section{Analysis of the Results}

\label{An}In the model discussion, we considered the mathematical structure
of perturbative amplitudes to establish identities at the integrand level.
Proper relations among Green functions (GF) should emerge with the
integration; however, the divergent character of calculations might affect
these expectations.

Verifying these relations requires performing momenta contractions with the
explicit form of $AV^{n}$-type amplitudes. Subsection (\ref{RAGF1}) develops
these operations for the $AVV$ triangle while highlighting tools and
patterns considered relevant to the more complex case. Afterward, Subsection
(\ref{RAGF2}) extends these explorations to the $AVVV$ box. Since
potentially violating terms emerge in this process, Subsection (\ref{Lin})
inquires about mathematical structures linked to them. Such analysis
elucidates the roles played by different trace expressions and vertex
configurations. Lastly, we study Ward identities (WIs) from their
association with relations among GF in Subsection (\ref{sym}). All mentioned
constraints depend on divergent objects materialized in surface terms;
therefore, our argumentation approaches their possible values and ensuing
implications.

\subsection{Relations Among Green Functions - $AVV$}

\label{RAGF1}This subsection aims to verify relations among GF derived for $%
AVV$ contractions (\ref{AVV}). The corresponding expectations are cast in
Equations (\ref{ragf1})-(\ref{ragf3}), so we transcribe them here:%
\begin{eqnarray}
\left( k_{1}-k_{3}\right) ^{\mu }T_{\mu \nu \alpha }^{AVV} &\rightarrow
&T_{\nu \alpha }^{AV}\left( k_{2},k_{3}\right) -T_{\alpha \nu }^{AV}\left(
k_{1},k_{2}\right) -2mT_{\nu \alpha }^{PVV}, \\
\left( k_{1}-k_{2}\right) ^{\nu }T_{\mu \nu \alpha }^{AVV} &\rightarrow
&T_{\mu \alpha }^{AV}\left( k_{2},k_{3}\right) -T_{\mu \alpha }^{AV}\left(
k_{1},k_{3}\right) , \\
\left( k_{2}-k_{3}\right) ^{\alpha }T_{\mu \nu \alpha }^{AVV} &\rightarrow
&T_{\mu \nu }^{AV}\left( k_{1},k_{3}\right) -T_{\mu \nu }^{AV}\left(
k_{1},k_{2}\right) .
\end{eqnarray}%
Our task consists of performing operations described on the left-hand side
of these equations, aiming to recognize the structures from the right. Since
contractions involving finite tensors and external momenta emerge throughout
this procedure, these primary ingredients are discussed in the sequence.

As anticipated in the $PVVV$ integration, connecting $J$-vector contractions
to reductions of finite functions is straightforward. That is transparent
when comparing the $J$-vector (\ref{J3}) with properties achieved in
Equations (\ref{p10}) and (\ref{q01}). After recognizing the $J$-scalar (\ref%
{J3}), we introduce the explicit results:%
\begin{eqnarray}
2p^{\mu }J_{3\mu } &=&p^{2}J_{3}-\frac{i}{\left( 4\pi \right) ^{2}}\left[
\xi _{0}^{\left( 0\right) }\left( p-q\right) -\xi _{0}^{\left( 0\right)
}\left( q\right) \right] ,  \label{pjz} \\
2q^{\mu }J_{3\mu } &=&q^{2}J_{3}-\frac{i}{\left( 4\pi \right) ^{2}}\left[
\xi _{0}^{\left( 0\right) }\left( p-q\right) -\xi _{0}^{\left( 0\right)
}\left( p\right) \right] .  \label{qjz}
\end{eqnarray}

That motivates us to pursue similar cases involving higher parameter powers,
following the condition $a+b=2$. From the definition of the 2nd-order tensor
(\ref{J3uv}), contracting the external momentum $p$ yields%
\begin{eqnarray}
p^{\mu }J_{3\mu \nu } &=&\frac{i}{\left( 4\pi \right) ^{2}}p_{\nu }\left[
p^{2}\xi _{20}^{\left( -1\right) }+\left( p\cdot q\right) \xi _{11}^{\left(
-1\right) }\right]  \notag \\
&&+\frac{i}{\left( 4\pi \right) ^{2}}q_{\nu }\left[ p^{2}\xi _{11}^{\left(
-1\right) }+\left( p\cdot q\right) \xi _{02}^{\left( -1\right) }\right] -%
\frac{i}{\left( 4\pi \right) ^{2}}\frac{1}{2}p_{\nu }\xi _{00}^{\left(
0\right) }.
\end{eqnarray}%
Combinations between brackets are the properties established in Equations (%
\ref{p20})-(\ref{p11}), whose replacement leads to the first reduction
within this category%
\begin{equation}
2p^{\mu }J_{3\mu \nu }=p^{2}J_{3\nu }-\frac{i}{\left( 4\pi \right) ^{2}}%
\frac{1}{2}\left[ \left( p+q\right) _{\nu }\xi _{0}^{\left( 0\right) }\left(
p-q\right) -q_{\nu }\xi _{0}^{\left( 0\right) }\left( q\right) \right] .
\label{pjza}
\end{equation}%
The second arises by using momentum $q$ and repeating this process:%
\begin{equation}
2q^{\mu }J_{3\mu \nu }=q^{2}J_{3\nu }-\frac{i}{\left( 4\pi \right) ^{2}}%
\frac{1}{2}\left[ \left( p+q\right) _{\nu }\xi _{0}^{\left( 0\right) }\left(
p-q\right) -p_{\nu }\xi _{0}^{\left( 0\right) }\left( p\right) \right] .
\label{qjza}
\end{equation}

Returning to the relations, we start with vector vertices, whose
manipulations must yield in pure surface terms since this is the structure
of the $AV$ amplitude (\ref{AV}). Let the contraction between $p=k_{1}-k_{2}$
and the first vector vertex be the outset of this discussion. Promptly,
several terms cancel out for being symmetric quantities multiplied by the
Levi-Civita symbol. Hence, we obtain the following expression after
relabeling some indices%
\begin{eqnarray}
p^{\nu }T_{\mu \nu \alpha }^{AVV} &=&2i\varepsilon _{\mu \nu \alpha \beta
}\left\{ \left[ \left( p-q\right) ^{\nu }p^{\rho }-p^{\nu }\left(
k_{1}+k_{3}\right) ^{\rho }\right] \Delta _{\rho }^{\beta }\right.  \notag \\
&&+4\left( p-q\right) ^{\nu }p_{\rho }J_{3}^{\rho \beta }-4p^{\nu }q^{\beta
}p^{\rho }J_{3\rho }-2\left( p^{2}-p\cdot q\right) p^{\nu }J_{3}^{\beta } 
\notag \\
&&\left. +p^{2}p^{\nu }q^{\beta }J_{3}+i\left( 4\pi \right) ^{-2}p^{\nu
}q^{\beta }\xi _{0}^{\left( 0\right) }\left( p-q\right) \right\} .
\end{eqnarray}

Obeying the hierarchy intrinsic to these calculations, we employ reduction (%
\ref{pjza}) to suppress the dependence on $a+b=2$ finite functions:%
\begin{eqnarray}
p^{\nu }T_{\mu \nu \alpha }^{AVV} &=&2i\varepsilon _{\mu \nu \alpha \beta
}\left\{ \left[ \left( p-q\right) ^{\nu }p^{\rho }-p^{\nu }\left(
k_{1}+k_{3}\right) ^{\rho }\right] \Delta _{\rho }^{\beta }\right.  \notag \\
&&-4p^{\nu }q^{\beta }p^{\rho }J_{3\rho }+2\left[ \left( p\cdot q\right)
p^{\nu }-p^{2}q^{\nu }\right] J_{3}^{\beta }+p^{2}p^{\nu }q^{\beta }J_{3} 
\notag \\
&&\left. +i\left( 4\pi \right) ^{-2}p^{\nu }q^{\beta }\left[ \xi
_{0}^{\left( 0\right) }\left( q\right) -\xi _{0}^{\left( 0\right) }\left(
p-q\right) \right] \right\} .
\end{eqnarray}%
Reducing $J$-vectors is necessary to cancel out all finite contributions;
however, there is a term where the corresponding contraction is disguised.
Symmetry properties allow us to uncover it through an index permutation 
\begin{equation}
\varepsilon _{\mu \nu \alpha \beta }\left[ \left( p_{ij}\cdot q\right)
p^{\nu }-\left( p_{ij}\cdot p\right) q^{\nu }\right] J_{3}^{\beta
}=\varepsilon _{\mu \nu \alpha \beta }q^{\nu }p^{\beta }p_{ij}^{\rho
}J_{3\rho }.  \label{idv}
\end{equation}%
This identity admits choices for the difference between routing $%
p_{ij}=k_{i}-k_{j}$, but we set the $p$ momentum for this particular
occurrence. These identifications concentrate $a+b=1$ contributions into
object (\ref{pjz}), reducing this sector and eliminating all finite parts.

The final step before concluding this demonstration is to recognize surface
terms as a difference between $AV$s. Thus, we reorganize coefficients to
achieve a transparent view 
\begin{equation}
p^{\nu }T_{\mu \nu \alpha }^{AVV}=2i\varepsilon _{\mu \nu \alpha \beta } 
\left[ \left( q-p\right) ^{\nu }\left( k_{2}+k_{3}\right) ^{\rho }-q^{\nu
}\left( k_{1}+k_{3}\right) ^{\rho }\right] \Delta _{\rho }^{\beta }.
\end{equation}%
Hence, a comparison with Equation (\ref{AV}) is enough to complete the proof
of this relation among GF: 
\begin{equation}
p^{\nu }T_{\mu \nu \alpha }^{AVV}=T_{\mu \nu }^{AV}\left( k_{2},k_{3}\right)
-T_{\mu \nu }^{AV}\left( k_{1},k_{3}\right) .
\end{equation}

Let us briefly describe the contraction between momentum $q-p=k_{2}-k_{3}$
and the index corresponding to the second vector vertex. It deals with a
difference between external momenta, which generates cancellations between
reductions. We emphasize this circumstance since it will simplify the box
analysis significantly. Again, only surface terms remain after contracting
the amplitude and employing \textit{all} reductions%
\begin{equation}
\left( q-p\right) ^{\alpha }T_{\mu \nu \alpha }^{AVV}=2i\varepsilon _{\mu
\nu \alpha \beta }\left[ p^{\alpha }\left( k_{1}+k_{2}\right) ^{\rho
}-q^{\alpha }\left( k_{1}+k_{3}\right) ^{\rho }\right] \Delta _{\rho
}^{\beta }.
\end{equation}%
Identifying $AV$ functions is straightforward for this particular case:%
\begin{equation}
\left( q-p\right) ^{\alpha }T_{\mu \nu \alpha }^{AVV}=T_{\mu \nu
}^{AV}\left( k_{1},k_{3}\right) -T_{\mu \nu }^{AV}\left( k_{1},k_{2}\right) .
\end{equation}%
Hence, we successfully verified the vector relations associated with
triangle contractions. Properties of finite tensors and algebraic operations
were the only resources necessary to achieve these results.

Lastly, we aim to perform the contraction between momentum $q=k_{1}-k_{3}$
and the index corresponding to the axial vertex. This operation must produce
surface terms corresponding to $AV$ amplitudes, similar to other cases.
Furthermore, even though this type of contribution is not visible at first
glance, finite functions proportional to the squared mass should arise. That
is a requirement to identify the amplitude $PVV$ (\ref{PVV}).

Once our expectations are clear, let us look closer at the expression
derived directly from the contraction:%
\begin{eqnarray}
q^{\mu }T_{\mu \nu \alpha }^{AVV} &=&2iq^{\mu }p^{\beta }\left( \varepsilon
_{\mu \nu \beta \rho }\Delta _{\alpha }^{\rho }-\varepsilon _{\mu \alpha
\beta \rho }\Delta _{\nu }^{\rho }\right) -2iq^{\mu }\left(
k_{1}+k_{3}\right) _{\rho }\varepsilon _{\mu \nu \alpha \beta }\Delta
^{\beta \rho }  \notag \\
&&+8iq^{\mu }p^{\beta }\left( \varepsilon _{\mu \nu \beta \rho }J_{3\alpha
}^{\rho }-\varepsilon _{\mu \alpha \beta \rho }J_{3\nu }^{\rho }\right)
-4i\left( p^{2}-p\cdot q\right) q^{\mu }\varepsilon _{\mu \nu \alpha \beta
}J_{3}^{\beta }  \notag \\
&&-2iq^{\mu }p^{\beta }\varepsilon _{\mu \nu \alpha \beta }\left\{
q^{2}J_{3}+i\left( 4\pi \right) ^{-2}\left[ \xi _{0}^{\left( 0\right)
}\left( p\right) +\xi _{0}^{\left( 0\right) }\left( p-q\right) \right]
\right\} .  \label{cite1}
\end{eqnarray}%
Unlike previous cases, there are no reductions of $J$-tensors since
contractions involve the Levi-Civita symbol instead of external momenta.
Besides, factorizing surface terms to recognize the required amplitudes is
not possible. That occurs because we chose a trace expression prioritizing
the $\mu $-index back in the integrand analysis, and now this feature
brought an inadequate index configuration that prevents identifications.
Therefore, our strategy is to exchange positions of indices to find known
ingredients.

Let us explore the 2nd-order $J$-tensor to illustrate this point. Following
the reasoning observed when discussing Dirac traces (\ref{Schouten}), we
construct a tensor with antisymmetry in five indices ($\rho $ fixed) through
the following Schouten identity%
\begin{equation}
\varepsilon _{\mu \nu \beta \rho }J_{3\alpha }^{\rho }-\varepsilon _{\mu
\alpha \beta \rho }J_{3\nu }^{\rho }=-\varepsilon _{\rho \alpha \mu \nu
}J_{3\beta }^{\rho }-\varepsilon _{\nu \beta \rho \alpha }J_{3\mu }^{\rho
}-\varepsilon _{\alpha \mu \nu \beta }J_{3\rho }^{\rho }.  \label{idS}
\end{equation}%
By replacing this result on the relation among GF, the first two terms on
the right-hand side generate momenta contractions. Hence, we must follow the
procedure established for vector contractions and reduce finite
contributions. These operations vanish most finite contributions, so the $%
AVV $ contraction assumes the form%
\begin{eqnarray}
q^{\mu }T_{\mu \nu \alpha }^{AVV} &=&2iq^{\mu }p^{\beta }\left( \varepsilon
_{\mu \nu \beta \rho }\Delta _{\alpha }^{\rho }-\varepsilon _{\mu \alpha
\beta \rho }\Delta _{\nu }^{\rho }\right) -2iq^{\mu }\left(
k_{1}+k_{3}\right) _{\rho }\varepsilon _{\mu \nu \alpha \beta }\Delta
^{\beta \rho }  \notag \\
&&-8iq^{\mu }p^{\beta }\varepsilon _{\mu \nu \alpha \beta }\left[ J_{3\rho
}^{\rho }+i\left( 4\pi \right) ^{-2}\xi _{0}^{\left( 0\right) }\left(
p-q\right) \right] .
\end{eqnarray}

The index permutation above also brought an additional term depending on
object $J_{3\rho }^{\rho }$. From definition (\ref{J3uv}), we take the $J$%
-tensor trace and identify reductions of finite structure functions (\ref%
{p20}) and (\ref{q02}):%
\begin{equation}
J_{3\rho }^{\rho }=\frac{i}{\left( 4\pi \right) ^{2}}\left\{ \left[ p^{2}\xi
_{20}^{\left( -1\right) }+\left( p\cdot q\right) \xi _{11}^{\left( -1\right)
}\right] +\left[ \left( p\cdot q\right) \xi _{11}^{\left( -1\right)
}+q^{2}\xi _{02}^{\left( -1\right) }\right] -2\xi _{00}^{\left( 0\right)
}\right\} .
\end{equation}%
Although this structure resembles those of momenta contractions, we stress
the presence of the finite function $\xi _{00}^{\left( 0\right) }$. By
replacing other reductions and expressing this contribution in terms of
elements belonging to the $\xi _{nm}^{\left( -1\right) }$-family (\ref{x00}%
), we obtain the following trace:%
\begin{equation}
J_{3\rho }^{\rho }=m^{2}J_{3}+\frac{i}{\left( 4\pi \right) ^{2}}\left[ \frac{%
1}{2}-\xi _{0}^{\left( 0\right) }\left( p-q\right) \right] .
\end{equation}

Both the term proportional to the squared mass and the numerical factor
remain when replacing this result within the $AVV$ contraction. A comparison
with Equation (\ref{PVV}) shows that the first corresponds to the $PVV$
amplitude\footnote{%
In general, subamplitudes within $AV^{n}$ might produce contributions
belonging to $PV^{n}$-type amplitudes. That does not transpire here due to
specific trace choices.}. With this identification, we finish explorations
about finite contributions for now:%
\begin{eqnarray}
q^{\mu }T_{\mu \nu \alpha }^{AVV} &=&2iq^{\mu }p^{\beta }\left( \varepsilon
_{\mu \nu \beta \rho }\Delta _{\alpha }^{\rho }-\varepsilon _{\mu \alpha
\beta \rho }\Delta _{\nu }^{\rho }\right) -2iq^{\mu }\left(
k_{1}+k_{3}\right) _{\rho }\varepsilon _{\mu \nu \alpha \beta }\Delta
^{\beta \rho }  \notag \\
&&-2mT_{\mu \nu }^{PVV}+\frac{1}{4\pi ^{2}}\varepsilon _{\mu \nu \alpha
\beta }q^{\mu }p^{\beta }.
\end{eqnarray}%
Alternatively, we could achieve this expression by making explicit the
content of $J$-tensors from the beginning. Such a perspective would make
calculations for this relation exceptionally simple. Even so, we chose to
preserve the elements given by the systematization and follow a longer path.
This reasoning established a routine, which will be fundamental to perform
box contractions.

Extending this discussion to divergent contributions is direct if we note
that the first structure of the equation above exhibits the same index
configuration observed for 2nd-order $J$-tensors. Therefore, if the Schouten
identity (\ref{idS}) applies to surface terms, index permutations produce
the organization required to recognize the remaining amplitudes%
\begin{eqnarray}
q^{\mu }T_{\mu \nu \alpha }^{AVV} &=&T_{\nu \alpha }^{AV}\left(
k_{2},k_{3}\right) -T_{\alpha \nu }^{AV}\left( k_{1},k_{2}\right) -2mT_{\nu
\alpha }^{PVV}  \notag \\
&&-2iq^{\mu }p^{\beta }\varepsilon _{\mu \nu \alpha \beta }\left[ \Delta
_{\rho }^{\rho }+\frac{i}{8\pi ^{2}}\right] ;  \label{cite5}
\end{eqnarray}%
see Equation (\ref{AV}). Once again, we have an additional object $\Delta
_{\rho }^{\rho }$ for this relation. We highlight that the only requirement
to obtain this result is the validity of the integral linearity.

Our objective was to perform the axial vertex contraction for the $AVV$
amplitude to verify the corresponding relation among GF; however, we found
an additional contribution in the second row of the equation above.
Differently from vector relations, this one is not automatic since it
depends on a condition over the value attributed to surface terms. Its
satisfaction occurs if the quantity in square brackets is null, which would
imply the ensuing values for the surface term and its trace:%
\begin{equation}
\Delta _{\rho \sigma }=-\frac{i}{32\pi ^{2}}g_{\rho \sigma }\text{, \ }%
\Delta _{\rho }^{\rho }=-\frac{i}{8\pi ^{2}}.  \label{cond1}
\end{equation}

We aim to extend these calculations to box contractions in the following
subsection. For both cases, Dirac traces admit different expressions because
they led to products involving the Levi-Civita symbol and metric tensors. We
expect that the reasoning developed for the triangle also applies in the box
context, so the chosen traces link to additional terms. Afterward, we
discuss the source of this mathematical structure and investigate its
implications.

\subsection{Relations Among Green Functions - $AVVV$}

\label{RAGF2}This subsection aims to verify relations among GF derived for $%
AVVV$ contractions (\ref{AVVV}). As the corresponding expectations are cast
in Equations (\ref{ragf4})-(\ref{ragf7}), we simply transcribe them here:%
\begin{eqnarray}
\left( k_{1}-k_{4}\right) ^{\mu }T_{\mu \nu \alpha \beta }^{AVVV}
&\rightarrow &T_{\nu \alpha \beta }^{AVV}\left( k_{2},k_{3},k_{4}\right)
-T_{\beta \nu \alpha }^{AVV}\left( k_{1},k_{2},k_{3}\right) -2mT_{\nu \alpha
\beta }^{PVVV}, \\
\left( k_{1}-k_{2}\right) ^{\nu }T_{\mu \nu \alpha \beta }^{AVVV}
&\rightarrow &T_{\mu \alpha \beta }^{AVV}\left( k_{2},k_{3},k_{4}\right)
-T_{\mu \alpha \beta }^{AVV}\left( k_{1},k_{3},k_{4}\right) , \\
\left( k_{2}-k_{3}\right) ^{\alpha }T_{\mu \nu \alpha \beta }^{AVVV}
&\rightarrow &T_{\mu \nu \beta }^{AVV}\left( k_{1},k_{3},k_{4}\right)
-T_{\mu \nu \beta }^{AVV}\left( k_{1},k_{2},k_{4}\right) , \\
\left( k_{4}-k_{3}\right) ^{\beta }T_{\mu \nu \alpha \beta }^{AVVV}
&\rightarrow &T_{\mu \nu \alpha }^{AVV}\left( k_{1},k_{2},k_{3}\right)
-T_{\mu \nu \alpha }^{AVV}\left( k_{1},k_{2},k_{4}\right) .
\end{eqnarray}%
Although they have different levels of complexity, triangle and box
calculations contain analogous ingredients. Notably, the systematization
through $J$-tensors establishes a clear link between both cases. That
strongly shapes the procedure adopted this time, so we introduce all
properties of these tensors beforehand.

Momenta contractions occur subsequently, starting with those involving
vector vertices. Observing the forms adopted for traces throughout this
investigation, we expect them to exhibit reductions from the outset. That
makes this context simpler even if numerous algebraic operations are
necessary. The axial contraction requires index permutations additionally;
thus, we approach this case carefully in the final subsubsection.

\subsubsection{Properties of Finite Tensors}

\label{FiniteTensorBox}One remarkable ingredient of the systematization
brought by IReg concerns structure functions used to describe the finite
part of amplitudes. Those functions typical of four-point integrals were
introduced in Subsection (\ref{Finite3}), where they receive integral
representations characterized by three Feynman parameters. Furthermore, we
derived reductions of these functions, in which case combinations
constrained by the same sum of parameter powers $a+b+c$ lead to structures
with decreased powers.

In Subsection (\ref{Fey4}), we computed four-point Feynman integrals and
projected their finite content through the mentioned functions. Nonetheless,
they did not appear randomly but grouped into symmetric objects following a
constraint regarding parameter powers, the so-called $J$-tensors. Reductions
appear inside momenta contractions and traces of them. Therefore, after
performing these operations, we cast properties that concern this
investigation below. Since the 4th-order tensor does not contribute to the
studied amplitudes, we omit the corresponding information. We recall the
notations for finite functions and tensors through the associations: $\xi
_{ab}=\xi _{ab}\left( p,q\right) $, $\xi _{ab}^{\prime }=\xi _{ab}\left(
p,r\right) $, $\xi _{ab}^{\prime \prime }=\xi _{ab}\left( q,r\right) $, $\xi
_{ab}^{\prime \prime \prime }=\xi _{ab}\left( q-p,r-p\right) $, and $\xi
_{abc}=\xi _{abc}\left( p,q,r\right) $.

$\bullet $ First-order tensor - reducing $a+b+c=1$%
\begin{eqnarray}
2p^{\mu }J_{4\mu } &=&p^{2}J_{4}+J_{3}^{\prime \prime \prime }-J_{3}^{\prime
\prime }  \label{c1} \\
2q^{\mu }J_{4\mu } &=&q^{2}J_{4}+J_{3}^{\prime \prime \prime }-J_{3}^{\prime
} \\
2r^{\mu }J_{4\mu } &=&r^{2}J_{4}+J_{3}^{\prime \prime \prime }-J_{3}
\label{c3}
\end{eqnarray}

$\bullet $ Second-order tensor - reducing $a+b+c=2$%
\begin{eqnarray}
2p^{\mu }J_{4\mu \nu } &=&p^{2}J_{4\nu }+J_{3\nu }^{\prime \prime \prime
}+p_{\nu }J_{3}^{\prime \prime \prime }-J_{3\nu }^{\prime \prime }
\label{c4} \\
2q^{\mu }J_{4\mu \nu } &=&q^{2}J_{4\nu }+J_{3\nu }^{\prime \prime \prime
}+p_{\nu }J_{3}^{\prime \prime \prime }-J_{3\nu }^{\prime } \\
2r^{\mu }J_{4\mu \nu } &=&r^{2}J_{4\nu }+J_{3\nu }^{\prime \prime \prime
}+p_{\nu }J_{3}^{\prime \prime \prime }-J_{3\nu } \\
J_{4\mu \mu } &=&m^{2}J_{4}+J_{3}^{\prime \prime \prime }  \label{c7}
\end{eqnarray}

$\bullet $ Third-order tensor - reducing $a+b+c=3$%
\begin{eqnarray}
2p^{\mu }J_{4\mu \nu \alpha } &=&p^{2}J_{4\nu \alpha }+J_{3\nu \alpha
}^{\prime \prime \prime }+p_{\nu }J_{3\alpha }^{\prime \prime \prime
}+p_{\alpha }J_{3\nu }^{\prime \prime \prime }+p_{\nu \alpha }J_{3}^{\prime
\prime \prime }-J_{3\nu \alpha }^{\prime \prime }  \label{c8} \\
2q^{\mu }J_{4\mu \nu \alpha } &=&q^{2}J_{4\nu \alpha }+J_{3\nu \alpha
}^{\prime \prime \prime }+p_{\nu }J_{3\alpha }^{\prime \prime \prime
}+p_{\alpha }J_{3\nu }^{\prime \prime \prime }+p_{\nu \alpha }J_{3}^{\prime
\prime \prime }-J_{3\nu \alpha }^{\prime } \\
2r^{\mu }J_{4\mu \nu \alpha } &=&r^{2}J_{4\nu \alpha }+J_{3\nu \alpha
}^{\prime \prime \prime }+p_{\nu }J_{3\alpha }^{\prime \prime \prime
}+p_{\alpha }J_{3\nu }^{\prime \prime \prime }+p_{\nu \alpha }J_{3}^{\prime
\prime \prime }-J_{3\nu \alpha } \\
J_{4\mu \mu \nu } &=&m^{2}J_{4\nu }+J_{3\nu }^{\prime \prime \prime }+p_{\nu
}J_{3}^{\prime \prime \prime }  \label{c11}
\end{eqnarray}

Although we already employed reductions of three-point functions,
introducing different momenta configurations is necessary. For such purpose,
recall the discussion developed when exploring 2nd-order standard tensors in
the box context (\ref{Second}). These properties are cast in the sequence.

$\bullet $ Denominator $D_{123}$ - $\xi _{ab}=\xi _{ab}\left( p,q\right) $%
\begin{eqnarray}
2p^{\mu }J_{3\mu } &=&p^{2}J_{3}-i\left( 4\pi \right) ^{-2}\left[ \xi
_{0}^{\left( 0\right) }\left( p-q\right) -\xi _{0}^{\left( 0\right) }\left(
q\right) \right] \\
2q^{\mu }J_{3\mu } &=&q^{2}J_{3}-i\left( 4\pi \right) ^{-2}\left[ \xi
_{0}^{\left( 0\right) }\left( p-q\right) -\xi _{0}^{\left( 0\right) }\left(
p\right) \right] \\
2p^{\mu }J_{3\mu \nu } &=&p^{2}J_{3\nu }-i\left( 4\pi \right) ^{-2}\frac{1}{2%
}\left[ \left( p+q\right) _{\nu }\xi _{0}^{\left( 0\right) }\left(
p-q\right) -q_{\nu }\xi _{0}^{\left( 0\right) }\left( q\right) \right] \\
2q^{\mu }J_{3\mu \nu } &=&q^{2}J_{3\nu }-i\left( 4\pi \right) ^{-2}\frac{1}{2%
}\left[ \left( p+q\right) _{\nu }\xi _{0}^{\left( 0\right) }\left(
p-q\right) -p_{\nu }\xi _{0}^{\left( 0\right) }\left( p\right) \right] \\
J_{3\rho }^{\rho } &=&m^{2}J_{3}+i\left( 4\pi \right) ^{-2}\left[ \frac{1}{2}%
-\xi _{0}^{\left( 0\right) }\left( p-q\right) \right]  \label{Trace0}
\end{eqnarray}

$\bullet $ Denominator $D_{124}$ - $\xi _{ab}^{\prime }=\xi _{ab}\left(
p,r\right) $%
\begin{eqnarray}
2p^{\mu }J_{3\mu }^{\prime } &=&p^{2}J_{3}^{\prime }-i\left( 4\pi \right)
^{-2}\left[ \xi _{0}^{\left( 0\right) }\left( p-r\right) -\xi _{0}^{\left(
0\right) }\left( r\right) \right] \\
2r^{\mu }J_{3\mu }^{\prime } &=&r^{2}J_{3}^{\prime }-i\left( 4\pi \right)
^{-2}\left[ \xi _{0}^{\left( 0\right) }\left( p-r\right) -\xi _{0}^{\left(
0\right) }\left( p\right) \right] \\
2p^{\mu }J_{3\mu \nu }^{\prime } &=&p^{2}J_{3\nu }^{\prime }-i\left( 4\pi
\right) ^{-2}\frac{1}{2}\left[ \left( p+r\right) _{\nu }\xi _{0}^{\left(
0\right) }\left( p-r\right) -r_{\nu }\xi _{0}^{\left( 0\right) }\left(
r\right) \right] \\
2r^{\mu }J_{3\mu \nu }^{\prime } &=&r^{2}J_{3\nu }^{\prime }-i\left( 4\pi
\right) ^{-2}\frac{1}{2}\left[ \left( p+r\right) _{\nu }\xi _{0}^{\left(
0\right) }\left( p-r\right) -p_{\nu }\xi _{0}^{\left( 0\right) }\left(
p\right) \right] \\
J_{3\rho }^{\prime \rho } &=&m^{2}J_{3}^{\prime }+i\left( 4\pi \right) ^{-2} 
\left[ \frac{1}{2}-\xi _{0}^{\left( 0\right) }\left( p-r\right) \right]
\end{eqnarray}

$\bullet $ Denominator $D_{134}$ - $\xi _{ab}^{\prime \prime }=\xi
_{ab}\left( q,r\right) $%
\begin{eqnarray}
2q^{\mu }J_{3\mu }^{\prime \prime } &=&q^{2}J_{3}^{\prime \prime }-i\left(
4\pi \right) ^{-2}\left[ \xi _{0}^{\left( 0\right) }\left( q-r\right) -\xi
_{0}^{\left( 0\right) }\left( r\right) \right] \\
2r^{\mu }J_{3\mu }^{\prime \prime } &=&r^{2}J_{3}^{\prime \prime }-i\left(
4\pi \right) ^{-2}\left[ \xi _{0}^{\left( 0\right) }\left( q-r\right) -\xi
_{0}^{\left( 0\right) }\left( q\right) \right] \\
2q^{\mu }J_{3\mu \nu }^{\prime \prime } &=&q^{2}J_{3\nu }^{\prime \prime
}-i\left( 4\pi \right) ^{-2}\frac{1}{2}\left[ \left( q+r\right) _{\nu }\xi
_{0}^{\left( 0\right) }\left( q-r\right) -r_{\nu }\xi _{0}^{\left( 0\right)
}\left( r\right) \right] \\
2r^{\mu }J_{3\mu \nu }^{\prime \prime } &=&r^{2}J_{3\nu }^{\prime \prime
}-i\left( 4\pi \right) ^{-2}\frac{1}{2}\left[ \left( q+r\right) _{\nu }\xi
_{0}^{\left( 0\right) }\left( q-r\right) -q_{\nu }\xi _{0}^{\left( 0\right)
}\left( q\right) \right] \\
J_{3\rho }^{\prime \prime \rho } &=&m^{2}J_{3}^{\prime \prime }+i\left( 4\pi
\right) ^{-2}\left[ \frac{1}{2}-\xi _{0}^{\left( 0\right) }\left( q-r\right) %
\right]
\end{eqnarray}

$\bullet $ Denominator $D_{234}$ - $\xi _{ab}^{\prime \prime \prime }=\xi
_{ab}\left( q-p,r-p\right) $%
\begin{eqnarray}
2\left( q-p\right) ^{\mu }J_{3\mu }^{\prime \prime \prime } &=&\left(
q-p\right) ^{2}J_{3}^{\prime \prime \prime }-i\left( 4\pi \right) ^{-2}\left[
\xi _{0}^{\left( 0\right) }\left( q-r\right) -\xi _{0}^{\left( 0\right)
}\left( r-p\right) \right] \\
2\left( r-p\right) ^{\mu }J_{3\mu }^{\prime \prime \prime } &=&\left(
r-p\right) ^{2}J_{3}^{\prime \prime \prime }-i\left( 4\pi \right) ^{-2}\left[
\xi _{0}^{\left( 0\right) }\left( q-r\right) -\xi _{0}^{\left( 0\right)
}\left( q-p\right) \right] \\
2\left( q-p\right) ^{\mu }J_{3\mu \nu }^{\prime \prime \prime } &=&\left(
q-p\right) ^{2}J_{3\nu }^{\prime \prime \prime }-\frac{1}{2}i\left( 4\pi
\right) ^{-2}\times  \notag \\
&&\times \left[ \left( q+r-2p\right) _{\nu }\xi _{0}^{\left( 0\right)
}\left( q-r\right) -\left( r-p\right) _{\nu }\xi _{0}^{\left( 0\right)
}\left( r-p\right) \right] \\
2\left( r-p\right) ^{\mu }J_{3\mu \nu }^{\prime \prime \prime } &=&\left(
r-p\right) ^{2}J_{3\nu }^{\prime \prime \prime }-\frac{1}{2}i\left( 4\pi
\right) ^{-2}\times  \notag \\
&&\times \left[ \left( q+r-2p\right) _{\nu }\xi _{0}^{\left( 0\right)
}\left( q-r\right) -\left( q-p\right) _{\nu }\xi _{0}^{\left( 0\right)
}\left( q-p\right) \right] \\
J_{3\rho }^{\prime \prime \prime \rho } &=&m^{2}J_{3}^{\prime \prime \prime
}+i\left( 4\pi \right) ^{-2}\left[ \frac{1}{2}-\xi _{0}^{\left( 0\right)
}\left( q-r\right) \right]  \label{Trace3}
\end{eqnarray}

\subsubsection{Vector Contractions}

Proceeding to the explicit computation of relations among GF of the $AVVV$
function (\ref{AVVV}), let us consider vector vertices first. For them, a
contraction with the corresponding momentum results in a difference between $%
AVV$ triangles. Hence, using the expression attributed to this amplitude (%
\ref{AVV}) gives hints for future calculations.

The most immediate implications concern terms whose index arrangements do
not find correspondence inside the triangle. For instance, the $AVPP$
function fits this category for still being proportional to the metric
tensor $g_{\alpha \beta }$ after contracting the index $\nu $. When
exploring other relations, this notion extends to similar amplitudes. Using
reductions of 2nd and 1st-order $J$-tensors, we prove that these products
indeed vanish%
\begin{eqnarray}
g_{\alpha \beta }p^{\nu }T_{\mu \nu }^{AVPP} &=&0, \\
g_{\nu \beta }\left( q-p\right) ^{\alpha }T_{\mu \alpha }^{APVP} &=&0, \\
g_{\nu \alpha }\left( q-r\right) ^{\beta }T_{\mu \beta }^{APPV} &=&0.
\end{eqnarray}

Subsequently, look at those structures proportional to the Levi-Civita
symbol having $\mu $ as the only free index. Comparing tensor (\ref{12})
with the adequate sectors from $APVP$ and $APPV$ functions (\ref{AVPP})
shows that these contributions cancel out identically for the first
contraction. Analogous structures arise for other contractions and cancel
out in the same way. Therefore, we cast these identities in the sequence%
\begin{eqnarray}
\varepsilon _{\mu XYZ}\left[ 4p^{Z}T_{XY\alpha \beta }^{\left( 12\right)
}+p_{\beta }F_{4\alpha XYZ}^{\left( +,+\right) }-p_{\alpha }F_{4\beta
XYZ}^{\left( +,-\right) }\right] &=&0, \\
\varepsilon _{\mu XYZ}\left[ 4\left( q-p\right) ^{Z}T_{XY\nu \beta }^{\left(
13\right) }-\left( q-p\right) _{\beta }F_{4\nu XYZ}^{\left( -,+\right)
}-\left( q-p\right) _{\nu }F_{4\beta XYZ}^{\left( +,-\right) }\right] &=&0,
\\
\varepsilon _{\mu XYZ}\left[ 4\left( q-r\right) ^{Z}T_{XY\nu \alpha
}^{\left( 14\right) }-\left( q-r\right) _{\alpha }F_{4\nu XYZ}^{\left(
-,+\right) }+\left( q-r\right) _{\nu }F_{4\alpha XYZ}^{\left( +,+\right) }%
\right] &=&0.
\end{eqnarray}%
Contractions assume the forms below when disregarding null objects:%
\begin{eqnarray}
p^{\nu }T_{\mu \nu \alpha \beta }^{AVVV} &=&\varepsilon _{\mu \alpha XY} 
\left[ 4p^{\nu }T_{XY\nu \beta }^{\left( 13\right) }+p_{X}T_{Y\beta
}^{VPPV}+p_{\beta }F_{4XY}\right]  \notag \\
&&+\varepsilon _{\mu \beta XY}\left[ 4p^{\nu }T_{XY\nu \alpha }^{\left(
14\right) }+p_{X}T_{Y\alpha }^{VPVP}-p_{\alpha }F_{4XY}\right]  \notag \\
&&+\varepsilon _{\mu \alpha \beta X}\left[ 2p_{X}T^{PPPP}-p^{\nu }T_{X\nu
}^{VVPP}\right] ,  \label{pT1}
\end{eqnarray}%
\begin{eqnarray}
\left( q-p\right) ^{\alpha }T_{\mu \nu \alpha \beta }^{AVVV} &=&\varepsilon
_{\mu \nu XY}\left[ 4\left( q-p\right) ^{\alpha }T_{XY\alpha \beta }^{\left(
12\right) }-\left( q-p\right) ^{X}T_{Y\beta }^{VPPV}-\left( q-p\right)
_{\beta }F_{4XY}\right]  \notag \\
&&+\varepsilon _{\mu \beta XY}\left[ 4\left( q-p\right) ^{\alpha }T_{XY\nu
\alpha }^{\left( 14\right) }+\left( q-p\right) ^{X}T_{Y\nu }^{VVPP}-\left(
q-p\right) _{\nu }F_{4XY}\right]  \notag \\
&&-\varepsilon _{\mu \nu \beta X}\left[ 2\left( q-p\right)
^{X}T^{PPPP}+\left( q-p\right) ^{\alpha }T_{X\alpha }^{VPVP}\right] ,
\end{eqnarray}%
\begin{eqnarray}
\left( q-r\right) ^{\beta }T_{\mu \nu \alpha \beta }^{AVVV} &=&\varepsilon
_{\mu \nu XY}\left[ 4\left( q-r\right) ^{\beta }T_{XY\alpha \beta }^{\left(
12\right) }-\left( q-r\right) ^{X}T_{Y\alpha }^{VPVP}-\left( q-r\right)
_{\alpha }F_{4XY}\right]  \notag \\
&&+\varepsilon _{\mu \alpha XY}\left[ 4\left( q-r\right) ^{\beta }T_{XY\nu
\beta }^{\left( 13\right) }-\left( q-r\right) ^{X}T_{Y\nu }^{VVPP}+\left(
q-r\right) _{\nu }F_{4XY}\right]  \notag \\
&&+\varepsilon _{\mu \nu \alpha X}\left[ 2\left( q-r\right)
^{X}T^{PPPP}-\left( q-r\right) ^{\beta }T_{X\beta }^{VPPV}\right] .
\end{eqnarray}

Our task becomes reducing all four-point finite functions, expecting that
only structures associated with the triangle remain. Although the number of
terms might bring complications, exploring each component separately is
possible since different tensor arrangements do not mix. Nevertheless, be
aware in this process that tensor subamplitudes carry contributions
proportional to the scalar one.

Once these operations are clear from the previous subsection, let us just
stress some details. The hierarchy associated with reductions must be
strictly followed; therefore, we start with the highest-order structure
function from four-point integrals $a+b+c=3$ and gradually decrease
parameter powers. With this stage complete, it is necessary to process
three-point structures using identity (\ref{idv}). That is possibly the most
intricate part of these calculations, so using the $AVV$ as a guide becomes
essential; consult Equation (\ref{AVV}). Meanwhile, reductions subtract each
other for contractions dealing with differences between external momenta.
That is a source of cancellations, decreasing our efforts when studying this
sector. To exemplify, we present the first contraction in its final
organization%
\begin{eqnarray}
\left[ p^{\nu }T_{\mu \nu \alpha \beta }^{AVVV}\right] _{fin}
&=&8\varepsilon _{\mu \alpha XY}\left\{ \left( q-p\right) _{X}J_{3Y\beta
}^{\prime \prime \prime }-q_{X}J_{3Y\beta }^{\prime \prime }-\left(
q-p\right) _{X}\left( r-p\right) _{\beta }J_{3Y}^{\prime \prime \prime
}\right.  \notag \\
&&\left. +q_{X}r_{\beta }J_{3Y}^{\prime \prime }\right\} +8\varepsilon _{\mu
\beta XY}\left\{ \left( r-q\right) _{X}\left( J_{3Y\alpha }^{\prime \prime
\prime }-J_{3Y\alpha }^{\prime \prime }\right) -q_{X}r_{Y}J_{3\alpha
}^{\prime \prime }\right.  \notag \\
&&\left. +\left( q-p\right) _{X}\left( r-p\right) _{Y}J_{3\alpha }^{\prime
\prime \prime }\right\} -2\varepsilon _{\mu \alpha \beta X}\left\{ \left(
q^{2}r_{X}-r^{2}q_{X}\right) J_{3}^{\prime \prime }\right.  \notag \\
&&-2q\cdot \left( q-r\right) J_{3X}^{\prime \prime }+2\left( q-p\right)
\cdot \left( q-r\right) J_{3X}^{\prime \prime \prime }  \notag \\
&&-\left[ \left( q-p\right) ^{2}\left( r-p\right) _{X}-\left( r-p\right)
^{2}\left( q-p\right) _{X}\right] J_{3}^{\prime \prime \prime }  \notag \\
&&\left. -i\left( 4\pi \right) ^{-2}\left[ \left( q-p\right) ^{X}\xi
_{0}^{\left( 0\right) }\left( q-p\right) -q^{X}\xi _{0}^{\left( 0\right)
}\left( q\right) \right] \right\} .
\end{eqnarray}

We still have to analyze divergent structures to complete this analysis. As
stated before, even though Feynman integrals depend on different standard
objects, only one type of surface term appears within the $AVVV$ box. Our
work summarizes into surveying substructures of this amplitude to find the
corresponding contributions and organize them through algebraic operations.
We exemplify this procedure for the first contraction:%
\begin{eqnarray}
\left[ p^{\nu }T_{\mu \nu \alpha \beta }^{AVVV}\right] _{div}
&=&2p^{\nu }\left( \varepsilon _{\mu \alpha \beta X}\Delta _{\nu
}^{X}+\varepsilon _{\mu \nu \alpha X}\Delta _{\beta }^{X}\right) \\
&=&-2\varepsilon _{\mu \beta XY}\left[ \left( q-r\right) -\left( q-r\right) %
\right] ^{X}\Delta _{\alpha }^{Y}-2\varepsilon _{\mu \nu \alpha X}\left[
\left( q-p\right) -q\right] ^{\nu }\Delta _{\beta }^{X}  \notag \\
&&-2\varepsilon _{\mu \alpha \beta X}\left[ \left( k_{2}+k_{4}\right)
-\left( k_{1}+k_{4}\right) \right] ^{\nu }\Delta _{\nu }^{X}.
\end{eqnarray}%
At this point, identifying divergent and finite parts as those belonging to
the triangle is straightforward (\ref{AVV}). That extends to all cases;
hence, all vector relations among GF apply regardless of the prescription
adopted to evaluate surface terms:%
\begin{equation}
p^{\nu }T_{\mu \nu \alpha \beta }^{AVVV}=T_{\mu \alpha \beta }^{AVV}\left(
k_{2},k_{3},k_{4}\right) -T_{\mu \alpha \beta }^{AVV}\left(
k_{1},k_{3},k_{4}\right) ,
\end{equation}%
\begin{eqnarray}
\left( q-p\right) ^{\alpha }T_{\mu \nu \alpha \beta }^{AVVV} &=&T_{\mu \nu
\beta }^{AVV}\left( k_{1},k_{3},k_{4}\right) -T_{\mu \nu \beta }^{AVV}\left(
k_{1},k_{2},k_{4}\right) , \\
\left( q-r\right) ^{\beta }T_{\mu \nu \alpha \beta }^{AVVV} &=&T_{\mu \nu
\alpha }^{AVV}\left( k_{1},k_{2},k_{3}\right) -T_{\mu \nu \alpha
}^{AVV}\left( k_{1},k_{2},k_{4}\right) .
\end{eqnarray}

\subsubsection{Axial Contraction}

The remaining box relation arises from the contraction between the momentum $%
r=k_{1}-k_{4}$ and the index corresponding to the axial vertex. Firstly,
following the route established for vector cases, observe that structures
associated with odd subamplitudes stand out from others. That is transparent
when comparing terms where the metric has exclusively free indices; consult
the final expressions for $AVVV$ (\ref{AVVV})\ and $PVVV$ (\ref{IntPVVV}).
Hence, our initial task is to verify the following expectation%
\begin{equation}
r^{\mu }\left[ g_{\alpha \beta }T_{\mu \nu }^{AVPP}+g_{\nu \beta }T_{\mu
\alpha }^{APVP}+g_{\nu \alpha }T_{\mu \beta }^{APPV}\right] =8im^{2}\left(
g_{\kappa \nu }g_{\alpha \beta }-g_{\kappa \alpha }g_{\nu \beta }+g_{\kappa
\beta }g_{\nu \alpha }\right) F_{4\kappa }  \label{part1}
\end{equation}

We resort to the information established in Subsubsection (\ref{AVPPexp}) to
accomplish this result. The first sector of the explored amplitudes features
a three-index contraction involving the Levi-Civita symbol; thus,
introducing another external momentum vanishes most contributions. Only the
2nd-order $J$-tensor (\ref{J4uv}) remains because it has terms on the metric
tensor:%
\begin{eqnarray}
&&-i\varepsilon _{\mu XYZ}r^{\mu }\left[ g_{\alpha \beta }F_{4\nu
XYZ}^{\left( -,+\right) }-g_{\nu \beta }F_{4\alpha XYZ}^{\left( +,+\right)
}+g_{\nu \alpha }F_{4\beta XYZ}^{\left( +,-\right) }\right]  \notag \\
&=&8i\varepsilon _{\mu XYZ}r^{\mu }p^{X}q^{Y}\left( g_{\alpha \beta }J_{4\nu
Z}-g_{\nu \beta }J_{4\alpha Z}+g_{\nu \alpha }J_{4\beta Z}\right)  \notag \\
&=&4ir^{\mu }p^{X}q^{Y}\left( g_{\alpha \beta }\varepsilon _{\mu \nu
XY}-g_{\nu \beta }\varepsilon _{\mu \alpha XY}+g_{\nu \alpha }\varepsilon
_{\mu \beta XY}\right) i\left( 4\pi \right) ^{-2}\xi _{000}^{\left(
-1\right) }.
\end{eqnarray}%
A finite function as $\xi _{000}^{\left( -1\right) }$ is typical of
higher-order Feynman integrals; therefore, not compatible with intended
identifications. Reduction (\ref{000}) handles this situation while bringing
the squared mass contribution necessary to find $F_{4\mu }$; we transcribe
this property here%
\begin{equation}
\xi _{000}^{\left( -1\right) }=2m^{2}\xi _{000}^{\left( -2\right) }-\left[
p^{2}\xi _{100}^{\left( -2\right) }+q^{2}\xi _{010}^{\left( -2\right)
}+r^{2}\xi _{001}^{\left( -2\right) }\right] +\left[ \xi _{00}^{\left(
-1\right) }\right] ^{\prime \prime \prime }.
\end{equation}

Notwithstanding that the situation is similar to the other sector, it leads
to a more complex expression due to the two-index contraction:%
\begin{eqnarray}
&&-ir^{\mu }\left( g_{\alpha \beta }\varepsilon _{\mu \nu XY}-g_{\nu \beta
}\varepsilon _{\mu \alpha XY}+g_{\nu \alpha }\varepsilon _{\mu \beta
XY}\right) F_{4XY}  \notag \\
&=&ir^{\mu }\left( g_{\alpha \beta }\varepsilon _{\mu \nu XY}-g_{\nu \beta
}\varepsilon _{\mu \alpha XY}+g_{\nu \alpha }\varepsilon _{\mu \beta
XY}\right) \times  \notag \\
&&\times \left\{ 4\left[ \left( q^{2}-q\cdot r\right) p_{X}-\left(
p^{2}-p\cdot r\right) q_{X}\right] J_{4Y}\right.  \notag \\
&&\left. +2p_{X}q_{Y}\left( r^{2}J_{4}-J_{3}^{\prime \prime \prime
}-J_{3}\right) \right\} .
\end{eqnarray}%
Even so, both parts fit perfectly since functions constrained by $a+b+c=1$
compound the vector reduction $r^{\mu }J_{4\mu }$. Such an object cancels
out all spare terms, completing the proof of relation (\ref{part1}). That
corresponds to the first row from $PVVV$ amplitude (\ref{PVVV}).

As in triangle calculations (\ref{cite1}), the remaining steps require index
permutations through the symmetry properties of tensors. A crucial feature
of these operations is that they generate additional contributions embodied
in traces, which generate the expected contributions proportional to the
squared mass.

To illustrate this procedure, we analyze finite functions whose parameter
powers follow the condition $a+b+c=3$. They compound 3rd-order $J_{4}$%
-tensors found inside tensor combinations belonging to the box amplitude:%
\begin{eqnarray}
\left[ r^{\mu }T_{\mu \nu \alpha \beta }^{AVVV}\right] _{a+b+c=3}
&=&16r^{\mu }p^{X}\left( \varepsilon _{\mu \nu XY}J_{4\alpha Y\beta
}-\varepsilon _{\mu \alpha XY}J_{4\nu Y\beta }\right)  \notag \\
&&+16r^{\mu }q^{X}\left( \varepsilon _{\mu \alpha XY}J_{4\beta Y\nu
}-\varepsilon _{\mu \beta XY}J_{4\alpha Y\nu }\right) .
\end{eqnarray}%
Our reasoning consists of building an object exhibiting antisymmetry in five
indices, a Schouten identity. Thus, considering only the first $J_{4}$-index
as changeable, let us rearrange indices accordingly to the expression 
\begin{eqnarray}
\left[ r^{\mu }T_{\mu \nu \alpha \beta }^{AVVV}\right] _{a+b+c=3}
&=&-16r^{\mu }p^{X}\left( \varepsilon _{\alpha \mu \nu X}J_{4YY\beta
}+\varepsilon _{Y\alpha \mu \nu }J_{4XY\beta }+\varepsilon _{\nu XY\alpha
}J_{4\mu Y\beta }\right)  \notag \\
&&-16r^{\mu }q^{X}\left( \varepsilon _{\beta \mu \alpha X}J_{4YY\nu
}+\varepsilon _{Y\beta \mu \alpha }J_{4XY\nu }+\varepsilon _{\alpha XY\beta
}J_{4\mu Y\nu }\right) .
\end{eqnarray}%
As all pieces are known, see Equations (\ref{c8})-(\ref{c11}), the adequate
replacements yield%
\begin{eqnarray}
&&\left[ r^{\mu }T_{\mu \nu \alpha \beta }^{AVVV}\right] _{a+b+c=3}  \notag
\\
&=&-16\varepsilon _{\alpha \mu \nu X}r^{\mu }p^{X}\left( m^{2}J_{4\beta
}+J_{3\beta }^{\prime \prime \prime }+p_{\beta }J_{3}^{\prime \prime \prime
}\right)  \notag \\
&&-8\varepsilon _{Y\alpha \mu \nu }r^{\mu }\left( p^{2}J_{4Y\beta
}+J_{3Y\beta }^{\prime \prime \prime }+p_{Y}J_{3\beta }^{\prime \prime
\prime }+p_{\beta }J_{3Y}^{\prime \prime \prime }+p_{Y\beta }J_{3}^{\prime
\prime \prime }-J_{3Y\beta }^{\prime \prime }\right)  \notag \\
&&-8\varepsilon _{\nu XY\alpha }p^{X}\left( r^{2}J_{4Y\beta }+J_{3Y\beta
}^{\prime \prime \prime }+p_{\beta }J_{3Y}^{\prime \prime \prime
}-J_{3Y\beta }\right)  \notag \\
&&-16\varepsilon _{\beta \mu \alpha X}r^{\mu }q^{X}\left( m^{2}J_{4\nu
}+J_{3\nu }^{\prime \prime \prime }+p_{\nu }J_{3}^{\prime \prime \prime
}\right)  \notag \\
&&-8\varepsilon _{Y\beta \mu \alpha }r^{\mu }\left( q^{2}J_{4Y\nu }+J_{3Y\nu
}^{\prime \prime \prime }+p_{Y}J_{3\nu }^{\prime \prime \prime }+p_{\nu
}J_{3Y}^{\prime \prime \prime }+p_{Y\nu }J_{3}^{\prime \prime \prime
}-J_{3Y\nu }^{\prime }\right)  \notag \\
&&-8\varepsilon _{\alpha XY\beta }q^{X}\left( r^{2}J_{4Y\nu }+J_{3Y\nu
}^{\prime \prime \prime }+p_{Y}J_{3\nu }^{\prime \prime \prime }+p_{\nu
}J_{3Y}^{\prime \prime \prime }+p_{Y\nu }J_{3}^{\prime \prime \prime
}-J_{3Y\nu }\right) .
\end{eqnarray}

The next step is to track all finite contributions under the restriction $%
a+b+c=2$. After rearrangements and other algebraic operations, we obtain
momenta contractions and traces of the 2nd-order $J_{4}$-tensor (\ref{c4})-(%
\ref{c7}). These traces contain terms proportional to the squared mass that
complete the content of four-point finite functions within $PVVV$ (there are
some missing pieces on $J_{3}$). Except for this sector, other structure
functions under this category disappear in the sequence through reductions
of $J_{4}$-vectors (\ref{c1})-(\ref{c3}). Although the process described in
this paragraph is notably extensive, all steps are transparent and easily
checked.

We must still explore those objects associated with three-point finite
functions to perform the remaining identifications, including the $AVV$ part
(\ref{AVV}). In addition to being quite extensive, this part also brings
complications due to the different momenta configurations associated with
the line notation. This discussion appears in detail when exploring
2nd-order standard tensors in the box context (\ref{Second}), while required
tensor properties are at the outset of this subsection (\ref{FiniteTensorBox}%
).

After fulfilling all reductions, we write for the finite sector%
\begin{eqnarray}
\left[ r^{\mu }T_{\mu \nu \alpha \beta }^{AVVV}\right] _{fin} &=&%
\left[ T_{\nu \alpha \beta }^{AVV}\left( k_{2},k_{3},k_{4}\right) -T_{\beta
\nu \alpha }^{AVV}\left( k_{1},k_{2},k_{3}\right) \right] _{fin%
}-2mT_{\nu \alpha \beta }^{PVVV}  \notag \\
&&-2\varepsilon _{\nu \alpha \beta X}\left( k_{1}-k_{2}+k_{3}-k_{4}\right)
^{X}\frac{i}{8\pi ^{2}}.
\end{eqnarray}%
Among all components, let us emphasize the role played by traces $J_{3\rho
}^{\rho }$ and $J_{3\rho }^{\prime \prime \prime \rho }$ from Equations (\ref%
{Trace0}) and (\ref{Trace3}). First, their terms on the squared mass led to
the missing pieces that completed the finite amplitude $PVVV$. Second,
numerical factors are additional terms if one considers the original
expectation for this relation. They correspond to the second line of the
equation above and will receive more attention soon enough.

Lastly, we pursue divergent objects that remain in even subamplitudes after
the axial vertex contraction:%
\begin{equation}
\left[ r^{\mu }T_{\mu \nu \alpha \beta }^{AVVV}\right] _{div%
}=2\varepsilon _{\mu \alpha \beta X}r^{\mu }\Delta _{X\nu }+2\varepsilon
_{\mu \nu \alpha X}r^{\mu }\Delta _{X\beta }.
\end{equation}%
Although that differs significantly from the organization expected for the
triangle (\ref{AVV}), performing algebraic manipulations and exchanging
index positions solve this situation. We add Schouten identities involving
routings $k_{2}$ and $k_{3}$ since they are absent in this equation. That
leads to the following structure%
\begin{eqnarray}
\left[ r^{\mu }T_{\mu \nu \alpha \beta }^{AVVV}\right] _{div}
&=&-2\varepsilon _{\mu \nu XY}\left( k_{4}-k_{3}\right) ^{X}\Delta _{\alpha
Y}+2\varepsilon _{\nu \alpha XY}\left( k_{2}-k_{3}\right) ^{X}\Delta _{\beta
Y}  \notag \\
&&-2\varepsilon _{\nu \alpha \beta X}\left( k_{2}+k_{4}\right) ^{Y}\Delta
_{XY}+2\varepsilon _{\beta \alpha XY}\left( k_{3}-k_{2}\right) ^{X}\Delta
_{\nu Y}  \notag \\
&&-2\varepsilon _{\beta \nu XY}\left( k_{1}-k_{2}\right) ^{X}\Delta _{\alpha
Y}+2\varepsilon _{\beta \nu \alpha X}\left( k_{1}+k_{3}\right) ^{Y}\Delta
_{XY}  \notag \\
&&-2\varepsilon _{\nu \alpha \beta X}\left( k_{1}-k_{2}+k_{3}-k_{4}\right)
^{X}\Delta _{YY},
\end{eqnarray}%
ultimately allowing the final identifications for the total amplitude%
\begin{eqnarray}
r^{\mu }T_{\mu \nu \alpha \beta }^{AVVV} &=&T_{\nu \alpha \beta
}^{AVV}\left( k_{2},k_{3},k_{4}\right) -T_{\beta \nu \alpha }^{AVV}\left(
k_{1},k_{2},k_{3}\right) -2mT_{\nu \alpha \beta }^{PVVV}  \notag \\
&&-2\varepsilon _{\nu \alpha \beta \sigma }\left( p-q+r\right) ^{\sigma } 
\left[ \Delta _{\rho }^{\rho }+\frac{i}{8\pi ^{2}}\right] .  \label{cite6}
\end{eqnarray}

We put additional terms together in the second line while writing their
coefficients in terms of external momenta. Satisfying the axial relation
among GF is not automatic since it requires the cancellations of these terms
as an extra condition; i.e., it depends on the prescription adopted to
evaluate the surface terms. Furthermore, note that the same condition was
acknowledged in the triangle analysis (\ref{cond1}).

\subsection{Further Explorations on Relations Among GF}

\label{Lin}Previously, we analyzed relations among GF emerging from
contractions involving amplitudes that are odd tensors. Relations obtained
for vector vertices were automatic, which means their achievement does not
depend on a prescription to evaluate divergent objects. In contrast, we
found that axial relations apply under a condition for the surface term and
its trace (\ref{cond1}). That works as a requirement for maintaining the
linearity of integration in this context.

Our first objective here is to understand the mechanisms that led to this
outcome. In Subsection (\ref{AVVint}), we discussed roles played by vertices
and Dirac traces. By endowing the $\mu $ index with a special role (\ref%
{form2})-(\ref{form1}), we shaped the tensor sector and fixed the $AVV$
integrand as (\ref{avv1}). Posteriorly, when evaluating the axial relation
among GF (also in $\mu $), index permutations brought additional
contributions to Equation (\ref{cite5}). We also computed traces found
inside the box amplitude by following the same strategy, and the
corresponding axial contraction produced a similar situation (\ref{cite6}).

Mathematical structures suggest a connection involving traces and the
acknowledged results. Let us propose other trace arrangements and inquire
about their implications over the triangle amplitude to clarify this
subject. From this point on, we explore three $AVV$ versions distinguished
through numerical subindices%
\begin{equation*}
\begin{tabular}{lllll}
$t_{1\mu \nu \alpha }^{AVV}\rightarrow $tr$\left( \gamma _{\mu 5A\nu B\alpha
C}\right) ,$ &  & $t_{2\mu \nu \alpha }^{AVV}\rightarrow $tr$\left( \gamma
_{\mu A\nu 5B\alpha C}\right) ,$ &  & $t_{3\mu \nu \alpha }^{AVV}\rightarrow 
$tr$\left( \gamma _{\mu A\nu B\alpha 5C}\right) .$%
\end{tabular}%
\end{equation*}%
These associations specify the position to replace the chiral matrix
definition, thus, prioritizing one free index among the options: $\mu $, $%
\nu $, and $\alpha $.

Take the first version as a guide since it corresponds to the former
integrand (\ref{AVV0}). Recognizing a Schouten identity with the prioritized
index fixed is possible for these versions, as it occurred in Equation (\ref%
{Schouten}). Even if one ignores this property, integrating the amplitudes
vanishes these sectors. Subsequently, our task is to organize integrands
through standard tensors and vector subamplitudes, namely, $VPP$, $SAP$, and 
$SPA$. We already verified some properties of antisymmetric objects (\ref%
{exp3})-(\ref{exp4}); therefore, using them leads to compact integrated
expressions%
\begin{eqnarray}
T_{1\mu \nu \alpha }^{AVV}=4i\varepsilon _{\mu \alpha XY}T_{3\nu
;XY}^{\left( -\right) }\left( k_{1};k_{2},k_{3}\right) +4i\varepsilon _{\mu
\nu XY}T_{3\alpha ;XY}^{\left( -\right) }\left( k_{3};k_{1},k_{2}\right)
-i\varepsilon _{\mu \nu \alpha \beta }T_{\beta }^{VPP}, &&  \label{AVV01} \\
T_{2\mu \nu \alpha }^{AVV}=4i\varepsilon _{\nu \mu XY}T_{3\alpha
;XY}^{\left( -\right) }\left( k_{2};k_{3},k_{1}\right) +4i\varepsilon _{\nu
\alpha XY}T_{3\mu ;XY}^{\left( -\right) }\left( k_{1};k_{2},k_{3}\right)
+i\varepsilon _{\mu \nu \alpha \beta }T_{\beta }^{SAP}, && \\
T_{3\mu \nu \alpha }^{AVV}=4i\varepsilon _{\alpha \nu XY}T_{3\mu
;XY}^{\left( -\right) }\left( k_{3};k_{1},k_{2}\right) +4i\varepsilon
_{\alpha \mu XY}T_{3\nu ;XY}^{\left( -\right) }\left(
k_{2};k_{1},k_{3}\right) -i\varepsilon _{\mu \nu \alpha \beta }T_{\beta
}^{SPA}. &&  \label{AVV03}
\end{eqnarray}

These equations show how traces link to additional terms emerging in
relations among GF. When prioritizing one vertex $\Gamma _{n}$, the
corresponding free index $\mu _{n}=\left\{ \mu ,\nu ,\alpha \right\} $
exclusively appears inside the Levi-Civita symbol for the tensor sector.
Hence, contracting this same index does not immediately lead to reductions.
Under these circumstances, we exchange index positions, and additional terms
emerge through traces of rank-2 objects: $J$-tensor and surface term.
Whereas other contractions are automatic, the $n$th relation among GF of the 
$n$th $AVV$ version is not; these specific cases come as follows:%
\begin{eqnarray}
q^{\mu }T_{1\mu \nu \alpha }^{AVV} &=&T_{\nu \alpha }^{AV}\left(
k_{2},k_{3}\right) -T_{\alpha \nu }^{AV}\left( k_{1},k_{2}\right) -2mT_{\nu
\alpha }^{PVV}  \notag \\
&&-2iq^{\mu }p^{\beta }\varepsilon _{\mu \nu \alpha \beta }\left[ \Delta
_{\rho }^{\rho }+\frac{i}{8\pi ^{2}}\right] , \\
p^{\nu }T_{2\mu \nu \alpha }^{AVV} &=&T_{\mu \alpha }^{AV}\left(
k_{2},k_{3}\right) -T_{\mu \alpha }^{AV}\left( k_{1},k_{3}\right) +2iq^{\nu
}p^{\beta }\varepsilon _{\mu \nu \alpha \beta }\left[ \Delta _{\rho }^{\rho
}+\frac{i}{8\pi ^{2}}\right] , \\
\left( q-p\right) ^{\alpha }T_{3\mu \nu \alpha }^{AVV} &=&T_{\mu \nu
}^{AV}\left( k_{1},k_{3}\right) -T_{\mu \nu }^{AV}\left( k_{1},k_{2}\right)
+2iq^{\alpha }p^{\beta }\varepsilon _{\mu \nu \alpha \beta }\left[ \Delta
_{\rho }^{\rho }+\frac{i}{8\pi ^{2}}\right] .
\end{eqnarray}

Integrated subamplitudes were necessary to inspect relations for new
triangle versions. If it interests the reader, follow the steps developed
for the $VPP$ (\ref{vpp}) to express them as combinations of Feynman
integrals. Posteriorly, the final forms emerge by replacing the necessary
ingredients; consult Equation (\ref{VPP}). Here, let us straightforwardly
introduce these quantities:%
\begin{eqnarray}
T_{\beta }^{SAP} &=&-2\left( k_{1}+k_{2}\right) ^{\rho }\Delta _{\beta \rho
}-2\left( p-2q\right) _{\beta }I_{\log }  \notag \\
&&-4\left( q^{2}-p\cdot q\right) J_{3\beta }-2\left[ p^{2}q_{\beta
}-q^{2}p_{\beta }+4m^{2}\left( q-p\right) ^{2}\right] J_{3}  \notag \\
&&-2i\left( 4\pi \right) ^{-2}\left[ \left( p-q\right) _{\beta }J_{2}\left(
q-p\right) -q_{\beta }J_{2}\left( q\right) \right] ,
\end{eqnarray}%
\begin{eqnarray}
T_{\beta }^{SPA} &=&2\left( k_{2}+k_{3}\right) ^{\rho }\Delta _{\beta \rho
}+2\left( p+q\right) I_{\log }  \notag \\
&&+4\left( p\cdot q\right) J_{3\beta }-2\left( p^{2}q_{\beta }+q^{2}p_{\beta
}+4m^{2}p^{2}\right) J_{3}  \notag \\
&&+2i\left( 4\pi \right) ^{-2}\left[ p_{\beta }J_{2}\left( p\right)
+q_{\beta }J_{2}\left( q\right) \right] .
\end{eqnarray}

This panorama concerns trace choices, having no strict relation with the
vertex content. That becomes even clearer by extending this argumentation to
all similar amplitudes ($AVV$, $VAV$, $VVA$, and $AAA$) since they all share
the same tensor structure:%
\begin{equation}
t_{\mu \nu \alpha }^{\Gamma \Gamma \Gamma }\rightarrow \text{tr}\left(
\gamma _{\mu }\gamma _{5}\gamma _{A}\gamma _{\nu }\gamma _{B}\gamma _{\alpha
}\gamma _{C}\right) \frac{K_{1}^{A}K_{2}^{B}K_{3}^{C}}{D_{123}}.
\end{equation}%
Regardless of its nature as an axial or a vector vertex, additional
contributions arise for a contraction if the contracted index links to the
vertex prioritized when taking the trace. For instance, prioritizing the $%
\mu $-index in the trace (\ref{form2})-(\ref{form1}) makes the first
relation among GF non-automatic for all four triangle amplitudes. Although
this situation is unavoidable, we still can choose the position of
additional terms by setting a specific trace expression.

Different integrands connect through algebraic operations, so one could
expect them to lead to identical results. Nevertheless, that was not
automatic after integration due to the divergent character of calculations.
After observing this feature in momenta contractions, it is reasonable to
compare different amplitude versions directly. With the aid of index
permutations and other algebraic operations, we evaluate differences between
versions%
\begin{equation}
T_{i\mu \nu \alpha }^{AVV}-T_{j\mu \nu \alpha }^{AVV}=i\varepsilon _{\mu \nu
\alpha \beta }P^{\beta }\left[ \Delta _{\rho }^{\rho }+\frac{i}{8\pi ^{2}}%
\right] ,  \label{diff}
\end{equation}%
where $i\neq j$ refers to Equations (\ref{AVV01})-(\ref{AVV03}) and $P$
represents a linear combination of the external momenta $p$ and $q$. The
term between square brackets equals the additional terms acknowledged in
contractions. Hence, opting for a prescription where the surface term
follows condition (\ref{cond1}) implies that all $AVV$ versions collapse
into one unique object while satisfying all relations among GF.

We still want to comment on the analysis regarding the box amplitude. Dirac
traces also admitted different expressions in this case because they led to
products involving the Levi-Civita symbol and the metric tensors. By
endowing the $\mu $ index with a prioritized role, the organization at the
integrand level puts this index exclusively in the Levi-Civita symbol while
other terms cancel out identically. Renaming indices within these traces
directly extends this notion to versions prioritizing other indices. That
applies to any amplitude under this category as they share the tensor
sector: $AVVV$, $AAAV$, and their permutations.

In general, for an amplitude version that prioritizes the index $\mu
_{n}=\left\{ \mu \text{, }\nu \text{, }\alpha \text{, }\beta \right\} $ in
the traces, the $n$th relation among GF requires index permutations to
identify momenta contractions and traces of 2nd-order tensors. Hence, using
analogous traces on the right-hand side of these relations produces the
additional term leading to condition (\ref{cond1}). Explorations considering
different trace versions on the left (box contraction) and on the right
(triangles) might bring further information, so this study remains a future
perspective. For this reason, we will not discuss the symmetry aspects of
box correlators.

The following subsection links the current discussion with WIs, so we can
inquire how the presence of surface terms reflects on the simultaneous
analysis of both types of constraints.

\subsection{Symmetries and Linearity}

\label{sym}In Subsection (\ref{Amplitudes}), we derived algebraic identities
among integrands of perturbative amplitudes. That suggests expectations
through relations among Green Functions (GF) that should apply as a direct
consequence of the linearity of integration. Hence, any violation of these
relations would imply linearity breaking. We tested them in Subsection (\ref%
{RAGF1}) for momenta contractions over the $AVV$ triangle, verifying part of
the cases without problems. Nonetheless, one relation among GF is not
automatic for containing an additional contribution depending on a surface
term.

We proved in Subsection (\ref{Lin}) that choosing a trace expression sets
the position of this additional contribution. By prioritizing one index when
taking the trace, its contraction automatically produces the mentioned
contributions. Although there are other trace possibilities, reference \cite%
{Arxiv22} shows that any other amplitude version combines those investigated
here. Consequently, it would carry potentially violating terms coming from
all combined parts. This overall situation has no relation with the vertex
nature as being axial or vector.

Let us return to the original prospects regarding triangle WIs (\ref{qAVV1}%
)-(\ref{qAVV3}) to continue this inspection. They are consequences of the
current algebra (\ref{dV})-(\ref{dA}) and comprise symmetry implications
over the complete amplitude. Hence, their verifications require symmetrizing
final states and summing up direct and crossed diagrams. We already obtained
the direct one (see Figure \ref{F}); thus, the crossed one arises by
changing the role of indices $\mu \leftrightarrow \nu $ and external momenta 
$p\leftrightarrow q$.

With that clear, consider in a preliminary argument that the satisfaction of
all relations among GF is automatic; i.e., they are valid without the need
for conditions over divergent objects. Under this hypothesis, canceling
differences between $AV$ amplitudes would be our sole concern regarding WIs.
Equations below follow the vertex order for $AVV$ contractions to cast these
structures:%
\begin{eqnarray}
&&T_{\nu \alpha }^{AV}\left( k_{2},k_{3}\right) -T_{\alpha \nu }^{AV}\left(
k_{1},k_{2}\right)  \notag \\
&=&2i\varepsilon _{\mu \nu \alpha \beta }\left[ \left( p-q\right) ^{\mu
}\left( k_{2}+k_{3}\right) ^{\rho }-p^{\mu }\left( k_{1}+k_{2}\right) ^{\rho
}\right] \Delta _{\rho }^{\beta },  \label{DAV1} \\
&&T_{\mu \alpha }^{AV}\left( k_{2},k_{3}\right) -T_{\mu \alpha }^{AV}\left(
k_{1},k_{3}\right)  \notag \\
&=&2i\varepsilon _{\mu \nu \alpha \beta }\left[ \left( q-p\right) ^{\nu
}\left( k_{2}+k_{3}\right) ^{\rho }-q^{\nu }\left( k_{1}+k_{3}\right) ^{\rho
}\right] \Delta _{\rho }^{\beta }, \\
&&T_{\mu \nu }^{AV}\left( k_{1},k_{3}\right) -T_{\mu \nu }^{AV}\left(
k_{1},k_{2}\right)  \notag \\
&=&2i\varepsilon _{\mu \nu \alpha \beta }\left[ p^{\alpha }\left(
k_{1}+k_{2}\right) ^{\rho }-q^{\alpha }\left( k_{1}+k_{3}\right) ^{\rho }%
\right] \Delta _{\rho }^{\beta }.  \label{DAV3}
\end{eqnarray}%
By eliminating surface terms $\Delta _{\rho }^{\beta }=0$, one disappears
with the $AV$ amplitudes and guarantees the satisfaction of all WIs.

There are some details to address about the equations above. Even though
similar structures arise for the crossed channel, combining these sectors is
not feasible. Energy-momentum conservation attributes a physical meaning to
differences of routings as external momenta, albeit not to routings
themselves. That means these quantities are different for each channel (let
us say $k_{i}$ and $k_{i}^{\prime }$), and there are no other connections
involving them.

Under these circumstances, the discussion about symmetry implications
applies channel by channel. Thus, we recall the referred WIs to cast \textit{%
Expectations} over the triangle amplitude below. Momenta contractions
associated with axial vertices should lead to a similar amplitude having a
pseudoscalar vertex $AVV\rightarrow PVV$, while vector contractions should
vanish $AVV\rightarrow 0$. Results different from these expressions
represent symmetry violations at the quantum level and carry anomalous
contributions.

\begin{itemize}
\item \textit{Expectations} - Ward identities (WIs) anticipated from current
algebra.%
\begin{eqnarray}
q^{\mu }T_{\mu \nu \alpha }^{AVV} &\rightarrow &-2mT_{\nu \alpha }^{PVV} \\
p^{\nu }T_{\mu \nu \alpha }^{AVV} &\rightarrow &0 \\
\left( q-p\right) ^{\alpha }T_{\mu \nu \alpha }^{AVV} &\rightarrow &0
\end{eqnarray}
\end{itemize}

It remains for us to evaluate the connection involving relations among GF
and WIs explicitly. Since no prescription was adopted to evaluate the
surface term up to this point, this analysis falls over the properties of
this object. We stress two lines of reasoning while doing so.

First, maintaining the linearity of integration occurs through a
prescription where the surface term assumes the finite non-zero value (\ref%
{cond1}). That occurs if one uses linearity to verify directly that the
surface term (\ref{delta3}) has a finite trace%
\begin{equation}
\Delta _{\rho }^{\rho }=4\lambda ^{2}\int \frac{d^{4}k}{\left( 2\pi \right)
^{4}}\frac{1}{D_{\lambda }^{3}}=-\frac{i}{8\pi ^{2}},
\end{equation}%
computed with the aid of integral (\ref{CC1}). This condition vanishes
additional contributions acknowledged before; hence, amplitude versions
obtained through different trace expressions coincide (\ref{diff}) and
satisfy all relations among GF. Nevertheless, that violates all symmetry
implications from WIs since the surface term itself is finite and non-zero.
After computing the differences involving $AV$s in Equations (\ref{DAV1})-(%
\ref{DAV3}), we cast these results in \textit{Condition I} below. Comparing
with the \textit{Expectations}, observe that all contractions exhibit an
anomalous contribution.

\begin{itemize}
\item \textit{Condition I} - Linearity of integration leads to the finite
non-zero value for the surface term $\Delta _{\rho \sigma }=-\frac{i}{32\pi
^{2}}g_{\rho \sigma }$ and $\Delta _{\rho }^{\rho }=-\frac{i}{8\pi ^{2}}$.%
\begin{eqnarray}
q^{\mu }T_{\mu \nu \alpha }^{AVV} &=&-2mT_{\nu \alpha }^{PVV}+\frac{1}{2\pi
^{2}}\varepsilon _{\nu \alpha \rho \sigma }k_{2}^{\rho }q^{\sigma }
\label{cont1} \\
p^{\nu }T_{\mu \nu \alpha }^{AVV} &=&-\frac{1}{2\pi ^{2}}\varepsilon _{\mu
\alpha \rho \sigma }k_{3}^{\rho }p^{\sigma } \\
\left( q-p\right) ^{\alpha }T_{\mu \nu \alpha }^{AVV} &=&\frac{1}{2\pi ^{2}}%
\varepsilon _{\mu \nu \rho \sigma }k_{1}^{\rho }\left( q-p\right) ^{\sigma }
\label{cont3}
\end{eqnarray}
\end{itemize}

On the other hand, it is possible to satisfy part of the WIs by adopting a
prescription that eliminates surface terms. As mentioned before, that occurs
in the case of Dimensional Regularization \cite{Bollini:1972bi,
tHooft:1972tcz, Ashmore:1972uj}. Non-automatic relations among GF are lost
since this value does not cancel out additional contributions in
contractions, characterizing a linearity violation. Meanwhile, canceling the 
$AV$ amplitude saves part of the symmetry relations; \textit{Condition II }%
below.

The possibility of changing the position of additional contributions by
adopting other trace versions has significant consequences within this
context. By eliminating surface terms, the first amplitude version preserves
vector implications while bringing an anomalous term to the axial WI. This
result is compatible with the usual perspective adopted in the literature
since it is necessary to explain the phenomenon of the neutral pion decay
into a pair of photons \cite{Cheng:1985bj}. Alternatively, vector identities
exhibit violations when it comes to the other two amplitude versions.

\begin{itemize}
\item \textit{Condition II} - Preserving part of the Ward identities (WIs)
leads to the null values $\Delta _{\rho \sigma }=0$ and $\Delta _{\rho
}^{\rho }=0$. This time, we only cast the violated implications for each
amplitude version.%
\begin{eqnarray}
q^{\mu }T_{1\mu \nu \alpha }^{AVV} &=&-2mT_{\nu \alpha }^{PVV}-\frac{1}{4\pi
^{2}}\varepsilon _{\nu \alpha \rho \sigma }p^{\rho }q^{\sigma }
\label{cont4} \\
p^{\nu }T_{2\mu \nu \alpha }^{AVV} &=&-\frac{1}{4\pi ^{2}}\varepsilon _{\mu
\alpha \rho \sigma }p^{\rho }q^{\sigma } \\
\left( q-p\right) ^{\alpha }T_{3\mu \nu \alpha }^{AVV} &=&\frac{1}{4\pi ^{2}}%
\varepsilon _{\mu \nu \rho \sigma }p^{\rho }q^{\sigma }  \label{cont12}
\end{eqnarray}
\end{itemize}

\newpage

\section{Final Remarks and Conclusions}

\label{Final}Throughout the third chapter, we investigated aspects of
fermionic amplitudes that are odd tensors. The $AVV$ triangle was our
primary target since its anomalous character is a recurrent subject in the
literature. We carefully examined its content and relations with other
amplitudes, thus understanding new aspects of anomalies while emphasizing
mathematical structures relevant to their discussion. We also extended this
analysis to the $AVVV$ box because it contains similar tensor structures.

Firstly, let us remark on the crucial role of traces having one chiral
matrix inside their argument in this context. They yield combinations of
monomials built through products between the Levi-Civita symbol and metric
tensors, in which case tensor properties allow different expressions.
Although they are identical at the integrand level, the connection among
corresponding versions for an integrated amplitude is not direct due to the
divergent character of calculations. This feature has motivated authors to
explore recipes for taking Dirac traces and study their implications \cite%
{Tsai:2009it, Tsai:2009hp, Bruque:2018bmy}.

To express this type of (odd) trace, one must suppress the dependence on the
chiral matrix and compute the ensuing (even) trace. Such an operation
requires employing one identity belonging to the set%
\begin{equation}
\gamma _{5}\gamma _{\left[ \mu _{1}\cdots \mu _{r}\right] }=\frac{%
i^{1+r\left( r+1\right) }}{\left( 4-r\right) !}\varepsilon _{\mu _{1}\cdots
\mu _{4}}\gamma ^{\left[ \mu _{r+1}\cdots \mu _{4}\right] },
\end{equation}%
where the notation $\gamma _{\left[ \mu _{1}\cdots \mu _{r}\right] }$
indicates antisymmetrized products of Dirac matrices. Reference \cite%
{Arxiv22} presents a broad discussion of this subject, approaching all
versions of the four-dimensional triangle and inquiring about analogous
cases in other space-time dimensions. Ultimately, the authors show that all
amplitude expressions coming from these identities are combinations of more
fundamental ones\footnote{%
That implies other versions carry anomalous terms in multiple vertices. For
instance, one form identified through the combination $\frac{1}{2}\left(
T_{1\mu \nu \alpha }^{AVV}+T_{2\mu \nu \alpha }^{AVV}\right) $ exhibits
violations for contractions with both first and second vertices.}, those
obtained through the chiral matrix definition (identity with $r=0$).

These ideas justify us targeting only these specific versions throughout
this work. In truth, we replaced the definition in all six positions
available to evaluate the trace containing six Dirac matrices plus the
chiral one. Comparing neighboring positions made evident the presence of
algebraic identities, which associate with null integrals when computing the
triangle. Despite this being almost a trivial example, it outlines a
strategy to pursue simplifications in more complex calculations. We used
this tool when computing the box amplitude, achieving a clear view of its
content and properties.

Replacing the chiral matrix definition in a particular position implies
prioritizing one vertex in the trace. By doing so, all contributions having
the corresponding index within metric tensors cancel out. Hence, this index
appears exclusively inside the Levi-Civita symbol, which is transparent by
the provided organization. Observe how the trace choice shapes tensor
contributions in the triangle versions from Equations (\ref{AVV01})-(\ref%
{AVV03}). Although we did not present other versions here, that also occurs
for the box amplitude. Posteriorly to the integration, index permutations
are necessary when performing momenta contractions with the prioritized
index. That is the mechanism inducing the presence of potentially violating
terms in relations among Green functions. This reasoning allows the reverse
way, choosing which index to prioritize aiming to position the additional
contributions.

We stress the generality of these concepts by commenting on triangle
amplitudes with similar tensor structures but different vertex
configurations, namely, $AVV$, $VAV$, $VVA$, and $AAA$. Since they share the
higher-order trace from Equation (\ref{AVV0}), opting for a trace expression
shapes the tensor sector of these amplitudes equally, and our conclusions
apply to all of them. When prioritizing the $n$th free index in the trace,
one induces potentially violating terms in the $n$th momenta contraction.
That does not depend on the character of the corresponding vertex as being
axial or vector. The same situation occurs for box amplitudes, i.e., $AVVV$, 
$VAAA$, and their permutations. Again, further explorations are necessary to
test the generality of the last statement.

Now, let us detail some aspects regarding integrated amplitudes. At the
beginning of this chapter, we mentioned that integrals exhibiting power
counting equal to or higher than linear are not translationally invariant.
That means performing shifts on the integration variable requires adequate
compensations to maintain the connection with the original expression. This
feature implies the presence of surface terms in perturbative calculations,
wholly expressed through the object $\Delta _{\mu \nu }$ in this
investigation.

Take the $AV$ bubble (\ref{AV}) as a preliminary study case. We observed a
priori that it should be a null object since it was impossible to build an
antisymmetric tensor exclusively using the external momentum. However,
two-point amplitudes exhibit quadratic power counting in the physical
dimension. Consequently, this amplitude admits the presence of a surface
term proportional to an ambiguous combination of arbitrary labels $%
k_{1}+k_{2}$. This type of contribution also arises for the $AVV$ triangle (%
\ref{AVV}), located inside the vector subamplitude (\ref{VPP}).

Albeit with non-ambiguous coefficients, the $AVVV$ box exhibits the same
surface term seen in the first two cases. Look into the complete amplitude (%
\ref{AVVV}) and its pertinent sectors (\ref{VVPPs}) to find these objects.
Their presence is characteristic of tensors with logarithmic power counting,
as observed in Feynman integrals (\ref{I3uv}) and (\ref{I4uvab}).

We also studied the implications of surface terms when exploring amplitude
versions. Since they differ in the index arrangement set through trace
choices, we had to permute indices to compare different possibilities. For
the $AVV$ triangle, this procedure emphasized the dependence on the surface
term value, represented by the structure on the right-hand side of Equation (%
\ref{diff}). Canceling this contribution occurs if one assumes the \textit{%
finite value} $\Delta _{\rho }^{\rho }=-i\left( 8\pi ^{2}\right) ^{-1}$. We
can interpret this constraint as a condition so all trace choices lead to
one unique expression for the amplitude. Although we did not extend this
argumentation, the involved tensors suggest that the box analysis is
analogous.

Next, let us comment on the results achieved when performing momenta
contractions. We identified the amplitudes from relations among Green
functions directly in part of the cases. Nevertheless, as mentioned in the
discussion about traces, potentially violating terms emerge in the $n$th
momenta contraction of an amplitude that prioritizes the $n$th free index in
the trace. Such additional contributions exhibit the same structure referred
to in the previous paragraph. At least one relation among Green functions is
not automatically satisfied but demands a condition over the surface term
value to do so. Hence, the amplitude expression considering the \textit{%
finite value} of the surface term satisfies all relations among Green
functions. This outcome breaks all symmetry implications through Ward
identities, which is transparent in the explicit values of these
contractions (\ref{cont1})-(\ref{cont3}). This part of the analysis also
applies to the box amplitude.

On the other hand, adopting a prescription that sets surface terms as zero $%
\Delta _{\mu \nu }=0$ preserves Ward identities for contractions that do not
produce additional contributions. We acknowledge violations in the
conditional relation among Green functions and the corresponding Ward
identity. That is consistent with the impossibility of preserving chiral and
gauge symmetry simultaneously. Furthermore, we clarify that it is possible
to choose the position of the violation by adopting the trace expression
accordingly. Equations (\ref{cont4})-(\ref{cont12}) illustrate these
possibilities for the triangle amplitude. Although we observed the same
situation in the box amplitude, there are more possibilities to study before
coming to a conclusion.

As a future perspective of this work, it is important to deepen the analysis
of symmetry aspects. Reference \cite{Arxiv22} is a work in progress from T.
J. Girardi, L. Ebani, and J. F. Thuorst and provides crucial information
regarding low-energy implications of anomalous amplitudes. Explorations on
the $AVV$ triangle are particularly detailed, but the authors also extend
this subject to analogous processes in other space-time dimensions.

Despite its similarities with the triangle, argumentations seem more
intricate for the box amplitude. We observed that versions differ in their
dependence on surface terms following the implications of trace choices.
This feature reflects on potentially violating terms in contractions when
prioritizing the first index in traces for all amplitudes within relations
among Green functions. Nonetheless, other choices are possible and require
further investigation.


\bigskip

\begin{thebibliography}{99}
\bibitem{Johnson:1963vz} K. Johnson, Phys. Lett. \textbf{5}, 253 (1963).

\bibitem{Adler:1969gk} S. Adler, Physical Review, \textbf{177}, 2426-2438
(1969).

\bibitem{Bell:1969ts} J. S. Bell, and R. Jackiw, Nuovo Cimento A, \textbf{60}%
, 47-61 (1969).

\bibitem{Jackiw-Johnson} R. Jackiw and K. Johnson, Phys.Rev. 182 (1969)
1459-1469.

\bibitem{Bardeen:1969md} W. A. Bardeen, Physical Review, \textbf{184},
1848-1857 (1969).

\bibitem{Cheng:1985bj} T. Cheng and L. Li, \textit{Gauge theory of
elementary particle physics}, Clarendon Press, Oxford (1982).

\bibitem{Jackiw} R. Jackiw, and R. Rajaraman, Phys. Rev. Lett. \textbf{54}
1219--1221 (1985).

\bibitem{Jackiw2} R. Jackiw, and R. Rajaraman, Phys. Rev. Lett. \textbf{55}
2224 (1985).

\bibitem{Faddeev} L.D. Faddeev, S.L. Shatashvili, Phys. Lett. B 167 (1986)
225--228.

\bibitem{Harada} K. Harada, I. Tsutsui, Phys. Lett. B 183 (311) (1987)
311--314.

\bibitem{Babelon} O. Babelon, F. Schaposnik, C. Viallet, Phys. Lett. B 177
(1986) 385--388.

\bibitem{Gabriel-Rafael-Tiao} G. L. S. Lima, R. C. S. Araujo, and S. A.
Dias, Annals Phys. 327 (2012) 1435-1449.

\bibitem{DEWITT} B. S. Dewitt; R. Stora. \textit{Relativity, Groups and
Topology II}\textbf{.} Course 3. Topological investigations of quantized
gauge theories, by R. Jackiw. Amsterdam: Elsevier Science Publishers B.V.,
1984.

\bibitem{Gabriel} G. L. S. Lima. \textit{Uma Abordagem Alternativa da Rela%
\c{c}\~{a}o entre Simetria de Calibre e Conserva\c{c}\~{a}o da Corrente.}%
\textbf{\ }Tese de Doutorado: CBPF, 2011.

\bibitem{Gabriel2} G. L. S. Lima, Annals of Physics 341 (2014) 183-194
(2013).

\bibitem{Rivers} R. J. Rivers. \textit{Path Integral Methods in Quantum
Field Theory}, Cambridge Monographs in Mathematical Physics, Cambridge
University Press, New York (1990).

\bibitem{Fujikawa} K. Fujikawa and H. Suzuki. \textit{Path Integrals and
Quantum Anomalies}, The International Series of Monographs on Physics,
Oxford University Press, New York (2004).

\bibitem{Fujikawa79} K. Fujikawa, Phys. Rev. Lett. \textbf{42} 1195 (1979).

\bibitem{Fujikawa80} K. Fujikawa, Phys. Rev. \textbf{D21} 2848 (1980). 
\textbf{D22} 1499 (E) (1980).

\bibitem{tHooft} G. 't Hooft, \textit{Nuclear Physics} \textbf{B33} (1971)
173-199; \textit{Nuclear Physics} \textbf{B35} (1971), 167-188.

\bibitem{refalgren} C. Becchi, A. Rouet and R. Stora, \textit{Ann. of Phys.} 
\textbf{98} (1976), 287-321.

\bibitem{GTS} G. de Lima e Silva, T. J. Girardi, and S. A. Dias. Universe 
\textbf{7}, 8 283 (2021).

\bibitem{Gnendiger:2017pys} C. Gnendiger, A. Signer, D. St\"{o}ckinger, A.
Broggio, A. L. Cherchiglia, F. Driencourt-Mangin, A. R. Fazio, B. Hiller, P.
Mastrolia, T. Peraro, R. Pittau, G. M. Pruna, G. Rodrigo, M. Sampaio, G.
Sborlini, W. J. Torres Bobadilla, F. Tramontano, Y. Ulrich, A. Visconti,
European Physical Journal C \textbf{77}, 471 (2017).

\bibitem{Pittau:2012zd} R. Pittau, JHEP \textbf{11} 141, (2012).

\bibitem{Bollini:1972bi} C. G. Bollini and J. J. Giambiagi, Phys. Lett. B 
\textbf{40,} 566 (1972).

\bibitem{tHooft:1972tcz} G.'t Hooft and M. Veltman, Nucl. Phys. B \textbf{44,%
} 189 (1972).

\bibitem{Ashmore:1972uj} J. F. Ashmore, Nuovo Cimento Lett. \textbf{4,} 289
(1972).

\bibitem{Treiman:1986ep} S. B. Treiman, R. Jackiw, B. Zumino, and E. Witten, 
\textit{Current algebra and anomalies}, World Scientific, Singapore (1985).

\bibitem{Bertlmann:1996xk} R. A. Bertlman, \textit{Anomalies in Quantum
Field Theory}, Clarendon Press, Oxford (1996).

\bibitem{Sterman:1994ce} G. Sterman, \textit{An Introduction to Quantum
Field Theory}, Cambridge University Press, Cambridge (1993).

\bibitem{Ferreira:2011cv} L. C. Ferreira, A. L. Cherchiglia, B. Hiller, M.
Sampaio, and M. C. Nemes, Phys. Rev. D \textbf{86}, 025016 (2012).

\bibitem{Vieira:2015fra} A. R. Vieira, A. L. Cherchiglia, and Marcos
Sampaio, Phys. Rev. D \textbf{93}, 025029 (2016).

\bibitem{Viglioni:2016nqc} A. C. D. Viglioni et al, Phys. Rev. D 94, 065023
(2016).

\bibitem{Tsai:2009it} E. Tsai, Phys. Rev. D \textbf{83}, 025020 (2011).

\bibitem{Tsai:2009hp} E. Tsai, Phys. Rev. D \textbf{83}, 065011 (2011).

\bibitem{Bruque:2018bmy} A. M. Bruque, A. L. Cherchiglia, and M. P\'{e}%
rez-Victoria, JHEP \textbf{08} 109, (2018).

\bibitem{ORIMAR-TESE} O. A. Battistel, Ph.D. Thesis, Universidade Federal de
Minas Gerais, Brazil, (1999).

\bibitem{Ebani2018:gre} O. A. Battistel, G. Dallabona, M. V. Fonseca, and L.
Ebani, Journal of Modern Physics \textbf{9}, 1153 (2018).

\bibitem{Battistel:2018rqe} O. A. Battistel, F. Traboussy, and G. Dallabona,
Int. J. Mod. Phys. A, \textbf{33,} 1850136 (2018).

\bibitem{Battistel:2012zz} O. A. Battistel, M. V. S. Fonseca, and G.
Dallabona, Phys. Rev. D \textbf{85}, 085007 (2012).

\bibitem{Battistel:2002dm} O. A. Battistel and G. Dallabona, Phys. Rev. D 
\textbf{65}, 125017 (2002).

\bibitem{Battistel:2002ve} O. A. Battistel and G. Dallabona, J. Phys. G:
Nucl. Part. Phys.\textbf{\ 28}, 2539 (2002).

\bibitem{FONSECA:2013eoa} M. V. S. Fonseca, T. J. Girardi, G. Dallabona, and
O. A. Battistel, Int. J. Mod. Phys. A \textbf{28, }1350135 (2013).

\bibitem{Battistel:2014vea} O. A. Battistel and G. Dallabona, Int. J. Mod.
Phys. A, \textbf{29}, 1450068 (2014).

\bibitem{Fonseca:2014cba} M. V. S. Fonseca, G. Dallabona, and O. A.
Battistel, Int. J. Mod. Phys. A \textbf{29}, 1450168 (2014).

\bibitem{Battistel:2006zq} O. A. Battistel and G. Dallabona, Eur. Phys. J. C%
\textbf{45}, 721 (2006).

\bibitem{Battistel:2012qpm} O. A. Battistel, and~G. Dallabona, Journal of
Modern Physics \textbf{3, }1408 (2012).

\bibitem{Arxiv22} L. Ebani, T. J. Girardi, and J. F. Thuorst,
arXiv:2212.03309, (2022).

\bibitem{Weinberg} S. Weinberg, \textit{The Quantum Theory of Fields, }%
Volume 1, Cambridge University Press, Cambridge (1995).
\end{thebibliography}
\end{document}